\documentclass[rmp,aps,twocolumn]{revtex4-1}

\pdfoutput=1

\usepackage{graphicx}
\usepackage{epsfig}
\usepackage{amsmath}
\usepackage{amssymb}
\usepackage{dcolumn}
\usepackage{color}
\usepackage{url}

\usepackage{ifpdf}

\newcolumntype{d}[1]{D{.}{.}{#1} }
\begin{document}

\title{Quantum Monte Carlo methods for nuclear physics}
\author{J. Carlson}
\affiliation{Theoretical Division, Los Alamos National Laboratory, Los Alamos, NM 87545}
\author{S. Gandolfi}
\affiliation{Theoretical Division, Los Alamos National Laboratory, Los Alamos, NM 87545 }
\author{F. Pederiva}
\affiliation{Dipartimento di Fisica, Universit\'{a} di Trento, I-38123 Trento, Italy}
\affiliation{INFN - Trento Institute for Fundamental Physics and Applications, I-38123 Trento Italy}
\author{Steven C. Pieper}
\affiliation{Physics Division, Argonne National Laboratory, Argonne, IL 60439}
\author{R. Schiavilla}
\affiliation{Theory Center, Jefferson Lab, Newport News, VA 23606}
\affiliation{Department of Physics, Old Dominion University, Norfolk, VA 23529}
\author{K.E. Schmidt}
\affiliation{Department of Physics, Arizona State University, Tempe, AZ 85287}
\author{R.B. Wiringa}
\affiliation{Physics Division, Argonne National Laboratory, Argonne, IL 60439 }

\begin{abstract}
Quantum Monte Carlo methods have proved very valuable to study
the structure and reactions of light nuclei and nucleonic matter starting
from realistic nuclear interactions and currents. These {\em ab-initio}
calculations reproduce many low-lying states, moments and transitions in light
nuclei, and simultaneously predict many properties of 
light nuclei and neutron matter over a rather wide range of energy and momenta.
We review the nuclear interactions and currents, and
describe the continuum Quantum Monte Carlo methods used in nuclear
physics. These methods are similar to those used in condensed matter and electronic 
structure but naturally include spin-isospin, tensor, spin-orbit, and 
three-body  interactions.
We present a variety of results including the low-lying spectra of light nuclei, 
nuclear form factors, and transition matrix elements. We also describe 
low-energy scattering techniques, studies of the electroweak 
response of nuclei relevant in electron and neutrino scattering, 
and the properties of dense nucleonic matter as found in neutron stars.
A coherent picture of nuclear structure and dynamics
emerges based upon rather simple but realistic interactions and currents.
\end{abstract}


\maketitle
\tableofcontents

\section{INTRODUCTION}
\label{sec:intro}

Nuclei are fascinating few- and many-body quantum systems,
ranging in size from the lightest nuclei formed in the big bang
to the structure of neutron stars with $\sim$10 km radii.
Understanding their structure and dynamics starting from realistic interactions
among nucleons has been a long-standing goal of nuclear physics.
The nuclear quantum many-body problem contains many features present in
other areas such as condensed matter physics, including pairing and superfluidity
and shell structure,
but also others that are less common including a very strong coupling
of spin and spatial degrees of freedom, clustering phenomena, and
strong spin-orbit splittings. The challenge is to describe diverse
physical phenomenon within a single coherent picture.

This understanding is clearly important to describe nuclear
properties and reactions, including reactions that synthesized
the elements and the structure of neutron-rich nuclei.
An accurate picture of interactions and currents at the
nucleonic level is critical to extend this understanding to the properties of dense 
nucleonic matter as occurs in neutron stars, and to use 
nuclei as probes of fundamental physics through, for example,
beta decay, neutrinoless double-beta decay, and neutrino-nucleus
scattering.

Over the last three decades it has become possible
using Quantum Monte Carlo (QMC) methods
to reliably compute the properties of light nuclei and neutron 
matter starting from realistic nuclear interactions.  While many of the
most basic properties of nuclei can be obtained from 
comparatively simple mean-field models, it has been a challenge to
relate the two- and three-nucleon interactions inferred from experiments 
to the structure and reactions of nuclei. This challenge arises
because the scale of the nuclear interactions obtained by examining
nucleon-nucleon phase shifts is of order (50-100) MeV or more,
significantly larger than a typical nuclear binding energy of 8 MeV per nucleon.

In addition, the nucleon-nucleon interaction is much more complex
than the Coulomb force used in molecular and atomic physics,
the van der Waals potential between atoms used, for example, in studies
of liquid helium systems, or the contact interaction that dominates dilute cold-atom physics.
The primary force carrier at large nucleon separations is the pion,
which couples strongly to both the spin and isospin of the nucleons
with a strong tensor component.
In addition there are significant spin-orbit forces. 
As a consequence, there is strong coupling between
the spin and isospin and spatial degrees of freedom.

These features lead to complex nuclear phenomena. The interactions
are predominantly attractive at low momenta, resulting in 
large pairing gaps in nuclei and associated superfluidity in matter.
In light nuclei, there is further clustering of neutrons and protons
into alpha-particle like configurations that are very evident in the
low-lying excitations of some nuclei.  At moderate nucleon separations, the
tensor character of the neutron-proton interaction produces 
significant high-momentum components in the nuclear wave function
that impact the electroweak
response observed in electron and neutrino scattering. 
The nuclear correlations also significantly quench the single-particle 
description of nucleon knockout and transfer reactions.
A major challenge
has been to include both the short-range high-momentum phenomena
and the long-range superfluid and clustering properties of nuclei and
matter in a consistent framework.

QMC methods based upon Feynman path integrals
formulated in the continuum have proved to
be very valuable in attacking these problems. The sampling of
configuration space in variational (VMC) and Green's function (GFMC)
Monte Carlo simulations gives access to many of the important properties
of light nuclei including spectra, form factors, transitions, low-energy
scattering and response.  Auxiliary Field Diffusion Monte Carlo (AFDMC)
uses Monte Carlo to also sample the spin-isospin degrees of freedom,
enabling studies of, for example, neutron matter that is so critical
to determining the structure of neutron stars.  In this review
we concentrate on continuum Monte Carlo methods. Lattice QMC
methods have also recently been employed to study both neutron 
matter~\cite{Muller:1999,Lee:2005,Seki:2006,Abe:2009,Wlazlowski:2014,Roggero:2014} 
and certain nuclei~\cite{lee2009,Epelbaum:2012}.
Other Monte Carlo methods combined with the use of effective interactions 
and/or space models like the shell model have been also developed to study properties
of larger systems; see for example~\cite{koonin1997,Bonett-Matiz:2013,Otsuka:2001,
Abe:2012,Bonnard:2013}. 

Other many-body methods, many of which have direct analogues in other fields of physics,
have also played important roles in the study of nuclei.
These include the coupled cluster method~\cite{Hagen:2014,Hagen:2014b},
the no core shell model~\cite{Barrett:2013}, the similarity renormalization
group~\cite{Bogner:2010}, and the Self Consistent Green's Function~\cite{Dickhoff:2004}. 
Each of these methods has distinct advantages, and
many are able to treat a wider variety of nuclear interaction models. Quantum
Monte Carlo methods, in contrast, are more able to deal with a wider range
of momentum and energy and to treat diverse phenomenon including superfluidity
and clustering.

Progress has been enabled by simultaneous advances in the
input nuclear interactions and currents, the QMC methods,
increasingly powerful computer facilities,
and the applied mathematics and computer science required to run efficiently these
calculations on the largest available machines~\cite{Lusk:2010}.
Each of these factors have been very important. QMC methods
have been able to make use of some of the most powerful computers available,
through extended efforts of physicists and computer scientists to scale the
algorithms successfully.  The codes have become much more efficient and also
more accurate through algorithmic developments.  The introduction of Auxiliary
Field methods paved the way to scale these results to much larger nuclear 
systems than would otherwise have been possible.
Equally important, advances in algorithms have allowed to expand the physics
scope of our investigations.  
Initial applications were to nuclear ground states, including energies and
elastic form factors. Later advances opened the way to study low-energy nuclear
reactions, the electroweak response of nuclei and infinite matter.

Combined, QMC and other computational methods in nuclear
physics have allowed us, for the first time, to directly connect the
underlying microscopic nuclear interactions and currents
with the structure and reactions of nuclei.  
Nuclear wave functions that contain the many-nucleon correlations
induced by these interactions are essential for
accurate predictions of many experiments.
QMC applications in nuclear physics span a wide range
of topics, including low-energy nuclear spectra and transitions,
low-energy reactions of astrophysical interest, tests of fundamental
symmetries, electron- and neutrino-nucleus scattering, and the
properties of dense matter as found in neutron stars.
In this review we briefly present the interactions and currents
and  the Monte Carlo methods, and then review results that have
been obtained to date across these different diverse and important
areas of nuclear physics.

\def\mb{many--body}
\def\Mb{Many--body}
\def\twb{two--body}
\def\Twb{Two--body}
\def\thb{three--body}
\def\Thb{Three--body}
\def\So{Spin--orbit}
\def\so{spin--orbit}
\def\pnm{pure neutron matter}
\def\snm{symmetric nuclear matter}
\def\Pnm{Pure neutron matter}
\def\Snm{Symmetric nuclear matter}
\def\QMC{QMC}
\def\ceft{chiral effective field theory}
\def\Ceft{Chiral effective field theory}
\def\Athpi{A_{3\pi}^{\Delta R}}
\newcommand{\boldtau}{\mbox{\boldmath$\tau$}}
\newcommand{\boldsigma}{\mbox{\boldmath$\sigma$}}

\section{HAMILTONIAN}
\label{sec:hamiltonian}
Over a substantial range of energy and momenta the structure and reactions
of nuclei and nucleonic matter can be studied with a
non-relativistic Hamiltonian with nucleons as the only active degrees
of freedom.  Typical nuclear binding energies are of order 10 MeV per nucleon
and Fermi momenta are around 1.35 fm$^{-1}$. Even allowing for substantial
correlations beyond the mean field, the nucleons are essentially non-relativistic.  
There is a wealth of nucleon-nucleon ($N\!N$) scattering data available that
severely constrains possible $N\!N$ interaction models. Nuclear 
interactions have been obtained that provide accurate fits to these data,
both in phenomenological models and in chiral effective field theory.
This is not sufficient to reproduce nuclear binding, however,
as internal excitations of the nucleon do have some impact.
The lowest nucleon excitation is the $\Delta(1232)$ resonance at $\sim$290 MeV. 
Rather than treat these excitations as dynamical degrees of freedom, however, it is more
typical to include them and other effects as three-nucleon ($3N$) interactions.

Therefore, in leading order approximation, one can integrate out nucleon
excitations and other degrees of freedom resulting in a
Hamiltonian of the form
\begin{align}
\label{eq:hamiltonian}
H = K + V  \,,
\end{align}
where $K$ is the kinetic energy and $V$ is an effective interaction, which, 
in principle, includes $N$-nucleon potentials, with $N\ge2$:
\begin{align}
\label{eq:potential}
 V = \sum_{i<j}v_{ij} + \sum_{i<j<k}V_{ijk} +\dots \,.
\end{align}
The $N\!N$ interaction term is the most studied of all, 
with thousands of experimental
data points at laboratory energies from essentially zero to hundreds of MeV.
Attempts are now being made to understand this interaction directly
through lattice QCD, though much more development will be required before
it can be used directly
in studies of nuclei~\cite{Beane:2013,Ishii:2007}.  Traditionally the $N\!N$ scattering data has been fit
with phenomenological interactions that require a rather complicated
spin-isospin structure because of the way the nucleon couples to the pion,
other heavier mesons, and nucleon resonances.
More recently, advances have been made using chiral effective field theory, 
which employs chiral symmetry and a set of low-energy constants to
fit the $N\!N$ scattering data. This has led to an 
understanding of why charge-independent
$N\!N$ terms are larger than isospin-breaking ones, why $3N$
interactions are a small fraction ($\sim 10\%$) of $N\!N$ interactions, 
and has provided a direct link between interactions and currents.

In what follows we will focus on potentials developed in coordinate space, 
which are particularly convenient for QMC calculations.
Many phenomenological models are primarily local interactions (although often 
specified differently in each partial wave) and local interactions can be obtained
within chiral effective theory, which is an expansion in the nucleon's momentum.
The interaction is predominantly local because of the nature of one-pion 
exchange, but at higher orders derivative (momentum) operators must be
introduced. Local interactions are simpler to treat in continuum
QMC methods because the $N\!N$ propagator is essentially positive definite,
a property that is not always true in non-local interactions.
The Monte Carlo sampling for such positive definite propagators is much
easier, reducing statistical errors in the simulation.

A number of very accurate $N\!N$ potentials constructed in the 1990s
reproduce the long-range one-pion-exchange part of the interaction and 
fit the large amount of empirical information about $N\!N$
scattering data contained in the Nijmegen database \cite{stoks1993b} with a
$\chi^2/N_{data}\sim1$ for lab energies up to $\sim$ 350 MeV. These include
the potentials of the Nijmegen group \cite{stoks1994}, the Argonne 
potentials \cite{wiringa1995,wiringa2002} and the CD-Bonn potentials
\cite{machleidt1996,Mac01}.
Of those potentials derived more recently by using \ceft, the most commonly
used is that of \citet{entem2002}. 
The most practical choice for QMC calculations is the Argonne $v_{18}$ potential
\cite{wiringa1995}, which is given in an r-space operator (non-partial
wave) format and has a very weak dependence on non-local terms. 
The latter are small and hence are tractable in QMC calculations. 
Another less sophisticated interaction that, apart from charge-symmetry breaking
effects, reproduces the gross features of Argonne $v_{18}$ is the Argonne $v_8'$.
These are the potentials adopted in most of the QMC calculations.

However all of these $N\!N$ interactions, when
used alone, underestimate the triton binding energy, indicating that
at least $3N$ forces are necessary to reproduce the physics of $^3$H
and $^3$He.
A number of semi-phenomenological $3N$ potentials, such as the Urbana
\cite{carlson1983,pudliner1996} series, were developed to fit three-
and four-body nuclear ground states. 
The more recent Illinois \cite{pieper2001,pieper2008} $3N$ potentials 
reproduce the ground state and
low-energy excitations of light $p$-shell nuclei ($A\le 12$). 
More sophisticated models may be required to treat nucleonic matter
at and above saturation density $\rho \gtrsim \rho_0$.
Particularly in isospin-symmetric nuclear matter, the many-body
techniques for realistic interactions also need to be improved.
Effective field theory techniques and QMC methods
may help to provide answers to these questions.

\subsection {The nucleon-nucleon interaction}
\label{sec:nn}

Among the realistic $N\!N$ interactions, 
the Argonne $v_{18}$ (AV18) $N\!N$ potential \cite{wiringa1995} is a finite, 
local, configuration-space potential that is defined in all partial waves.
AV18 has explicit charge-independence breaking (CIB) terms, so it should be 
used with a kinetic energy operator that keeps track of the proton-neutron
mass difference by a split into charge-independent (CI) and charge-symmetry 
breaking (CSB) pieces:
\begin{align}
 K &= \sum_i K^{\rm CI}_{i} + K^{\rm CSB}_{i} \\
   &\equiv -\frac{\hbar^2}{4} \sum_i
    \left[ \left(\frac{1}{m_{p}} + \frac{1}{m_{n}}\right)
    + \left(\frac{1}{m_{p}} - \frac{1}{m_{n}}\right) \tau_{z_i} \right] \nabla^{2}_{i}  \nonumber \, ,
\end{align}
where $m_p$ and $m_n$ are the proton and neutron mass, and $\tau_{z_i}$ is the 
operator that selects the third component of the isospin.
AV18 is expressed as a sum of electromagnetic and one-pion-exchange (OPE) 
terms and phenomenological intermediate- and short-range parts, which can
be written as an overall operator sum
\begin{equation}
 v_{ij} = v^{\gamma}_{ij} + v^{\pi}_{ij} + v^{I}_{ij} + v^{S}_{ij}
        = \sum_{p} v_{p}(r_{ij}) O^{p}_{ij} \ .
\label{eq:vijgen}
\end{equation}
The electromagnetic term $v^\gamma_{ij}$ has one- and two-photon-exchange
Coulomb interaction, vacuum polarization, Darwin-Foldy, and magnetic moment
terms, with appropriate form factors that keep terms finite at $r_{ij}$=0.
The OPE part includes the charge-dependent (CD) terms due to the difference
in neutral and charged pion masses:
\begin{equation}
    v^{\pi}_{ij} = f^{2} \left[ X_{ij} \boldtau_{i} \cdot \boldtau_{j}
                       + \tilde{X}_{ij} T_{ij} \right] \ ,
\label{eq:vpi}
\end{equation}
where the coupling constant is $f^2\!=\!0.075$, 
$\boldtau$ are the Pauli matrices that operate over 
the isospin of particles, and
$T_{ij} = 3\tau_{z_i}\tau_{z_j}\!-\!\boldtau_{i}\cdot\boldtau_{j}$ is the
isotensor operator.
The radial functions are
\begin{align}
   & X_{ij} = \case{1}{3} \left( X^{0}_{ij} + 2 X^{\pm}_{ij} \right) , \\
   & \tilde{X}_{ij} = \case{1}{3} \left( X^{0}_{ij} - X^{\pm}_{ij}
                                      \right) , \\
   & X^{m}_{ij} =  \left(\frac{m}{m_{s}}\right)^{2} \case{1}{3} mc^{2}
                    \left[ Y(\mu r_{ij}) \boldsigma_{i} \cdot \boldsigma_{j} +
                           T(\mu r_{ij}) S_{ij}  \right] \,,
\end{align}
where $m\!=\!m_{\pi^{\pm}}$ or $m_{\pi^{0}}$, $\mu\!=\!m/\hbar c$, 
the scaling mass $m_s\!=\!m_{\pi^{\pm}}$, 
$\boldsigma$ are Pauli matrices that operate over the spin of nucleons,
and $S_{ij}\!=\!3\boldsymbol\sigma_i\cdot\hat r_{ij}
\boldsymbol\sigma_j\cdot\hat r_{ij}\!-\!\boldsymbol\sigma_i\cdot\boldsymbol\sigma_j$ is the tensor operator.
The $Y(x)$ and $T(x)$ are the normal Yukawa
 $Y(x) = \frac{e^{-x}}{x} \  \xi(r)$ and tensor
 $T(x) = \left( 1 + \frac{3}{x} + \frac{3}{x^2} \right) Y(x) \ \xi(r)$
functions with a short-range cutoff $\xi(r) = 1-\exp(-cr^2)$ with $c=2.1$ fm$^{-2}$.

The intermediate- and short-range strong-interaction terms have eighteen 
operators and are given the functional forms
\begin{align}
  v^{I}_{ij} &= \sum_{p=1}^{18} I^{p} T^2(\mu r_{ij}) O^{p}_{ij} \ , \\
  v^{S}_{ij} &= \sum_{p=1}^{18} \left[ P^p + Q^p r + R^p r^2 \right] W(r) O^p_{ij} \ ,
\end{align}
where $T^2$ is constructed with the average pion mass,
$\mu=(\case{1}{3}m_{\pi^0}+\case{2}{3}m_{\pi^{\pm}})/\hbar c$, and
$W(r)$ is a Woods-Saxon potential with radius $r_0 = 0.5$ fm and diffuseness
$a = 0.2$ fm.
Thus the former has two-pion-exchange (TPE) range, while the short-range
part remains finite and is constrained to have zero slope at the origin,
except for tensor terms which vanish at the origin. 
The first fourteen operators are CI terms:
\begin{align}
\label{eq:vci}
  O^{\rm CI}_{ij} &= \left[1, \boldsymbol{\sigma}_{i}\cdot\boldsymbol{\sigma}_{j}, S_{ij},
    {\bf L\cdot S},{\bf L}^{2},{\bf L}^{2}(\boldsymbol{\sigma}_{i}\cdot\boldsymbol{\sigma}_{j}),
    ({\bf L\cdot S})^{2}\right] \nonumber \\
               &\otimes \left[1, \boldsymbol{\tau}_{i}\cdot\boldsymbol{\tau}_{j}\right] \,,
\end{align}
where $\boldsymbol L_{ij}=\frac{1}{2i}(\boldsymbol r_i-\boldsymbol r_j)
\times(\boldsymbol\nabla_i-\boldsymbol\nabla_j)$ is the relative angular momentum of
the pair $ij$, and $\boldsymbol S_{ij}=\frac{1}{2}(\boldsymbol\sigma_i+\boldsymbol\sigma_j)$
is the total spin.
The remaining operators include three CD and one CSB terms:
\begin{align}
  O^{\rm CD}_{ij} &= [1, \boldsymbol{\sigma}_{i}\cdot\boldsymbol{\sigma}_{j}, S_{ij}]
    \otimes T_{ij} \ , \\
  O^{\rm CSB}_{ij} &= \tau_{z_i}+\tau_{z_j} \ .
\end{align}
The maximum value of the central ($p$=1) potential is $\sim 2$ GeV.

The AV18 model has a total of 42 independent parameters $I^p$, $P^p$, $Q^p$ 
and $R^p$.
A simplex routine \cite{nelder1965} was used to make an initial fit to
the phase shifts of the Nijmegen PWA93 analysis \cite{stoks1993}, followed by
a final fit direct to the data base, which contains 1787 $pp$ and
2514 $np$ observables for $E_{lab} \leq 350$ MeV.
The $nn$ scattering length and deuteron binding energy were also fit.
The final $\chi^2/N_{data}$ = 1.1 \cite{wiringa1995}.
While the fit was made up to 350 MeV, the phase shifts are qualitatively
good up to much larger energies $\geq 600$ MeV~\cite{Gandolfi:2014}.

The CD and CSB terms are small, but there is clear evidence for their presence.
The CD terms are constrained by the long-range OPE form and the
differences between $pp$ and $np$ scattering in the $^1S_0$ channel.
The CSB term is short-ranged and constrained by the difference in
$pp$ and $nn$ scattering lengths, and is necessary to obtain the correct
$^3$He--$^3$H mass difference.

Direct GFMC and AFDMC calculations with the full AV18 potential are
not practical because the spin-isospin-dependent terms which involve
the square of the orbital momentum operator have very large statistical errors.
However, these terms in AV18 are fairly weak and can be treated as
a first-order perturbation.
Using a wave function of good isospin also significantly reduces the
cost of calculations in GFMC. 
Hence it is useful to define a simpler isoscalar AV8$^\prime$ potential
with only the first eight (central, spin, isospin, tensor and spin-orbit) 
operators of Eq.~(\ref{eq:vci}); details are given in \cite{pudliner1997,wiringa2002}.
The AV8$^\prime$ is not a simple truncation of AV18, but a reprojection
that preserves the isoscalar average of the strong interaction in all $S$
and $P$ partial waves as well as the deuteron.
It has been used in benchmark calculations of $^4$He by seven different
many-body methods, including GFMC~\cite{kamada2001}.

It has proved useful to define even simpler reprojections of AV8$^\prime$,
particularly an AV6$^\prime$ potential without spin-orbit terms that is 
adjusted to preserve deuteron binding. 
The AV6$^\prime$ has the same CI OPE potential as AV8$^\prime$ and
preserves deuteron binding and $S$-wave and $^1P_1$ partial wave phase shifts, 
but $^3P_{0,1,2}$ 
partial waves are no longer properly differentiated.
Details are given in \citet{wiringa2002}, where the evolution of nuclear
spectra with increasing realism of the potentials was investigated.

\subsection {Three-body forces}
\label{sec:TNI}

\begin{figure}
\includegraphics[scale=0.35]{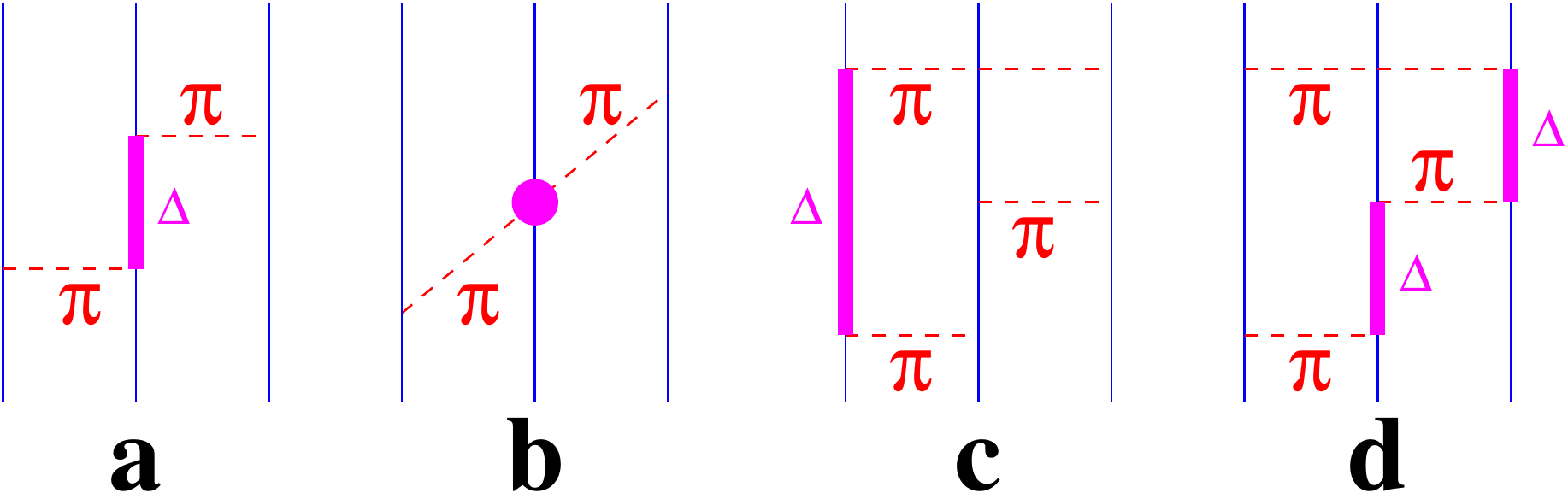}
\caption{Three-nucleon force diagrams for (a) two-pion $P$-wave, (b) two-pion
$S$-wave, and (c--d) three-pion ring terms.}
\label{fig:vijk}
\end{figure}

The Urbana series of $3N$ potentials \cite{carlson1983} 
is written as a sum of
two-pion-exchange $P$-wave and remaining shorter-range phenomenological terms,
\begin{equation}
   V_{ijk} = V^{2\pi,P}_{ijk} + V^{R}_{ijk} ~.
\end{equation}
The structure of the two-pion $P$-wave exchange term with an intermediate
$\Delta$ excitation (Fig.~\ref{fig:vijk}a)
was originally written down by \citet{Fujita1957}; it
can be expressed simply as
\begin{align}
   V^{2\pi,P}_{ijk} &=& \sum_{cyc} A^P_{2\pi}
     \{X^{\pi}_{ij},X^{\pi}_{jk}\}
     \{\boldtau_{i}\cdot\boldtau_{j},\boldtau_{j}\cdot\boldtau_{k}\}
     \nonumber \\
   && \ + \ C^P_{2\pi} [X^{\pi}_{ij},X^{\pi}_{jk}]
     [\boldtau_{i}\cdot\boldtau_{j},\boldtau_{j}\cdot\boldtau_{k}] \ ,
\end{align}
where $X^{\pi}_{ij}$ is constructed with the average pion mass and $\sum_{cyc}$
is a sum over the three cyclic exchanges of nucleons $i,j,k$.
For the Urbana models $C^P_{2\pi} = \case{1}{4}A^P_{2\pi}$, as in the original
Fujita-Miyazawa model, while other potentials like the
Tucson-Melbourne~\cite{coon1979} and Brazil~\cite{coelho1983} models, 
have a ratio slightly larger than $\case{1}{4}$.
The shorter-range phenomenological term is given by
\begin{equation}
   V^{R}_{ijk} = \sum_{cyc} A_R T^2(\mu r_{ij}) T^2(\mu r_{jk}) \ .
\label{eq:vrijk}
\end{equation}
For the Urbana IX (UIX) model \cite{pudliner1995}, the two parameters 
$A^P_{2\pi}$ and $A_R$ were determined by fitting the binding energy of 
$^3$H and the density of nuclear matter in conjunction with AV18.

While the combined AV18+UIX Hamiltonian reproduces the binding energies 
of $s$-shell nuclei, it somewhat underbinds light $p$-shell nuclei.
A particular problem is that the two-parameter Urbana form is not flexible 
enough to fit both $^8$He and $^8$Be at the same time.
A new class of $3N$ potentials, called the Illinois models, has
been developed to address this problem~\cite{pieper2001}.
These potentials contain the Urbana terms and two additional terms, resulting
in a total of four strength parameters that can be adjusted to fit the data.
The general form of the Illinois models is
\begin{equation}
V_{ijk}=V^{2\pi,P}_{ijk}+V^{2\pi,S}_{ijk}+V^{3\pi,\Delta R}_{ijk}+V^{R}_{ijk} ~.
\end{equation}
One term, $V^{2\pi,S}_{ijk}$, is due to $\pi N$ $S$-wave scattering as
illustrated in Fig.~\ref{fig:vijk}b and is parametrized with a strength
$A^S_{2\pi}$.
It has been included in a number of $3N$ potentials like the
Tucson-Melbourne and Brazil models.
The Illinois models use the form recommended in the latest Texas 
model \cite{friar1999},
where chiral symmetry is used to constrain the structure of the interaction.
However, in practice, this term is much smaller than the $V^{2\pi,P}_{ijk}$
contribution and behaves similarly in light nuclei, so it is difficult to
establish its strength independently just from calculations of energy levels.

A more important addition is a simplified form for three-pion rings
containing one or two $\Delta$s (Fig.~\ref{fig:vijk}c,d).  As discussed
by~\citet{pieper2001}, these diagrams result in a large number of terms,
the most important of which are used to construct the Illinois models:
\begin{equation}
V^{3\pi,\Delta R}_{ijk} = \Athpi \left[ \case{50}{3}S^I_{\tau}S^I_{\sigma}+
\case{26}{3}A^I_{\tau}A^I_{\sigma} \right] \ .
\end{equation}
Here the $S^I_x$ and $A^I_x$ are operators that are symmetric or antisymmetric
under any exchange of the three nucleons, and the subscript $\sigma$ or
$\tau$ indicates that the operators act on, respectively, spin 
or isospin degrees of freedom.

The $S^I_\tau$ is a projector onto isospin-$\case{3}{2}$ triples:
\begin{equation}
S^I_{\tau} = 2 + \case{2}{3}\left(\boldtau_i \cdot \boldtau_j
+\boldtau_j \cdot \boldtau_k + \boldtau_k \cdot \boldtau_i \right)
= 4 P_{T=3/2}  \ .
\label{eq:sitau}
\end{equation}
To the extent isospin is conserved, there are no such triples in the $s$-shell
nuclei, and so this term does not affect them.  It is also zero for $Nd$
scattering.  However, the $S^I_{\tau}S^I_{\sigma}$ term is
attractive in all the $p$-shell nuclei studied.  The $A^I_\tau$ has the
same structure as the isospin part of the anticommutator part of $V^{2\pi,P}$,
but the $A^I_{\tau}A^I_{\sigma}$ term is repulsive in all nuclei studied so far.
In $p$-shell nuclei, the magnitude of the $A^I_{\tau}A^I_{\sigma}$ term is
smaller than that of the $S^I_{\tau}S^I_{\sigma}$ term, so the net effect
of the $V^{3\pi,\Delta R}_{ijk}$ is slight repulsion in $s$-shell nuclei and
larger attraction in $p$-shell nuclei.
The reader is referred to the appendix of \citet{pieper2001} for
the complete structure of $V^{3\pi,\Delta R}_{ijk}$.

The first series of five Illinois models (IL1-5) explored different
combinations of the parameters $A^P_{2\pi}$, $A^S_{2\pi}$, 
$A^{\Delta R}_{3\pi}$, and $A_R$, and also variation of the OPE cutoff
function $\xi(r)$.
One drawback of these models is that they appear to provide too much attraction
in dense neutron matter calculations~\cite{Sarsa:2003}.
To help alleviate this problem, the latest version Illinois-7 (IL7)
introduced an additional repulsive term with the isospin-$\case{3}{2}$
projector:
\begin{equation}
   V^{R,T=3/2}_{ijk} = \sum_{cyc} A_{R,T=3/2} T^2(\mu r_{ij}) 
                       T^2(\mu r_{jk}) P_{T=3/2} \ .
\label{eq:vrijk3/2}
\end{equation}
After fixing $A^S_{2\pi}$ at the Texas value, and taking $\xi(r)$ from AV18,
the four parameters $A^P_{2\pi}$, $A^{\Delta R}_{3\pi}$, $A_R$, and
$A_{R,T=3/2}$ were searched to obtain a best fit, in conjunction with AV18,
for energies of about 20 nuclear ground and low-lying excited states in $A\leq 10$ nuclei
\cite{pieper2008}.

\subsection{Nuclear Hamiltonians from chiral effective field theory}
  
Chiral effective field theory ($\chi$EFT) has witnessed 
much progress during the two decades since 
the pioneering papers by \citet{Wei90a,Wei90b,Wei90c}.
In $\chi$EFT, the symmetries 
of quantum chromodynamics (QCD), in particular its approximate chiral symmetry,
are employed to systematically constrain classes of Lagrangians describing,
at low energies, the interactions of baryons (in particular, nucleons and $\Delta$-isobars)
with pions as well as the interactions of these hadrons with electroweak
fields.  Each class is characterized by a given power of the pion mass and/or
momentum, the latter generically denoted by $P$, and can therefore be thought
of as a term in a series expansion in powers of $P/\Lambda_\chi$, where
$\Lambda_\chi\simeq 1$ GeV specifies the chiral-symmetry breaking scale.
Each class also involves a certain number of unknown coefficients, called
low-energy constants (LEC's), which are determined by fits to experimental data.
See, for example, the review papers~\citet{Bed02} and~\citet{Epelbaum:2009},
and references therein.  Thus $\chi$EFT provides a direct connection between QCD
and its symmetries and the strong and electroweak interactions in nuclei.
From this perspective, it can be justifiably argued to have put
low-energy nuclear physics on a more fundamental basis.  Just as importantly,
it yields a practical calculational scheme, which can, at least in principle, be
improved systematically.

Within the nuclear $\chi$EFT approach, a variety of studies have been carried out
in the strong-interaction sector dealing with the derivation of $N\!N$ and $3N$
potentials~\cite{Ord95,Epe98,Ent03,Mac11,Nav07,Epe02,van94,Ber11,Gir11}
and accompanying isospin-symmetry-breaking corrections~\cite{Fri99,Epe99,Fri04,Fri05}.
In the electroweak sector additional studies have been made
dealing with the derivation of parity-violating
$N\!N$ potentials induced by hadronic weak interactions~\cite{Hax13,Zhu05,Gir08,Viv14}
and the construction of nuclear electroweak currents~\cite{Par93,Pas09,Pas11,Pia13,Koe09,Koe11}.

Recently chiral nuclear interactions have been developed that are local
up to next-to-next-to-leading order (N$^2$LO)~\cite{Gezerlis:2013}.
These interactions employ a different regularization scheme from previous
chiral interactions, with a cutoff in the relative $N\!N$ momentum $q$.
They are therefore fairly simple to treat with standard QMC techniques
to calculate properties of nuclei and neutron matter~\cite{Lynn:2014,Gezerlis:2013}.

As explained in \citet{Gezerlis:2014}, up to N$^2$LO, the momentum-dependent contact interactions can be
completely removed by choosing proper local operators. For example,
at LO there are several operators that are equivalent for contact 
interactions: $\boldsymbol{1}$, $\boldsymbol{\sigma}_1\cdot\boldsymbol{\sigma}_2$, 
$\boldsymbol{\tau}_1\cdot\boldsymbol{\tau}_2$, and
$\boldsymbol{\sigma}_1\cdot\boldsymbol{\sigma}_2\boldsymbol{\tau}_1\cdot\boldsymbol{\tau}_2$.
Similarly, interactions at NLO and N$^2$LO can be constructed by adding
extra operators that include the $S_{12}$, $S_{12}\boldsymbol{\tau}_1\cdot\boldsymbol{\tau}_2$,
and ${\bf L\cdot S}$.
The short-range regulators are also chosen to be local, i.e., $f_{\rm cut}=[1-\exp(-r/R_0)^4]$.
In this way, by fitting the low-energy constants, the chiral potentials are completely 
local up to N$^2$LO. At the next order N$^3$LO non-local operators start to appear,
but their contributions are expected to be very small~\cite{Piarulli:2015}.

\newcommand{\bfsigma}{\boldsymbol{\sigma}}
\newcommand{\bftau}{ \boldsymbol{\tau}}
\newcommand{\bfr}{ \boldsymbol{r}}
\newcommand{\deltau}{ \delta \tau }
\newcommand{\bfR}{ \boldsymbol{R} }
\newcommand{\bfRp}{ \boldsymbol{R}' }
\section{Quantum Monte Carlo methods}

There is a large variety of Quantum Monte Carlo algorithms, and it would
be out of the scope of this review to cover all of them.
We will limit ourselves to describing a specific subset of QMC algorithms
that has been consistently applied to the many nucleon problem,
namely algorithms that are based on a coordinate representation of the
Hamiltonian, and that are based on recursive
sampling of a probability density or of a propagator.
This set of methods includes the standard Variational Monte Carlo (VMC),
Green's Function Monte Carlo (GFMC) and Diffusion Monte Carlo methods.

These methods have been successfully applied to a broad class of problems.
The major fields of application of this set of algorithms
are quantum chemistry and materials science~\cite{Hammond:1994,Nightingale:1999,Foulkes:2001}, where QMC is
a natural competitor of methods such as Coupled Cluster theory and
standard Configuration Interaction methods that are very accurate
for problems where the uncorrelated or Hartree-Fock state provides 
already a good description of the many-body ground state. 
In these fields several software packages have been developed 
with the aim of making the
use of QMC methods more and more widespread across the community.
Other applications in condensed matter theory concern the physics
of condensed helium systems, both $^4$He and $^3$He~\cite{Schmidt:1992,ceperley1995}.
Several QMC calculations have been extensively performed
to investigate properties of both bosonic and fermionic
ultracold gases; see for example \citet{Carlson:2003,Giorgini:2008}.

Because of the strong correlations induced by nuclear Hamiltonians, 
QMC methods have proved to be very valuable
in understanding properties of nuclei and nucleonic matter.
Variational Monte Carlo methods were introduced for use with
nuclear interactions in the early 1980s~\cite{lomnitz1981monte}.
VMC requires an accurate understanding of the structure
of the system to be explored.  Typically, 
a specific class of trial wave functions is considered, and using
Monte Carlo quadrature to evaluate the multidimensional integrals, the
energy with respect to changes in a set of variational parameters is minimized.

GFMC was introduced in nuclear physics for 
spin- isospin-dependent Hamiltonians 
in the late 1980s~\cite{carlson1987,carlson1988}.  It involves the projection
of the ground state from an initial trial state with an evolution in
imaginary time in terms of a path integral,
using Monte Carlo techniques to sample the paths.  GFMC works best when
an accurate trial wave function is available, often developed through
initial VMC calculations.  
This method is very accurate for light nuclei, but becomes increasingly more
difficult moving toward larger systems.  The growth in computing time is
exponential in the number of particles because of the number of spin and isospin
states. The largest nuclear GFMC calculations to
date are for the $^{12}$C nucleus~\cite{lovato2013,Lovato:2014,Lovato:2015}, and for systems
of 16 neutrons~\cite{Gandolfi:2011,Maris:2013} (540,672 and 65,536 spin-isospin states, respectively).

    Auxiliary Field Diffusion Monte Carlo (AFDMC) was introduced in 
1999~\cite{Schmidt:1999}.  In this algorithm the spin- and isospin-dependence
is treated using auxiliary fields. These fields are sampled using
Monte Carlo techniques, and the coordinate-space diffusion in GFMC
is extended to include a diffusion in the spin and isospin states of
the individual nucleons as well.  This algorithm is much more efficient
at treating large systems.  It has been very successful in studying
homogeneous and inhomogeneous neutron matter, and recently has been shown to
be very promising for calculating properties of heavier nuclei,
nuclear matter~\cite{Gandolfi:2014b}, and systems including 
hyperons~\cite{Lonardoni:2013,Lonardoni:2014,Lonardoni:2014b}.
It does require the 
use of simpler trial wave functions, though, and is not yet quite as flexible 
in the complexity of nuclear Hamiltonians that can be employed.
Extending the range of interactions that can be treated with AFDMC
is an active area of research.

\subsection{Variational Monte Carlo}

In VMC, one assumes a form for the trial wave function $\Psi_T$
and optimizes variational parameters, typically by minimizing the energy
and/or the variance of the energy with respect to variations in the
parameters. The energy of the variational wave function $E_V$
\begin{equation}
E_V = \frac{ \langle \Psi_T |  H |\Psi_T \rangle }{\langle \Psi_T | \Psi_T \rangle} \geq E_0,
\label{eq:ev}
\end{equation}
is greater than or equal to the ground-state energy with the same
quantum numbers as $\Psi_T$. Monte Carlo methods can be used to
calculate $E_V$ and to minimize the energy with respect to changes
in the variational parameters.

For nuclear physics, the trial wave function $| \Psi_T \rangle$
has the generic form:
\begin{equation}
| \Psi_T \rangle \  = \ {\cal F} | \Phi \rangle.
\label{eq:psitrial}
\end{equation}

With this form, a factorization of the wave function
into long-range low-momentum components and short-range high-momentum
components is assumed.  The short-range behavior of the wave
functions is controlled by the correlation operator
${ \cal F}$, and the quantum numbers of the system and the long-range behavior
by $| \Phi \rangle $.  In nuclei the separation between the
short-distance correlations and the low-momentum structure of the
wave function is 
less clear than in some systems. For example, alpha particle clusters
can be very important in light nuclei, and their structure is of the
order of the interparticle spacing. Also the pairing
gap can be a nontrivial fraction of the Fermi energy, and hence
the coherence length may be smaller than the system.
Nevertheless this general form has proved to be extremely
useful in both light nuclei and nuclear matter.

\subsubsection{Short-range structure:  {\cal F}}
The correlation operator is dominated by Jastrow-like correlations between 
pairs and triplets of particles:
\begin{equation}
{\cal F} = \left({\cal S} \prod_{i<j<k} (1+F_{ijk})\right) \ \left({\cal S} \prod_{i<j} F_{ij}\right) \, ,
\label{eq:ftrial}
\end{equation}
where ${\cal S}$ is the symmetrization operator, $F_{ij}$ is a 
two-body and $F_{ijk}$ is a three-body correlation.
The two-body correlation operator can include a strong dependence upon spin 
and isospin, and is typically taken as:
\begin{equation}
\label{eq:fij}
F_{ij} = \sum_p f^p ( r_{ij} )  O_{ij}^p \, ,
\end{equation}
where
\begin{equation}
O_{ij}^p = { 1, \bftau_i \cdot \bftau_j, \bfsigma_i \cdot \bfsigma_j, 
             (\bfsigma_i \cdot \bfsigma_j)(\bftau_i \cdot \bftau_j), 
             S_{ij}, S_{ij} \bftau_i \cdot \bftau_j } \, ,
\end{equation}
and the  $f^p$ are functions of the distance $r_{ij}$ 
between particles $i$ and $j$.  The pair functions $f_{ij}^p$
are usually obtained as the solution of Schr\"{o}dinger-like equations
in the relative distance between two particles:
\begin{equation}
\left[- \ \frac{\hbar^2}{2 \mu} \nabla^2 + v_{S,T} (r) + \lambda_{S,T} (r) \right] f_{S,T} (r) = 0 \, .
\end{equation}
The pair functions are obtained by solving this equation in different
spin and isospin channels, for example $S=0$, $T=1$, and can then be recast
into operator form.  For $S$=1 channels the tensor force enters
and this equation becomes two
coupled equations for the components with $L = J -1$ and $L = J+1$.

The $\lambda_{S,T} (r)$ are functions designed to encode the variational
nature of the calculation, mimicking the effect of other particles
on the pair in the many-body system. Additional variational choices
can be incorporated into boundary conditions on the $f_{S,T} (r)$.
For example, in nuclear and neutron matter the pair functions are typically
short-ranged functions and the boundary condition that $f^{p=1} = 1$
and $f^{p>1} = 0$ at some distances $d$, which may be different in different channels, is enforced.  
Usually it is 
advantageous for the tensor correlation to be finite out to longer distances
because of the one-pion-exchange interaction.
The distances $d$ are variational
parameters,  and the equations for the pair correlations are 
eigenvalue equations;
the eigenvalues are contained in the $\lambda(r)$.
See \citet{pandharipande1979} for complete details.

For the lightest $s$-shell nuclei (A= 3 and 4), on the other hand,
the asymptotic properties of the wave function are encoded in the
pair correlation operators $f^p$. To this end the $\lambda (r)$
are determined by requiring the product of pair correlations 
${\cal S} \prod_j F_{ij}$ to have the correct asymptotic behavior
as particle $i$ is separated from the system.
These boundary conditions are described in \citet{schiavilla1986}
and \citet{wiringa1991}.

It has been found advantageous to reduce the strength of the 
spin- and isospin-dependent pair correlation functions $F_{ij}$
when other particles are nearby, with the simple form above
altered to 
\begin{equation}
F_{ij} = \sum_p f^p ( r_{ij} ) \prod_k q^p (\bfr_{ij}, \bfr_{ik}, \bfr_{jk}) O_{ij}^p,
\end{equation}
where the central (spin-isospin independent) quenching factor
$q^{p=1}$ is typically 1,
while for other operators it is parametrized so as to reduce
the pair correlation when another particle $k$ is near the pair $ij$~\cite{pudliner1997}.

The $F_{ijk}$ becomes particularly important when the Hamiltonian includes 
a $3N$ force.
A good correlation form is:
\begin{equation}
     F_{ijk} = \sum_x \epsilon_x V^x_{ijk}(\tilde{ r}_{ij},
                     \tilde{r}_{jk}, \tilde{ r}_{ki}) \ ,
\label{eq:bestfijk}
\end{equation}
with $\tilde{r}=y_xr$, $y_x$ a scaling parameter, and $\epsilon_x$ a (small
negative) strength parameter.
The superscript $x$ denotes various pieces of the $3N$ force like
$(2\pi,P)$ and $R$, so Eq.~(\ref{eq:bestfijk}) brings in all the spin-isospin
dependence induced by that piece of the $3N$ potential.
In practice the ${\cal S} \prod_{i<j<k} (1+F_{ijk})$ in Eq.~(\ref{eq:ftrial}) is
usually replaced with a sum $(1+\sum_{i<j<k} F_{ijk})$ which is significantly
faster and results in almost as good a variational energy.  
For three- and four-body nuclei and nuclear matter, pair spin-orbit correlations
have also been included in Eq.~(\ref{eq:ftrial}), but they are expensive to
compute and not used in the work reviewed here.

The typical number of variational parameters for s-shell nuclear wave functions
is about two dozen for a two-body potential like AV18, as shown in
\citet{wiringa1991} and \citet{pudliner1997}.
Another four to six parameters are added if a three-body potential
is included in the Hamiltonian.
One can also add a few additional parameters to break charge independence,
e.g., to generate $T=\frac{3}{2}$ components in the trinucleon wave functions,
but these are generally used only for studies of isospin violation.
For $p$-shell nuclei, the alpha-particle pair and triplet correlations
are varied only minimally, and most optimization is done with the long-range
correlations discussed below.

The variational parameters have generally been optimized by hand.
Variational wave functions with significantly larger numbers of
parameters and more sophisticated optimization have since been
developed \cite{usmani2009,usmani2012}, but are not in general use.
However, they have provided useful insight for improving the simpler
parameter sets.
The calculation of light nuclei is now sufficiently fast that automated
optimization programs might be profitably employed in the future.

\subsubsection{Long-Range Structure: ${ | \phi \rangle }$}

The quantum numbers and long-range structure of the wave function are generally 
controlled by the $| \Phi \rangle$ term in Eq.~(\ref{eq:psitrial}).
For nuclear and neutron matter this has often been taken to
be an uncorrelated Fermi gas wave function.  Recently, the
crucial role of superfluidity has been recognized, particularly
in low-density neutron matter.  In such cases the trial wave function
includes a $| \Phi \rangle$ of BCS form.  For the $s$-wave pairing
relevant to low-density neutron matter, this can be written:
\begin{equation}
| \Phi \rangle = {\cal A} \left[\phi(r_{11'}),\phi(r_{22'}),\phi(r_{33'}), ...\right],
\end{equation}
where the finite particle number projection of the
BCS state has been taken, with $\phi (r)$ the individual pair functions,
and the unprimed and primed indices refer to spin-up and spin-down 
particles respectively.
  These
pair states are functions of the distance between the two nucleons
in the pair.  
The operator ${\cal A}$ is an anti-symmetrization operator~\cite{Carlson:2003c,Gezerlis:2008}. For a more general pairing, a Pfaffian wave 
function is needed (see for example~\citet{Gandolfi:2008b,Gandolfi:2009b} and
references therein).

For light nuclei, the simplest $| \Phi \rangle $ can be written as the sum of
a few Slater determinants, essentially those arising from a very modest
shell-model treatment of the nucleus.  The single-particle orbitals
in such calculations are written in relative coordinates so as to
avoid introducing any spurious center-of-mass (CM) motion. 
An explicit antisymmetrization of the wave function summing over particles
in $s$-wave, $p$-wave, etc., orbitals is required to compute $| \Phi \rangle$.

Improved wave functions can be obtained by considering the
significant cluster structures present in light nuclei.
For example the ground
state of $^8$Be has a very large overlap with two well-separated
alpha particles. Alpha-cluster structures are important in many light
nuclei, for example states in helium and carbon. 
To this end, it is useful to use a ``Jastrow'' wave function $ | \Phi_J \rangle$ 
which includes spin-isospin independent two- and three-body correlations and the cluster-structure for the
$| \Phi \rangle$:
\begin{align}
| \Phi_J \rangle \  & =  \  {\cal A}
\prod_{i<j<k} f_{ijk}^c 
\prod_{i<j \leq 4} f_{ss} (r_{ij})
\prod_{k\leq 4 < l \leq A} f_{sp} (r_{kl}) \nonumber \\
& 
\times\sum_N \prod_{4 < l < m \leq A} f_{pp} (r_{lm}) |\Phi_N (1234:56...A) \rangle.
\label{eq:phij}
\end{align}

This wave function must be explicitly antisymmetrized as it is written
in a particular cluster structure, with particles $1\dots 4$ being in an
alpha-particle cluster, summed over the $N=\binom{A}{4}$ possible partitions.  
The spin-isospin independent two-body correlations
$f_{ss}$, $f_{pp}$, and $f_{sp}$ are different for pairs of particles where
both are in the $s$-shell, both in the $p$-shell, or one in each. The
$f_{ss}$ comes from the structure of an alpha particle, the $f_{sp}$
is constructed to go to unity at large distances.  The $f_{pp}$ is set
to give the appropriate cluster structure outside the $\alpha$-particle
core, for example it is similar to a deuteron for $^6$Li and to a triton
for $^7$Li; see \citet{pudliner1997} for more details.

Except for closed-shell nuclei, the complete trial wave function is constructed
by taking a linear set of states of the form in Eq.~(\ref{eq:phij}) with the 
same total angular momentum and parity.
Typically these correspond to the lowest shell-model states of the system.
QMC methods are then used to compute the Hamiltonian
and normalization matrix elements in this basis.
These coefficients are often similar in magnitude to those produced by a 
very small shell-model calculation of the same nucleus.
In light nuclei $LS$ coupling is most efficient; examples of the 
diagonalization may be found in \citet{pudliner1997,Wiringa:2000,pieper2002}
and compared to traditional shell model studies such as \citet{kumar1974}.
The VMC calculations give very good descriptions of inclusive observables 
including momentum distributions, but the energies and other observables 
can then be improved, using  the results of the VMC diagonalization
to initiate the GFMC calculations.

\subsubsection{ Computational Implementation}

The spatial integrals in Eq.~(\ref{eq:ev}) are evaluated using Metropolis Monte
Carlo techniques \cite{metropolis1953}.
A weight function $W(\boldsymbol{R} )$ is first defined to 
sample points in 3A-dimensional coordinate space.
The simplest choice is $W (\boldsymbol{R}) = \langle \Psi_T (\boldsymbol{R}) | \Psi_T (\boldsymbol{R}) \rangle$, where the brackets indicate a sum over all the spin isospin parts
of the wave function.
For spin-isospin independent interactions the $A$-particle  wave function is
a function of the $3A$ coordinates of the system only, and the weight function
$W$ is the square of the wave function.
The Metropolis method allows one to sample points in large-dimensional spaces
with probability proportional to any positive 
function W through a suitable combination of proposed 
(usually local) moves and an acceptance or rejection of the proposed move based
upon the ratio of the function W at the original or proposed points.  Iterating
these steps produces a set of points in 3A dimensional space with probability
proportional to W({\bf R}).

For spin-isospin dependent interactions, the wave function 
$| \Psi_T (\boldsymbol{R}) \rangle $ is a sum of complex amplitudes for each
spin-isospin state of the system:
\begin{equation}
| \Psi_T(\boldsymbol{R}) \rangle = \sum_{s \leq 2^A, t \leq 2^A} \phi_{s,t} ({\bf R}) \ 
\chi_s (\sigma) \ \chi_t (\tau),
\label{eq:psit}
\end{equation}
and the spin states $\chi_s$ are:
\begin{align}
\chi_1 & =  {| \downarrow_1 \downarrow_2 ... \downarrow_A} \rangle, \nonumber \\
\chi_2 & =  {| \uparrow_1 \downarrow_2 ... \downarrow_A} \rangle, \nonumber \\
\chi_3 & =  {| \downarrow_1 \uparrow_2 ... \downarrow_A} \rangle, \nonumber \\
& ... \nonumber \\
\chi_{2^A} & =  {| \uparrow_1 \uparrow_2 ... \uparrow_A} \rangle, 
\end{align}
and similarly for the isospin states with $n$ and $p$ instead of $\downarrow$ and
$\uparrow$.  The $2^A$ isospin states can be reduced
by using charge conservation to $A! / (N! Z!)$ states and,
by assuming the nucleus has good isospin $T$, further
reduced to
\begin{equation}
   I(A,T) = \frac{2T+1}{\case{1}{2}A+T+1} 
            \left(\begin{array}{c} A \\ \case{1}{2}A+T \end{array}\right) 
\label{eq:numiso}
\end{equation}
components.
The weight function in this case is the sum of the squares
of the individual amplitudes: $W (\boldsymbol{R}) = \sum_{s,t} |\phi_{s,t}(\boldsymbol{R})|^2$.

Given a set of coordinates $ \{ \boldsymbol{R} \}$, to calculate the wave function
one must first populate the various amplitudes in the trial state
by calculating the Slater determinant, BCS state, or Jastrow wave function $| \Phi \rangle$.
Spin-isospin independent operators acting on $ | \Phi \rangle $ are
simple multiplicative constants for each amplitude $\phi_{s,t}$.
Pair correlation operators then operate on the $\Phi$; these are
sparse matrix multiplications for each pair. The sparse matrices
are easily computed on-the-fly 
using explicitly coded subroutines~\cite{pieper2008b}.
The product over pair correlations is built up by successive operations
for each pair. For example, the effect of the operator 
$\boldsymbol{\sigma}_1\cdot\boldsymbol{\sigma}_2$ on the wave function of 
three-particles can be written as follows 
(The notation $a(\uparrow_1\downarrow_2\downarrow_3)$ means the amplitude
for nucleon 1 being spin up and nucleons 2 and 3 being spin down;
the isospin components have been omitted for simplicity):
\begin{align}
\boldsymbol{\sigma}_1 \cdot \boldsymbol{\sigma}_2  \left(
\begin{array}{c}
a(\downarrow_1\downarrow_2\downarrow_3) \\ a(\uparrow_1\downarrow_2\downarrow_3)\\
a(\downarrow_1\uparrow_2\downarrow_3) \\ a(\uparrow_1\uparrow_2\downarrow_3)\\
a(\downarrow_1\downarrow_2\uparrow_3) \\ a(\uparrow_1\downarrow_2\uparrow_3)\\
a(\downarrow_1\uparrow_2\uparrow_3) \\ a(\uparrow_1\uparrow_2\uparrow_3)
\end{array} \right)  
= \left(
\begin{array}{c}
a(\downarrow_1\downarrow_2\downarrow_3)\\
2a(\downarrow_1\uparrow_2\downarrow_3)-a(\uparrow_1\downarrow_2\downarrow_3)\\
2a(\uparrow_1\downarrow_2\downarrow_3)-a(\downarrow_1\uparrow_2\downarrow_3)\\
a(\uparrow_1\uparrow_2\downarrow_3)\\
a(\downarrow_1\downarrow_2\uparrow_3)\\
2a(\downarrow_1\uparrow_2\uparrow_3)-a(\uparrow_1\downarrow_2\uparrow_3)\\
2a(\uparrow_1\downarrow_2\uparrow_3)-a(\downarrow_1\uparrow_2\uparrow_3)\\
a(\uparrow_1\uparrow_2\uparrow_3)
\end{array} \right).
\end{align}

Metropolis Monte Carlo is used to sample points in the 3$A$-dimensional
space by accepting and rejecting trial moves of the particles. Enforcing
detailed balance ensures that the asymptotic distribution of such
points will be distributed according to the weight $W (\boldsymbol{R})$.
The energy can then be computed as the average over the $N$ points in the
random walk:
\begin{equation}
\label{eq:ener}
E_V = \frac{1}{N}  \sum_{i=1}^N \frac{\langle \Psi_T (\boldsymbol{R}_i)  | H | \Psi_T (\boldsymbol{R}_i)\rangle}
{W(\boldsymbol{R}_i)},
\end{equation}
where the angled brackets imply the sum over spin and isospin states for
each set of spatial coordinates $\boldsymbol{R}_i$.
The matrix elements of the Hamiltonian are evaluated using the same techniques
as those used for the pair correlation operators.

The computational time for the VMC method
scales exponentially with the particle number. At first glance,
this may seem to be because of the explicit sums over exponentially
large number of spin-isospin
amplitudes calculated from the trial wave function.  If that were the
only reason, it would be trivial to sample the spin-isospin state
and evaluate the trial wave function's amplitude for that sampled spin-isospin
state. This sampling
can in fact be done but the fundamental problem remains that
good trial wave functions constructed as described in
Eqs.~(\ref{eq:psitrial}--\ref{eq:fij}),
require exponential in the particle number operations to evaluate either
a single spin-isospin amplitude or all of them. Evaluating a single
amplitude provides negligible savings, so the computational time is
reduced by explicitly summing over the amplitudes, which removes any
variance that would occur from sampling. If trial wave functions could
be constructed which
capture the important physics, while requiring computational time that
scales polynomially with particle number for a single spin-isospin
amplitude, VMC calculations
would be straightforward for all nuclei.

In reality one does not usually compute the full wave function with all
orders of pair operators implied by the symmetrization operator ${\cal S}$
in the definition of the wave function.  One can sample the orders of the
pairs independently for the left and right (bra and ket) wave functions 
of Eq.~(\ref{eq:ener}), and define a slightly
more complicated positive definite form for the weight function $W$
in terms of the two sets of amplitudes $\phi_{s,t,l}$  and $\phi_{s,t,r}$
for the order of pair operators $l$ and $r$ in the left- and right-hand
wave functions.
From several thousand to several tens of thousands of points are sufficient
for a typical evaluation of the energy, and statistical errors are obtained 
using standard techniques.

To search for optimal variational parameters embedded in $\Psi_T$, it is 
very useful to first generate a Monte Carlo walk with configurations 
${\bf R}_i$ and weights $W({\bf R}_i)$ for a given parameter set.
Then one can change one or more parameters and reuse the same set of
configurations to evaluate the change in the energy.
The correlated energy difference will have a much smaller statistical error
than differencing two large energies obtained from independent random walks.
In this manner, a chain of small incremental improvements can be developed
that leads to a lower variational energy.
When the norm of the improved wave function starts to differ significantly
from the original walk, a new reference walk can be made and the search
continued from that set.

One way to overcome the exponential growth in computational requirements 
and access larger nuclei is to use a cluster expansion.
Cluster expansions in terms of the operator correlations in
the variational wave function were developed more than two decades 
ago and used in the first QMC calculations of $^{16}$O~\cite{pieper1992}.
In these calculations a full $3A$-dimensional integral was
done for the Jastrow part of the wave function while
up to four-nucleon linked-clusters were used for the
operator terms.  Earlier versions of the Argonne $N\!N$ and Urbana $3N$ 
interactions were used.
Given the tremendous increase in computer power since
then, this method might profitably be reconsidered for
calculations of much bigger nuclei.

\subsection{Green's function Monte Carlo}
\label{sec:gfmc}

GFMC methods 
 are used to project out the ground
state with a particular set of quantum numbers.
GFMC methods were invented in the 1960s~\cite{Kalos:1962} and have been applied 
to many different problems
in condensed matter, chemistry, and related fields.  
They are closely related to finite-temperature algorithms which
calculate the density matrix \cite{ceperley1995}, 
but they use trial wave functions
on the boundaries of the paths to project out the quantum
numbers of specific states.

GFMC typically starts from a trial wave function $| \Psi_T \rangle$
and projects:
\begin{equation}
| \Psi_0 \rangle \propto \lim_{\tau \rightarrow \infty} \exp [ - (H - E_0) \tau ] | \Psi_T \rangle,
\end{equation}
where $E_0$ is a parameter used to control the normalization.
For strongly-interacting systems one cannot compute 
$\exp [ - (H - E_0) \tau ]$
directly,  however one can compute the high-temperature or short-time
propagator, and insert complete sets of
states between each short-time propagator,
\begin{align}
| \Psi_0(R_N) \rangle & = \prod_{1..N} \langle \bfR_N | 
\exp [ - (H - E_0) \deltau ] 
| \bfR_{N-1} \rangle 
\nonumber \\
&  ...
\langle \bfR_1 |
\exp [ - (H - E_0) \deltau ] 
| \bfR_0 \rangle
 | \Psi_T(R_0) \rangle \, ,
\label{eq:gfiter}
\end{align}
and then use Monte Carlo techniques to sample the paths ${ \bfR_i }$ in the
propagation. The method is accurate for small values of 
the time step $\deltau$, and the accuracy 
can be determined by simulations using several different values of the
time step and extrapolating to zero.
In the GFMC method, Monte Carlo is used to sample the coordinates $\bfR$; 
Eq.~(\ref{eq:gfiter}) also has an implied sum over spin and isospin states at 
each step of the walk which is calculated explicitly.

\subsubsection{Imaginary-Time Propagator}
In the simplest approximation the propagator:
\begin{align}
G_{\deltau} &( \bfRp, \bfR )  \equiv 
\langle\bfRp|\exp ( - H  \deltau )|\bfR\rangle \\
& \approx \langle\bfRp| \exp ( -V \deltau / 2) \exp ( -T \deltau ) \exp ( -V \deltau / 2 )|\bfR\rangle  \nonumber \, ,
\label{eq:gsimple}
\end{align}
where $T$ is the non-relativistic kinetic energy:
\begin{align} 
G^0 ( {\bfR', \bfR}) & =
\langle \bfR' | \exp [ - T \deltau ] | \bfR \rangle
\nonumber \\
& =
\left[\frac{1}{\lambda^3 \pi^{3/2}}\right]^A
\exp [ -  (\bfR-\bfR') ^2 / \lambda^2 ] \, ,
\end{align}
with $\lambda^2 = 4 \frac{\hbar^2}{2m} \deltau$,
yielding a Gaussian diffusion for the paths.
The matrix $V$ is the spin- and 
isospin-dependent interaction: 
\begin{equation}
\langle \bfR | \exp ( - V \deltau ) | \bfR \rangle \approx
{\cal S}\prod_{i<j} \exp [ - V_{ij} ({\bfr_{ij}})  \deltau ] \, ,
\end{equation}
where ${\cal S}$ indicates a symmetrization over orders of pairs.
Each pair interaction can be 
simply evaluated as the exponent of a small spin-isospin
matrix.
This treatment is adequate for static spin-dependent $N\!N$ interactions.

In practice one needs to include momentum-dependent spin-orbit $N\!N$
interactions as well as $3N$ interactions. It is more efficient
to calculate the $N\!N$ propagator explicitly, storing the radial
and spin-isospin dependence on a grid for each initial and final $N\!N$
state. This is done by calculating the propagator independently in each
partial wave and then summing them to create the full $N\!N$ 
propagator.  This was first done in studies of liquid Helium~\cite{ceperley1995,Schmidt:1995} and then adapted to the nuclear 
physics case ~\cite{pudliner1997}.
This has the advantage of summing all $N\!N$ interactions for each
pair explicitly, allowing for larger time steps in the path-integral
simulation.  The $N\!N$ propagator $g_{ij}$ is defined:
\begin{align}
\langle \chi'_\sigma \chi'_\tau |   
& g_{ij}  ( \bfr'_{ij}, \bfr_{ij}; \deltau)
| \chi_\sigma \chi_\tau \rangle  = \nonumber \\
& \langle \chi'_\sigma \chi'_\tau \bfr'_{ij} |   \exp [ - H_{ij} \deltau ]
| \chi_\sigma \chi_\tau \bfr_{ij} \rangle,
\end{align}
where $\bfr_{ij}$ and $\bfr'_{ij}$ are the initial and final $N\!N$
relative coordinates, $H_{ij}$ is the $N\!N$ Hamiltonian including
relative kinetic energy and the $N\!N$ interaction, and $\chi'_\sigma,
\chi_\sigma$ and $\chi'_\tau, \chi_\tau$ are $N\!N$ initial and final
spin and isospin states, respectively. The pair propagator is calculated
for the AV8$'$ Hamiltonian, denoted as $g_{ij}^{v8}$.
At present higher order terms in the momenta $( {\bf p}^2, {\bf L}^2,
({\bf L \cdot S})^2, ...)$ are treated perturbatively. Though the pair
propagator can be calculated for these interactions, the Monte Carlo
sampling can lead to large variance~\cite{Lynn2012}.

The pair propagators are then combined to produce the full propagation
matrix for the system.
The $3N$ interaction $V_{ijk}$ is included symmetrically,
and the full propagation matrix for each step 
$ G_{\deltau} ( \bfRp, \bfR ) $
can then be written as:
\begin{align}
 &G_{\deltau} ( \bfRp, \bfR )   =  
\langle\bfRp| \left( 1 - \sum_{i<j<k}V_{ijk} \deltau / 2 \right) |\bfRp\rangle \,
G^0 (\bfRp,\bfR) 
\nonumber \\
&\times{\cal S} \prod_{i<j} \frac{ g_{ij}^{v_8}(\bfRp,\bfR) }
{g_{ij}^{0}(\bfRp,\bfR)}  
\,\langle\bfR| \left( 1 - \sum_{i<j<k}V_{ijk} \deltau / 2 \right)|\bfR\rangle .
\label{eq:fullprop}
\end{align}
The spin-orbit interaction in the
product of propagators with the full v$_8$ interaction yields
spurious interactions resulting from quadratic terms in the
difference ${\bfR'} - {\bfR}$ from different pairs.  One can
correct for this but 
in practice the effect is not significant.
Using the calculated $N\!N$ propagators allows for a factor of 5-10
larger time steps $\deltau$ than the simple approximation in Eq.(\ref{eq:gsimple})
~\cite{pudliner1997}.

\subsubsection{Implementation}
\label{sec:gfmcimpl}
Once the propagator for each step is specified, an algorithm must be
chosen to sample over all possible paths. A branching
random walk algorithm very similar to that used in standard diffusion
Monte Carlo (DMC)~\cite{Foulkes:2001} is used. This random walk does not sample the entire path at once;
it uses Markov Chain Monte Carlo to perform each step given the present
coordinates and amplitudes in the propagated wave function.
One difference with standard DMC is that the importance
sampled Green's function is explicitly sampled rather than using a small time-step extrapolation
for the wave functions.

A positive definite ``weight'' $W (\Psi_T, \Psi (\tau))$ is first
defined
as a function of the trial function $\Psi_T$ and the propagated wave function
$\Psi (\tau)$. Typically the form used is
\begin{align}
W & =  \left| \sum_{s,t}  
\langle \Psi_T | \chi_s \chi_t \rangle
\langle \chi_s \chi_t  | \Psi (\tau) \rangle \right| \nonumber \\
& + \epsilon \sum_{s,t} \left|
\langle \Psi_T | \chi_s \chi_t \rangle
\langle \chi_s \chi_t  | \Psi (\tau) \rangle \right| ,
\label{eq:gfmcw}
\end{align}
where $\epsilon$ is a small parameter.
Sampling of the paths and branching for the importance function are then
implemented with the scalar function $W$.
Given the present position ${\bfR}$, several different possible
final states ${\bfR'} = \bfR + \delta \bfR$ are sampled from the free propagator $G^0$.
For each sample of $\delta \bfR$ the corresponding $- \delta \bfR$ configuration is included in the 
sample.  The weight function $W_i$ is then calculated for each of the possible
new points $\bfR'_i$, and the final point is chosen according to the relative weights and
scaled with the ratio of the average $W_i$ to the 
actual $W_i$. Branching is performed with the ratio of weight functions
after and before the step, or typically after several steps.
The weights of different paths used to calculate observables will eventually diverge,
yielding the entire contribution from only a few paths that dominate. This is commonly avoided
by using the branching technique, in which the configurations are redistributed 
by killing or making $N$ copies of each one according to
\begin{equation}
N_i=\left[W_i+\xi\right] \,,
\end{equation}
where $W_i$ is the weight of the i-th configuration
obtained by multiplying the weight of Eq.~(\ref{eq:gfmcw}) by
$\exp([E_0-V(R)]\delta\tau)$ ($V$ is the spin/isospin independent part of
the potential), $\xi$ is a random number with uniform distribution
between 0 and 1, and in the above equation $[...]$ means the truncated integer
number of the argument. 
Different random number seeds are given to new
copies generated from the same walker. This procedure guarantees that
the configurations with small weight, contributing by generating only
noise to the observables, are dropped. The full procedure is described
in~\citet{pudliner1997}.

\begin{figure}
\includegraphics[scale=0.33]{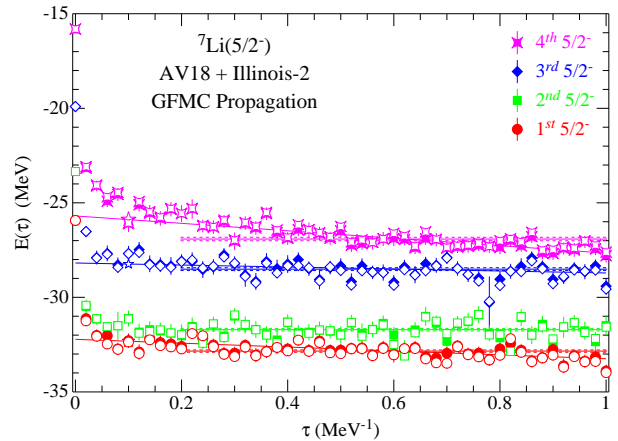}
\caption{GFMC energies of four $\frac{5}{2}^-$ states in $^7$Li vs. imaginary
time $\tau$.  The solid symbols show the computed energies at each $\tau$,
open symbols show the results of rediagonalization.}
\label{fig:mult_ex_a7}
\end{figure}

After every typically 20 to 40 steps,
the energy as a function of imaginary time $\tau$ is calculated as:
\begin{align}
E ( \tau )  & = \frac{\langle \Psi_T | H | \Psi (\tau) \rangle}
{\langle \Psi_T | \Psi (\tau) \rangle} \nonumber \\
& = 
\frac{\sum_i \langle \Psi_T (i) | H | \Psi (\tau,i) \rangle / W_i}
{\sum_i \langle \Psi_T (i) | \Psi (\tau,i) \rangle / W_i},
\label{eq:evstau}
\end{align}
where the sum over $i$ indicates the sum over samples of the wave function.
The brackets in the numerator and denominator of the last expression
indicate sums over spins and
isospins for each sample.  
The $E(\tau)$ initially decrease rapidly from the VMC ($\tau=0$) energy
but then stabilizes and just fluctuates within the statistical errors;
examples of this are shown in Fig.~\ref{fig:mult_ex_a7}, discussed below,
and also in Sec.~\ref{sec:Hoyle}.
These stable values are averaged to get the converged GFMC results.

In principle, the GFMC propagation should converge to 
the lowest-energy state of given quantum numbers $J^\pi;T$.
The nuclei considered here may have a few particle-stable and multiple
particle-unstable excited states of the same quantum numbers.
In practice, GFMC propagation can obtain good energy estimates for many
of these additional states.

First, a set of orthogonal VMC trial functions are generated that are
diagonalized in the small single-particle $p$-shell basis of differing
$LS$ and spatial symmetry combinations that can make a given $J^\pi;T$.
These are pseudo-bound wave functions that fall off exponentially
at long range, with matter radii not much larger than the ground state.
Then independent GFMC propagations are carried out starting from each of these 
trial functions.
An example is shown in Fig.~\ref{fig:mult_ex_a7} for the four $\frac{5}{2}^-$ $p$-shell
states in $^7$Li, all of which are particle-unstable \cite{pieper2004}.
The GFMC propagations stay nearly orthogonal to fairly large $\tau \sim 1$ 
MeV$^{-1}$, as shown by the solid symbols in the figure.
The overlaps between different states can be evaluated, and an explicit
reorthogonalization made, shown by the open symbols.
The states remain well-separated in energy.

The first $\frac{5}{2}^-$ state in Fig.~\ref{fig:mult_ex_a7} is physically wide
($\sim 900$ keV) because it has the spatial symmetry of alpha plus triton
and is several MeV above the threshold for breakup into separated clusters.
Consequently, a GFMC propagation is expected to eventually drop
to that threshold energy, and the figure shows, after a rapid initial
drop from $-$26 to $-$32 MeV by $\tau = 0.1$ MeV$^{-1}$, a slowly 
decreasing energy as $\tau$ increases, reaching $-$33 MeV at $\tau = 1$ 
MeV$^{-1}$.
In cases like this, the energy is quoted at the small value of $\tau$
where the rapid initial improvement over the variational starting point has 
saturated.
The second $\frac{5}{2}^-$ state in Fig.~\ref{fig:mult_ex_a7} is physically narrow
($\sim 80$ keV) because it has a spatial symmetry like $^6$Li+$n$ and is
only 20 keV above that breakup threshold.
The GFMC propagation shows the same rapid initial drop in energy, and then
no appreciable further decline, allowing us to identify a clear energy
for this state.
The third and fourth $\frac{5}{2}^-$ states are not experimentally identified,
but from the GFMC propagation behavior we would expect the third state to be 
physically narrow, and the fourth to be fairly broad.
An alternative approach to calculate systems in the continuum by imposing 
specific boundary conditions is presented in sec.~\ref{sec:scat}.

In general the GFMC method suffers from the fermion sign problem, in that
the numerator and denominator of Eq.~(\ref{eq:evstau}) tend to 
have an increasing ratio of error to signal for a finite 
sample size and large imaginary times $\tau$. Other than for a few 
special cases such as purely attractive interactions, Hubbard models at
half-filling, or lattice QCD at zero chemical potential, QMC methods
typically all have this difficulty. 
This is basically because when $\Psi_T$ is not exact it contains contamination
from the Bosonic ground-state that will be unavoidably sampled.
For scalar potentials, or in any case where a real wave function can be 
used, the sign problem is avoided by using the fixed-node approximation,
and the problem is solved in a restricted (Bosonic) sub-space, where the 
trial wave function always maintains the same sign. In this case the 
problem would be exactly solved if the nodes of the true ground-state were
known. Because this is not the case, the solution obtained is a rigorous
upperbound to the true ground-state energy~\cite{Moskowitz:1982}.
For spin-isospin dependent Hamiltonians a complex wave function must be used, 
and the general fixed-node approximation does not apply.
Instead the sign problem is circumvented by
using a `constrained path' algorithm, essentially limiting the original
propagation to regions where the propagated and trial wave functions have
a positive overlap.  This approximation, like the fixed-node algorithm
for spin-independent interaction, involves discarding configurations that
have zero overlap with the trial wave function. As such, they are exact
for the case when the trial wave function is exact and are therefore
variational. However, unlike the fixed-node case, the constrained path
method does not provide upper bounds~\cite{Wiringa:2000}.

To address the possible bias introduced by the constraint,
all the configurations (including those that would be discarded) for
a previous number of steps $N_{uc}$ are used when evaluating energies and other
expectation values.  $N_{uc}$ is chosen to be 
as large a number of time steps as feasible with reasonable
statistical error (again typically 20 to 40 steps).  
Tests using different trial functions and very long runs indicate that 
energies in $p$-shell nuclei are accurate to around one per cent using these
methods. This has been tested in detail in~\citet{Wiringa:2000}, where the
use of different wave functions is discussed.

Expectation values other than the energy are typically calculated from
``mixed'' estimates; for diagonal matrix elements this is: 
\begin{equation}
\label{eq:mixed}
\langle {\cal O(\tau)} \rangle \ \approx \ 2 \,\frac{\langle \Psi_T | {\cal O} | \Psi (\tau)  \rangle}
{\langle \Psi_T|\Psi (\tau)  \rangle}
 - \frac{\langle \Psi_T | {\cal O} | \Psi_T \rangle}{\langle \Psi_T|\Psi_T \rangle} \,.
\end{equation}
The above equation can be verified by assuming that the true ground state is
well represented by the variational wave function and a small perturbation,
i.e., $|\Psi(\tau)\rangle\approx|\Psi_T\rangle+\lambda|\Psi\rangle$, and $\lambda$ is a
small parameter.
Since the variational wave functions are typically very good the extrapolation
is quite small. This can be further tested by using different trial wave
functions to extract the same observable, or using the Hellman-Feynman theorem. 
For the case of simple static
operators, improved methods are available that propagate both before and
after the insertion of the operator ${\cal O}$~\cite{Liu:1974}, i.e. directly
calculating operators with $\Psi(\tau)$ on both sides. However these techniques
might be very difficult to apply for non-local operators.

Because a Hamiltonian commutes with itself, the total energy 
of the Hamiltonian used to construct the propagator [Eq.~(\ref{eq:fullprop})] is not extrapolated;
thus this total energy is not the sum of its extrapolated pieces,
rather the sum differs by the amount the $\Psi_T$ energy was improved.
As was noted above, the full AV18 $N\!N$ potential cannot be used in the
propagator; rather an $H^\prime$ containing the AV8$^\prime$ approximation to AV18 is used.  
In practice AV8$^\prime$ gives slightly more binding than AV18 so the
the repulsive part of the $3N$ potential is increased to make
$\langle H^\prime \rangle \approx \langle H \rangle$.  The difference
$\langle H-H^\prime \rangle$ must be extrapolated by Eq.~(\ref{eq:mixed}).
The best check of the systematic error introduced by this procedure
is given by comparing GFMC calculations of $^3$H and $^4$He energies
with results of more accurate few-nucleon methods; this suggests
that the error is less than 0.5\%~\cite{pudliner1997}.

In the case of off-diagonal matrix elements, e.g., in transition matrix
elements between initial $\Psi^i$ and final $\Psi^f$ wave functions, 
Eq.~(\ref{eq:mixed}) generalizes to:
\begin{eqnarray}
   \nonumber
\langle {\cal O(\tau)} \rangle \ &\approx&
   \frac{\langle \Psi^f_T | {\cal O} | \Psi^i(\tau)  \rangle}
        {\langle \Psi^i_T|\Psi^i(\tau) \rangle} 
        \frac{|\Psi^i_T|}{|\Psi^f_T|}
 + \frac{\langle \Psi^f(\tau) | {\cal O} | \Psi^i_T  \rangle}
        {\langle \Psi^f(\tau)|\Psi^f_T \rangle}
        \frac{|\Psi^f_T|}{|\Psi^i_T|}  \\
&& - \frac{\langle \Psi^f_T | {\cal O} | \Psi^i_T \rangle}
        {|\Psi^f_T||\Psi^i_T|} \ .
\label{eq:transmix}
\end{eqnarray}
Technical details can be found in \citet{pervin2007}.

Recently, the capability to make correlated GFMC propagations
has been added~\cite{Lovato:2015}.  In these calculations, the values of 
$\boldsymbol{R}$ for every $\delta \tau$ time step,
the corresponding weights $W$, and other quantities are saved during
an initial propagation.  Subsequent propagations for different initial
$\Psi_T$ or different nuclei (such as isobaric analogs) then follow
the original propagation and correlated differences of expectation values
can be computed with much smaller statistical errors than for the individual
values.

\subsection{Auxiliary Field Diffusion Monte Carlo}

The GFMC method works very well for calculating the low lying states of nuclei
up to $^{12}$C. Its major limitation is that the computational
costs scale exponentially with the number of particles, because of the
full summations of the spin-isospin states. 
An alternative approach is to use a basis given by the outer product of nucleon position
states, and the outer product of single nucleon spin-isospin spinor states.
An element of this overcomplete basis is given by specifying the
$3A$ Cartesian coordinates for the $A$ nucleons, and specifying four complex
amplitudes for each nucleon to be in a $|s\rangle=|p\uparrow, p\downarrow, n\uparrow,
n\downarrow\rangle$ spin-isospin state. A basis state is then defined
\begin{equation}
|\boldsymbol{R}\, S\rangle =
|\boldsymbol{r}_1 s_1\rangle \otimes |\boldsymbol{r}_2 s_2 \rangle \dots \otimes 
|\boldsymbol{r}_n s_n\rangle \,.
\label{eq:basis}
\end{equation}

The trial functions must be antisymmetric under interchange. The only such
functions with polynomial scaling are Slater
determinants or Pfaffians (BCS pairing functions), for example,
\begin{equation}
\langle \boldsymbol{R}\, S|\Phi\rangle = 
{\cal A} \left [
\langle \boldsymbol{r}_1 s_1|\phi_1\rangle 
\langle \boldsymbol{r}_2 s_2|\phi_2\rangle 
\dots
\langle \boldsymbol{r}_A s_A|\phi_n\rangle  \right ]
\label{eq:wf1}
\end{equation}
or linear combinations of them. Operating on these with the product of
correlation operators, Eq.~(\ref{eq:ftrial}), again gives a state with
exponential scaling with nucleon number. In most of the AFDMC calculations,
these wave functions include a state-independent, or
central, Jastrow correlation:
\begin{equation}
\langle \boldsymbol{R}\,S|\Psi_T\rangle=\langle\boldsymbol{R}\,S|\left[\prod_{i<j}f^c(r_{ij})\right]
\Phi\rangle \,.
\label{eq:wfsimple}
\end{equation}
Calculations of the Slater determinants
or Pfaffians scale like $A^3$ when using standard dense matrix methods,
while the central Jastrow requires $A^2$ operations if its range is
the same order as the system size.
These trial functions capture only the physics of the
gross shell structure of the nuclear
problem and the state-independent part of the two-body interaction.
Devising trial functions that are both computationally efficient to
calculate and that capture the state-dependent two- and three-body
correlations that are important would greatly improve both
the statistical and systematic errors of QMC methods
for nuclear problems.

The trial wave functions above can be used for variational calculations.
However, the results are poor since the functions miss the
physics of the important tensor interactions. 
More recently the improved form
\begin{align}
&\langle \boldsymbol{R}\,S|\Psi_T\rangle =
\nonumber \\
&\langle \boldsymbol{R}\,S|
\left[\prod_{i<j}f^c(r_{ij})\right] 
\left [1+\sum_{i<j} F_{ij}
+\sum_{i<j<k} F_{ijk} \right ]|\Phi\rangle \,,
\label{eq:wfbetter}
\end{align}
has been employed,
where $f^c$ are spin-isospin independent correlations, and 
the correlations $F$ have a form similar to those discussed in the 
previous sections.
These wave functions can be
used as importance functions for AFDMC calculations where they have been
found adequate for this purpose in a variety of problems.

Using the basis state as in Eq.~(\ref{eq:basis}) requires the use of a different
propagator, with at most linear spin-isospin operators. The propagator
can be rewritten using the Hubbard-Stratonovich transformation:
\begin{equation}
e^{-O^2/2} = \frac{1}{\sqrt{2\pi}}
\int_{-\infty}^\infty dx\, e^{-x^2/2} e^{x O} \,,
\label{eq:hs}
\end{equation}
where the variables $x$ are called auxiliary fields, and 
$O$ can be any type of operator included in the propagator.

It is helpful to apply the auxiliary field formalism to derive
the well known central potential
diffusion Monte Carlo algorithm~\cite{anderson1976}.
The Hamiltonian is
\begin{equation}
H = \sum_{n}^A \frac{\boldsymbol{p}_{n}^2}{2m} + V(\boldsymbol{R}) \,,
\quad V(\boldsymbol{R})=\sum_{i<j} v(r_{ij}) \,,
\end{equation}
and $v(r_{ij})$ is a generic potential whose form depends on the system. 
Making the short-time approximation, the propagator can be written
as
\begin{equation}
e^{-(H-E_0)\delta \tau} \approx \exp\left(-\sum_{n}^A
\frac{\boldsymbol{p}_{n}^2}{2m} \delta \tau\right)
\exp\left[-(V(\boldsymbol{R})-E_0)\delta \tau\right] \,.
\end{equation}
Since the Hamiltonian does not operate on the spin, the
spin variables can be dropped
from the walker expressions to leave just a position basis 
$|\boldsymbol{R}\rangle$.
Operating with the local-potential term gives just a weight factor:
\begin{equation}
e^{-[V(\boldsymbol{R})-E_0]\delta \tau}|\boldsymbol{R}\rangle = 
W|\boldsymbol{R}\rangle \,.
\end{equation}
The kinetic energy part of the propagator can be applied by
using the Hubbard-Stratonovich transformation:
\begin{align}
\label{eq:hskinetic}
\exp&\left(-\sum_n\frac{\boldsymbol{p}_{n}^2}{2m} \delta \tau\right)
\approx
\prod_n \exp\left(-\frac{\boldsymbol{p}_{n}^2}{2m} \delta \tau\right)
\\
&=\prod_{n} \frac{1}{(2\pi)^{3/2}} \int d x_{n}
e^{-x_{n}^2/2}
\exp\left({-\frac{i}{\hbar} \boldsymbol{p}_{n}
x_{n} \sqrt{\frac{\hbar^2 \delta \tau}{m}}}\right) \nonumber \,.
\end{align}
This propagator applied to a walker $|\boldsymbol{R}\rangle$
generates a new position $|\boldsymbol{R}+\Delta \boldsymbol{R}\rangle$, where
each particle position is shifted as
\begin{equation}
\boldsymbol{r}_{n}' = \boldsymbol{r}_{n} + \frac{\hbar^2 \delta \tau}{m} x_n \,.
\end{equation}
This is identical to the standard diffusion Monte Carlo algorithm without
importance sampling. 
Each particle is moved with a Gaussian distribution of variance
$\hbar^2 \delta \tau /m$, and a weight of
$\exp[-(V(\boldsymbol{R})-E_0)\delta \tau]$ is included. The branching on the weight
is then included to complete the algorithm.

The $N\!N$ potential in the general form of Eq.~(\ref{eq:vijgen}) can be written as 
\begin{equation}
V = \sum_{i<j}v_p(r_{ij})O^{p}_{ij} =
\frac{1}{2} \sum_{i,j} O_i^\alpha A_{i\alpha,j\beta} O_j^\beta=
\frac{1}{2} \sum_n \lambda_n {\cal O}_n^2
\end{equation}
where $O_i^\alpha$ are $\boldsymbol{\sigma}_i$, $\boldsymbol{\tau}_i$ or similar combinations;
see~\citet{Gandolfi:2007c} for more details.
The new operators ${\cal O}$ are defined 
\begin{equation}
{\cal O}_n = \sum_{j\beta} \psi_{j\beta}^{(n)} O_j^\beta \,.
\label{eq:odiag}
\end{equation}
Here $\psi_{j\beta}^{(n)}$ and $\lambda_n$ are the eigenvectors and eigenvalues 
obtained by diagonalizing the matrix $A_{i\alpha,j\beta}$. 

It is easy to see that applying the Hubbard-Stratonovich transformation 
consists in a rotation of the spin-isospin states of nucleons:
\begin{align}
\label{eq:hsspin}
&\prod_{i<j}e^{-V_{ij}\delta\tau}\,|\boldsymbol{R}\,S\rangle= 
\\
& \prod_n \frac{1}{(2\pi)^{3/2}}  \int d x_{n} e^{-x_{n}^2/2} \,
e^{\sqrt{-\lambda_n\delta\tau}x_n{\cal O}_n}
|\boldsymbol{R}\,S\rangle = |\boldsymbol{R}\,S'\rangle \nonumber \,,
\end{align}
The propagation is performed by sampling the auxiliary fields from the 
probability distribution $\exp (-x_n^2/2)$, and applying the rotations to
the nucleon spinors. 
At order $\delta\tau$ the above propagator is the same as that described
in the previous sections. The advantage of this procedure is that a wave function 
with the general spin-isospin structure of Eq.~(\ref{eq:wf1}) can be used, at
a much cheaper computational cost than that of including all the spin-isospin states
of Eq.~(\ref{eq:psit}). However, one must then solve the integral in Eq.~(\ref{eq:hs}),
which is done by Monte Carlo sampling of the auxiliary fields $x$.

The inclusion of importance sampling within the auxiliary fields formalism 
is straightforward, and is currently done as described in Sec.~\ref{sec:gfmcimpl}.
At each time-step a random vector $\Delta\boldsymbol{R}$ for the 
spatial coordinates, and the required auxiliary fields $X$ are sampled.
The four weights corresponding to these samples are
\begin{equation}
W_i=\frac{\langle\Psi_I|\boldsymbol{R}\pm\Delta\boldsymbol{R}\,S'(\pm X)\rangle}
{\langle\Psi_I|\boldsymbol{R}\,S)\rangle}\exp\left[-V_c(\boldsymbol{R})\delta\tau\right] \,,
\end{equation}
where $\Psi_I$ is used for the importance sampling, 
$S'(X)$ are obtained 
by rotating the spinors $S$ of the previous time-step using the auxiliary 
fields $X$, and $V_c$ includes all the spin-isospin independent terms 
of the interaction. 
The procedure is then completed as done in GFMC: one of the above configurations
is taken according to the probabilities, and the branching is done by considering
the cumulative weight.
This procedure lowers the variance as the "plus-minus" sampling cancels
the linear terms coming from the exponential of Eqs.~(\ref{eq:hskinetic},\ref{eq:hsspin}). 
Note that in the example of the kinetic energy presented 
above, the effect of sampling using $\pm\Delta\boldsymbol{R}$ is identical to sampling 
configurations using $\nabla\Psi_I/\Psi_I$ commonly adopted in standard
diffusion Monte Carlo~\cite{Foulkes:2001}.

The importance function $\Psi_I$ must be real and positive, and an efficient algorithm to deal
with complex wave functions has been proposed by~\citet{Zhang:2003}, i.e., consider 
$\langle\Psi_I|\boldsymbol{R}\,S\rangle  =|\langle\Psi_T|\boldsymbol{R}\,S\rangle|$, 
and multiply the weight terms $W_i$ by $\cos\Delta\theta$, where 
$\Delta\theta$ is the phase of 
$\langle\Psi_T|\boldsymbol{R}'\,S'\rangle/\langle\Psi_T|\boldsymbol{R}\,S\rangle$,
and for each $W_i$, $|\boldsymbol{R}'\,S'\rangle$ is the corresponding 
configuration obtained from the corresponding $\pm\Delta\boldsymbol{R}$ 
and $\pm X$ sampling. This method 
samples configurations with a very low variance.

Previous applications of the AFDMC method used a somewhat different
importance sampling, using $\nabla\Psi_I/\Psi_I$ for the kinetic energy,
and the strategy described by~\citet{Sarsa:2003} and~\citet{Gandolfi:2009} for the spin;
the two methods become the same in the limit of $\delta\tau\rightarrow 0$. 
In \citet{Gandolfi:2014b} it has been found that the procedure described above is much less
time-step dependent for calculations including protons. This is due to
the strong tensor force in the $np$ channel that in the case of pure neutron
systems is very weak.  The two algorithms give very similar results.

The energy and other observables are calculated after a block of steps 
in imaginary time, where each block comprises a number of steps that is chosen to 
be large enough (typically around 100-500) such that the configurations 
are statistically uncorrelated. This is done to save computing time in
calculating observables for data that are not useful to reduce the statistical errors.

While the Hubbard-Stratonovich transformation is the most common, there are
many other possibilities. For example, the propagator for the
relativistic kinetic energy can be sampled by using
\begin{equation}
\exp\left[-\left (\sqrt{p^2 c^2 + m^2 c^4}-mc^2 \right ) \delta \tau\right] =
\int d^3x f(x) e^{-i \boldsymbol{p} \cdot \boldsymbol{x} / \hbar}
\end{equation}
with
\begin{align}
f(x) &= \int \frac{d^3\boldsymbol{p}}{(2\pi)^3} \, e^{i \boldsymbol{p} \cdot \boldsymbol{x} / \hbar} \,
e^{-\left (\sqrt{p^2 c^2 + m^2 c^4}-mc^2 \right ) \delta \tau}
\nonumber\\
&= e^{mc^2 \delta \tau} \,
K_2\left( \frac{m c}{\hbar} \sqrt{x^2+c^2 \delta \tau^2} \right)
\end{align}
where $K_2$ is the modified Bessel function of order 2~\cite{carlson1993}.

\newcommand{\nuc}[2] {$^{#1}$#2}

\section{Light nuclei}

\subsection{Energy spectra}
\label{subsec:nuclei}

\begin{figure*}
\includegraphics[height=4.0in]{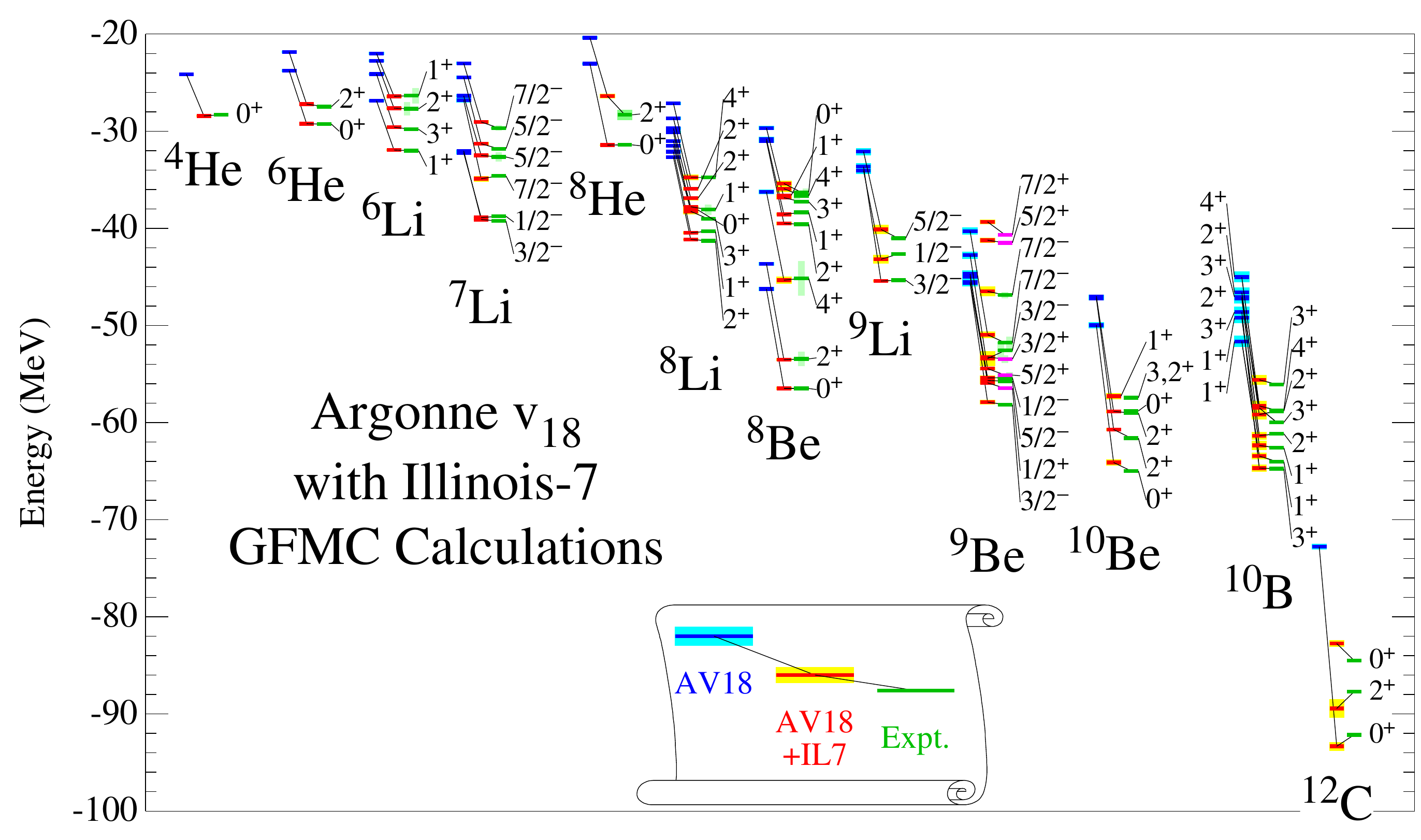}
\caption{GFMC energies of light nuclear ground and excited states for the
AV18 and AV18+IL7 Hamiltonians compared to experiment.
See Table~\ref{tab:energies} for references.}
\label{fig:spectrum}
\end{figure*}

Results of GFMC calculations for light nuclei using the AV18+IL7 Hamiltonian are compared 
to experiment in Fig.~\ref{fig:spectrum} and Table~\ref{tab:energies}
\cite{brida2011,mccutchan2012,pastore2013,lovato2013,wiringa2013,pastore2014,Pieper:2015}.
Results using just AV18 with no $3N$ potential are also shown in the figure.
Figure~\ref{fig:spectrum} shows the absolute energies of more than 50 ground 
and excited states.
The experimental energies of the 21 ground states shown in the table are
reproduced with an rms error of 0.36 MeV and an average signed error of only
0.06 MeV. The importance of the three-body interaction is confirmed by 
the large corresponding numbers for AV18 with no $3N$ potential, namely
10.0 and 8.8 MeV.
About sixty additional isobaric analog states also have been evaluated but are
not shown here.

Table~\ref{tab:energies} gives the ground state energies $E$, proton (neutron) 
point radii $r_p$ ($r_n$), magnetic moments $\mu$ (including two-body current 
contributions, see Sec.~\ref{sec:currents}), and quadrupole moments $Q$ for all the particle-stable ground 
states of $A\leq 10$ nuclei, plus $^{12}$C and the resonant ground states of 
$^7$He and $^8$Be.
Many of these results were obtained in recent studies of spectroscopic
overlaps, electromagnetic transitions and sum rules, and isospin mixing.
The energies, radii, and electromagnetic moments are in generally good
agreement with experiment.

\begin{table*}
\caption{AV18+IL7 GFMC results for $A\leq12$ nuclear ground 
states~\cite{brida2011,mccutchan2012,pastore2013,lovato2013,wiringa2013,pastore2014,Pieper:2015},
compared to experimental values~\cite{NNDC-data,Purcell10,Tilley02,Tilley04,
Amroun94,Shiner94,Nortershauser11,Nortershauser09}.
Numbers in parentheses are statistical errors for the GFMC calculations
or experimental errors; errors of less than one in the last decimal place
are not shown.}
\label{tab:energies}
\begin{ruledtabular}
 \begin{tabular}
{l d{3.3} d{3.1} d{3.1} d{3.1} d{3.2} d{3.3} d{3.3} d{3.1} d{3.3} @{\hspace{12pt}}}
   $^AZ(J^\pi;T)$                &
   \multicolumn{2}{c}{$E$ (MeV)} &
   \multicolumn{3}{c}{$r_p$~[$r_n$] (fm)} &
   \multicolumn{2}{c}{$\mu$ ($\mu_N$)} &
   \multicolumn{2}{c}{$Q$ (fm$^2$)}
   \\ \cline{2-3} \cline{4-6} \cline{7-8} \cline{9-10}
                            &
   \multicolumn{1}{c}{GFMC} &
   \multicolumn{1}{c}{Expt.}&
   \multicolumn{2}{c}{GFMC} &
   \multicolumn{1}{c}{Expt.}&
   \multicolumn{1}{c}{GFMC} &
   \multicolumn{1}{c}{Expt.}&
   \multicolumn{1}{c}{GFMC} &
   \multicolumn{1}{c}{Expt.} \\
   \hline
   \nuc{2}{H}$(1^{+};0)$                    &  -2.225   &  -2.2246& 1.98&     & 1.96    & 0.8604   & 0.8574 & 0.270   &  0.286
   \\
   \nuc{3}{H}$(\frac{1}{2}^+;\frac{1}{2})$  &  -8.47(1) &  -8.482& 1.59&[1.73]& 1.58    & 2.960(1) & 2.979
   \\
   \nuc{3}{He}$(\frac{1}{2}^+;\frac{1}{2})$ &  -7.72(1) &  -7.718& 1.76&[1.60]& 1.76    &-2.100(1) &-2.127
   \\
   \nuc{4}{He}$(0^+;0)$                     & -28.42(3) & -28.30 & 1.43&      & 1.462(6)
   \\ 
   \nuc{6}{He}$(0^+;1)$                     & -29.23(2) & -29.27 & 1.95(3)&[2.88] & 1.93(1) 
   \\
   \nuc{6}{Li}$(1^+;0)$                     & -31.93(3) & -31.99 & 2.39   &   & 2.45(4) & 0.835(1) & 0.822  & 0.1(2)  &-0.082(2)
   \\
   \nuc{7}{He}$(\frac{3}{2}^-;\frac{3}{2})$ & -28.74(3) & -28.86 & 1.97&[3.32(1)]
   \\   
   \nuc{7}{Li}$(\frac{3}{2}^-;\frac{1}{2})$ & -39.15(3) & -39.25 & 2.25&[2.44]& 2.31(5) & 3.24(1)  & 3.256  &-3.9(2)  &-4.06(8)
   \\   
   \nuc{7}{Be}$(\frac{3}{2}^-;\frac{1}{2})$ & -37.54(3) & -37.60 & 2.51&[2.32]& 2.51(2) &-1.42(1)  &-1.398(15) &-6.6(2)
   \\
   \nuc{8}{He}$(0^+;2)$                     & -31.42(3) & -31.40 & 1.83(2)&[2.73]    & 1.88(2)
   \\
   \nuc{8}{Li}$(2^+;1)$                     & -41.14(6) & -41.28 & 2.10&[2.46]& 2.20(5) & 1.48(2)  & 1.654 & 2.5(2) & 3.27(6)
   \\
   \nuc{8}{Be}$(0^+;0)$                     & -56.5(1)  & -56.50 & 2.40(1)
   \\
   \nuc{8}{B}$(2^+,1)$                      & -37.51(6) & -37.74 & 2.48&[2.10]&         & 1.11(2)  & 1.036 & 5.9(4) & 6.83(21)
   \\
   \nuc{8}{C}$(0^+;2)$                      & -24.53(3) & -24.81 & 2.94&[1.85] 
   \\
   \nuc{9}{Li}$(\frac{3}{2}^-,\frac{3}{2})$ & -45.42(4) & -45.34 & 1.96&[2.33]& 2.11(5) & 3.39(4)& 3.439 &-2.3(1) & -2.74(10)
   \\
   \nuc{9}{Be}$(\frac{3}{2}^-,\frac{1}{2})$ & -57.9(2)  & -58.16 & 2.31&[2.46]& 2.38(1) &-1.29(1)&-1.178 & 5.1(1) & 5.29(4)
   \\
   \nuc{9}{C}$(\frac{3}{2}^-,\frac{3}{2})$  & -38.88(4) & -39.04 & 2.44&[1.99]&         &-1.35(4)&-1.391 &-4.1(4)
   \\
   \nuc{10}{Be}$(0^+;1)$                    & -64.4(2)  & -64.98 & 2.20&[2.44]& 2.22(2)
   \\
   \nuc{10}{B}$(3^+;0)$                     & -64.7(3)  & -64.75 & 2.28&      & 2.31(1) & 1.76(1)  & 1.801 & 7.3(3) & 8.47(6)
   \\
   \nuc{10}{C}$(0^+;1)$                     & -60.2(2)  & -60.32 & 2.51&[2.25]
   \\
   \nuc{12}{C}$(0^+;0)$                     & -93.3(4)  & -92.16 & 2.32&      & 2.33
   \\
\end{tabular}
\end{ruledtabular}
\end{table*}

\begin{figure*}
\includegraphics[height=4.0in]{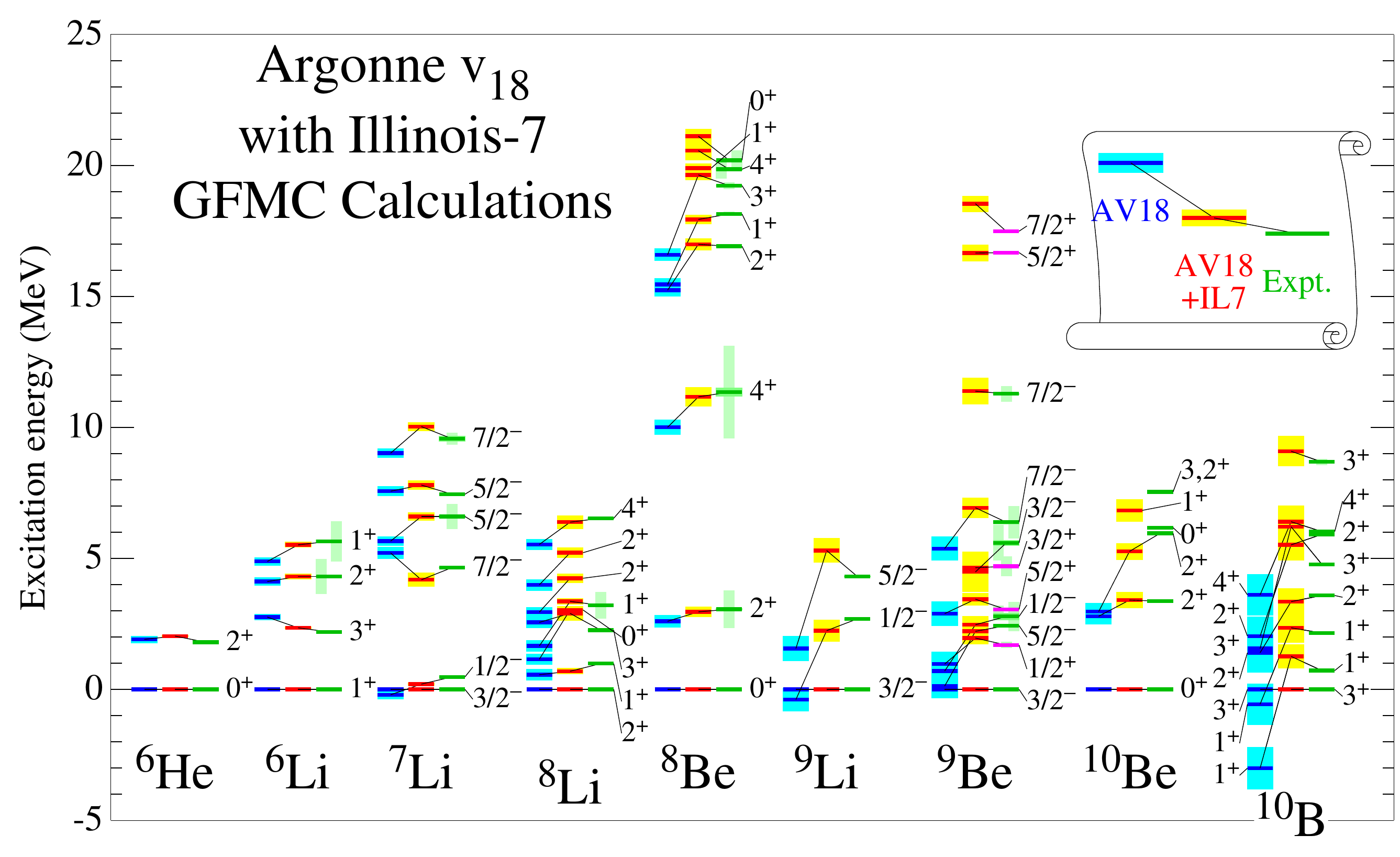}
\caption{GFMC excitation energies of light nuclei for the
AV18 and AV18+IL7 Hamiltonians compared to experiment.
See Table~\ref{tab:energies} for references.}
\label{fig:excited}
\end{figure*}

A detailed breakdown of the AV18+IL7 energies into various pieces
for some of the nuclear ground states is shown in Table~\ref{tab:breakdown}.
The components include the total kinetic energy $K$, 
the contribution $v_{18}$ of the strong interaction part of AV18,
the full electromagnetic potential $v_{ij}^{\gamma}$, 
the two-pion-exchange parts of IL7 $V_{ijk}^{2\pi}$,
the three-pion-ring parts $V_{ijk}^{3\pi}$, 
and the short-range repulsion $V_{ijk}^{R}$.
In the last column, $\delta v_{ij}$ is the expectation value of the
difference between $v_{18}$ and $v_{8^\prime}$, which is the part of
the $N\!N$ interaction that is treated perturbatively because $v_{8^\prime}$
is used in the propagation Hamiltonian.
The sum of the six contributions $K$ through $V_{ijk}^{R}$ does not quite
match the total energy reported in Table~\ref{tab:energies} because they
have been individually extrapolated from the mixed energy expression
Eq.~(\ref{eq:mixed}).

Several key observations can be drawn from Table~\ref{tab:breakdown}.
First, there is a huge cancellation between kinetic and two-body terms.
Second, the net perturbative correction $\delta v_{ij}$ is tiny
($< 2\%$) compared to the full $v_{18}$ expectation value.
Third, the total $V_{ijk}$ contribution is $\sim 5\%$ of $v_{ij}$, 
suggesting good convergence in many-body forces, but it is not negligible
compared to the binding energy.
Finally, the $V_{ijk}^{3\pi}$ contribution that is unique to the Illinois
potentials is a small fraction of the $V_{ijk}^{2\pi}$ in $T=0$ states, 
but does get as large as 35\% in $T=2$ states.

\begin{table}
\caption{Breakdown of GFMC energy contributions for AV18+IL7, in MeV.
See Table~\ref{tab:energies} for references.}
\begin{ruledtabular}
\begin{tabular}{l d{2} d{2} @{\hspace{12pt}} d{2} d{1} d{2} d{1} d{2}}
   \multicolumn{1}{c}{$^AZ(J^\pi;T)$} &
   \multicolumn{1}{c}{$K$} &
   \multicolumn{1}{c}{$v_{18}$} &
   \multicolumn{1}{c}{$v_{ij}^{\gamma}$} &
   \multicolumn{1}{c}{$V_{ijk}^{2\pi}$} &
   \multicolumn{1}{c}{$V_{ijk}^{3\pi}$} &
   \multicolumn{1}{c}{$V_{ijk}^{R}$} &
   \multicolumn{1}{c}{$\delta v_{ij}$} \\
\hline
\nuc{2}{H}$(1^+;0)$  &  19.81 &  -22.05 & 0.02&       &       &      &  0.09\\
\nuc{3}{H}$(\frac{1}{2}^+;\frac{1}{2})$ 
                     &  50.9  &  -58.5  & 0.04&  -1.8 &  -0.03&  0.7 &  0.18\\
\nuc{4}{He}$(0^+;0)$ & 112.(1)& -136.(1)& 0.9 &  -9.8 &  -0.3 &  3.9 &  1.4 \\
\nuc{6}{He}$(0^+;1)$ & 141.(1)& -167.(1)& 0.9 & -11.5 &  -1.5 &  5.1 &  1.8 \\
\nuc{6}{Li}$(1^+;0)$ & 154.(1)& -184.(1)& 1.7 & -11.4 &  -1.0 &  4.9 &  1.8 \\
\nuc{7}{He}$(\frac{3}{2}^-;\frac{3}{2})$ 
                     & 160.(1)& -185.(1)& 0.9 & -13.3 &  -2.9 &  6.4 &  2.3 \\
\nuc{7}{Li}$(\frac{3}{2}^-;\frac{1}{2})$ 
                     & 196.(1)& -231.(1)& 1.8 & -15.4 &  -2.0 &  7.1 &  2.6 \\
\nuc{8}{He}$(0^+;2)$ & 208.(1)& -235.(1)& 0.9 & -17.1 &  -6.8 &  9.0 &  3.6 \\
\nuc{8}{Li}$(2^+;1)$ & 236.(2)& -274.(2)& 2.0 & -19.0 &  -4.7 &  9.4 &  3.7 \\
\nuc{8}{Be}$(0^+;0)$ & 238.(2)& -290.(2)& 3.2 & -20.1 &  -1.4 &  8.8 &  3.3 \\
\nuc{9}{Li}$(\frac{3}{2}^-;\frac{3}{2})$ 
                     & 283.(1)& -322.(1)& 2.1 & -25.1 & -10.3 & 13.6 &  5.9 \\
\nuc{9}{Be}$(\frac{3}{2}^-;\frac{1}{2})$ 
                     & 282.(2)& -336.(2)& 3.5 & -25.0 &  -4.7 & 11.9 &  4.9 \\
\nuc{10}{Be}$(0^+;1)$& 331.(2)& -391.(1)& 3.7 & -31.1 &  -8.3 & 15.7 &  6.6 \\
\nuc{10}{B}$(3^+;0)$ & 339.(2)& -405.(2)& 5.7 & -32.7 &  -8.8 & 16.0 &  6.9 \\
\nuc{12}{C}$(0^+;0)$ & 437.(3)& -534.(2)& 8.3 & -45.0 & -14.1 & 23.9 & 10.9 \\
\end{tabular}
\label{tab:breakdown}
\end{ruledtabular}
\end{table}

In describing the structure of the light nuclei, it is convenient to 
characterize specific $J^\pi;T$ states by their dominant orbital and spin 
angular momentum and spatial symmetry $^{2S+1}L_J[n]$
where $[n]$ denotes the Young diagram for spatial symmetry \cite{wiringa2006}.
(This classification is essentially a modern update of the discussion in
\citet{feenberg1937}.)
For example, $^4$He is a $^1S_0$[4] state, and the ground state of $^6$Li
is predominantly $^3S_1$[42], with admixtures of $^3D_1$[42] and
$^1P_1$[411].
Because $N\!N$ forces are strongly attractive in relative $S$-waves, and
repulsive in $P$-waves, ground states of given $J^\pi;T$ have the maximum 
spatial symmetry allowed by the Pauli exclusion principle.
For the same spatial symmetry, states of higher $L$ are higher in the spectrum.
Further, due to the effect of $N\!N$ spin-orbit forces, iterated tensor forces
and also $3N$ forces, the spin doublets, triplets, etc., are split, with the 
maximum $J$ value for given $[n]$ lying lowest in the spectrum (up to mid
$p$-shell).
These features are evident in the excitation spectra discussed next.

The excitations relative to the ground state energies for many states 
are shown in Fig.~\ref{fig:excited} and tabulated in Table~\ref{tab:excited}.
These excitation energies are each the difference of two independent GFMC
calculations; the quoted statistical errors are the uncorrelated combination
of the errors of each calculation.
In general, the excitation energies are quite satisfactory with an rms error 
of 0.5 MeV  for 58 $A \leq 10$ states using AV18+IL7 compared to 1.8 MeV
using just AV18.
Thus we see that AV18 alone does a much better job on excitation energies
than it does for absolute binding, and that the addition of IL7 greatly
improves both aspects.

The $^6$He ground state is a $^1S_0$[42] combination, with a $^1D_2$[42] 
first excited state; the AV18+IL7 Hamiltonian gets an excitation in fair
agreement with experiment.
The first three $T=0$ excited states in $^6$Li constitute a $^3D_J$[42] triplet,
and the spin-orbit splitting between the $3^+$, $2^+$, and $1^+$ states 
is also reproduced very nicely.
The first two states in $^7$Li are a narrowly split $^2P_J$[43] pair,
while the next two are a $^2F_J$[43] pair, followed by the lowest
member of a $^4P_J$[421] triplet, all with a reasonably good reproduction
of experiment.
The $^8$Be nucleus exhibits a strong $2\alpha$ rotational spectrum, with 
a $^1S_0$[44] ground state and widely spaced $^1D_2$[44] and $^1G_4$[44] 
excited states, also with excitation energies in excellent agreement with
experiment.
Above this rotational band are $^3P_2$[431], $^3P_1$[431], and 
$^3D_3$[431] $T=0$ states that isospin mix with the $T=1$ isobaric analogs of
the $^8$Li ground and first two excited states.

The $A=10$ nuclei, which are mid $p$-shell nuclei, have the interesting 
feature of having two linearly independent ways of constructing 
$^{2S+1}D_J$[442] states.
In $^{10}$Be, the ground state is $^1S_0$[442] (much like $^6$He with an
added $\alpha$) followed by two $^1D_2$[442] excited states.
In $^{10}$B, the lowest state might be expected to be a $^3S_1$[442] state
similar to $^6$Li ground state plus an $\alpha$, but there are also two
$^3D_J$[442] triplets, one of which is so widely split by the effective 
one-body spin-orbit force that one $^3D_3$[442] component becomes the ground
state leaving the $^3S_1$[442] state as the first excited state
\cite{kurath1979}.

The IL7 $3N$ force plays a key role in getting these spin-orbit splittings
correctly.
The AV18 $N\!N$ force alone splits the $^6$Li $^3D_J$[42] states in the 
correct order, but with insufficient spacing.
It leaves the $^7$Li $^2P_J$[43] doublet degenerate, as well as the two 
$^1D_2$[442] states in $^{10}$Be, and the $^3S_1$[442] state in
$^{10}$B is predicted to be the ground state.
IL7 not only splits the two $2^+$ states in $^{10}$Be by about the correct
amount, but splits them in the correct direction, making the predicted $E2$ 
transitions to the ground state significantly different in size as 
experimentally observed \cite{mccutchan2012}.
By increasing the splitting of the $^3D_J$[442] states in $^{10}$B, IL7
also gives the correct $3^+$ ground state for $^{10}$Be.
Addition of the older Urbana $3N$ potentials fixes some, but not all of
these problems.
The superior behavior of the Illinois $3N$ interactions is also seen in $^5$He, i.e., $\alpha n$
scattering, as discussed in Sect.~\ref{sec:scat}.
The importance of $3N$ interactions is also observed in no-core shell model
calculations~\cite{navratil2007}.

\begin{center}
\begin{table}
\caption{GFMC excitation energies in MeV for the AV18+IL7 Hamiltonian 
compared to experiment~\cite{Tilley04} for selected $A\leq 12$ states;
those marked with a * are the empirical isospin-unmixed values.
See Table~\ref{tab:energies} for references.}
\label{tab:excited}
\begin{ruledtabular}
\begin{tabular}{l d{1} d{2} @{\hspace{18pt}} }
$^AZ(J^\pi;T)$ & \multicolumn{1}{c}{GFMC} & \multicolumn{1}{c}{Expt.} \\
\hline
\nuc{6}{He}$(2^+;1)$                       &  2.0(1) &  1.80 \\
\nuc{6}{Li}$(3^+;0)$                       &  2.3(1) &  2.19 \\
\nuc{6}{Li}$(2^+;0)$                       &  4.1(1) &  4.31 \\
\nuc{6}{Li}$(1^+;0)$                       &  5.4(1) &  5.37 \\
\nuc{7}{Li}$(\frac{1}{2}^-;\frac{1}{2})$   &  0.2(1) &  0.48 \\
\nuc{7}{Li}$(\frac{7}{2}^-;\frac{1}{2})$   &  5.0(1) &  4.65 \\  
\nuc{7}{Li}$(\frac{5}{2}^-;\frac{1}{2})$   &  6.6(2) &  6.60 \\
\nuc{7}{Li}$(\frac{5}{2}^-_2;\frac{1}{2})$ &  7.8(2) &  7.45 \\
\nuc{8}{He}$(2^+;2)$                       &  4.7(3) &  3.1(4) \\
\nuc{8}{Li}$(1^+;1)$                       &  1.4(3) &  0.98 \\
\nuc{8}{Li}$(3^+;1)$                       &  3.0(5) &  2.26 \\
\nuc{8}{Be}$(2^+;0)$                       &  3.2(2) &  3.03(1) \\
\nuc{8}{Be}$(4^+;0)$                       & 11.2(3) & 11.35(15) \\
\nuc{8}{Be}$(2^+_2;0)$                     & 16.8(2) & 16.75* \\
\nuc{8}{Be}$(1^+;0)$                       & 18.0(2) & 18.13* \\
\nuc{8}{Be}$(3^+;0)$                       & 19.9(2) & 19.21* \\
\nuc{9}{Li}$(\frac{1}{2}^-;\frac{3}{2})$   &  2.0(5) &  2.69 \\
\nuc{9}{Be}$(\frac{1}{2}^+;\frac{1}{2})$   &  1.5(3) &  1.68 \\
\nuc{9}{Be}$(\frac{5}{2}^-;\frac{1}{2})$   &  2.4(3) &  2.43 \\
\nuc{10}{Be}$(2^+;1)$                      &  3.4(3) &  3.37 \\
\nuc{10}{Be}$(2^+_2;1)$                    &  5.3(3) &  5.96 \\
\nuc{10}{B}$(1^+;0)$                       &  1.3(4) &  0.72 \\
\nuc{10}{B}$(1^+_2;0)$                     &  2.4(5) &  2.15 \\
\nuc{10}{B}$(2^+;0)$                       &  3.3(5) &  3.59 \\
\end{tabular}
\end{ruledtabular}
\end{table}
\end{center}

\subsection{Isospin breaking}

Energy differences among isobaric analog states are probes of the
charge-independence-breaking parts of the Hamiltonian.
The energies for a given isospin multiplet can be expanded as
\begin{equation}
   E_{A,T}(T_z) = \sum_{n\leq 2T} a_n(A,T) Q_n(T,T_z)
\end{equation}
where $Q_0 = 1$, $Q_1 = T_z$, $Q_2 = \frac{1}{2}(3T_z^2-T^2)$, and
$Q_3 = \frac{1}{2}(5T_z^3 - 3T^2 + T_Z)$
are orthogonal isospin polynomials~\cite{peshkin1960}.
GFMC calculations of the coefficients $a_n(A,T)$ for a number of isobaric sequences
and various contributions for the AV18+IL7 Hamiltonian are shown in
Table~\ref{tab:imme} along with the experimental values.
The contributions are the CSB component of the kinetic energy $K^{\rm CSB}$,
all electromagnetic interactions $v^{\gamma}$,
and the strong CIB interactions,  $v^{\rm CIB} = v^{\rm CSB}+v^{\rm CD}$.
The experimental values were computed using ground-state energies
from \cite{NNDC-data} and excitation energies from \cite{TUNL-data}.
By using the correlated GFMC propagations described in Sec.~\ref{sec:gfmc},
it is possible to extract statistically significant values for some of
the $a_3(A,T)$.
An additional contribution is the second-order perturbation correction
to the CI part of the Hamiltonian $\delta H^{CI}$ due to differences in
the wave functions. 
Although this term is small, it is the difference between two large energies
and has the greatest Monte Carlo statistical error of any of the contributions;
again correlated GFMC propagations make its extraction possible.

\begin{table}
\caption{GFMC isovector and isotensor energy coefficients $a_n(A,T)$
computed using AV18+IL7, in keV, compared to experiment~\cite{wiringa2013,Pieper:unpub}.
}
\begin{ruledtabular}
\begin{tabular}{lcrrcrr}
   \multicolumn{1}{c}{$a_n(A,T)$} & 
   \multicolumn{1}{c}{$K^{\rm CSB}$} &
   \multicolumn{1}{c}{$v^{\gamma}$} &
   \multicolumn{1}{c}{$v^{\rm CIB}$} &
   \multicolumn{1}{c}{$\delta H^{CI}$} &
   \multicolumn{1}{c}{Total} &
   \multicolumn{1}{c}{Expt.} \\
\hline
$a_1(3,\case{1}{2})$ & 14 &  670(1) & ~~65(0)~ &  8(1) &  755(1)~ &  764 \\
$a_1(6,1)$           & 18 & 1056(1) &  44(0)   & 68(3) & 1184(4)~ & 1174 \\
$a_1(7,\case{1}{2})$ & 23 & 1478(2) & ~~83(1)~ & 27(10)& 1611(10) & 1644 \\
$a_1(7,\case{3}{2})$ & 17 & 1206(1) & 45~~~~  & 85(4) & 1358(3)~ & 1326 \\
$a_1(8,1)$           & 25 & 1675(1) &   77~~~~  & 43(6) & 1813(6)~ & 1770 \\
$a_1(8,2)$           & 22 & 1557(1) &   63~~~~~ & 104(4)& 1735(3)~ & 1651 \\
$a_1(9,\case{1}{2})$ & 19 & 1713(6) & ~~55(1)~ &       & 1786(7)~ & 1851 \\
$a_1(9,\case{3}{2})$ & 26 & 1976(1) &   91(0)~ &   84(7)& 2176(7)~ & 2102 \\
$a_1(10,1)$          & 25 & 2155(7) & ~~85(1)~ &       & 2170(8)~ & 2329 \\
\hline                              
$a_2(6,1)$           &    &  153(1) & 112(2)   &  5(4) &  270(5)~ &  223 \\
$a_2(7,\case{3}{2})$ &    &  106(0) & ~~34(1)~ & 13(2) &  158(5)~ &  137 \\
$a_2(8,1)$           &    &  136(1) & $-$3(2)~ & 10(5) &  139(5)~ &  127 \\
$a_2(8,2)$           &    &  130(0) &   38(0)~ &  9(2) &  178(1)~ &  151 \\
$a_2(9,\case{3}{2})$ &    &  150(1) & 44(1)    &  4(5) &  200(4)~ &  176 \\
$a_2(10,1)$          &    &  178(1) & ~119(18) &       &  297(19) &  241 \\
\hline                              
$a_3(7,\case{3}{2})$ &    &$-$3(0)  &  0(0)    &  0(2) & $-$3(1) & $-$20(8) \\
$a_3(8,2)$           &    &$-$1(0)  &  0(0)    &$-$1(1)& $-$2(1) &  $-$3(1) \\
$a_3(9,\case{3}{2})$ &    &$-$1(1)  &  0(0)    &$-$0(4)& $-$1(3) &  $-$2(5) \\
\end{tabular}
\end{ruledtabular}
\label{tab:imme}
\end{table}
The dominant piece in all these terms is the Coulomb interaction
between protons, giving 85-95\% (70-100\%) of the experimental isovector 
(isotensor) total.  
However the strong CSB and CD interactions give important corrections,
and the other terms are not negligible.
In particular, the $v^{CSB}$ contribution is just the right size to fix the 
$^3$He -- $^3$H mass difference and is a strong constraint on the 
difference of $nn$ and $pp$ scattering lengths.
Overall, the isoscalar terms are in good agreement with experiment, while
the isotensor terms are perhaps a little too large.
One can understand the negative values of $a_3(A,T)$ as coming from the 
increasing Coulomb repulsion as $T_z$ increases; this expands the nucleus
and reduces $v_{C1}(pp)$.

Another place that CSB interactions play a role is in the isospin mixing
of nearby states with the same spin and parity but different isospins
\cite{wiringa2013}.
A classic case is the appearance in the $^8$Be excitation spectrum of
three pairs of states with $J^\pi$ of $2^+$ (at 16.6--16.9 MeV),
$1^+$ (at 17.6--18.2 MeV) and $3^+$ (at 19.0--19.2 MeV).
The unmixed states come from three $T=0$ states, including the second $2^+$
excitation and first $1^+$ and $3^+$ states in the $^8$Be spectrum and
three $T=1$ states that are the isobaric analogs of $^8$Li ground state
and its first two excited states.
These states have the same dominant [431] spatial symmetry, so it is not
surprising that their energies are closely paired.
The CSB components of the Hamiltonian have $\sim 100$ keV off-diagonal
(in isospin) matrix elements $H_{01}$ leading to significant isospin mixing..
Experimentally this is observed in the two-alpha decay of the
$2^+$ states, which have comparable widths and which can only go via the 
$T=0$ component of the wave functions.
The mixing of the $1^+$ doublet is apparent in their $M1$ decays
\cite{pastore2014}.

GFMC calculations of the isospin-mixing matrix elements are shown in 
Table~\ref{tab:imixing}.
The table includes a small contribution from class IV CSB terms $v^{\rm IV}$
that can connect $T=0$ and $T=1$ $np$ pairs \cite{henley1979}.
The theoretical total provides about 90\% of the inferred experimental
values in the $2^+$ and $1^+$ doublets, but is too large for the (poorly
determined) $3^+$ case.

\begin{table}
\caption{GFMC isospin mixing matrix elements $H_{01}$ in $^8$Be spin doublets
computed using AV18+IL7 (augmented by class IV CSB contributions) in keV, 
compared to experiment~\cite{wiringa2013}.}
\begin{ruledtabular}
\begin{tabular}{lccccrrr}
   \multicolumn{1}{c}{$H_{01}(J^\pi)$} & 
   \multicolumn{1}{c}{$K^{\rm CSB}$} &
   \multicolumn{1}{c}{$v^{\gamma}$} &
   \multicolumn{1}{c}{$v^{\rm CSB}$} &
   \multicolumn{1}{c}{$v^{\rm IV}$} &
   \multicolumn{1}{c}{Total} &
   \multicolumn{1}{c}{Expt.} \\
\hline
$H_{01}(2^+)$  & -4 & -99(1) & -23 & -2(1) & -128(2) & -145(3)~~\\
$H_{01}(1^+)$  & -3 & -74(1) & -19 & ~3(1) &  -93(2) & -103(14)\\
$H_{01}(3^+)$  & -3 & -87(1) & -17 & -6(2) & -113(3) &  -59(12)\\
\end{tabular}
\end{ruledtabular}
\label{tab:imixing}
\end{table}

\subsection{Densities}

The one- and two-nucleon density distributions of light nuclei
are interesting in a variety of experimental settings.
They are evaluated as the expectation values
\begin{align}
\rho_N(r) &= \frac{1}{4 \pi r^2}
\langle \Psi | \sum_i 
P_{N_i} \delta ( r - | {\bf r}_i - {\bf R}_{cm} | ) | \Psi \rangle \ , \\
\rho_{N\!N} (r) &= \frac{1}{4 \pi r^2}
\langle \Psi | \sum_{i<j} P_{N_i} P_{N_j} \delta ( r - |{\bf r}_i - {\bf r}_j|)|
\Psi \rangle \ ,
\end{align}
where $P_N$ is a proton or neutron projector.

\begin{figure*}
\includegraphics[scale=0.6]{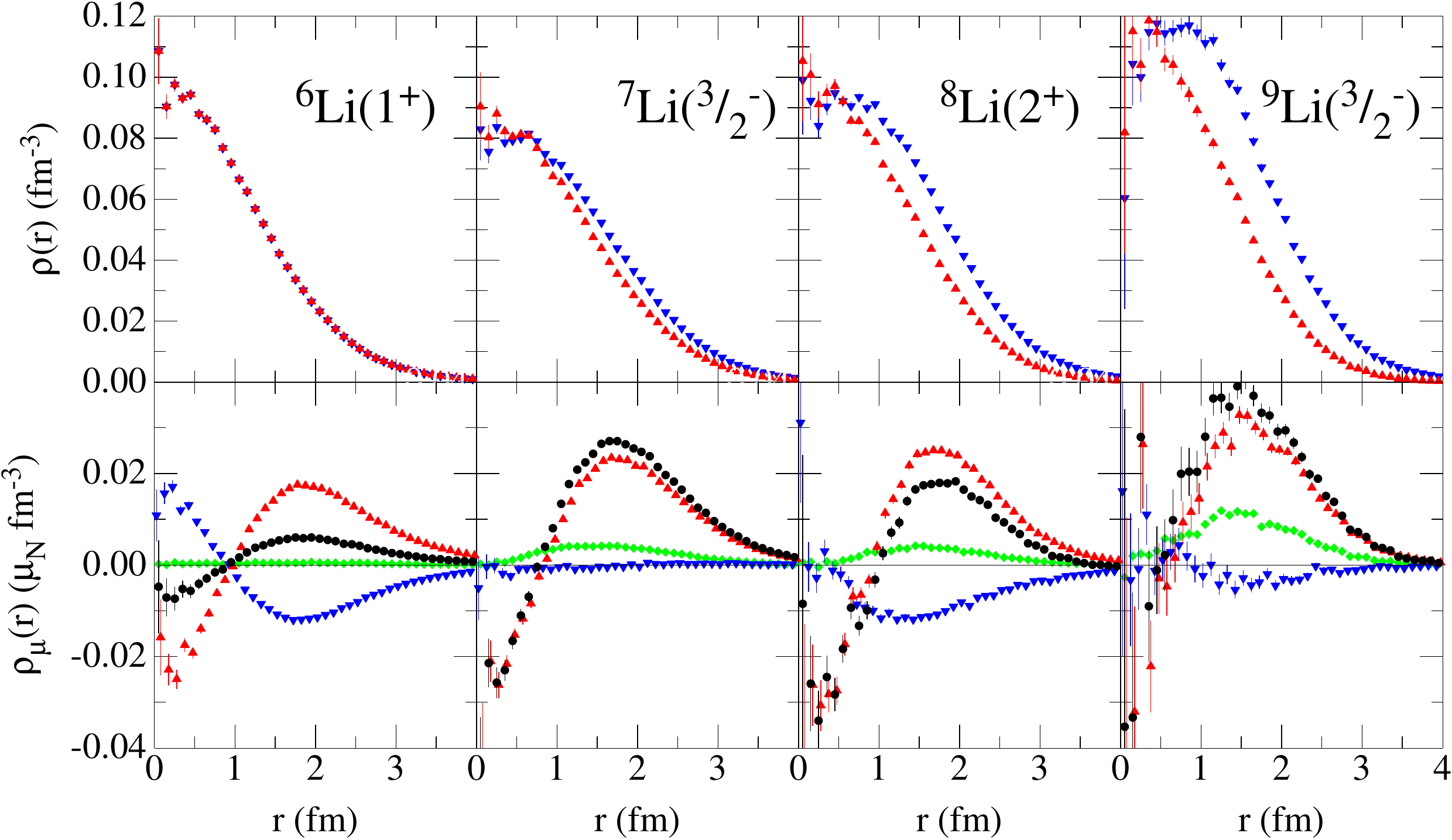}
\caption{GFMC point proton (red up triangles) and neutron (blue down triangles) 
densities (upper panel) and magnetic spin densities (lower panel) for the 
chain of lithium isotopes; also shown are proton magnetic orbital density 
(green diamonds), and total magnetic density in IA (black circles)
\cite{Wiringa:unpub}. }
\label{fig:density}
\end{figure*}

Ground state proton and neutron rms point 
radii are tabulated in Table~\ref{tab:energies}.
These can be related to the charge radii, which have been measured very 
accurately for the helium, lithium, and beryllium isotopic chains in
recent years by a combination of electron scattering from stable nuclei
and isotopic differences by atomic spectroscopy on rare isotopes.
A recent review~\cite{lu2013} discusses these developments
and the conversion between point and charge radii and presents
figures for the GFMC one- and two-body densities of the helium isotopes.

The proton and neutron one-body densities for the lithium
isotope chain are shown, as red up triangles and blue down 
triangles, respectively, in the upper panels of Fig.~\ref{fig:density}.
As the binding energy increases with $A$, the central proton 
density increases, even though the number of protons is constant.
Consequently, the proton point radius decreases by 0.4~fm in going from $^6$Li
to $^9$Li, in fair agreement with the experimentally observed reduction 
of 0.34~fm.
In contrast, the neutron point radius is relatively constant, even though
neutrons are being added, varying only 0.15~fm over the same range.

The magnetic moments of $A \leq 9$ nuclei have been calculated in GFMC
\cite{pervin2007,marcucci2008,pastore2013} including contributions from
two-body meson-exchange currents (MEC), as discussed in Sec.~\ref{sec:currents}.
The MEC can give 20--40\% contributions over the impulse approximation (IA)
values, resulting in very good agreement with experiment as shown in
Table~\ref{tab:energies}.

The origin of the IA contributions from the proton and neutron spin
densities and proton orbital density are illustrated in the bottom
panels of Fig.~\ref{fig:density}, also for the lithium isotope chain.
Here, the proton spin contribution 
$\mu_p[\rho_{p\uparrow}(r)\!-\!\rho_{p\downarrow}(r)]$ is shown by red
upward-pointing triangles, the neutron spin contribution by blue 
downward-pointing triangles, the proton orbital contribution by green diamonds,
and the total by black circles.
The proton spin density, due to one unpaired $p$-shell proton, is similar in 
all cases, with a negative region at short distance from the core and a 
positive peak near 2~fm that gradually shifts inward as the binding increases.
The neutron spin density has the opposite sign and alternates between a 
significant unpaired neutron contribution in $^{6,8}$Li and a very small 
paired contribution in $^{7,9}$Li.
The proton orbital piece gets progressively larger as $A$ increases.
The MEC contributions are discussed in more detail below, but come largely 
from pion exchange and are primarily isovector in character, ranging from 
2\% in $^6$Li to 10\% in $^7$Li and 20\% in $^{8,9}$Li.

\begin{figure}
\includegraphics[scale=0.33]{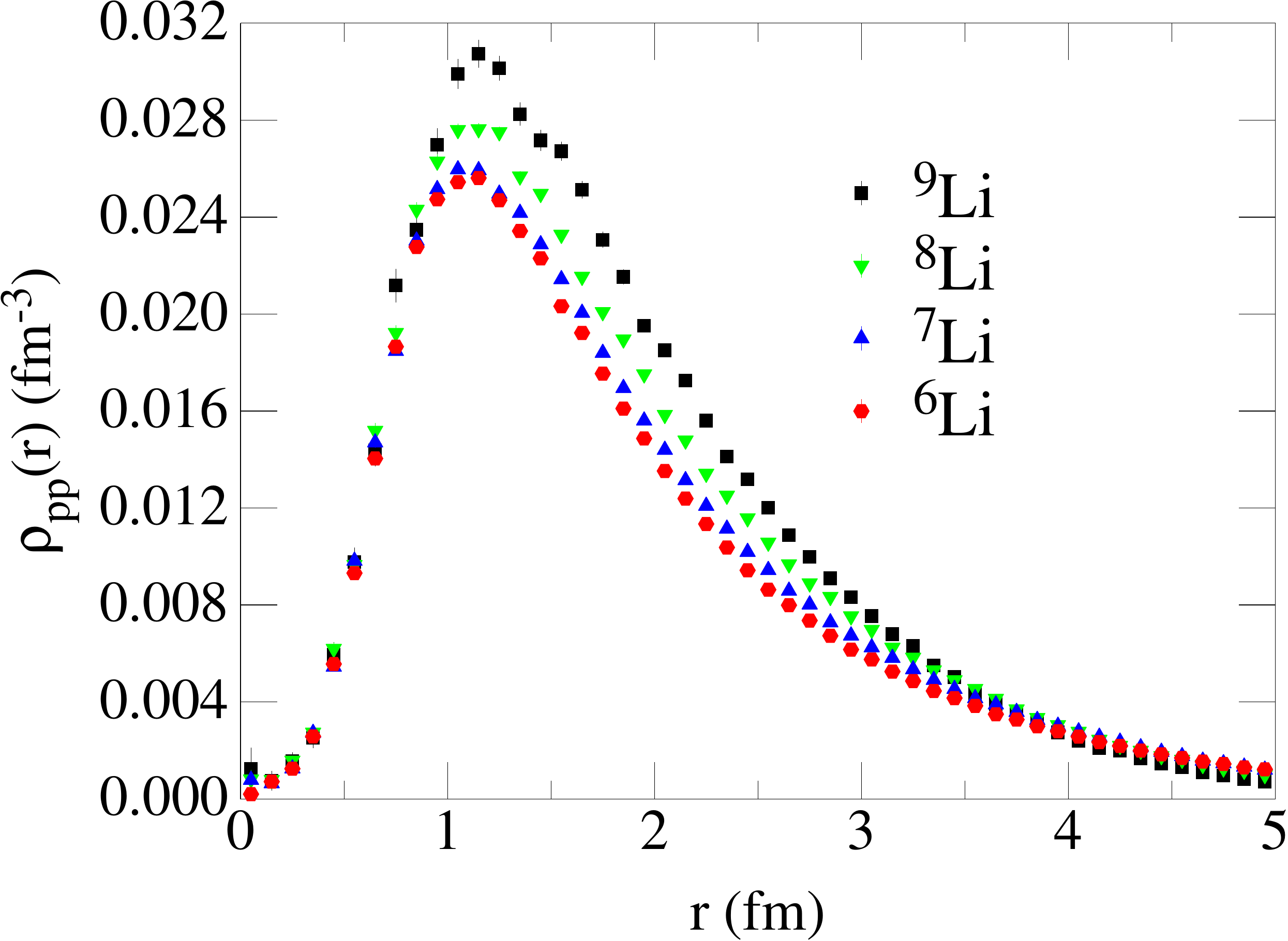}
\caption{GFMC $pp$ densities for the chain of lithium isotopes~\cite{Wiringa:unpub}.}
\label{fig:rhor_pp}
\end{figure}

The two-nucleon density for $pp$ pairs in the lithium isotopes is shown 
in Fig.~\ref{fig:rhor_pp} and all four curves integrate to three pairs.
Because the third proton is in the $p$-shell, the behavior of $\rho_{pp}(r)$
is rather different from the one $pp$ pair in the core of the helium 
isotopes shown in Fig.~12 of \citet{lu2013}.
In that case, there is a slight decrease in the peak value as $A$ increases
because the $p$-shell neutrons in $^{6,8}$He tug the core protons out a little.
In lithium the peak value of $\rho_{pp}(r)$ gets progressively larger with 
increasing $A$ due to the increasing binding, so the pair rms radius decreases 
from 4.03~fm in $^6$Li to 3.20~fm in $^9$Li.

\subsection{Momentum distributions}
\label{sec:momenta}

\begin{figure}
\includegraphics[scale=0.36]{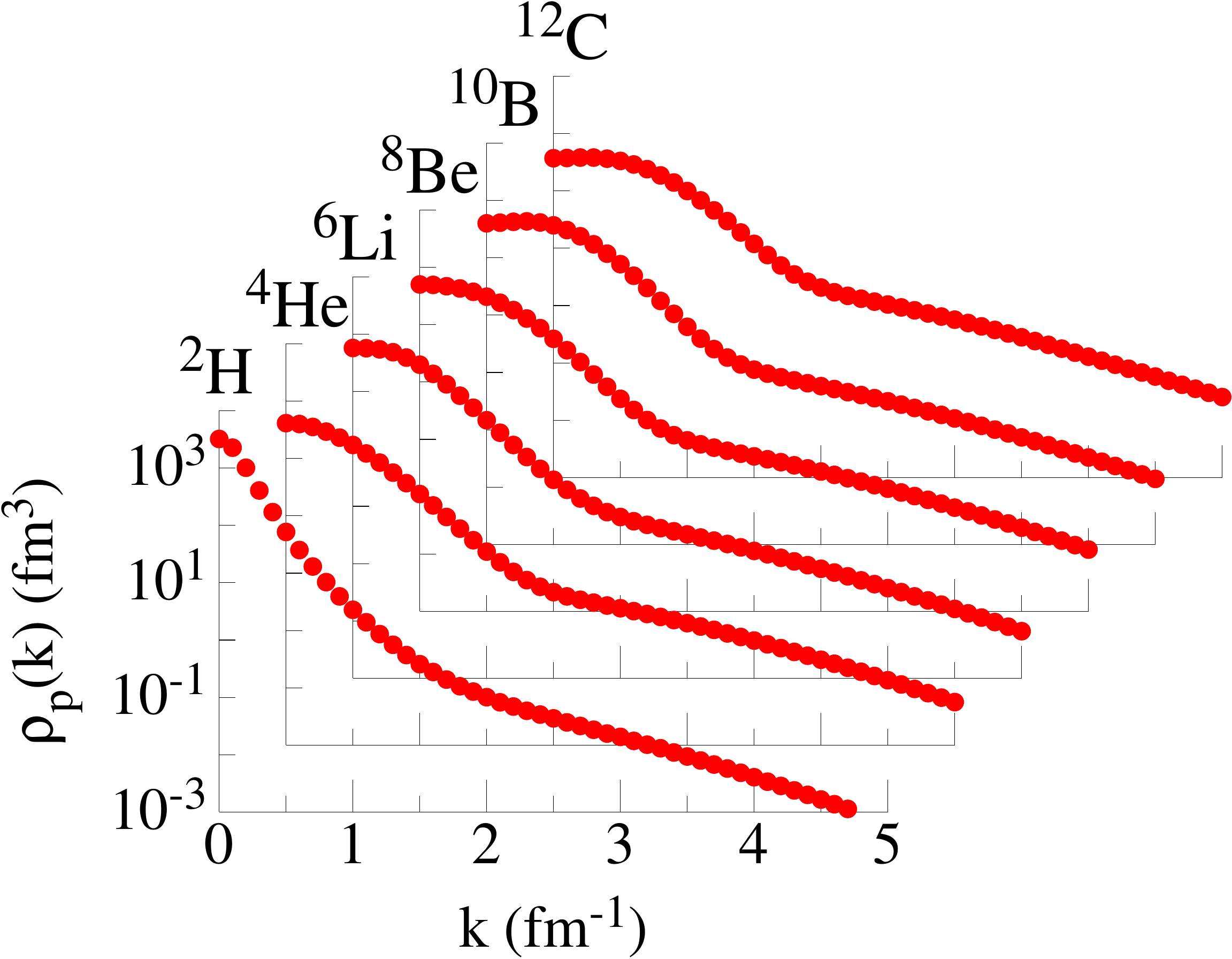}
\caption{VMC proton momentum distributions in $T=0$ light nuclei~\cite{wiringa2014}.}
\label{fig:rhokp}
\end{figure}

Momentum distributions of individual nucleons, nucleon pairs, and nucleon
clusters reflect features of the short-range structure of nuclei.
They can provide useful insight into various reactions on nuclei, such as
$(e,e^\prime p)$ and $(e,e^\prime pp/pn)$ electrodisintegration processes
or neutrino-nucleus interactions.

The probability of finding a nucleon with momentum $k$ and spin-isospin
projection $\sigma$,$\tau$ in a given nuclear state
is proportional to the density
\begin{align}
\label{eq:rhok}
\rho_{\sigma\tau}({\bf k})\!\!&=\!\!
\int d{\bf r}^\prime_1\, d{\bf r}_1\, d{\bf r}_2 \cdots d{\bf r}_A\,
\psi^\dagger_{JM_J}({\bf r}_1^\prime,{\bf r}_2, \dots,{\bf r}_A)\,  \nonumber \\
& \times \, e^{-i{\bf k}\cdot ({\bf r}_1-{\bf r}^\prime_1)}
\, P_{\sigma\tau}(1) \,
\psi_{JM_J} ({\bf r}_1,{\bf r}_2, \dots,{\bf r}_A) \, .
\end{align}
$P_{\sigma\tau}(i)$ is the spin-isospin projection operator for nucleon $i$,
and $\psi_{JM_J}$ is the nuclear wave function with total spin $J$ and
spin projection $M_J$.  The normalization is
\begin{equation}
\label{eq:nst}
  N_{\sigma \tau} =
    \int \frac{d{\bf k}}{(2\pi)^3}\,\,\rho_{\sigma\tau}({\bf k}) \ ,
\end{equation}
where $N_{\sigma \tau}$ is the number of spin-up or spin-down protons
or neutrons.

Early variational calculations of few-nucleon momentum distributions 
\cite{schiavilla1986} evaluated Eq.~(\ref{eq:rhok}) by following a Metropolis 
Monte Carlo walk in the $d{\bf r}_1\, d{\bf r}_2 \cdots d{\bf r}_A$ space 
and one extra Gaussian integration over $d{\bf r}^\prime_1$ at each Monte 
Carlo configuration.
This was subject to large statistical errors originating from the rapidly
oscillating nature of the integrand for large values of $k$.

\begin{figure}
\includegraphics[scale=0.36]{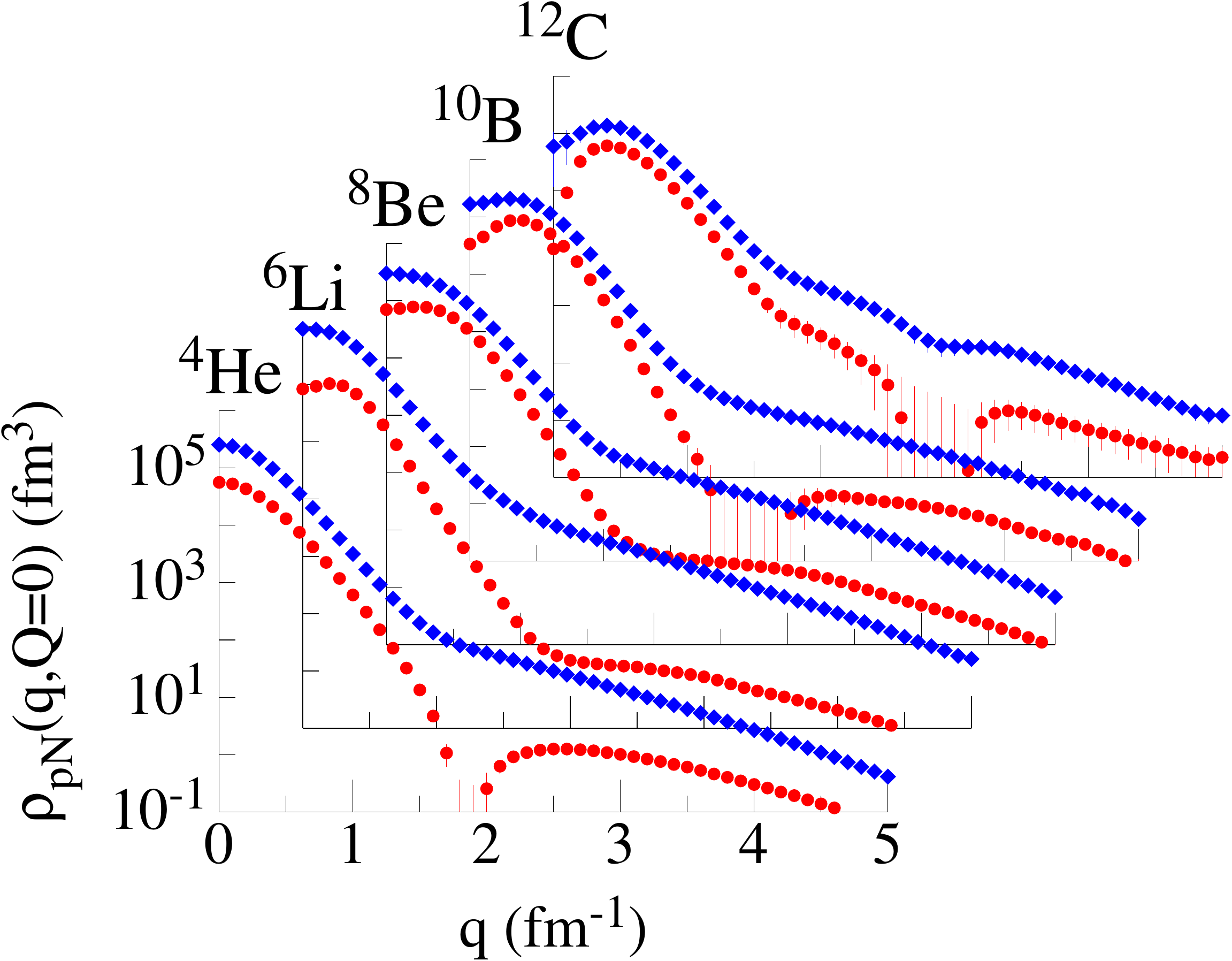}
\caption{VMC $pn$ (blue diamonds) and $pp$ (red circles) back-to-back ($Q=0$)
pair momentum distributions for $T=0$ nuclei~\cite{wiringa2014}.}
\label{fig:rhokqQ0}
\end{figure}

A more efficient method is to rewrite Eq.~(\ref{eq:rhok}) as 
\begin{align}
\label{eq:actual}
  \rho_{\sigma\tau}({\bf k}) &= \frac{1}{A}\sum_i
  \int d{\bf r}_1 \cdots d{\bf r}_i \cdots d{\bf r}_A \int d{\bf \Omega}_x
  \int_0^{x_{\rm max}} x^2 dx \nonumber \\
  & \times \psi^\dagger_{JM_J}
  ({\bf r}_1,\dots,{\bf r}_i+{\bf x}/2,\dots,{\bf r}_A) \,
  e^{-i{\bf k}\cdot {\bf x} }  \nonumber \\
  &\times \,  P_{\sigma\tau}(i) \,
  \psi_{JM_J}
  ({\bf r}_1,\dots,{\bf r}_i-{\bf x}/2,\dots,{\bf r}_A) \, .
\end{align}
and perform the Gaussian integration over ${\bf x}$.
However, this requires re-evaluating both initial and final wave functions 
at multiple configurations, which limits the present calculations to VMC.
A comprehensive set of single-nucleon momentum distributions for
$A\leq 12$ nuclei, evaluated with the AV18+UX Hamiltonian, has been 
published \cite{wiringa2014} with figures
and tables available on-line \cite{webmomenta1}.

The overall evolution of the proton momentum distribution in light $T=0$
nuclei is shown in Fig.~\ref{fig:rhokp}.
The shape of the distributions shows a smooth progression as nucleons are
added.  
As $A$ increases, the nuclei become more tightly bound, and the fraction
of nucleons at zero momentum decreases.
As nucleons are added to the $p$-shell, the distribution at low momenta
becomes broader, and develops a peak at finite $k$.
The sharp change in slope near $k=2$ fm$^{-1}$ to a broad shoulder is
present in all these nuclei and is attributable to the strong tensor
correlation induced by the pion-exchange part of the $N\!N$ potential,
further increased by the two-pion-exchange part of the $3N$ potential.
Above $k=4$ fm$^{-1}$, the bulk of the momentum density appears to come from
short-range spin-isospin correlations.

Two-nucleon momentum distributions, i.e., the probability of finding
two nucleons in a nucleus with relative momentum
${\bf q}=({\bf k}_1-{\bf k}_2)/2$ and total center-of-mass momentum
${\bf Q}={\bf k}_1+{\bf k}_2$, provide insight into the short-range
correlations induced by a given Hamiltonian.
They can be formulated analogously to Eqs.~(\ref{eq:rhok},\ref{eq:actual}),
and projected with total pair spin-isospin $ST$, or as $pp$, $np$, and 
$nn$ pairs.
Again, a large collection of VMC results has been published \cite{wiringa2014}
and figures and tables are available on-line \cite{webmomenta2}.

Experiments to search for evidence of short-range correlations have
been a recent focus of activity at Jefferson Laboratory.
In an $(e,e^\prime pN)$ experiment on $^{12}$C at JLab, a very large ratio 
$\sim 20$ of $pn$ to $pp$ pairs was observed at momenta $q$=1.5--2.5~fm$^{-1}$
for back-to-back ($Q=0$) pairs~\cite{Subedi:2008}.
VMC calculations for $\rho_{pN}(q,Q=0)$ are shown in Fig.~\ref{fig:rhokqQ0}
as blue diamonds for $pn$ pairs and red circles for $pp$ pairs for $T=0$
nuclei from $^4$He to $^{12}$C \cite{schiavilla2007,wiringa2014}.
The $pp$ back-to-back pairs are primarily in $^1S_0$ states and have
a node near 2~fm$^{-1}$, while the $pn$ pairs are in deuteron-like
$^3S_1-^3D_1$ states where the $D$-wave fills in the $S$-wave node.
Consequently, there is a large ratio of $pn$ to $pp$ pairs in this region.
This behavior is predicted to be universal across a wide range of nuclei.

\begin{figure}
\includegraphics[scale=0.36]{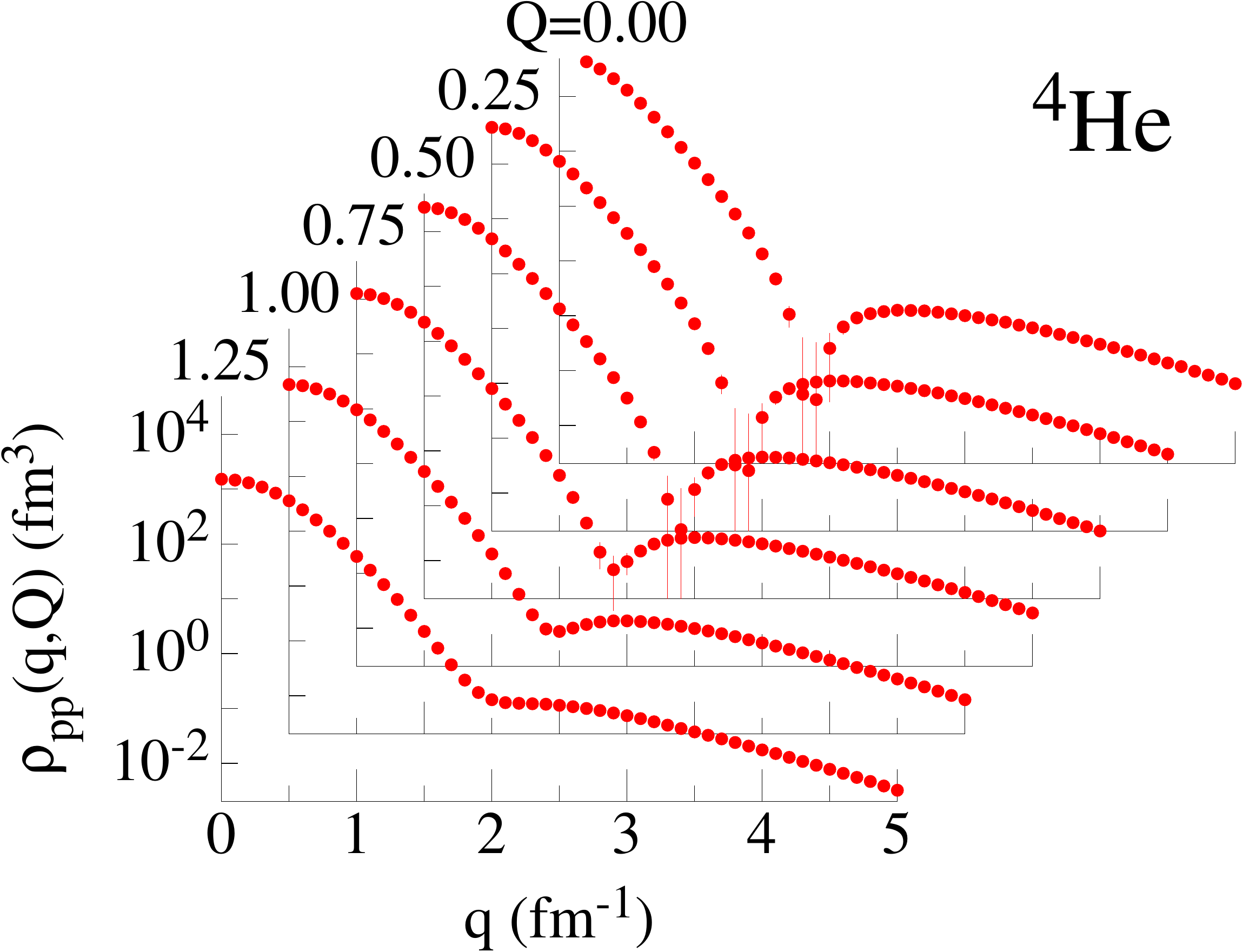}
\caption{VMC proton-proton momentum distributions in $^4$He averaged over 
the directions of ${\bf q}$ and ${\bf Q}$ as a function of $q$ for several 
fixed values of $Q$ from 0 to 1.25 fm$^{-1}$~\cite{wiringa2014}.}
\label{fig:rhokpp}
\end{figure}

\begin{figure}
\includegraphics[scale=0.36]{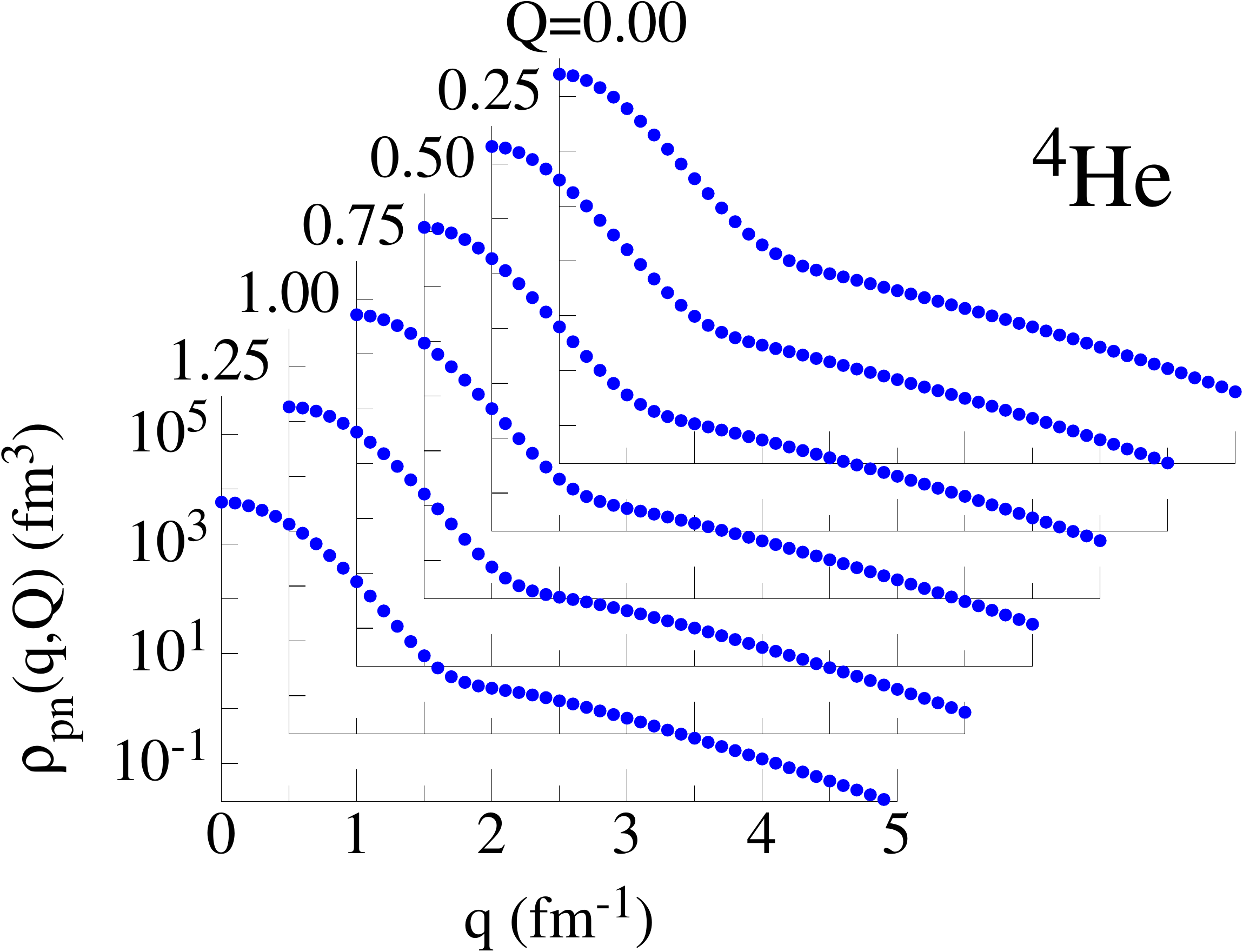}
\caption{VMC proton-neutron momentum distributions in $^4$He averaged over 
the directions of ${\bf q}$ and ${\bf Q}$ as a function of $q$ for several 
fixed values of $Q$ from 0 to 1.25 fm$^{-1}$~\cite{wiringa2014}.}
\label{fig:rhokpn}
\end{figure}

As $Q$ increases, the $S$-wave node in $pp$ pairs will gradually fill in,
as illustrated for $^4$He in Fig.~\ref{fig:rhokpp}, where $\rho_{pp}(q,Q)$ 
is shown as a function of $q$ for several fixed values of $Q$,
averaged over all directions of ${\bf q}$ and ${\bf Q}$.
In contrast, the deuteron-like distribution in $pn$ pairs is maintained
as $Q$ increases, as shown in Fig.~\ref{fig:rhokpn}, with only a gradual 
decrease in magnitude because there are fewer pairs at high total $Q$.
Recently, these momentum distributions for $^4$He have been tested in
new JLab experiments and found to predict the ratio of $pp$ to $pn$
pairs at higher missing momentum very well \cite{korover2014}.

\subsection{Spectroscopic overlaps, spectroscopic factors, and ANCs}

Determining the influence of nuclear structure on nuclear reactions is a 
challenging subject.
One source of theoretical input is the calculation of spectroscopic overlaps,
spectroscopic factors (SFs), and asymptotic normalization coefficients (ANCs).
They are steps on the way to calculating reaction cross sections in direct 
nuclear reactions, like nucleon knockout or radiative capture.

A one-nucleon spectroscopic overlap is the expectation value of the nucleon
removal operator between states of nuclei differing by one particle.
It can be written as
\begin{eqnarray}
\label{eq:overlap}
   R(\beta,\gamma,\nu;r) &=& \sqrt{A} \left\langle \left[
\Psi_{A-1}(\gamma) \otimes \mathcal{Y}(\nu;r_{Cv}) \right]_{J_A,T_A}
\right. \nonumber \\
  &&  \left. \left| \frac{ \delta( r - r_{Cv}) } { r_{Cv}^2 } \right|
      \Psi_A(\beta) \right\rangle
\end{eqnarray}
where $\beta \equiv \{A,J_A^{\pi},T_A,T_{z,A} \}$ denotes the quantum 
numbers of a parent $A$-body nucleus, 
$\gamma \equiv \{C,J_C^{\pi},T_C,T_{z,C} \}$ specifies an $(A-1)$-body core,
and $\nu \equiv \{v,l,s, j,t,t_z\}$ specifies the valence nucleon.
Here $r_{Cv}$ is the distance between the valence nucleon and the center
of mass of the core, and 
$\mathcal{Y}(\nu;r) \equiv \left[ Y_l(\hat{r}) \otimes \chi_s (\sigma_v) \right]_j \chi_{t,t_z}(\tau_v)$ 
is the valence angle-spin-isospin function.
The SF is then defined  as the norm of the overlap:
\begin{equation}
\label{eq:SF}
  S(\beta,\gamma,\nu) = \int \left| R(\beta,\gamma,\nu;r) \right|^2 
  r^2 \mathrm{d}r.
\end{equation}
In standard shell model calculations \cite{cohen1967}, the SFs obey 
various sum rules \cite{macfarlane1960}, including that for a given
state of the parent nucleus, the SFs to all possible final states of
the core plus valence nucleon add up to the parent's number of such
nucleons.  For example, 
$\sum_{\gamma,\nu}\langle ^6{\rm He}(\gamma)+p(\nu)|^7{\rm Li}(\beta)\rangle = 1$
because $^7$Li has one $p$-shell proton.

Overlap functions $R(r)$ satisfy a one-body Schr\"{o}dinger equation with
appropriate source terms \cite{Pinkston1965}. 
Asymptotically, at $r \rightarrow \infty$, these source terms contain 
core-valence Coulomb interaction at most, and hence the long-range part of 
overlap functions for parent states below core-valence separation thresholds 
is proportional to a Whittaker function $W_{-\eta,l+1/2}$:
\begin{equation}
\label{eq:ANC}
   R(\beta,\gamma,\nu;r) \xrightarrow{r \rightarrow \infty}
   C(\beta,\gamma,\nu) \frac{ W_{-\eta,l+1/2} (2kr) }{r},
\end{equation}
where $\eta = Z_C Z_\nu \alpha \sqrt{ \mu c^2 / 2|B|}$ depends on 
proton numbers $Z_C$ and $Z_\nu$, the fine-structure constant $\alpha$, 
and the core-valence reduced mass $\mu$ and the separation energy $B$ 
(negative for parent states below core-valence separation thresholds). 
The wave number $k$ is defined as $\sqrt{2\mu |B|}/\hbar$, and $l$ is the 
orbital momentum in $\mathcal{Y}(\nu)$.
The proportionality constant $C(\beta,\gamma,\nu)$ in Eq.~(\ref{eq:ANC})
is the ANC.

VMC calculations of overlaps and SFs for $s$-shell nuclei were first
reported in \cite{schiavilla1986}, followed by calculations in various 
$p$-shell nuclei for application to $(e,e^\prime p)$ experiments 
\cite{lapikas1999}, transfer reactions like $(d,p)$ and $(d,^3{\rm He})$
\cite{wuosmaa2005,wuosmaa2008}, and single-neutron knockout reactions 
\cite{grinyer2011,grinyer2012}.
The first GFMC calculations for $A \leq 7$ nuclei were reported in 
\citet{brida2011}.
These are off-diagonal calculations, as in Eq.~(\ref{eq:transmix}), so the 
final GFMC result is extrapolated from two different mixed estimates, one 
where $\Psi(\tau)$ is propagated for the $A$-body nucleus and one where it
is propagated for the $(A-1)$-body nucleus.
A large collection of VMC and GFMC results can be found on-line
\cite{weboverlaps}.

\begin{figure}
\includegraphics[scale=0.30]{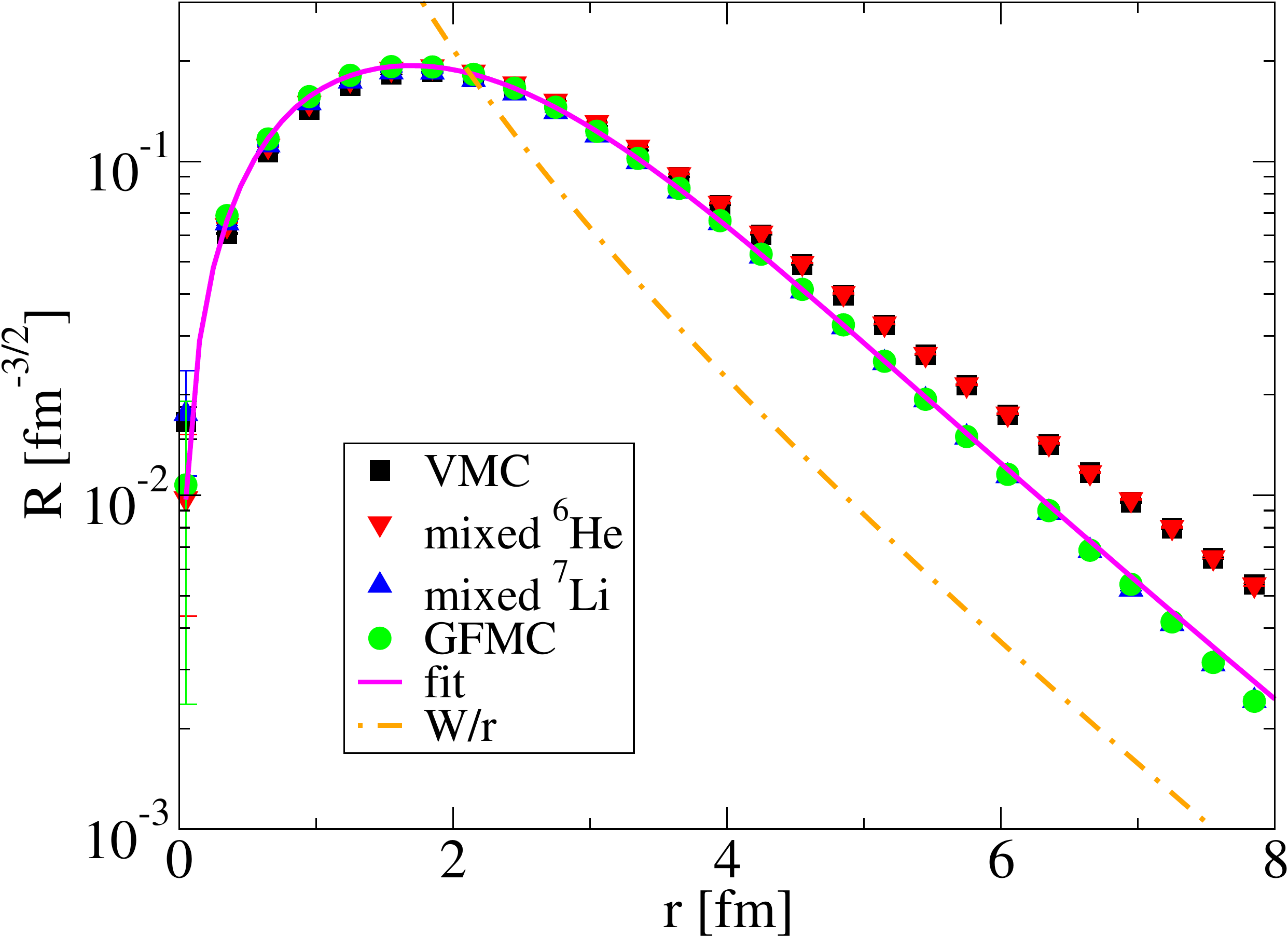}
\caption{VMC and GFMC calculations of the
$\langle ^6{\rm He}(0^+)+p(p_{3/2})|^7{\rm Li}(\frac{3}{2}^-)\rangle$ overlap~\cite{brida2011};
see text for details.}
\label{fig:overlap}
\end{figure}

For the $s$-shell nuclei, VMC energies and densities are very close to the
exact GFMC results, so VMC and GFMC overlaps $R(r)$ for cases like 
$\langle ^3{\rm H}+p(s_{1/2})|^4{\rm He}\rangle$
are in excellent agreement, both in the peak values at small $r$ and in
the asymptotic regime.
This translates into very similar SF and ANC predictions.
However, for $p$-shell nuclei, the VMC energies are progressively smaller in magnitude
relative to GFMC as $A$ increases, although the one-body densities
remain fairly close.
Consequently the overlaps have similar peak values but different asymptotic
behavior.

An example of $p$-shell overlap calculations is shown in Fig.~\ref{fig:overlap} 
for $\langle ^6{\rm He}(0^+)+p(p_{3/2})|^7{\rm Li}(\frac{3}{2}^-)\rangle$.
The VMC calculation is shown by black squares, the two GFMC mixed estimates
by red down (blue up) triangles for GFMC propagation of the $^6$He ($^7$Li) 
states, and the final GFMC result by green circles.
In this case, the VMC overlap and the GFMC mixed estimate when $^6$He is
propagated give virtually identical results, so the GFMC mixed estimate
when $^7$Li is propagated coincides with the final result.
The smooth fit to the GFMC result shown by the solid purple line is parallel at 
large $r$ to the Whittaker function $W/r$ (constructed with the experimental 
separation energy) shown by the dot-dash orange line.
The integrated VMC and GFMC SFs for this case are 0.44 and 0.41, respectively.
These values are consistent with experiment \cite{lapikas1999,wuosmaa2008}
but much smaller than the standard shell model value (corrected for 
center of mass) of 0.69 \cite{cohen1967}.

\begin{figure}
\includegraphics[scale=0.30]{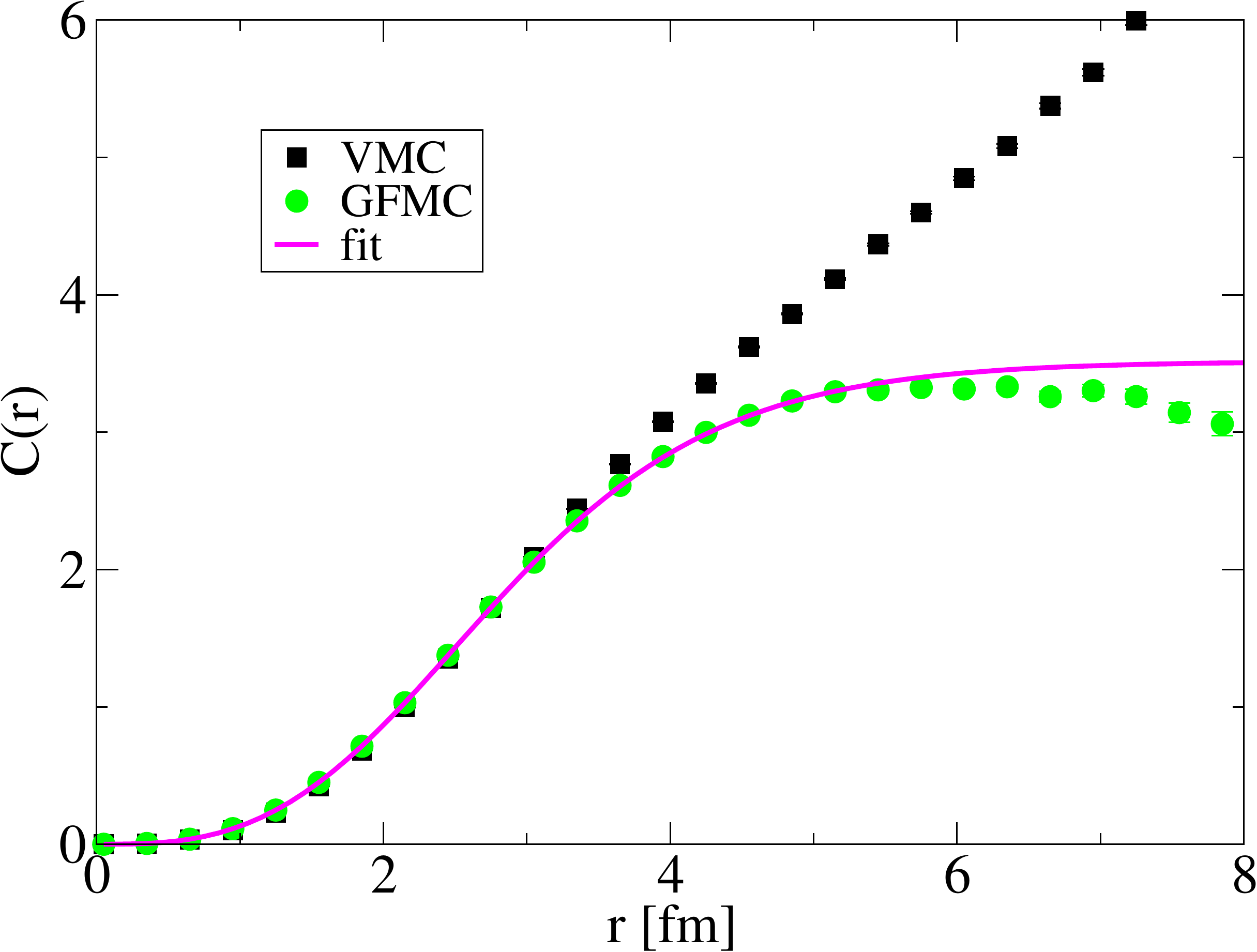}
\caption{The ratios $C(r)$ of the VMC and GFMC 
$\langle ^6{\rm He}(0^+)+p(p_{3/2})|^7{\rm Li}(\frac{3}{2}^-)\rangle$ overlaps
to the asymptotic Whittaker function~\cite{brida2011}; see text for details.}
\label{fig:anc}
\end{figure}

\begin{figure*}[t]
\includegraphics[scale=0.50,angle=270]{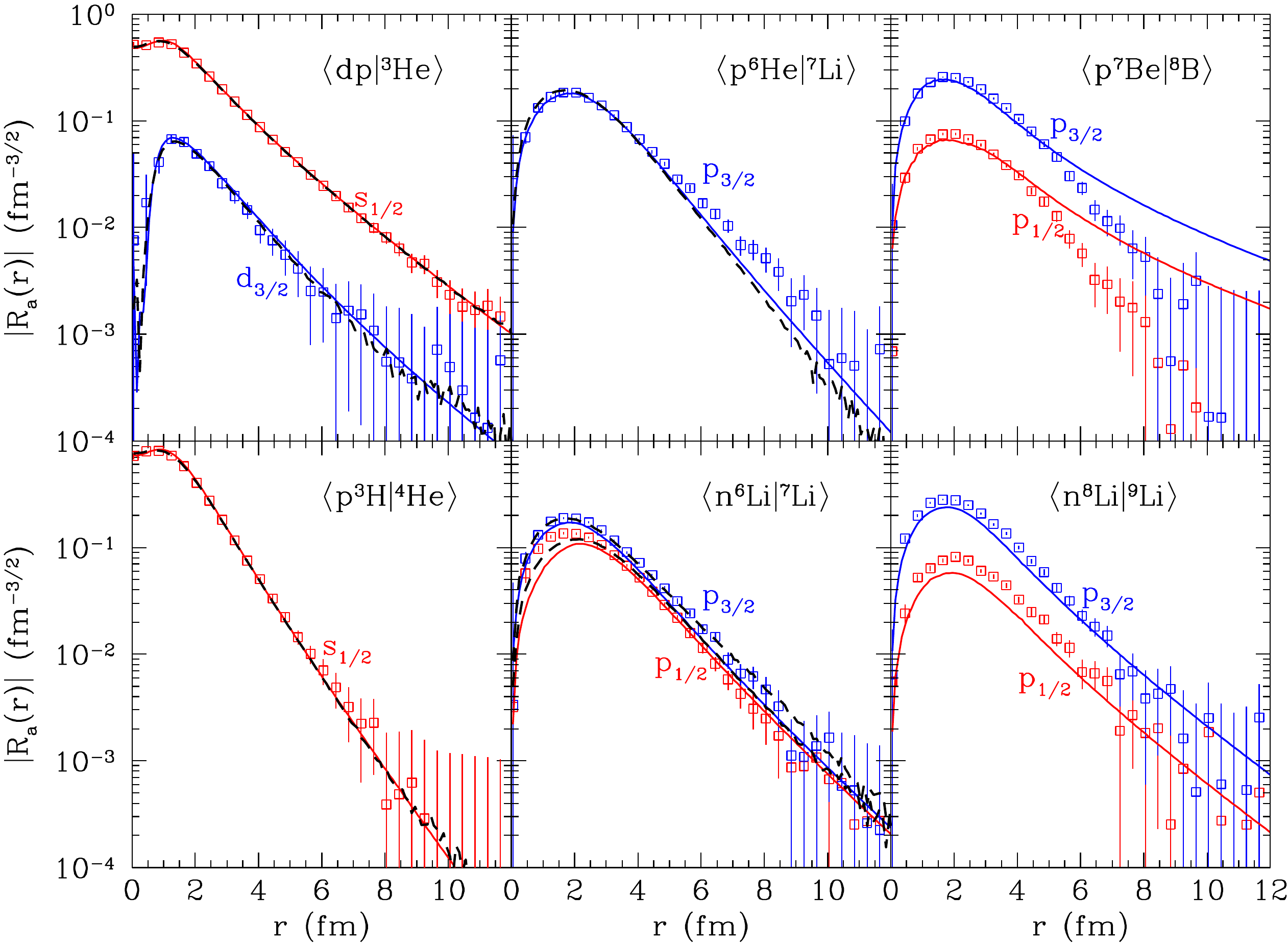}
\caption{Overlaps for various bound states as computed by 1) VMC sampling
(points with error bars), 2) a bound-state integral relation with the
VMC as input but imposing experimental separation energies (solid curves)
evaluated by \citet{nollett2012}, and 3), GFMC overlaps (dashed curves)
from \citet{brida2011}.}
\label{fig:unified}
\end{figure*}

In general, the SFs predicted by the VMC and GFMC calculations show a
significant quenching relative to standard shell model estimates which
are based on notions of independent particle motion.
The low-energy states of light nuclei can be interpreted as having
quasiparticles in single-particle orbitals \cite{pandharipande1997}.    
The difference between physical particles and quasiparticles is the 
consequence of the correlations in the system, which push a significant
fraction of nucleons above the nominal Fermi sea, as noted in the 
momentum distribution calculations of Sec.~\ref{sec:momenta}.
The SF is the quasihole strength, i.e., the probability of the
quasiparticle being a physical particle.
A variety of experiments find that, for a broad range of nuclei from 
$^4$He to $^{208}$Pb, SFs are quenched $\sim 0.5$ relative
to standard shell model, consistent with the VMC and GFMC calculations
\cite{kay2013}.

The ratio $C(r)$ of the VMC and GFMC overlaps with the Whittaker function 
constructed with the experimental separation energy are shown in 
Fig.~\ref{fig:anc}.
The incorrect asymptotic behavior of the VMC calculation means the $C(r)$ does
not reach a constant value and precludes extracting a reasonable ANC from
this ratio.
However the GFMC calculation, with its much better asymptotic behavior,
does go to a constant at large $r$, as indicated by the purple line fit.

There is an alternative method to obtain overlaps, ANCs, and estimates of
widths from variational wave functions
using integral constraints that are insensitive to their asymptotic behavior
\cite{Barletta:2009,nollett2011,nollett2012}.
As an example, the ANC is given by a sort of modified overlap integral with
a finite-range potential insertion:
\begin{eqnarray}
\label{eq:anc-integral}
 C(\beta,\gamma,\nu) &=&  \frac{2\mu}{k \hbar^2 w} \, \times 
 \mathcal{A} \int \frac{M_{-\eta ,l+\frac{1}{2}}(2kr)}{r} \\
&& { \Psi_{A-1}^\dag(\gamma) \mathcal{Y}^\dag(\nu;r) }
\left(U_\mathrm{rel}-V_C\right) \Psi_A(\beta) \, d^3r \, . \nonumber
\end{eqnarray}
The integral extends over all particle coordinates, $\mathcal{A}$ is an
antisymmetrization operator for the core and valence particle,
$M_{-\eta ,l+\frac{1}{2}}$
is the Whittaker function that is irregular at infinity, and $w$ is its
Wronskian with the regular Whittaker function $W_{-\eta ,l+\frac{1}{2}}$.
The $U_\mathrm{rel}$ is a sum of two- and three-body potentials involving 
the last nucleon
\begin{equation}
U_\mathrm{rel} = \sum_{i<A} v_{iA} + \sum_{i<j<A} V_{ijA}\,,
\end{equation}
where we have labeled the last nucleon $A$.
The point-Coulomb potential between the residual nucleus and last nucleon
is $V_C = Z_{A-1}Z_\nu \alpha \hbar c/r$ and in the limit of large separation,
typically $r > 7$~fm, $(U_\mathrm{rel}-V_C)$ vanishes.
This provides a natural cutoff to the integral of Eq.~(\ref{eq:anc-integral}).

This integral method has been implemented, using VMC wave functions obtained
for the AV18+UIX Hamiltonian, for 19 one-nucleon removals from nuclear
states with $3 \leq A \leq 9$.
Detailed tables are given in \citet{nollett2011}, as well as comparisons
to available experimental determinations and previous theoretical work.
In general, when the experimental binding energy $B_{\rm expt}$ is used in 
the wave number $k$, the ANCs derived from VMC wave functions through
Eq.~(\ref{eq:anc-integral}) are in excellent agreement with experiment.
The results also agree with the GFMC determinations discussed above at 
$\sim 10\%$ level, e.g., the GFMC ANC for 
$\langle ^6{\rm He}(0^+)+p(p_{3/2})|^7{\rm Li}(\frac{3}{2}^-)\rangle$
from Fig.~\ref{fig:anc} is 3.5, while the VMC integral value is 3.7.
Of particular note, the astrophysical $S$-factor for 
$^8{\rm B}\rightarrow p + ^7{\rm Be}$ is related to the ANCs by
$S_{17}(0) = [38.7~\rm eV~b~fm] \sum_j |C(2^+,\frac{3}{2}^-,j)|^2$
\cite{esbensen2004}.
Inserting the VMC ANC values gives the result 20.8~eV~b, which is
exactly the current recommended value from the Solar Fusion II
analysis \cite{adelberger2011}.

Relations similar to Eq.~(\ref{eq:anc-integral}) can be used to generate 
overlaps and also to estimate the widths of resonant states \cite{nollett2012}.
Examples of overlaps evaluated in this way are shown in Fig.~\ref{fig:unified},
where they are compared to the VMC input and the GFMC overlaps of
\citet{brida2011}.
Many widths in $5 \leq A \leq 9$ nuclei have also been evaluated, using as
input VMC pseudo-bound wave functions from the AV18+UIX Hamiltonian.
Detailed tables are given in \citet{nollett2012}.
The agreement with experiment is generally satisfactory when the physical
states are narrow, but the method fails for broad states; the overlaps
can help differentiate these cases.
For broad states, true scattering wave functions need to be developed, as
discussed below.

While the preceding discussion has focused on single-nucleon spectroscopic
overlaps, SFs, ANCs, and widths, the techniques involved are readily
adaptable to other cluster-cluster pairings, e.g., with deuterons or
$\alpha$s as the valence cluster.
Spectroscopic overlaps for $dd$ in $^4$He, $\alpha d$ in $^6$Li, and
$\alpha t$ in $^7$Li are included in the on-line overlap tabulations of
\citet{weboverlaps} and spectroscopic factors can be obtained from the 
cluster-cluster momentum distribution tables in \citet{webmomenta1}.
It should be possible in future to evaluate $\alpha$ ANCs and widths
from the VMC wave functions and generalized integral relations.

\subsection{Low-Energy Scattering}
\label{sec:scat}

   Quantum Monte Carlo methods can also be used to treat low-energy scattering
in nuclear systems~\cite{Carlson:1987,Nollett:2007}. The methods employed are similar to bound-state methods,
and are easily applicable at low energies where the combined system breaks
up into at most two clusters. 
One enforces one or more boundary conditions on the
asymptotic wave function at large cluster separations and then solves for the
energy levels with these boundary conditions.  
The resulting energies can be used with the boundary conditions to
determine the elements of the $S$-matrix for those energies.

   The simplest example is for a one-channel case with only elastic scattering,
for example $n-\alpha$ scattering. The asymptotic wave function for the
relative motion of the neutron and the alpha particle is given by:
\begin{equation}
\label{eq:asymptotic}
\Psi \propto \left\{\Phi_{c1}\Phi_{c2} Y_{L}\right\}_J
\left[\cos \delta_{JL} j_L(kr) 
- \sin\delta_{JL} n_L(kr)\right] \ ,
\end{equation}
where $\Phi_{c1}$ and $\Phi_{c2}$ are the internal wave functions of the
two clusters, $k$ and $r$ are the relative momentum and spatial separation
between the two clusters, and
$\delta_{JL}$ is the phase shift in the $JL$ partial wave. For problems
with Coulomb interactions between the clusters the relative wave function 
will contain Coulomb rather than Bessel functions. 

   The original QMC scattering calculations required the wave function
to be zero at a specified cluster separation~\cite{Carlson:1987}, while
in recent work the logarithmic derivative
$\gamma$ of the relative wave function at a boundary $r=R_0$ is specified~\cite{Nollett:2007}:
\begin{equation}
\gamma =  \frac{\nabla_r \Psi }{  \Psi } \Bigg |_{r=R_0}.
\end{equation}
In VMC calculations this is enforced within the form of the trial wave function,
which is required at large distances to go like Eq.~(\ref{eq:asymptotic}).
The radius $R_0$ should be large enough so that there is no strong interaction
between the clusters at that separation.
The scattering energy and hence the relative momentum between clusters is
unknown initially, but these are obtained by variationally solving for
states confined within the boundary $r=R_0$. Knowledge of the energy
and the boundary condition is then sufficient to determine the phase
shift at that energy.  The method for GFMC is very
similar, except that the logarithmic derivative of the wave function must
also be enforced in the propagator.  This can be incorporated through an
image method.  For each point $\bfR$ near the boundary $r=R_0$
reached during the random walk, the contribution to the internal 
wave function from points originally outside the boundary are added.
Consider an image at a cluster separation ${\bf r_e} = {\bf r} (R_0/r)^2$;
simple manipulations yields
\begin{align}
\Psi_{n+1} ({\bf R}^\prime)& =  \int_{|{\bf r}| < R_0}
                      d{\bf R}_{c1}\, d{\bf R}_{c2}\,   d{\bf r} 
           \ G ({\bf R}^\prime, {\bf R}) \\
& \times \left[ \Psi_n ({\bf R}) + \gamma \ 
     \frac{G ({\bf R}^\prime, {\bf R}_e)} {G ({\bf R}^\prime,{\bf R})} 
  \  \left( \frac{{ r}_e}{{ r}} \right)^3
  \  \Psi_n ({\bf R}_e) \right] \ , \nonumber
\end{align}
where ${\bf R}$ and ${\bf R'}$ are the initial and final points of all 
the particles, ${\bf R}_{c1} $ and ${\bf R}_{c2}$ are the internal
coordinates of the clusters, and ${\bf r}$ is the separation between
clusters. The image point for all the particles is denoted by ${\bf R}_e$,
and ${\bf r}_e$ is its cluster separation.
The image contribution ensures the correct logarithmic derivative of
the wave function at the boundary is preserved in the propagation.

   The $n-\alpha$ system is interesting as it is the lightest system
where $T= 3/2$ triplets play a significant role. QMC methods have been
used to study low-energy scattering in $n-\alpha$, including the two
low-lying $P$-wave resonances and $S$-wave scattering \cite{Nollett:2007}.
The spin-orbit splitting is especially interesting, as it can be examined
by comparing the $^3P_{1/2}$ and $^3P_{3/2}$ partial waves.

\begin{figure}[]
\includegraphics[width=3.25in,angle=0]{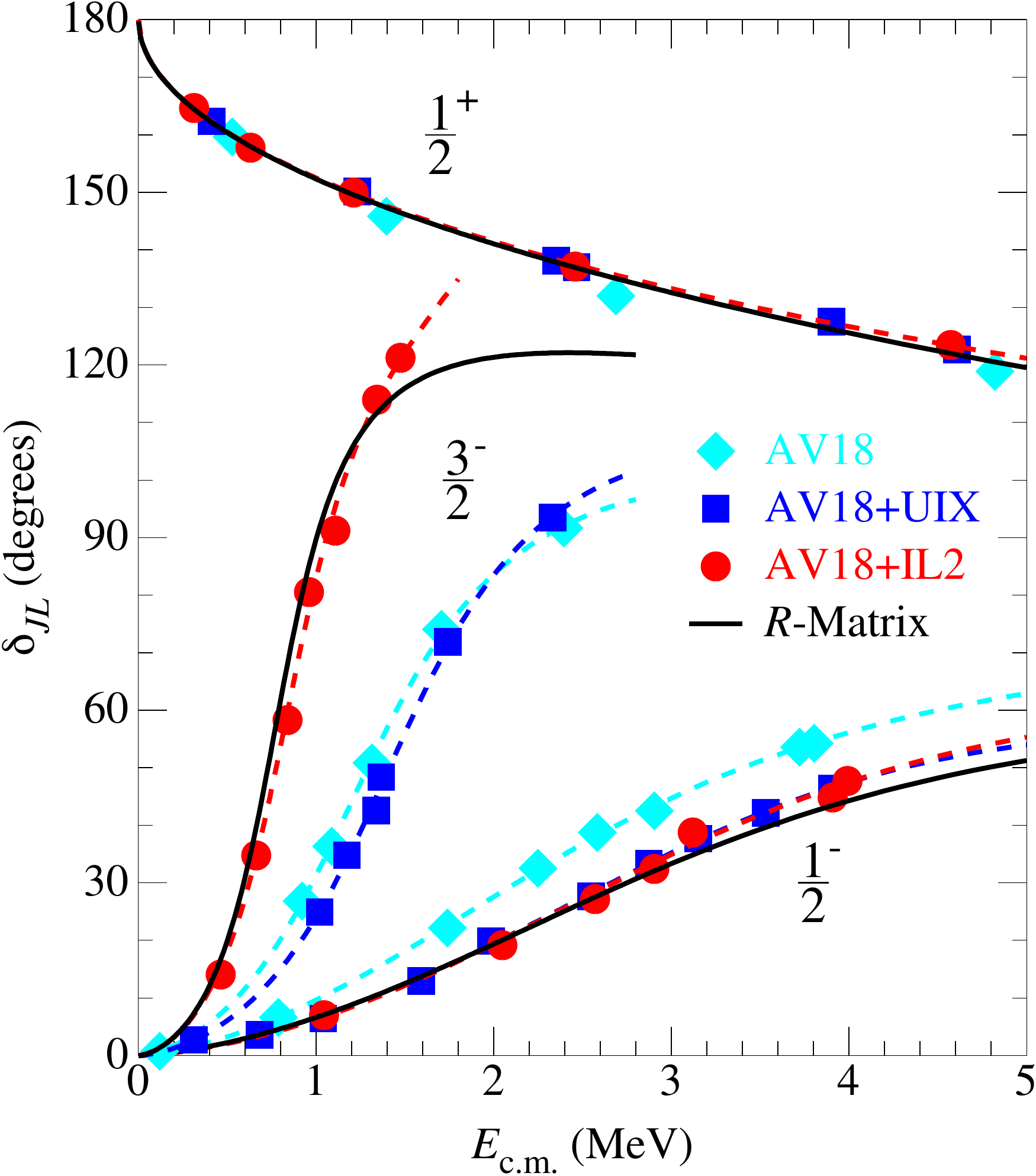}
\caption{(Color online) Phase shifts for $n$-$\alpha$ scattering.
Filled symbols (with statistical errors smaller
than the symbols) are GFMC results, dashed curves are polynomial fits, and 
solid curves are from an $R$-matrix fit to data \cite{Nollett:2007}.}
\label{fig:shifts}
\end{figure}

   The results of calculations with the AV18 $N\!N$ interaction and with
different $3N$ interactions 
are shown in Fig.~\ref{fig:shifts}.
The various calculations are also compared with an $R$-matrix analysis of the
experimental data. As is evident from the figure, the AV18 interaction alone
significantly underpredicts the spin-orbit splitting. The two-pion-exchange
in the UIX $3N$ interaction increases the splitting,
but not enough to agree with the experimental data. The IL2 model of the
$3N$ interaction results in good agreement with the experimental spin-orbit
splitting.

   These scattering methods have many possible applications. They can be
extended to inelastic multichannel processes in a fairly straightforward
manner. In this case there are multiple independent solutions for a given
scattering energy, hence one must study the energy as a function of the
boundary conditions in each channel and obtain multiple independent
solutions for the same energy. From the boundary conditions, the energy,
and the relative asymptotic magnitude of the wave functions, one can obtain
the full multichannel $S$-matrix.  It should be possible to treat a variety
of low-energy strong reactions, as well as electroweak transitions 
involving scattering states using these methods. In addition, hadronic parity violation in few-nucleon systems is an important application.

\subsection{Chiral Interactions}

Local $N\!N$ potentials derived within chiral effective field
theory have been used to calculate properties of A=3,4 nuclei with GFMC
by~\citet{Lynn:2014}.
Although the calculations do not yet include $3N$ interactions that also appear
at N$^2$LO, they are nevertheless interesting, showing the order-by-order
results for the binding energies and also the range of results
for different cutoffs. Also the question of perturbative treatments
of higher-order corrections has been investigated, as well as one- and
two-nucleon distributions.

\begin{figure}
\includegraphics[width=3.5in]{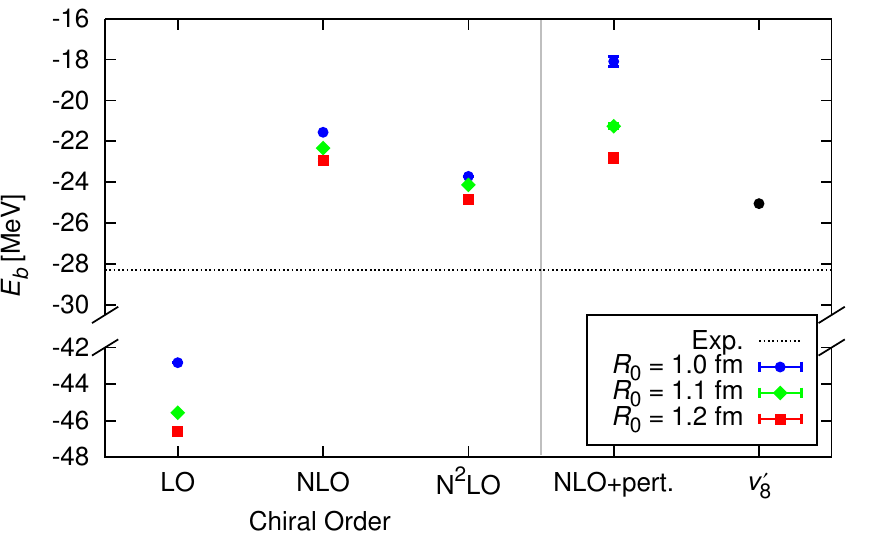}
\caption{
GFMC $^4$He binding energies at LO, NLO, and N$^2$LO compared with experiment 
(dashed line)
and with the Argonne AV8$'$ energy.
Also shown is a first-order perturbation-theory calculation of 
the N$^2$LO binding energy using the NLO wave function~\cite{Lynn:2014}.}
\label{fig:chiralspectrum}
\end{figure}

Figure~\ref{fig:chiralspectrum} shows results at various orders and
for different values of the cutoff $R_0$ used to regulate the small-$r$
behavior of the pion-exchange potentials, $f_{\rm long}=1-\exp[-(r/R_0)^4]$.
The LO result is extremely overbound, whereas the NLO and N$^2$LO results
are underbound as expected because of the lack of the $3N$ interaction.
The NLO interaction includes pion-exchange diagrams, and the N$^2$LO
two-pion exchange terms.
On the right, the column labeled `NLO+pert' shows the results for the
N$^2$LO binding energy using the NLO wave function plus the perturbative
contribution of the difference between the two interactions. The perturbative
treatment is reasonable, but the spread of energies is significantly larger
and of course the binding is less than in the full nonperturbative calculation.
The role of chiral $3N$ interactions in light nuclei and in matter 
are currently being investigated.

It should be noted that lattice  QMC approaches 
to the study of chiral interactions have been pursued~\cite{Lee:2004, Epelbaum:2011, Epelbaum:2012, Epelbaum:2014}.  
These methods have also been used to study,
for example, the Hoyle state in $^{12}$C and the ground state and 
excitations in $^{16}$O.  While the lattices used to
date are rather coarse, using a lattice spacing near 2 fm corresponding
to a maximum momenta of $\sim$ 1.5 fm$^{-1}$, they obtain
very good results for the energies of the Hoyle state and for other nuclei
with alpha particle substructure.
Comparisons for different forms of chiral interactions and for a variety
of observables could prove very valuable.

\section{Electroweak currents}
\label{sec:currents}

\subsection{Conventional nuclear electroweak currents} 
\label{subsec:2bcurrent}
A fundamental aspect in the description of electroweak processes
in nuclei is the construction of a realistic set of nuclear electroweak
currents.  The electromagnetic current is denoted by $j^\mu_\gamma$,
and the neutral and charge-changing weak currents
as $j^\mu_{NC}$ and $j^\mu_{CC}$, respectively.  In the Standard
Model of particle physics, the latter consist of polar-vector ($j^\mu_\gamma$ or $j^\mu$)
and axial-vector ($j^{\mu5}$) parts, and read
\begin{align}
\label{eq:jwnc}
j^\mu_{NC}&=-2\, {\rm sin}^2\theta_W\, j^\mu_{\gamma, S} 
+ (1-2\, {\rm sin}^2\theta_W) \, j^\mu_{\gamma, z} +\, j^{\mu5}_z \ , 
\nonumber \\
j^\mu_{CC}&=j^\mu_{\pm}+j^{\mu5}_{\pm} \, \qquad j_\pm = j_x \pm i\, j_y \ ,
\end{align}
where $\theta_W$ is the Weinberg angle (${\rm sin}^2\theta_W=0.2312$~\citet{PDG}),
$j^\mu_{\gamma,S}$ and $j^\mu_{\gamma,z}$ are, respectively, the isoscalar and
isovector pieces of the electromagnetic current, and the
subscript $b$ with $b$=$x$, $y$, or $z$ on $j^\mu_{\gamma,b}\,$,
$j^\mu_b$, and $j^{\mu5}_b$ denotes components in isospin space.  The
conserved-vector-current constraint relates the polar-vector components
$j^\mu_b$ of the charge-changing weak current to the isovector component
$j^\mu_{\gamma,z}$ of the electromagnetic current via
\begin{equation}
\left[ \, T_a \, , \, j^\mu_{\gamma,z} \, \right]=i\, \epsilon_{azb}\, j^\mu_b \ ,
\end{equation}
where $T_a$ are isospin operators, the implication being that
$(j^\mu_x,j^\mu_y,j^\mu_{\gamma,z})$ form a vector in isospin
space.  There are in principle isoscalar contributions
to $j^\mu_{NC}$ associated with strange quarks, but they are ignored in Eq.(\ref{eq:jwnc}),
since experiments at Bates~\cite{Spa00,Bei05} and Jefferson
Lab~\cite{Ahm12,Ani04,Ach07} have found them to be very small.

The leading terms in $j^\mu_\gamma$ and $j^\mu_{NC/CC}$ are expected to be those associated
with individual nucleons.  A single nucleon absorbs the momentum and
energy of the external electroweak field, and can later share
this momentum and energy with other nucleons via two- and three-body
interactions.  These interactions determine the final state of the nucleus, and
are not part of the current operator.  They are known as final state interactions in
approaches based on perturbation theory.  Interactions between nucleons
that take place before the absorption of the external field momentum and energy
are known as initial state interactions.  Nonperturbative approaches, such as those
discussed in this review, use eigenstates of the
nuclear Hamiltonian as initial and final states, and treat only the
interaction with the external field, described by the above currents, as a weak
perturbation.  The nuclear eigenstates contain all the effects of
nuclear forces including those of the electroweak interaction
between nucleons in the nucleus.

The one-body electroweak operators follow from a nonrelativistic expansion
of the single-nucleon covariant currents.  By retaining terms proportional to
$1/m^2$ in this expansion, one finds in the electromagnetic case the
following time-like (charge) and space-like (current) components
 \begin{align}
 \label{eq:chargenr}
j^0_{\gamma}({\bf q};i)&=\Bigg[ \frac{1}{\sqrt{1+Q^2/(2\,m)^2}} \, \epsilon_i(Q^2)
\nonumber \\
&- \frac{i}{4\,m^2} \left[2\, \mu_i(Q^2)-\epsilon_i(Q^2)\right]{\bf q}
 \cdot\left( {\boldsymbol \sigma}_i\times {\bf p}_i\right)\Bigg]
  {\rm e}^{{\rm i} {\bf q}\cdot{\bf r}_i}\ , 
\\
 {\bf j}_\gamma({\bf q};i)&=\frac{\epsilon_i(Q^2)}{2\,m}\,
\left\{ {\bf p}_i\, ,\, {\rm e}^{{\rm i}{\bf q}\cdot{\bf r}_i}\right\}
-\frac{\rm i}{2\,m}\, \mu_i(Q^2)\, {\bf q}\times{\boldsymbol \sigma}_i\,
{\rm e}^{{\rm i}{\bf q}\cdot{\bf r}_i} \,
\label{eq:1bcurr}
\end{align}
where ${\bf q}$ and $\omega$ are the momentum and energy transfers (due to the
external field) with $Q^2=q^2-\omega^2$,
${\bf p}_i$ is the momentum
operator of nucleon $i$ with its charge and magnetization
distributions described by the form factors $\epsilon_i(Q^2)$ and $\mu_i(Q^2)$,
 \begin{align}
  \epsilon_i(Q^2)&=\frac{1}{2}\left[G_E^S(Q^2) 
 +G_E^V(Q^2)\, \tau_{i,z}\right]\ ,\label{eq:ei} \\
  \mu_i(Q^2)&=\frac{1}{2}\left[G_M^S(Q^2) 
 +G_M^V(Q^2)\, \tau_{i,z}\right]\ .
 \label{eq:mui}
\end{align}
Here $G_E^S(Q^2)$ and $G_M^S(Q^2)$, and $G_E^V(Q^2)$ and $G_M^V(Q^2)$,
are, respectively, the  isoscalar electric and magnetic, and isovector electric
and magnetic, combinations of the proton and neutron form factors,
normalized as $G_E^S(0)=G_E^V(0)=1$,
$G_M^S(0)=\mu^S$, and $G_M^V(0)=\mu^V$, with $\mu^S$ and $\mu^V$
denoting the isoscalar and isovector combinations of the proton and neutron
magnetic moments, $\mu^S=0.880$ and $\mu^V=4.706$ in units of
nuclear magnetons $\mu_N$.  These form factors are obtained from fits
to elastic electron scattering data off the proton and deuteron; for
a recent review see \citet{Hyd04}.

The isoscalar $j^\mu_{\gamma,S}$ and isovector $j^\mu_{\gamma,z}$ pieces
in $j^\mu_{NC}$ are easily identified as the terms proportional to $G^S_{E/M}$ and
$G^V_{E/M}$ in the expressions above, while the isovector components $j^{\mu5}_z$ are
given by
\begin{align}
\label{eq:rho5}
j^{05}_z({\bf q};i) &= -\frac{1}{4\, m}\, \tau_{i,z}\Bigg[ G_A(Q^2)\,{\boldsymbol \sigma}_i\,
\cdot \left\{\, {\bf p}_i\, , \,  {\rm e}^{i\, {\bf q}\cdot {\bf r}_i} \right\} 
\nonumber \\
&+{G_{PS}(Q^2)\over{m_\mu}}\, \omega \, {\boldsymbol \sigma}_i\cdot{\bf q} \, 
\, {\rm e}^{ i  {\bf q} \cdot {\bf r}_i }\Bigg] ,
\\
\label{eq:j5}
{\bf j}^{5}_z({\bf q};i)&= -\frac{G_A(Q^2)}{2}\, \tau_{i,z}\, 
\Bigg [ {\boldsymbol \sigma}_i\,  {\rm e}^{i\, {\bf q}\cdot {\bf r}_i} 
\nonumber \\
&-\frac{1}{4\, m^2} \Big( {\boldsymbol \sigma}_i \left\{ \,{\bf p}_i^2 \, , \, {\rm e}^{i\, {\bf q}\cdot {\bf r}_i} \right\}
- \left\{ {\boldsymbol \sigma}_i \cdot {\bf p}_i\,\, {\bf p}_i \, , \, {\rm e}^{i\, {\bf q}\cdot {\bf r}_i} \right\} 
\nonumber \\
&-\frac{1}{2} {\boldsymbol \sigma}_i \cdot {\bf q} \left\{ \, {\bf p}_i \, , \, {\rm e}^{i\, {\bf q}\cdot {\bf r}_i} \right\}
-\frac{1}{2} {\bf q} \left\{ \,{\boldsymbol \sigma}_i \cdot  {\bf p}_i \, , \, {\rm e}^{i\, {\bf q}\cdot {\bf r}_i} \right\}
\nonumber \\
+i\,&{\bf q}\times {\bf p}_i\,  {\rm e}^{i\, {\bf q}\cdot {\bf r}_i}\Big)  \Bigg] 
-{G_{PS}(Q^2)\over{4\,m\,m_\mu}}\,\tau_{i,z} \, {\bf q}  \, {\boldsymbol \sigma}_i\cdot{\bf q} \, 
\, {\rm e}^{ i  {\bf q} \cdot {\bf r}_i } \ , 
\end{align}
where $G_A$ and $G_{PS}$ are the nucleon axial and induced pseudoscalar
form factors.  The former is obtained from analysis of pion electro-production
data~\cite{Amaldi:1979} and measurements of the reaction $n(\nu_\mu,\mu^-)p$ in the
deuteron at quasi-elastic kinematics~\cite{Bak81,Mil82,Kit83} and
of $\nu_\mu\, p$ and $\overline{\nu}_\mu \,p$ elastic scattering~\cite{Ahr87}.
It is normalized as $G_A(0)=g_A$, where $g_A$ is the nucleon axial coupling
constant, $g_A=1.2694$ \cite{PDG}.   The form factor
$G_{PS}$ is parametrized as 
\begin{equation}
G_{PS}(Q^2) = -{2\,m_\mu\,m \over m_\pi^2 + Q^2 }\, G_{A}(Q^2) \ ,
\label{eq:gps}
\end{equation}
where $m_\mu$ and $m_\pi$ are the muon and pion masses, respectively.
This form factor is not well known; see \citet{Gor03} and \citet{Kam10} for 
recent reviews.
The parametrization
above is consistent with values extracted~\cite{Cza07,Mar12} from precise
measurements of muon-capture rates on hydrogen~\cite{And07} and $^3$He~\cite{Ack98},
as well as with the most recent  theoretical predictions based on chiral perturbation theory~\cite{Ber94}. 
Lastly, the polar-vector $j^\mu_{\pm}$ and axial-vector $j^{\mu 5}_{\pm}$ components in $j^\mu_{CC}$
follow, respectively, from $j^\mu_{\gamma,z}$ and $j^{\mu5}_{z}$ by
the replacements $\tau_{i,z}/2 \longrightarrow \tau_{i,\pm}=(\tau_{i,x} \pm \tau_{i,y})/2$.

In a nucleus, these one-body (1b) contributions lead to the impulse approximation (IA)
electroweak current
\begin{equation}
j^\mu_{\rm 1b}({\bf q})=\sum_{i\leq A} j^\mu({\bf q};i) \ .
\end{equation}
In the limit of small momentum transfers $q^\mu$, and ignoring
relativistic corrections proportional to $1/m^2$ and neutron charge contributions,
it is easily seen that $j^\mu_{\gamma,{\rm 1b}}$ reduces to the
charge and convection current operators of individual protons, and to
the magnetization current operator of individual protons and neutrons,
while the time-like $j^0_{\pm}$ and space-like ${\bf j}^5_\pm$ components
in $j^\mu_{CC}$ reduce, respectively, to the familiar Fermi and Gamow-Teller
operators.

There is ample evidence for the inadequacy of the IA currents
to provide a quantitatively satisfactory description of electroweak
observables at low and intermediate values of energy and
momentum transfers, especially in light s- and p-shell nuclei
with $A \leq 12$, for which essentially exact calculations can be carried
out.  This evidence is particularly striking in the case of electromagnetic
isovector transitions.  Well-known illustrations are, among others, the
10\% underestimate of the $np$ radiative capture cross section at thermal
neutron energies, which in fact provided the initial impetus to consider
two-body terms in the nuclear electromagnetic current
operator~\cite{Ris72}, the 15\% underestimate
of the isovector magnetic moment of the trinucleons
and the large discrepancies
between the experimental and calculated magnetic and charge form factors
of the hydrogen and helium isotopes~\cite{Had83,Str87,Schiavilla:1989,Sch90},
particularly in the first diffraction region
at momentum transfers in the range of  (3.0--3.5) fm$^{-1}$, the large underprediction,
by respectively about 50\% and 90\%, of the $nd$ and $n\,^3$He radiative
capture cross sections~\cite{Marcucci:2005,Gir10a}, and, finally, the significant underestimate,
in some cases as large as 40\%, of magnetic moments and $M_1$ radiative
transition rates in $A$=7--9 nuclei~\cite{pastore2013}.

In the case of charge-changing weak transitions, discrepancies between
experimental data and theoretical results obtained with the IA operators
are not as large and are all limited to the low momentum and energy transfers
of interest in $\beta$ decays and electron- and muon-capture processes.
They are nevertheless significant.  Examples of these in the few-nucleon
systems are the few \% underestimate of the Gamow-Teller matrix element in tritium $\beta$
decay~\cite{Sch98} and the 10\% underprediction~\cite{Mar12} of the
precisely measured~\cite{Ack98} $^3$He($\mu^-,\nu_\mu$)$^3$H rate.

Many-body terms in the nuclear electroweak current operators
arise quite naturally in the conventional meson-exchange picture
as well as in more modern approaches based on chiral effective field
theory.  
Below we provide a brief review of both frameworks; a recent review
on reactions on electromagnetic reactions in 
light nuclei~\cite{bacca2014} is also available.
\subsubsection{Two- and three-body electromagnetic currents}
We first discuss electromagnetic operators.
There is a large body of work dealing with the problem of their
construction from meson-exchange theory, crystallized in a
number of reviews of the 1970s and 1980s, e.g.,
\citet{Che71,Tow87,Ris89,Mat89}.  Here we describe an
approach, originally proposed by Riska~\cite{Ris85a,Ris85b,Ris85c},
that leads to conserved currents, even in the presence of $N\!N$ and
$3N$ potentials, not necessarily derived from meson-exchange
mechanisms (as is the case for the AV18 and UIX or IL7 models).
This approach has been consistently used to study many
photo- and electro-nuclear observables, and has proved to be quite
successful in providing predictions systematically in close agreement
with experiment.

Leading electromagnetic two-body charge and current operators are derived
from the static (that is, momentum-independent) components of the $N\!N$
potential, consisting of the isospin-dependent central, spin, and tensor terms.
These terms are assumed to be due to exchanges of effective pseudo-scalar
($PS$ or $\pi$-like) and vector ($V$ or $\rho$-like) mesons, and the corresponding
charge and current operators are constructed from nonrelativistic reductions
of Feynman amplitudes with the $\pi$-like and $\rho$-like effective propagators.
For the $\pi$-like case (we defer to \citet{carlson1998} and
\citet{Marcucci:2005} for a complete listing) they are given in 
momentum space by
\begin{align}
j_{\gamma}^{0,PS}({\bf k}_i,{\bf k}_j)&= \left[ F_1^S(Q^2)   \, {\boldsymbol \tau}_i \cdot {\boldsymbol \tau}_j 
+F_1^V(Q^2) \,\tau_{j,z} \right]\times
\nonumber \\
&\frac{v_{PS}(k_j)}{2\,m}
 {\boldsymbol \sigma}_i \cdot {\bf q} \,\,   {\boldsymbol \sigma}_j \cdot {\bf k}_j  + (i \rightleftharpoons j) \ ,
\label{eq68a}  \\
{\bf j}_{\gamma}^{PS}({\bf k}_i,{\bf k}_j)
&=i\,G_{E}^{V}(Q^2)
   ({\boldsymbol\tau}_i \times {\boldsymbol\tau}_j)_z\times 
\nonumber \\
v_{PS}(k_j)\, 
&  \bigg[  {\boldsymbol\sigma}_i 
 -{ {\bf k}_i - {\bf k}_j \over k_i^2 -k_j^2 }\,
  {\boldsymbol\sigma}_i \cdot {\bf k}_i \bigg] {\boldsymbol\sigma}_j \cdot {\bf k}_j + (i \rightleftharpoons j) \ .
  \label{eq:jps}
\end{align}
Here ${\bf k}_i$ and ${\bf k}_j$ are the fractional momenta delivered to 
nucleons $i$ and $j$, with ${\bf q}={\bf k}_i+{\bf k}_j$, and
$v_{PS}(k)$ is projected out of the (isospin-dependent) spin
and tensor components of the potential \cite{Marcucci:2005}.
The Dirac nucleon electromagnetic form factors $F^{S/V}_1$ are related
to those introduced previously via $F_1^{S/V}=\left(G_E^{S/V} +\eta 
\, G_M^{S/V}\right)/(1+\eta)$ with $\eta=Q^2/(4\,m^2)$, and therefore
differ from $G^{S/V}_E$ by relativistic corrections proportional to $\eta$.
The representation of these operators in coordinate space follows from
\begin{align}
j_\gamma^{\mu, PS}({\bf q};ij)&=
\int \frac{{\rm d}{\bf k}_i}{(2\pi)^3}
\frac{{\rm d}{\bf k}_i}{(2\pi)^3}
\, (2\pi)^3 \times \\
&\delta({\bf k}_i+{\bf k}_j-{\bf q})\, {\rm e}^{i{\bf k}_i\cdot{\bf r}_i}\,
{\rm e}^{i{\bf k}_j\cdot{\bf r}_j} \, j^{\mu,PS}_\gamma({\bf k}_i,{\bf k}_j) \ 
\nonumber ,
\end{align}
and explicit expressions for them can be found in \citet{Schiavilla:1989}.

By construction, the longitudinal components of the resulting
${\bf j}_{\gamma}^{PS}$ and ${\bf j}_{\gamma}^V$ currents satisfy current
conservation with the static part of the potential $v_{ij}({\rm static})$,
\begin{align}
{\bf q}\cdot &\left[ {\, \bf j}_\gamma^{PS}({\bf q};ij)
+{\bf j}_\gamma^V({\bf q};ij)\right]= 
\nonumber \\
&\left[\, v_{ij}({\rm static})\, ,\, j^0_{\gamma}({\bf q};i)
   +j^0_{\gamma}({\bf q};j)\, \right]
 \ , \label{eq:ccr2} 
 \end{align}
where $j^0_\gamma({\bf q};i)$ is the one-body charge operator of
Eq.~(\ref{eq:chargenr}) to leading order in an expansion in powers of $1/m$.
The continuity equation requires that the
form factor $G_E^V(Q^2)$ be used in the longitudinal components
of the $PS$ and $V$ currents.  However, it poses no
restrictions on their transverse components, in particular on the
electromagnetic hadronic form factors that may be used
in them.  Ignoring this ambiguity, the choice $G_E^V$
has been made for both longitudinal {\it and} transverse components.

Additional conserved currents follow from minimal substitution in the
momentum-dependent part of the potential $v_{ij}$(nonstatic). 
In a realistic potential like the AV18,
this momentum dependence enters explicitly via the spin-orbit,
quadratic orbital angular momentum, and quadratic spin-orbit
operators, and implicitly via ${\boldsymbol \tau}_i\cdot {\boldsymbol \tau}_j$,
which for two nucleons can be expressed in terms of space- and
spin-exchange operators as
\begin{equation}
  {\boldsymbol \tau}_i\cdot{\boldsymbol \tau}_j = -1-
  (1+{\boldsymbol \sigma}_i\cdot{\boldsymbol \sigma}_j) \, {\rm e}^{-i\, {\bf r}_{ij}\cdot\left( {\bf p}_i
  - {\bf p}_j\right)} \ .
  \label{eq:tt}
\end{equation}
Both the explicit and implicit (via ${\boldsymbol \tau}_i\cdot {\boldsymbol \tau}_j$) momentum-dependent 
terms need to be gauged with ${\bf p}_i \longrightarrow {\bf p}_i -
\epsilon_i(Q^2)\, {\bf A}({\bf r}_i)$, where ${\bf A}({\bf r})$ is the
vector potential, in order to construct exactly conserved currents with 
$v_{ij}({\rm non}$-static)~\cite{Sac48}.  The procedure, including the non-uniqueness inherent
in its implementation, is described in \citet{Marcucci:2005} and \citet{Sac48}.
In contrast to the purely
isovector ${\bf j}^{PS}_{\gamma}$ and ${\bf j}^V_{\gamma}$, the currents from
$v_{ij}({\rm non}$-static) have both isoscalar and isovector terms, which,
however, due to their short-range nature lead to contributions that are typically
much smaller (in magnitude) than those generated by ${\bf j}^{PS}_{\gamma}$ and
${\bf j}^V_{\gamma}$.

Conserved three-body currents associated with the $V^{2\pi}_{ijk}$ term of
the $3N$ potential have also been derived by assuming that this term originates
from the exchange of effective $PS$ and $V$ mesons with excitation of
an intermediate $\Delta$ isobar.  However, their
contributions have been found to be generally negligible,
except for some of the polarization observables, like $T_{20}$ and $T_{21}$,
measured in proton-deuteron radiative capture at low energy~\cite{Marcucci:2005}.

It is important to stress that the two- and three-body charge and current operators
discussed so far have no free parameters, and that their short-range behavior
is consistent with that of the potentials---for the $N\!N$ potential, in particular,
this behavior is ultimately constrained by scattering data.  It is also worthwhile
noting that in a nucleus $^AZ$ global charge conservation requires that
\begin{equation}
\langle ^AZ\!\mid \int{\rm d}{\bf x}\, j^0_\gamma({\bf x}) \mid\! ^AZ\rangle = Z\ .
\end{equation}
This condition is obviously satisfied by $j^0_{\gamma,{\rm 1b}}({\bf q}\!\!=\!\!0)$
(equivalent to the volume integral of the charge density above); it implies that
two-body (and many-body) charge operators must vanish at ${\bf q}$=0, to which
both $j^{0,PS}_\gamma$ and $j^{0,V}_\gamma$ conform.  As emphasized by
\citet{Fri77}, a proper derivation of the leading two-body charge
operator $j^{0,PS}_\gamma$ necessarily entails the study of nonstatic corrections
to the OPE potential.  However, these corrections are neglected in the AV18, and in
fact in most modern realistic potentials.  These issues have recently been re-examined
(and extended to the two-pion-exchange potential and charge operator) within
the context of chiral effective field theory \cite{Pas11}.
 
There are many-body currents arising from magnetic-dipole excitation
of $\Delta$ resonances.  They have been derived in a number of different
approaches, the most accurate of which is based on the explicit inclusion
of $\Delta$-isobar degrees of freedom in nuclear wave functions.  In this
approach, known as the transition-correlation-operator (TCO) method and
originally developed by \citet{Sch92}, the nuclear wave function is
written as
\begin{align}
\Psi_{N+\Delta}=\left[ {\cal S} \prod_{i <j }\left( 1+U^{\rm TR}_{ij} \right) \right] \Psi
\simeq \left( 1 +\sum_{i <j }U^{\rm TR}_{ij} \right)\Psi
\end{align}
where $\Psi$ is the purely nucleonic component and ${\cal S}$ is the symmetrizer,
and in the last expression on the r.h.s.~only admixtures with one and two
$\Delta$'s are retained.  The transition operators $U_{ij}^{\rm TR}$ convert
$N\!N$ into $N\Delta$ and $\Delta\Delta$ pairs and are obtained from two-body
bound and low-energy scattering solutions of the full $N$+$\Delta$ coupled-channel
problem, including transition potentials  $v^{\rm TR}_{ij}(N\!N \rightarrow N\Delta)$
and $v^{\rm TR}_{ij}(N\!N \rightarrow \Delta\Delta)$; see \citet{wiringa1984}.  
The simpler perturbative
treatment of $\Delta$-isobar degrees of freedom, commonly used in estimating the
$\Delta$-excitation current contributions, uses the approximation
\begin{align}
U^{\rm TR, PT}_{ij}&=\frac{1}{m-m_\Delta}\left[ v^{\rm TR}_{ij}(N\!N \rightarrow N\Delta)
+(i\rightleftharpoons j)\right] + 
\nonumber \\
&\frac{1}{2\,(m-m_\Delta)} \,v^{\rm TR}_{ij}(N\!N 
\rightarrow \Delta\Delta)  \ ,
\end{align}
and $m_\Delta$ (1232 MeV) is the $\Delta$ mass.
This perturbative treatment has been found to overestimate $\Delta$-isobar
contributions~\cite{Sch92}, since $U^{\rm TR, PT}_{ij}$ ignores the repulsive
core in the $N\Delta \rightleftharpoons N\Delta$ and
$\Delta\Delta \rightleftharpoons \Delta\Delta$ interactions as well as
the significant kinetic energies of the $\Delta$'s in these channels. 

In the presence of an electromagnetic field, $N \rightleftharpoons \Delta$
and $\Delta \rightleftharpoons \Delta$ couplings need to be accounted for.
For the first process, the coupling and associated electromagnetic form
factor are taken from $N(e,e^\prime)$ data in the resonance region~\cite{Car86},
while for the second, experimental information on the
magnetic moment $\mu_{\gamma\Delta\Delta}$ comes from soft-photon analysis of pion-proton
bremsstrahlung data near the $\Delta$ resonance~\cite{Lin91}.
The associated currents give important contributions to isovector
transitions, comparable to those from the $PS$ current.  In particular, the leading $N \rightarrow \Delta$
current is parametrized as
\begin{equation}
\label{eq:ndelta}
{\bf j}_\gamma({\bf q};i, N\rightarrow \Delta) = 
\frac{i}{2\, m}
\, G_{\gamma N\Delta}(Q^2) \, {\bf S}_i \times {\bf q} \, T_{i,z}\, {\rm e}^{i{\bf q}\cdot r_i} \ ,
\end{equation}
where ${\bf S}_i$ and ${\bf T}_i$ are spin and isospin transition operators
converting a nucleon into a $\Delta$.  The $\Delta\rightarrow N$ current follows
from the expression above by replacing ${\bf S}_i$ and ${\bf T}_i$ by their
adjoints ${\bf S}^\dagger_i$ and ${\bf T}^\dagger_i$.  The electromagnetic
$\gamma N\Delta$ form factor, obtained from fits of $\gamma N$ data
at resonance, is normalized as $G_{\gamma N\Delta}(0)=\mu_{\gamma N\Delta}$
with $\mu_{\gamma N\Delta}\simeq 3 \mu_N$ \cite{Car86}.
There can also be an electric quadrupole
transition between the $N$ and $\Delta$ states.  However, this coupling is very
weak compared to the magnetic dipole, and has typically been neglected.
In the perturbative approach above, the $N\rightleftarrows\Delta$ current in 
Eq.(\ref{eq:ndelta})
leads to a two-body current given by
\begin{align}
{\bf j}_\gamma^{\Delta,{\rm PT}}&({\bf q};ij)=
\nonumber \\
&\left[ v^{\rm TR}_{ij}({N\!N \rightarrow\Delta N})\right]^\dagger\, \frac{1}{m_N-m_\Delta} \,
{\bf j}_\gamma({\bf q}; {i,N\rightarrow\Delta})\nonumber\\
&+\,\, {\bf j}_\gamma({\bf q};{i,\Delta \rightarrow N}) \,\frac{1}{m_N-m_\Delta}\,
v^{\rm TR}_{ij}({N\!N \rightarrow\Delta N}) 
\nonumber \\
&+ (i \rightleftharpoons j) \ .
\label{eq:jndc}
\end{align}
This current is obviously transverse,
and hence unconstrained by current conservation.

The $\Delta$-excitation currents in either perturbation theory or in the
nonperturbative TCO approach can be reduced to effective two- and
many-body operators depending on $U^{\rm TR}_{ij}$, but acting only
on the nucleonic component $\Psi$ of the full wave function.  This is
accomplished by making use of standard identities which allow one to 
express products of spin and isospin transition operators in terms of Pauli
spin and isospin matrices.  Both perturbation theory and the TCO method
have been used to obtain results reported in the present review.

Finally, additional short-range isoscalar and isovector two-body charge and
(purely transverse) current operators follow from, respectively, the $\rho\pi\gamma$
and $\omega\pi\gamma$ transition mechanisms.  The coupling constants
and hadronic and electromagnetic form factors at the $\rho N\!N$,
$\omega N\!N$, $\rho\pi\gamma$, and $\omega\pi\gamma$ vertices are
poorly known~\cite{carlson1998}.  In reference to the $\rho\pi\gamma$ current,
it is important to note that, because of the large tensor coupling of the
$\rho$-meson to the nucleon, a nonrelativistic expansion of
$j_\gamma^{\mu, \rho\pi}$ which only
retains the leading order is not accurate~\cite{Schiavilla:2002}.  The inadequacy
of this approximation becomes especially apparent in the
deuteron magnetic form factor at high momentum transfers.
However, with the exception of this observable,
these transition currents typically lead to very
small corrections to charge and magnetic form factors of light nuclei,
in the momentum transfer range where data
are available.

\subsubsection{Two- and three-body weak currents in the conventional approach}
\label{subsec:2bweak}

Among the axial current operators, the leading
terms are those associated with the excitation of $\Delta$ resonances.
The $N\rightarrow \Delta$ axial
current is
\begin{equation}
{\bf j}^5_a({\bf q};i, N\rightarrow \Delta) = 
-\frac{G_{A N\Delta}(Q^2)}{2} \, {\bf S}_i  \, T_{i,a}\, {\rm e}^{i{\bf q}\cdot r_i} \ ,
\end{equation}
where the (unknown) $N$ to $\Delta$ axial form factor is parametrized as
\begin{equation}
 G_{AN\Delta} (Q^2) = \frac{g_{A N\Delta}}{\left(1+Q^2/\Lambda_A^2\right)^2} \ ,
\end{equation}
and the cutoff $\Lambda_A$ is taken of the order 1 GeV (as in the case of
the nucleon).  The coupling constant $g_{A N\Delta}$ is not known.  In the
static quark model, it is related to the nucleon axial coupling constant via
$g_{A N\Delta}= (6\sqrt{2}/5)g_A$.  This value has often been used in the
literature in the calculation of $\Delta$-induced axial current contributions
to weak transitions~\cite{Car91,Sai90}.  However, in view of the
uncertainty in the naive quark model predictions, a more reliable estimate
of $g_{A N\Delta}$ is obtained by determining it phenomenologically in
the following way.  It is well established that the one-body axial current
leads to a 3--4\% under-prediction of the measured Gamow-Teller matrix
element of tritium $\beta$-decay~\cite{Sch98}, the relatively small
spread depending on the particular realistic Hamiltonian adopted
to generate the trinucleon wave functions.  Since the contributions
due to $\Delta \rightarrow \Delta$ currents~\cite{Sch92}, and
to the other mechanisms discussed below, have been found to be
numerically small, this 3--4\% discrepancy can be used to determine
$g_{AN\Delta}$.  Of course, the resulting value depends on how the
$\Delta$ degrees of freedom are treated in nuclear wave functions,
whether perturbatively as in Eq.~(\ref{eq:jndc}) or nonperturbatively
in the full TCO approach~\cite{Sch92,Mar00a}.  In any case, this value is
typically significantly smaller than the quark-model estimate.

There are additional axial two-body currents due to $\pi$-
and $\rho$-meson exchange and $\rho \pi$ transition; explicit
expressions have been listed most recently in \citet{Shen:2012}.
They are derived from nonrelativistic reduction of Feynman
amplitudes \cite{Tow87}.
However, the contributions of these two-body operators
to weak transitions in light nuclei have been found to be numerically far
less important than those from $\Delta$ degrees of freedom~\cite{Car91,Sch92}.

Finally, in the axial charge there is a two-body operator
of pion range, whose model-independent structure and strength are determined by
soft-pion theorem and current algebra arguments~\cite{Kub78} and it arises
naturally in chiral effective field theory:
\begin{align}
j^{05,\pi}_a({\bf k}_i,{\bf k}_j)&=
-i\,\frac{G_A(Q^2)}{4 \,f_\pi^2}\,\frac{h_\pi^2(k_i) }{ k_i^2+ m_\pi^2 }\, 
({\boldsymbol \tau}_i \times{\boldsymbol \tau}_j)_a\times
\nonumber \\
&{\boldsymbol \sigma}_i \cdot{\bf k}_i +(i \rightleftharpoons j)  \ .
\label{eq:rho2pi}
\end{align}
Here $f_\pi$ is pion decay amplitude ($f_\pi \simeq 93$ MeV),
the $Q^2$ dependence of the form factor $G_A$ is assumed to be the
same as in the nucleon, and the hadronic form factor $h_\pi$ is parametrized as
\begin{equation}
h_\pi(k)=\frac{\Lambda_\pi^2 -m_\pi^2}{\Lambda_\pi^2+k^2} \  .
\label{eq:hff}
\end{equation}
The $\Lambda_\pi$ is in the range (1.0--1.5) GeV, consistent with
values inferred from the OPE component of realistic $N\!N$
potentials.  Because of the absence of $J_i^{\pi_i}=0^+ \rightarrow J_f^{\pi_f}=0^-$
weak transitions in light nuclei, it does not play a significant role in these systems.
\subsection{Electromagnetic currents in chiral effective field theory}
Electromagnetic charge and current operators were derived up to one loop originally by
\citet{Par96} in the heavy-baryon formulation of covariant perturbation theory.
More recently, however, two independent derivations, based
on time-ordered perturbation theory (TOPT), have appeared in the literature, one
by \citet{Pas09,Pas11,Pia13} and the other by 
\citet{Koe09,Koe11}.  In the following, we only discuss briefly
the electromagnetic current operator, since it has been used recently
in QMC calculations of magnetic moments and $M1$ transition rates
in light p-shell nuclei~\cite{pastore2013,pastore2014}.  For a derivation of this as well
as of the electromagnetic charge operator, we refer the reader to the
original papers.
  
\begin{figure}
\centerline{\includegraphics[width=0.35\textwidth]{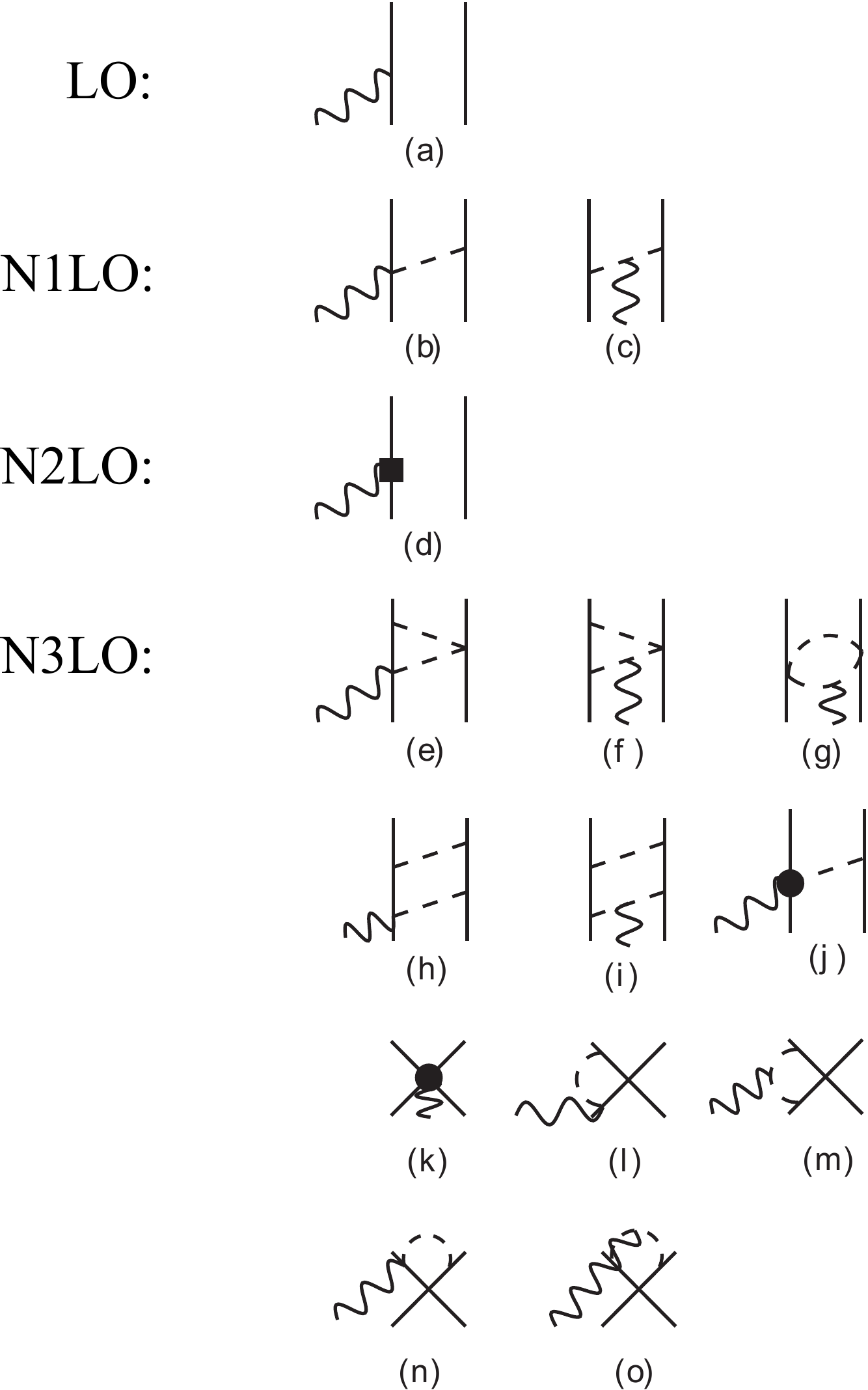}}
\vspace{8pt}
\caption{Diagrams illustrating one- and two-body electromagnetic current operators at
$\left(P/\Lambda_\chi\right)^{-2}$ (LO),
$\left(P/\Lambda_\chi\right)^{-1}$ (N1LO),
$\left(P/\Lambda_\chi\right)^{0}$ (N2LO),
and $\left(P/\Lambda_\chi\right)^{1}$ (N3LO).  Nucleons, pions,
and photons are denoted by solid, dashed, and wavy lines, respectively.  The square
in panel (d) represents the relativistic correction
to the LO one-body current, suppressed relative to it by an additional
$\left(P/\Lambda_\chi\right)^2$ factor;
the solid circle in panel (j) is associated with a $\gamma \pi N$ vertex in $H_{\gamma \pi N}$
involving the low-energy constants (LECs) $d_8^\prime$, $d_9^\prime$, and $d_{21}^\prime$;
the solid circle in panel (k) denotes two-body contact terms of minimal and nonminimal
nature, the latter involving the LECs $C_{15}^\prime$ and $C_{16}^\prime$.
Only one among all possible time orderings is shown for the N1LO and N3LO currents,
so that both direct- and crossed-box contributions are retained.}
\label{fig:vvchi}
\end{figure}

The contributions to the current operators up to one loop
are illustrated diagrammatically in Fig.~\ref{fig:vvchi}, where
the N$n$LO terms correspond to the power
counting $\left(P/\Lambda_\chi\right)^n \times \left(P/\Lambda_\chi\right)^{\rm LO}$. 
The electromagnetic currents from LO, N1LO, and N2LO terms and from N3LO
loop corrections depend only on the known parameters $g_A$ and $f_\pi$ (N1LO and N3LO),
and the nucleon magnetic moments (LO and N2LO).  Note that the LO and N1LO currents
are the same as the conventional ones, while the N2LO current consists
of relativistic corrections to the LO one.
Unknown low-energy constants (LECs) enter the N3LO OPE contribution
involving a $\gamma \pi N$ vertex from a higher order chiral Lagrangian ${\cal L}^{(3)}_{\pi N}$
(proportional to the LECs $d_i^\prime$) and
contact currents implied by nonminimal couplings~\cite{Pas09,Pia13}.
They are given by
\begin{align}
\label{eq:cdlt}
{\bf j}^{\rm N3LO}_{\gamma\pi N}({\bf k}_i,{\bf k}_j)&=
i\, \frac{g_A}{F_\pi^2} 
\frac{{\bm \sigma}_j \cdot {\bf k}_j}{\omega_{k_j}^2}\,
\bigg[  d_8^\prime\, \tau_{j,z} \,{\bf k}_j +d^\prime_9 \, 
 {\bm \tau}_i\cdot{\bm \tau}_j\,{\bf k}_j \nonumber\\
&-d_{21}^\prime ({\bm \tau}_i\times{\bm \tau}_j)_z\, {\bm \sigma}_i\times {\bf k}_j  \bigg] \times {\bf q} 
+ (i\rightleftharpoons j) \ ,\\
\label{eq:nmcounter}
{\bf j}^{\rm N3LO}_{\gamma,{\rm nm}}({\bf k}_i,{\bf k}_j)&= - i\,e \Big[  C_{15}^\prime\, {\bm \sigma}_i 
+  C_{16}^\prime
 (\tau_{i,z} - \tau_{j,z})\,{\bm \sigma}_i  \Big]\times {\bf q}  
\nonumber \\
&+ (i \rightleftharpoons j)\ .
\end{align}
Before discussing the determination of these LECs, we note that
the loop integrals in the N3LO diagrams of Fig.~\ref{fig:vvchi}
are ultraviolet divergent and are regularized using dimensional regularization.
The divergent parts of these loop integrals are reabsorbed by the LECs
multiplying contact terms.  Finally, the resulting renormalized
electromagnetic operators
have power-law behavior for large momenta, and must be further regularized
before they can be sandwiched between nuclear wave functions.  This is accomplished by
the inclusion of a momentum-space cutoff of the type $C_\Lambda(k)={\rm exp}(-k^4/\Lambda^4)$
with $\Lambda$ in the range $\simeq (500$--700) MeV/c.  The expectation is that observables,
like magnetic moments and $M1$ transitions in light nuclei
are fairly insensitive to variations of $\Lambda$ in this range.  

The $d_i^{\, \prime}$, entering the OPE N3LO current, could be
fitted to pion photo-production data on a single nucleon or related
to hadronic coupling constants by resonance saturation arguments~\cite{Pas09,Pia13}.
Both procedures have drawbacks.  While the former achieves consistency
with the single-nucleon sector, it nevertheless relies on single-nucleon
data involving photon energies much higher than those relevant to the
threshold processes under consideration and real (in contrast to virtual)
pions.  The second procedure is questionable because of poor knowledge
of some of the hadronic couplings, such as $g_{\rho N\!N}$.  Alternative strategies
have been investigated for determining the LECs $d_i^\prime$ as well as $C_{15}^\prime$
and $C^\prime_{16}$~\cite{Pia13}.  In this respect, it is convenient
to define the dimensionless LECs $d_i^{S,V}$ (in units
of the cutoff $\Lambda$) related to the original set via
\begin{align}
C_{15}^\prime&=d_1^S/\Lambda^4  \ , \quad d_9^\prime=d_2^S/\Lambda^2 , \nonumber \\
C_{16}^\prime&=d_1^V/\Lambda^4 \ , \quad d_8^\prime=d_2^V/\Lambda^2\ ,
\quad d_{21}^\prime=d_3^V/\Lambda^2 \ ,
\end{align}
where the superscript $S$ or $V$ on the $d^{S,V}_i$ characterizes the isospin of the
associated operator.

The isoscalar $d_1^S$ and $d_2^S$ have been fixed by reproducing the experimental
deuteron magnetic moment $\mu_d$ and isoscalar combination $\mu^S$ of the
trinucleon magnetic moments.  It turns out that in calculations based on the AV18
and AV18+UIX Hamiltonians the LEC $d_1^S$ multiplying the contact current assumes
reasonable values, $d_1^S\simeq 2.5$ and 5.2 corresponding to $\Lambda=500$ MeV and
600 MeV, while the LEC $d_2^S$ values are quite small $\simeq-0.17$ and --0.20
for the same range of cutoff $\Lambda$ \cite{Pia13}.

Three different strategies, referred to as I, II, and III, have been investigated
to determine the isovector LECs $d_1^V$, $d_2^V$, and $d_3^V$.  In all
cases I-III, $d_3^V/d_2^V=1/4$ is assumed as suggested by $\Delta$
dominance in a resonance saturation picture of the N3LO OPE current of panel
(j) in Fig.~\ref{fig:vvchi}. In set I, $d_1^V$ and $d_2^V$ have been constrained
to reproduce the experimental values of the $np$ radiative capture cross section
$\sigma_{np}$ at thermal neutron energies and the isovector combination $\mu^V$
of the trinucleon magnetic moments.  This, however, leads to unreasonably
large values for both LECs, and is clearly unacceptable~\cite{Pia13}.
In sets II and III, the LEC $d_2^V$ is fixed by assuming $\Delta$ dominance
while the LEC $d_1^V$ multiplying the contact current is fitted to reproduce
either $\sigma_{np}$ in set II or $\mu^V$ in set III.  
Both alternatives still lead
to somewhat large values for this LEC: $d_1^V\simeq -9.3$ and --11.6 in set II
and $d_1^V\simeq -5.2$ and --1.0 in set III.
There are no three-body currents at N3LO~\cite{Pas09},
and therefore it is reasonable to fix the strength of the $N\!N$ contact operators by
fitting a $3N$ observable such as $\mu^S$ and $\mu^V$.

\label{sec:form}
\subsection{Elastic and inelastic form factors}

The longitudinal $F_L$ and transverse $F_T$ form factors
for elastic and inelastic transitions are extracted from
electron scattering data by measuring the cross section~\cite{Don84}
\begin{equation}
\frac{{\rm d}\sigma}{{\rm d}\Omega}=4\, \pi\, \sigma_M \, f_{\rm rec}^{-1}\, \left[ \frac{Q^4}{q^4}\, F_L^{\, 2}
+\left( \frac{Q^2}{2\, q^2}+{\rm tan}^2\theta_{\rm e}/2\right)\, F_T^{\,2}\right]\ ,
\end{equation}
where $\sigma_M$ is the Mott cross section, $q$ and $Q$ are the
electron three- and four-momentum transfers, $f_{\rm rec}$ is the
recoil correction $f_{\rm rec}=1+(2\, \epsilon/m_A)\, {\rm sin}^2\theta_{\rm e}/2$,
$\epsilon$ and $\theta_{\rm e}$ are the electron initial energy and scattering angle in
the laboratory, and $m_A$ is the mass of the target nucleus.  In the case of elastic
scattering, the electron energy transfer is $\omega_{\rm el}=\sqrt{q^2+m_A^2} -m_A$ and
the four-momentum transfer $Q^2_{\rm el}=2\, m_A\, \omega_{\rm el}$.
The form factors $F_L$ and $F_T$ are expressed in terms of reduced
matrix elements (RMEs) of charge ($C_L$), magnetic ($M_L$), and electric
($E_L$) multipole operators, defined below, as
\begin{align}
F_L^{\,2}(q)=\frac{1}{2\, J_i+1} &\sum_{L=0}^\infty \mid \langle J_f\mid\mid C_L(q)\mid\mid J_i\rangle\mid^2\ ,\\
F_T^{\,2}(q)=\frac{1}{2\, J_i+1} &\sum_{L=1}^\infty\Big[\, \mid \langle J_f\mid\mid M_L(q)\mid\mid J_i\rangle\mid^2 
\nonumber \\
&+\mid \langle J_f\mid\mid E_L(q)\mid\mid J_i\rangle\mid^2\,  \Big]\ .
\end{align}
We note that for elastic scattering $J_i=J_f=J$ and the $E_L$ RMEs vanish 
because of time reversal invariance.

Standard techniques~\cite{Wal95} are used to carry out
the multipole expansion of the electromagnetic charge $j^0_\gamma({\bf q})$
and current ${\bf j}_\gamma({\bf q})$ operators in a reference frame
in which the $\hat{\bf z}$ axis defines the spin-quantization axis,
and the direction $\hat{\bf q}$ is specified by the angles $\theta$ and $\phi$:
\begin{align}
\label{eq:rcl}
\j^0_\gamma({\bf q})&= \int{\rm d}{\bf x}\, {\rm e}^{i\, {\bf q}\cdot{\bf x}}\, j^0_\gamma({\bf x}) \nonumber\\
&=\sum_{LM_L} 4\,\pi\,i^L\, Y^*_{LM_L}(\hat{\bf q}) \, C_{LM_L}(q) \ , \\
j_{\gamma, q\lambda}({\bf q})&=\int{\rm d}{\bf x}\, {\rm e}^{i\, {\bf q}\cdot{\bf x}}\, 
\hat{\bf e}_{q\lambda}\cdot {\bf j}_\gamma({\bf x})\nonumber\\
&=-\!\!\!\sum_{LM_L ( L\ge 1)}\sqrt{2\,\pi\,(2\, L+1)}\, i^L\, D^L_{M_L,\,\lambda}(-\phi,-\theta,\phi)\,
\nonumber \\
&\times\left[ \lambda\, M_{LM_L}(q)+E_{LM_L}(q)\right]\ ,
\label{eq:jeml}
\end{align}
where $\lambda=\pm 1$, the $Y_{LM_L}$ are spherical harmonics, and the $D^L_{M_L,\, \lambda}$
are rotation matrices~\cite{Edm57}.
The unit vectors $\hat{\bf e}_{q\lambda}$ denote
the linear combinations
\begin{equation}
\hat{\bf e}_{q\pm 1}=\mp\frac{1}{\sqrt{2}}\left( \hat{\bf e}_{q1} \pm i\, \hat{\bf e}_{q2}\right) \ ,
\label{eq:eu}
\end{equation}
with $\hat{\bf e}_{q3}=\hat{\bf q}$, $\hat{\bf e}_{q2}=\hat{\bf z}\times {\bf q}/\mid\!\hat{\bf z}\times {\bf q}\! \mid$,
and $\hat{\bf e}_{q1}=\hat{\bf e}_{q2}\times \hat{\bf e}_{q3}$.  These relations are used
below to isolate the contributing RMEs to elastic transitions in nuclei with $A \le 12$.
%
%
%
%
The ground states of nuclei in the mass range $6 \leq A \leq 12$
have spins ranging from $J=0$ (as in $^{12}$C) to $J=3$
(as in $^{10}$B), and are described by VMC or GFMC wave
functions.  For reasons of computational efficiency, it is convenient to determine
the RMEs of charge and magnetic multipoles contributing to a specific
transition by evaluating the matrix elements of $j^0_\gamma({\bf q})$ and
${\bf j}_{\gamma}({\bf q})$ between states having a given spin projection $M_J$,
usually the stretched configuration with $M_J=J$, for a number of
different $\hat{\bf q}$ directions.  The matrix
element of the charge operator can then be written as
\begin{equation}
\langle JJ ;{\bf q}|
 j^0_\gamma({\bf q}) |JJ\rangle = \sum^\infty_{L=0} \sqrt{4\,\pi}\, i^L \, c_{LJ} \, P_L({\rm cos}\,\theta)\, 
 \langle J|| C_L(q)||J\rangle
\end{equation}
where $\theta$ is the angle that $\hat{\bf q}$ makes with the $\hat{\bf z}$
spin-quantization axis, the $P_L$ are Legendre polynomials, and $c_{LJ}$ is 
the Clebsch-Gordan coefficient
$\langle JJJ-\!J| L0\rangle$.
Generally, for a nucleus of spin $J$ the number of contributing (real)
RMEs of charge multipole operators is $[J]+1$  (here $[J]$ denotes the integer
part of $J$)  and the allowed $L$ are the even integers between 0 and $2\, J$.
Thus, it is possible to select $[J]+1$ independent $\hat{\bf q}$ directions, evaluate the matrix
element of the charge operator for each of these different $\hat{\bf q}$, and then
determine the RMEs by solving a linear system.  For example, for a nucleus of spin
$J=1$ (like $^6$Li) 
\begin{align}
\langle 11 ;q \,\hat{\bf z}| j^0_\gamma(q\, \hat{\bf z}) |11\rangle &= \sqrt{\frac{4\, \pi}{3}}
\left( C_0 - \frac{1}{\sqrt{2}}\, C_2 \right) \, , \\
\langle 11;q\, \hat{\bf x}| j^0_\gamma(q\, \hat{\bf x}) |11\rangle &= \sqrt{\frac{4\, \pi}{3}}
\left( C_0 + \frac{1}{2\,\sqrt{2}}\, C_2 \right) \ ,
\end{align}
where $C_L$ is a short-hand notation for $\langle 1|| C_L(q)||1\rangle$.
%

For the transverse elastic form factor, it is possible to proceed in a similar fashion.
Since electric multipoles do not contribute in elastic scattering
\begin{align}
\langle&JJ; {\bf q}\mid \hat{\bf e}_{{\bf q} \lambda} \cdot {\bf j}_\gamma({\bf q})\mid JJ\rangle
\nonumber \\
&=-\lambda \sum_{L \ge 1}i^L \, \sqrt{2\pi}\, c_{LJ}\, D^L_{0,\,\lambda}(-\phi,-\theta,\phi)\,
\langle J|| M_L(q)||J\rangle \ ,
\end{align}
where the unit vectors $\hat{\bf e}_{{\bf q}\lambda}$, $\lambda =\pm 1$, have been
defined in Eq.~(\ref{eq:eu}). Using the identity~\cite{Edm57}
\begin{equation}
 D^L_{0,\,\lambda}(-\phi,-\theta,\phi) = - \sqrt{\frac{4\,\pi}{2\, L+1}} \, Y_{L\,\lambda}(\theta,\phi) \ ,\qquad \lambda=\pm 1 \ , 
\end{equation}
and, rather than considering the spherical components $j_{q\lambda}({\bf q})$ of the current,
it is possible to work with its component along the unit vector ${\bf e}_{q2}$ defined earlier;
further, ${\bf q}$ can be taken in the $xz$-plane ($\phi=0$), in which case ${\bf e}_{q2}$ is along the
$\hat{\bf y}$ axis, leading to
\begin{align}
\langle JJ;{\bf q} \mid & j_{\gamma,y}({\bf q})\mid JJ\rangle
= \sqrt{4\,\pi} \sum_{L \ge 1}i^{L+1} \,\frac{c_{LJ}}{\sqrt{L\left(L+1\right)}}
\nonumber \\
&\times P_L^1({\rm cos}\, \theta)\,
\langle J|| M_L(q)||J\rangle \ ,
\end{align}
where $P_L^1(x)$ are associated Legendre functions.
For a nucleus of spin $ J > 0$, the number of contributing (purely
imaginary) RMEs of magnetic multipole operators is $[J-1/2]+1$,
and the allowed $L$ are the odd integers between 0 and $2\, J$.  In the
case of a $J=1$ nucleus, for example, it is possible to take
${\bf q}$ along the $\hat{\bf x}$ axis ($\theta=\pi/2$), and determine
$M_1\equiv \langle 1|| M_1(q)||1\rangle$ from
\begin{equation}
\langle 11;q\, \hat{\bf x} \mid  j_y(q\, \hat{\bf x})\mid 11\rangle=\sqrt{\pi}\, M_1 \ .
\end{equation}

Finally, the small $q$ behavior of the charge monopole and quadrupole, and
magnetic dipole RMEs is given by:
\begin{align}
&\langle J||C_0(q=0)|| J\rangle  = \sqrt{\frac{2\, J+1}{4\, \pi}} \, Z \ , \\
&\langle J||C_2(q)|| J\rangle \simeq \frac{1}{12\,  \sqrt{\pi}\, c_{2J}}\,q^2\, Q \ , \quad J \ge 1 \ ,\\
&\langle J||M_1(q)|| J\rangle \simeq\frac{i}{\sqrt{2\, \pi}\, c_{1J}}\,\frac{q}{2\, m} \, \mu \ , \quad J \ge 1/2 \ ,
\end{align} 
where $Q$ and $\mu$ are the quadrupole moment and magnetic moment, defined in terms
of matrix elements of the charge and current density operators $j^0_\gamma({\bf x})$ and ${\bf j}_\gamma({\bf x})$
respectively as
\begin{eqnarray}
Q &=& \langle JJ \mid \int{\rm d}{\bf x} \, j^0_\gamma({\bf x})\, (3\, z^2-{\bf x}^2 ) \mid JJ\rangle \ , \\
\frac{\mu}{2\, m} &=& \langle JJ \mid \frac{1}{2} \int{\rm d}{\bf x} \,\left[ {\bf x}\times 
{\bf j}_\gamma({\bf x})\right]_z \mid JJ\rangle \ .
\end{eqnarray}
They are determined by extrapolating to
zero a polynomial fit (in powers of $q^2$) to the calculated $C_2/q^2$ and
$M_1/q$ on a grid of small $q$ values.  Consequently, the longitudinal
form factor at $q=0$ is normalized as
\begin{equation}
F_L^{\, 2}(q=0) = \frac{Z^2}{4\, \pi} \ ,
\end{equation}
while the transverse form factor $F_T^{\,2}(q)$ vanishes at $q=0$.  Note that
experimental data for $F_L^{\, 2}(q)$ are often reported in the literature as normalized
to one at $q=0$.

\begin{figure}
\includegraphics[width=0.45\textwidth]{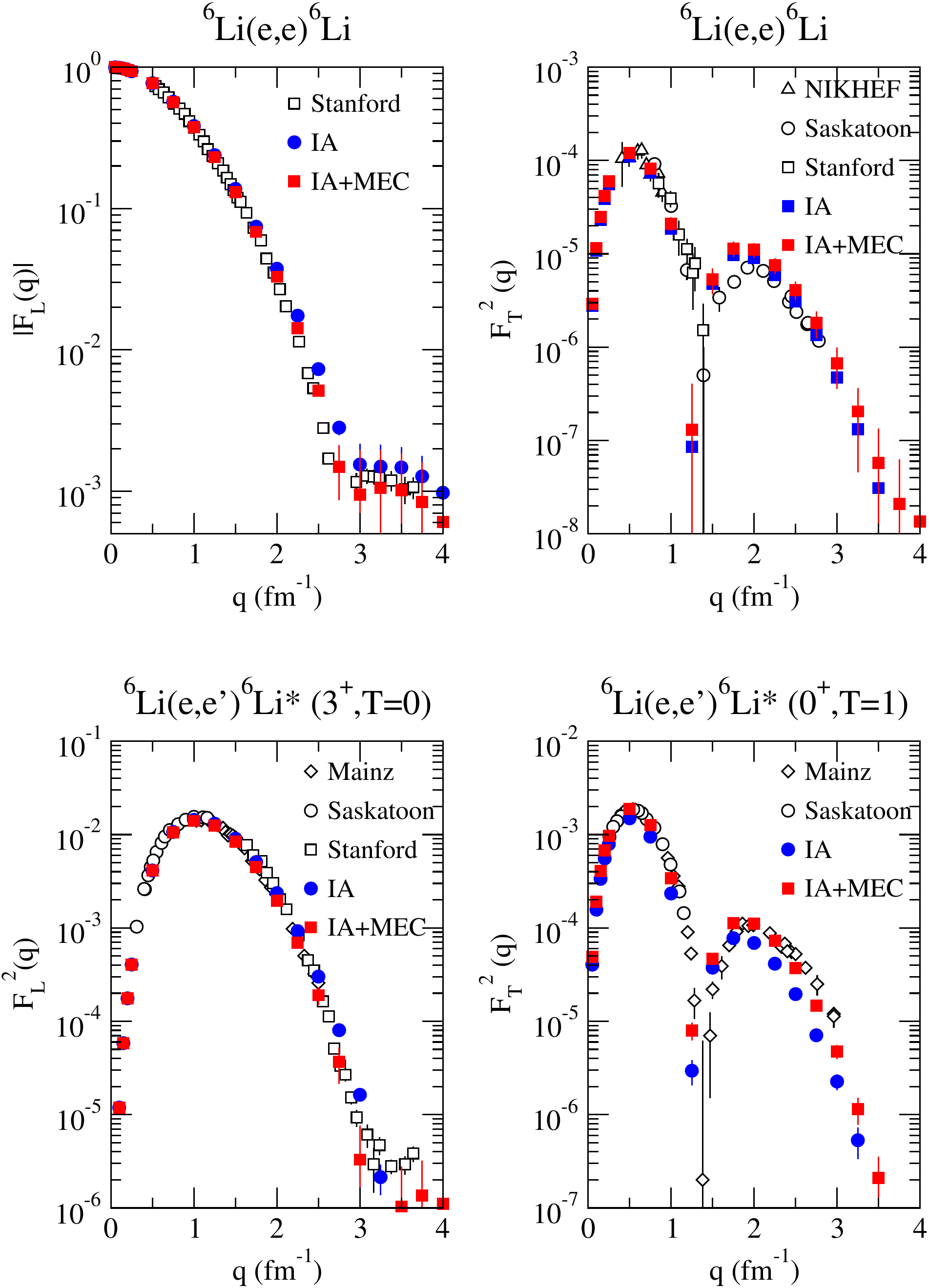}
\caption{The $^6$Li longitudinal elastic (upper left panel), inelastic (bottom left),
and transverse elastic (upper right), and inelastic (bottom right) calculated with VMC
in the impulse approximation (IA), and with the addition of MEC contributions~\cite{Wiringa:1998}.
The results are compared to the experimental data indicated in the legend.
See \citet{Wiringa:1998} and references therein.}
\label{fig:li6ff}
\end{figure}

\begin{figure}
\includegraphics[width=0.45\textwidth]{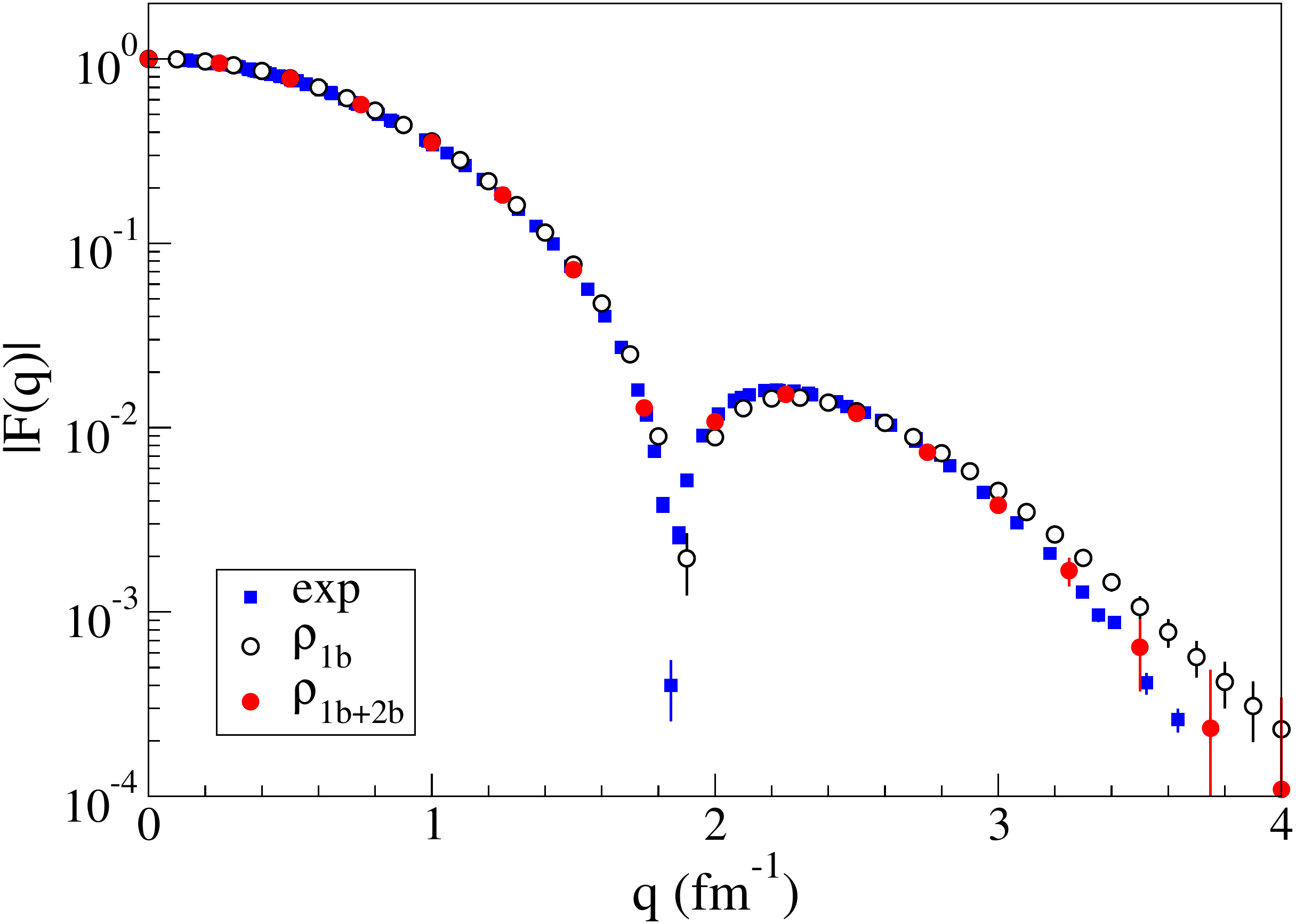}
\caption{The longitudinal elastic form factor of $^{12}$C including one- 
(empty circles) and one- plus two-body operators (red filled circles) calculated 
with GFMC. The results are compared to the experimental data~\cite{lovato2013}.}
\label{fig:c12ff}
\end{figure}

In QMC, matrix elements are evaluated as described in
Sec.~\ref{sec:gfmcimpl}.  The results of elastic and
inelastic electromagnetic form factors for $^6$Li are shown
in Fig.~\ref{fig:li6ff}. The calculations were performed
within the impulse approximation (IA), and two-body operators added
(IA+MEC). Overall, the agreement with the experimental data is excellent.
The contribution of MEC is generally small but its inclusion improves
the agreement between theory and data. In particular, it
shifts the longitudinal elastic and inelastic form factors to slightly 
lower values, and sensibly increases the transverse inelastic form factor.

The longitudinal form factor of $^{12}$C is shown in Fig.~\ref{fig:c12ff}.
The calculation has been performed including only one-body operators (empty 
symbols), and one- plus two-body operators~\cite{lovato2013}.
The experimental data are from a compilation by \citet{Sick:2013,Sick:1982}, and are well
reproduced by theory over the whole range of momentum transfers.
The two-body contributions are negligible at low $q$, and become
appreciable only for $q > 3$ fm$^{-1}$, where they interfere destructively
with the one-body contributions, bringing theory into closer agreement with
experiment.

\subsection{Second $0^+$ state of $^{12}$C: Hoyle state}
\label{sec:Hoyle}

The second $0^+$ state of $^{12}$C is the famous Hoyle state, the gateway
for the triple-alpha burning reaction in stars.  It is a particularly
difficult state for shell model calculations as it is predominantly a
four-particle four-hole state.  However the flexible nature of the
variational trial functions allows us to directly describe this aspect
of the state.  

To do this~\cite{Pieper:2015} two different types of single-particle
wave functions have been used in the $|\Phi_N\rangle$ of Eq.~(\ref{eq:phij}): 1) the five conventional
$0^+$ $LS$-coupled shell model states and 2) states that have an
explicit three-alpha structure.  The first alpha is in the $0s$ shell,
the second in the $0p$ shell and the third in either the $0p$ or
$1s0d$ shells.  The latter can have four nucleons in $1s$ or four in $0d$
or two in $1s$ and two in $0d$.  In addition we allow the third alpha
to have two nucleons in $0p$ and two in $1s0d$ (a two-particle two-hole excitation).
This gives us a total of 11 components in $|\Phi_N\rangle$; a diagonalization
gives the $\Psi_T$ for the ground and excited $0^+$ states.  

\begin{figure}
\includegraphics[width=0.45\textwidth]{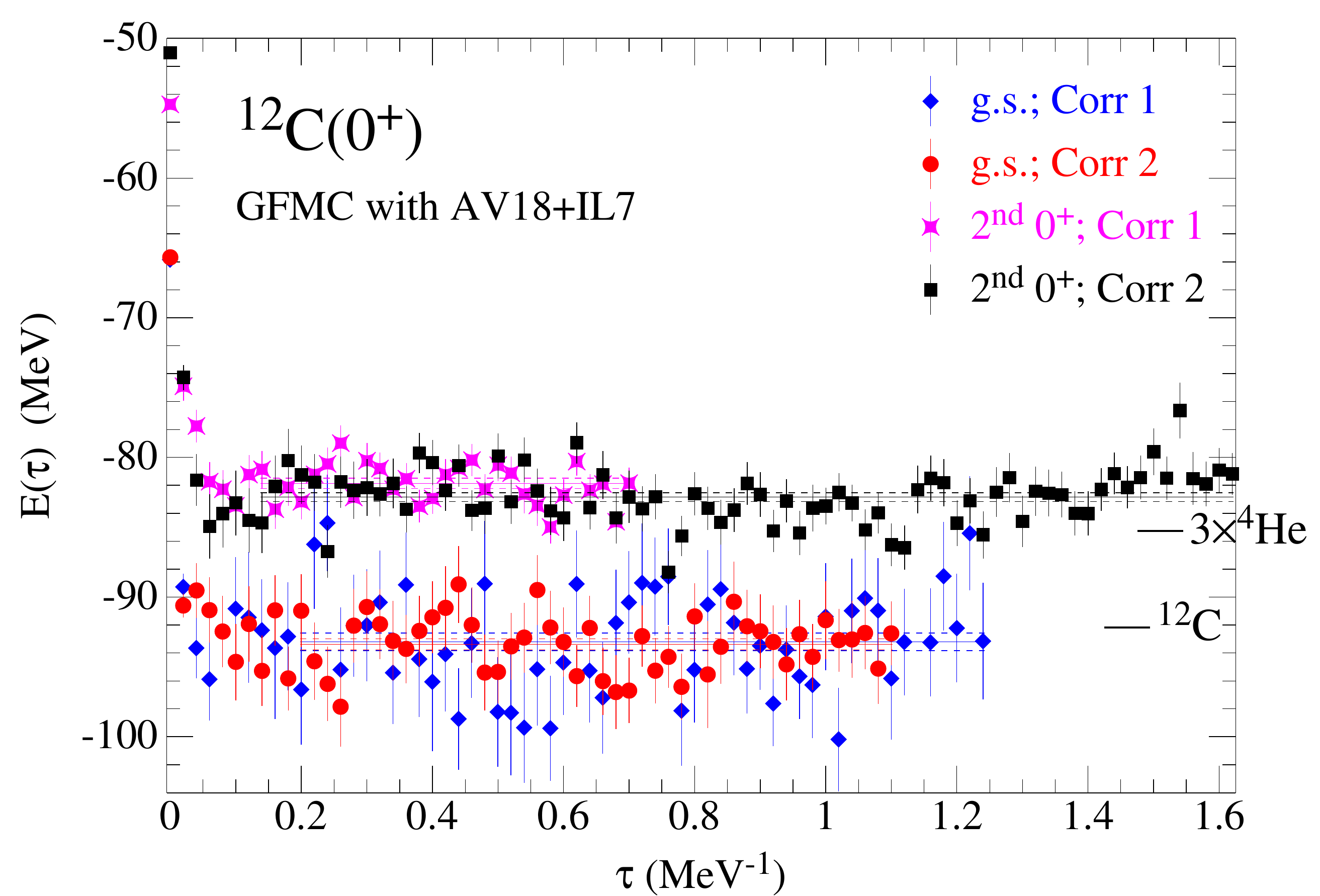}
\caption{GFMC propagated energy versus imaginary time for the first
two $0^+$ states of $^{12}$C.}
\label{fig:c12_e_vs_tau}
\end{figure}

The resulting ground state has less than 1\% of its $\Psi_T$ in the $1s0d$ shell
while the second state has almost 70\% in the $1s0d$ shell.
The GFMC propagation is then done for the first two states; the resulting
energies are shown as a function of imaginary time $\tau$ in Fig.~\ref{fig:c12_e_vs_tau}
which has results for two different initial sets of $\Psi_T$.  The
GFMC rapidly improves the variational energy and then produces
stable (except for Monte Carlo fluctuations) results to large $\tau$.
The resulting ground state energy is very good, $-93.3(4)$ MeV versus the experimental
value of $-92.16$ MeV.  However the Hoyle state excitation energy is
somewhat too high, 10.4(5) versus 7.65 MeV.

\begin{figure}
\includegraphics[width=0.45\textwidth]{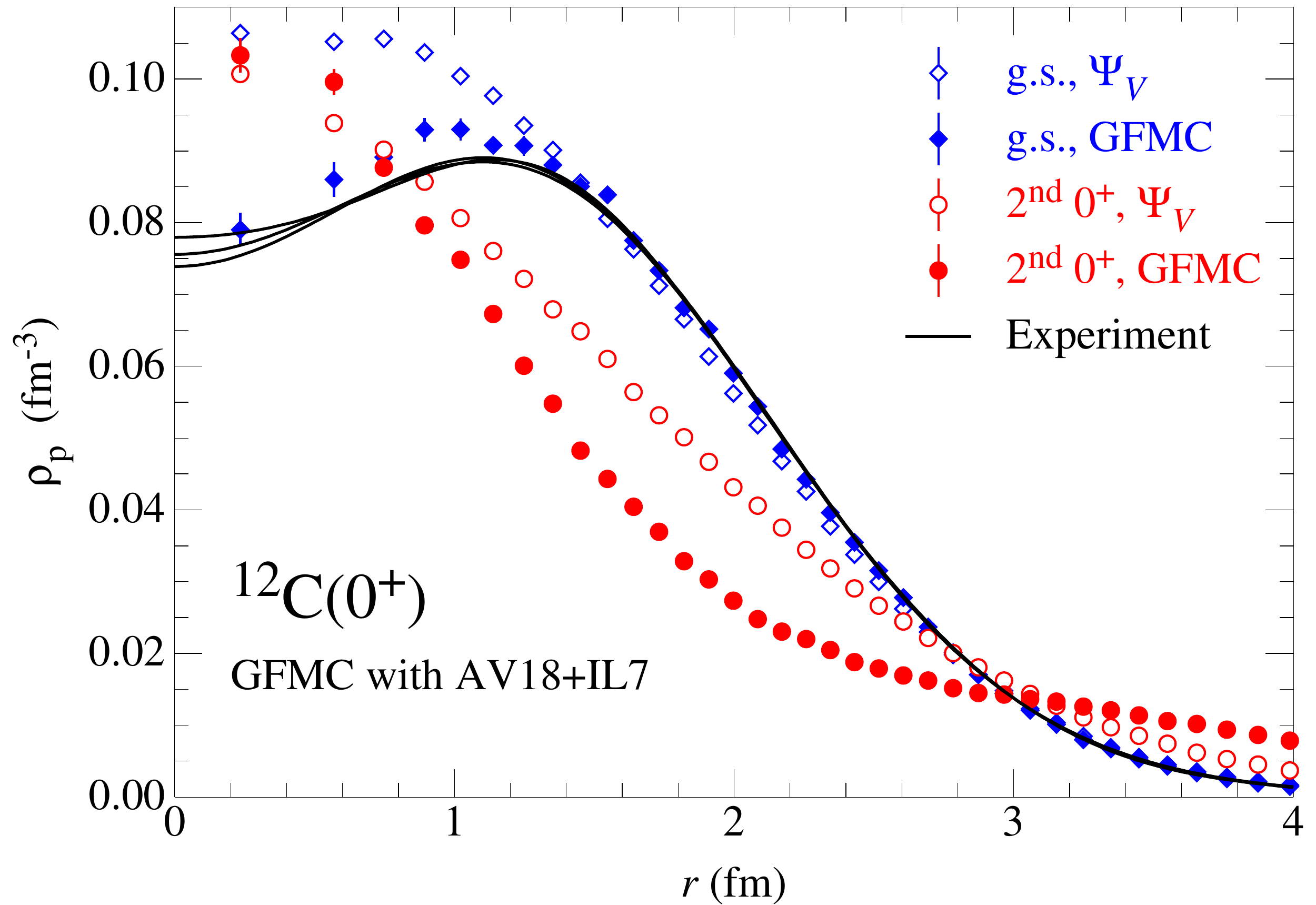}
\caption{VMC and GFMC point-proton densities for the first two $0^+$ states of $^{12}$C.
The experimental band was unfolded from electron scattering data in Ref.~\cite{DeVries:1987}}
\label{fig:c12_densities}.
\end{figure}

Figure~\ref{fig:c12_densities} shows the resulting VMC and GFMC densities
for one of the sets of $\Psi_T$.  The GFMC propagation builds a dip at $r=0$
into the ground-state density which results in good agreement with the experimental
value.  However the Hoyle-state density is peaked at $r=0$ in both the
VMC and GFMC calculations. A possible interpretation of these results is
that the ground state is dominated by an approximately equilateral distribution
of alphas while the Hoyle state has an approximately linear distribution.

\begin{figure}
\includegraphics[width=0.45\textwidth]{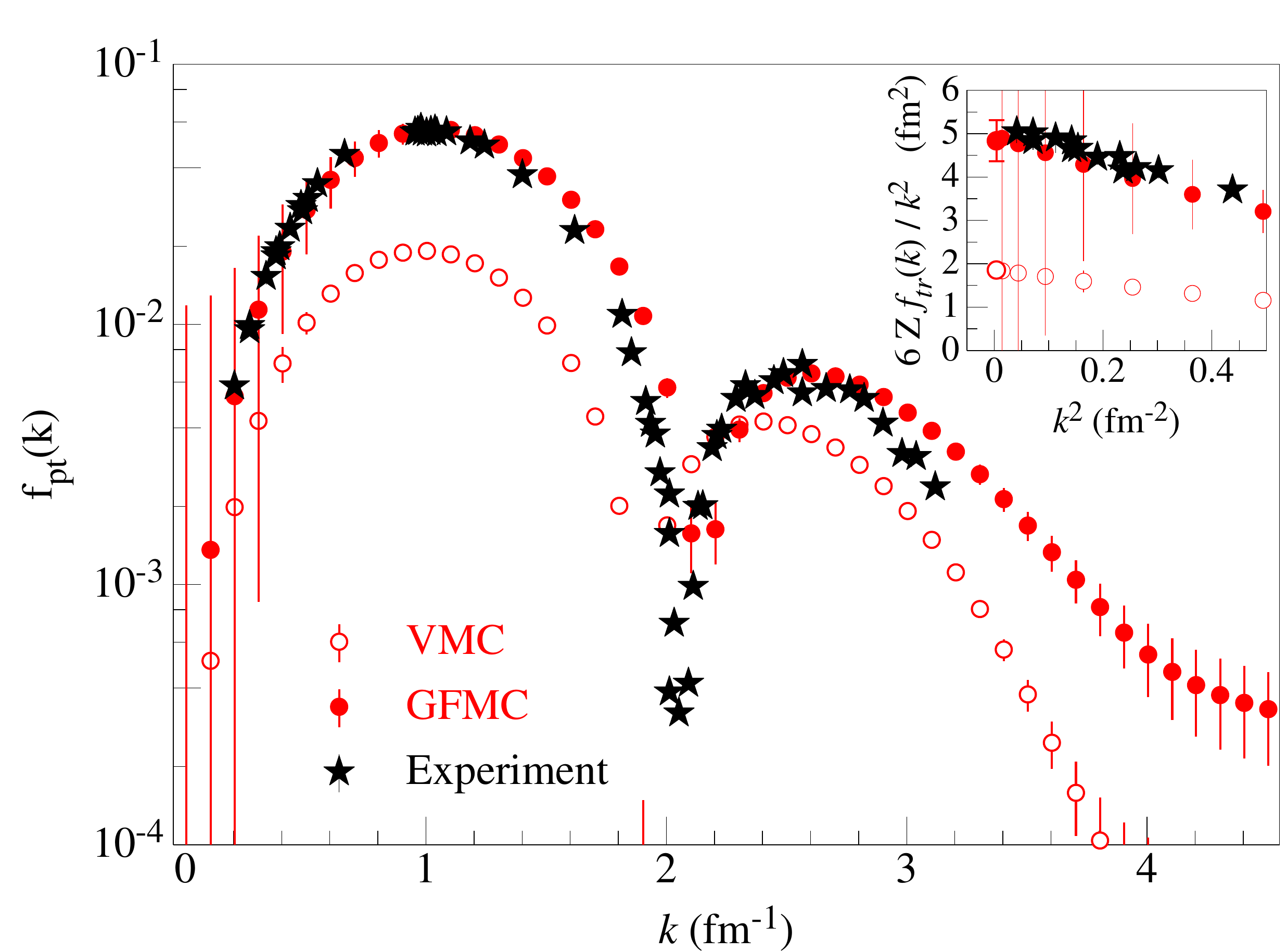}
\caption{VMC and GFMC $E0$ transition form factor between the first two $0^+$ 
states of $^{12}$C in the impulse approximation. The data is from ~\citet{Chernylch:2010}}
\label{fig:c12_e0}
\end{figure}

The calculated impulse $E0$ transition form factor is compared to the experimental
data in Fig.~\ref{fig:c12_e0}.  The insert is scaled such that (linear)
extrapolation to $k^2=0$ gives the $B(E0)$.  The GFMC more than doubles the
VMC result and gives excellent agreement with the data.

\subsection{Magnetic moments and electroweak transitions}

In the impulse approximation (IA), magnetic moments are calculated as
\begin{equation}
\mu^{IA}=\sum_i(e_{N,i}\boldsymbol{L}_i+\mu_{N,i}\boldsymbol{\sigma}_i) \,,
\end{equation}
where $e_{N,i}=(1+\tau_{i,z})/2$, $\mu_N=e_N+\kappa_N$, 
$\kappa_N=(\kappa_S+\kappa_V\tau_{i,z})/2$, and $\kappa_S=-0.120$ 
and $\kappa_V=3.706$ are the isoscalar and isovector combinations of the 
anomalous magnetic moment of the proton and neutron.
The magnetic moment corrections associated with the two-body operators
discussed in the previous sections are obtained from diagonal nuclear matrix elements
\begin{equation}
\mu^{MEC}=-i\lim_{q\rightarrow 0}\frac{2m}{q}\langle 
J^\pi,M_J;T|j_y^{MEC}(q\boldsymbol{\hat x})|J^\pi,M_J;T\rangle\,,
\end{equation}
where the nuclear wave function is taken with $M=J$, the momentum
transfer $q$ is taken along $\hat{\boldsymbol{x}}$, $m$ is the nucleon mass, and the
extrapolation to determine $\mu$ is done from calculations performed at
several small values of $q$.  

\begin{figure}
\includegraphics[width=0.475\textwidth]{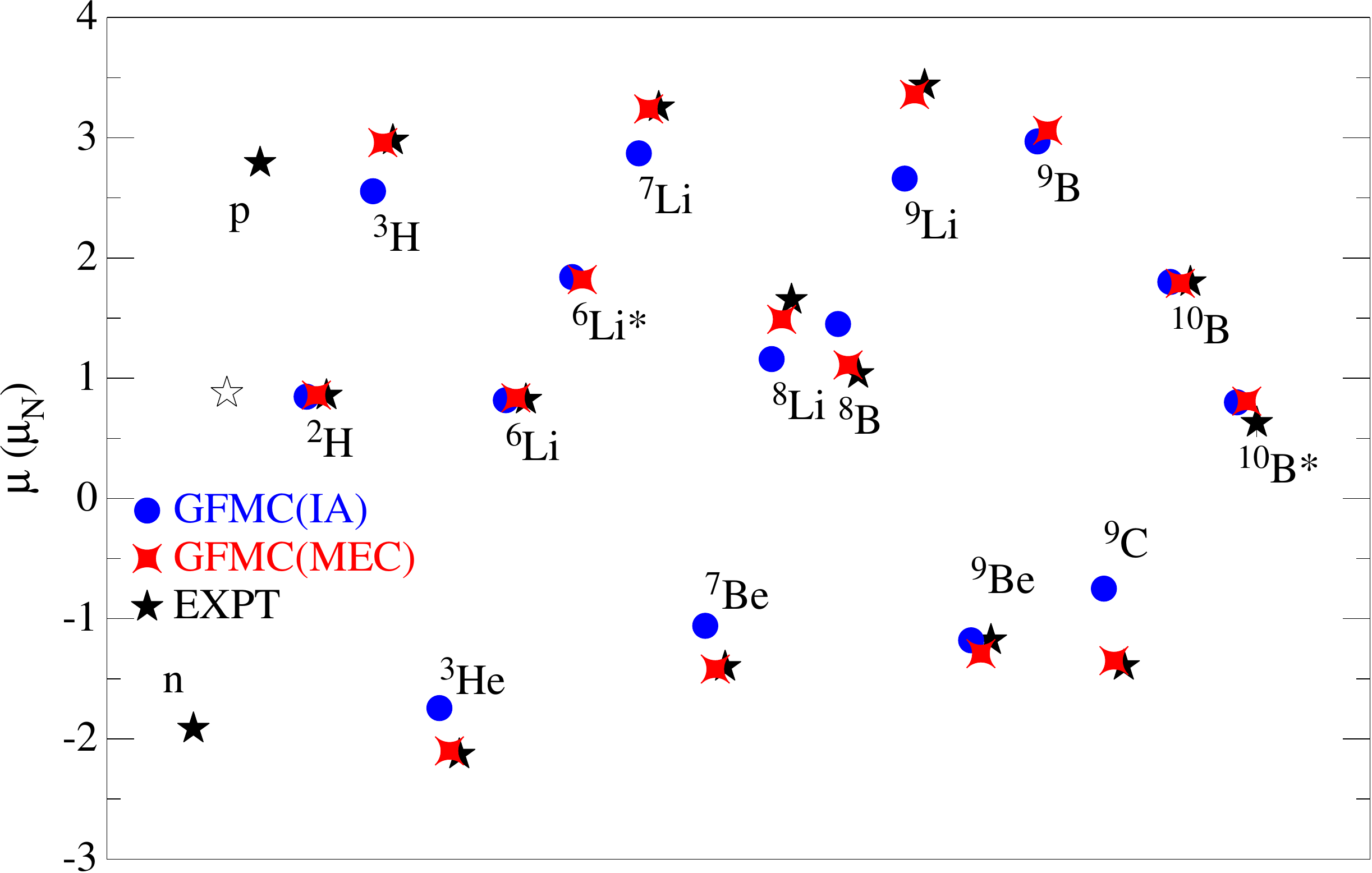}
\caption{Magnetic moments in nuclear magnetons for $A\leq10$ nuclei. 
Black stars indicate
the experimental values~\cite{Tilley02,Tilley04}, while blue dots (red diamonds)
represent GFMC calculations which include the IA one-body EM current
(full $\chi$EFT current up to N3LO); asterisks denote first excited states.
Results are from~\cite{pastore2013,pastore_pc}.}
\label{fig:magmom}
\end{figure}

The total magnetic moments, including MEC derived within $\chi$EFT, 
have been presented in Table~\ref{tab:energies} of Sec.~\ref{subsec:nuclei}. 
Results obtained using MEC derived in the conventional approach and
within $\chi$EFT are very similar, and have been discussed in detail
in~\citet{pastore2013}. 
Here it is interesting to discuss the role of MEC compared to the IA. 
GFMC calculations using AV18+IL7 and chiral two-body currents 
of the magnetic moments are shown in Fig.~\ref{fig:magmom}.
The experimental magnetic moments of $A=2,3$ nuclei
were used to constrain the LECs of the $\chi$EFT; all the results
for heavier nuclei are predictions.

In many cases the two-body currents significantly change the IA results
and in all of these much better agreement with experiment is achieved.
The contribution of MEC is generally larger for even-odd and odd-even nuclei, 
in particular for $^9$Li and $^9$C.
The exceptions are $^9$Be and $^9$B, which with their [441] spatial symmetry
are essentially single nucleons outside a $^8$Be($0^+)$ core; on average, 
these have no OPE interaction with the core and therefore no significant
MEC contribution. 
For odd-odd isoscalar nuclei, the IA results are already in good agreement 
with experimental data; only for the $T=1$ nuclei
$^8$Li and $^8$B are the MEC contributions significant.

$M1$ and $E2$ electromagnetic transitions for $A$=6--9 nuclei have been calculated 
with GFMC. The one-body part of these operators are given by
\begin{align} 
M1&=\mu_N\sum_i (L_i+g_p S_i)(1+\tau_{i,z})/2+g_n S_i(1-\tau_{i,z})/2 \,,
\nonumber \\
E2&=e\sum_i\left[r_i^2 Y_2(\hat r_i)\right](1+\tau_{i,z}) \,
\end{align}
where $Y$ is a spherical harmonic, $L$ and $S$ the orbital and spin angular momentum
operators, and $g_p$ and $g_n$ the gyromagnetic ratio of protons and neutrons.
MEC are also included in the $M1$ transitions.
The nuclear matrix elements can be compared with the experimental widths.
In units of MeV, they are given by \cite{preston1962}
\begin{align}
\Gamma_{M1}=\frac{16\pi}{9} \, \left(\frac{\Delta E}{\hbar c}\right)^3\,B(M1) \,,
\nonumber \\
\Gamma_{E2}=\frac{4\pi}{75} \, \left(\frac{\Delta E}{\hbar c}\right)^5\,B(E2) \,,
\end{align}
where $\Delta E$ is the energy difference between the 
final and initial state and 
$B(M1) = \langle J_F||M1||J_I\rangle^2/(2J_I+1)$ is in units of $\mu_N^2$ and
$B(E2) = \langle J_F||E2||J_I\rangle^2/(2J_I+1)$ is in units of $e^2$ fm$^4$.

\begin{figure}
\includegraphics[width=0.475\textwidth]{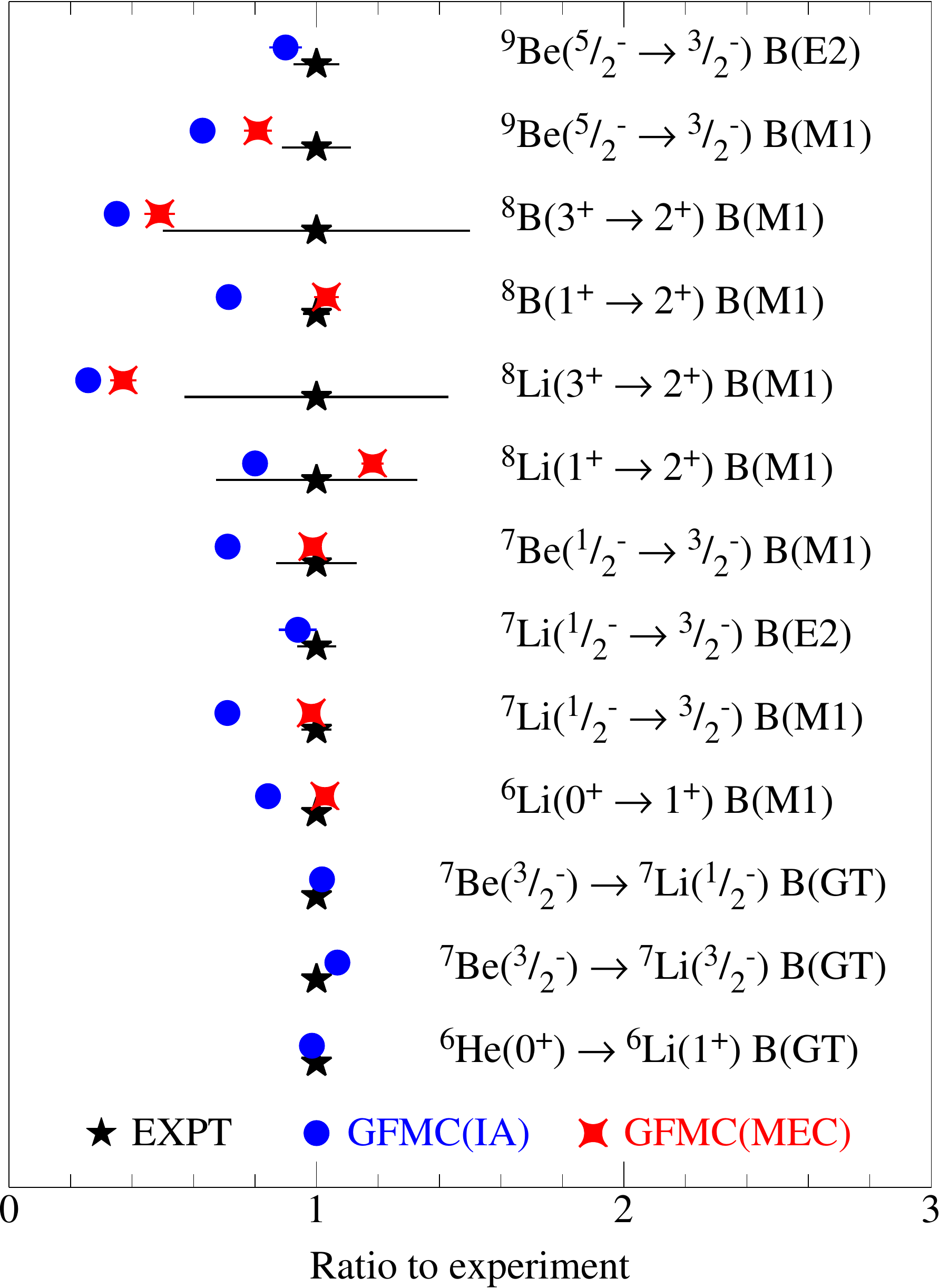}
\caption{Ratio of calculated to experimental $M1$, $E2$~\cite{pastore2013}, 
and GT reduced transition probabilities \cite{pervin2007} in $A\leq9$ nuclei.  
Symbols are as in Fig.~\ref{fig:magmom}.}
\label{fig:EW_transitions}
\end{figure}

A number of calculated electromagnetic transition strengths are compared to
experiment in Fig.~\ref{fig:EW_transitions}.  
Many additional transitions within $^8$Be are reported in \citet{pastore2014}.
Again GFMC calculations were made using AV18+IL7 and chiral two-body currents.  
The two-body currents make large corrections to the IA results for the 
$M1$ transitions; these often result in excellent agreement with experiment.

Weak decays of $A$=6, 7 nuclei have been evaluated using QMC
but much more needs to be done in the future.
In IA, the weak Fermi (F) and Gamow-Teller (GT) operators 
to be evaluated are:
\begin{eqnarray}
 {\rm F}&=&{\sum_{i}} \tau_{i\pm} \, , \nonumber \\
 {\rm GT}&=&{\sum_{i}} {\bf \sigma}_{i}\tau_{i\pm} \, .
\end{eqnarray}
A first calculation for the weak decays $^6$He$(\beta^-)^6$Li and
$^7$Be$(\varepsilon)^7$Li was made by \citet{Schiavilla:2002b} using
VMC wave functions for the AV18+UIX Hamiltonian, and incorporating conventional 
MEC as discussed in Sec.~\ref{subsec:2bweak}.
Parameters in the MEC were fixed to reproduce $^3$H $\beta$-decay \cite{Sch98}.

The $^6$He $\beta$-decay is a pure GT transition, while the $^7$Be
electron capture is a mixed F+GT transition to the ground state, and a
GT transition to the first excited state of $^7$Li.
These are superallowed decays where the dominant spatial symmetry of the
parent and daughter states is the same, e.g., [42]$\rightarrow$[42] in
$A$=6 and [43]$\rightarrow$[43] in $A$=7.
In these cases, the F and GT matrix elements are of order 1--2 and
the MEC contributions are only a 2--4\% correction.

Subsequently, a GFMC calculation for these transitions was made by
\citet{pervin2007} based on the AV18+IL2 Hamiltonian, but only in
the IA.
The GFMC results for these three B(GT) reduced transition probabilities are 
shown at the bottom of Fig.~\ref{fig:EW_transitions}.
These are already in fairly good agreement with experiment, and small
MEC corrections will not shift the results by much.

Weak decays in the $A$=8,9 nuclei pose a much bigger challenge.
For example, $^8$He$(\beta^-)^8$Li goes from a predominantly [422] symmetry
state to multiple $1^+$ excited states, but primarily to the first excited 
state in $^8$Li.  
The latter is predominantly a [431] symmetry state with only a small [422] 
component, so the allowed GT matrix element is of order 0.1--0.2.
Similarly, the $^8$Li$(\beta^-)^8$Be and $^8$B$(\beta^+)^8$Be decays
are transitions from large to small components, with the added complication
that the final $2^+$ state in $^8$Be is a moderately broad resonant state.
GFMC calculations in impulse approximation underpredict the $A$=8 experimental
matrix elements by a factor of two \cite{pastore_pc}.
It is possible that GFMC does an inadequate job of accurately determining small
components in the final state wave functions, or that the specific Hamiltonian
does not induce the required correlations.
However, if the magnitude of the MEC corrections is comparable to that
in the $A$=6,7 superallowed decays, then the MEC will be relatively much
more important in the allowed decays and may resolve the problem.
This is an important task for future QMC studies.

\subsection{Electroweak Response of Light Nuclei}
\label{sec:ewrespon}
The response to electroweak probes provides direct information
on dynamics in the nucleus.  The rich structure of nuclear interactions
and currents, combined with the availability of different probes,
offers the opportunity to study many intriguing aspects of nuclear
dynamics.  Here we describe theoretical approaches
for describing inclusive scattering of electrons and neutrinos
from a nucleus, including both sum-rule techniques and direct
computations of response functions, as well as comparisons to
available experimental data.  In
the last few years inclusive neutrino scattering from nuclear targets
has seen a surge in interest, spurred by the excess, at relatively low
energy, of measured cross section relative to theoretical calculations 
observed in recent neutrino quasi-elastic scattering data on
$^{12}$C~\cite{Aguilar:2008,Butkevich10}.
Analyses based on these calculations have led to speculations that our
present understanding of the nuclear response to charge-changing weak
probes may be incomplete~\cite{Benhar:2010}.
However, it should be emphasized that the calculations on which these
analyses are based use rather crude models of nuclear structure---Fermi
gas or local density approximations of the nuclear matter spectral function---and
simplistic treatments of the reaction mechanism, and
should therefore be viewed with some skepticism.
%
%
\label{sec:eucl}
The differential cross section for neutrino $\nu$ and
antineutrino $\overline{\nu}$ inclusive scattering off a nucleus, specifically
the processes $A(\nu_l,\nu_l)$ and $A(\overline{\nu}_l,\overline{\nu}_l)$
induced by neutral weak currents (NC), and the processes
$A(\nu_l,l^-)$ and $A(\overline{\nu}_l,l^+)$ induced by charge-changing weak
currents (CC), can be expressed in terms of five response functions
$R_{\alpha\beta}$ as
\begin{align}
\left(\frac{ {\rm d}\sigma}{ {\rm d}\epsilon^\prime 
{\rm d}\Omega}\right)_{\nu/\overline{\nu}}&= \frac{G^2}{2\pi^2}\, 
k^\prime \epsilon^\prime \,F(Z,k^\prime)\, {\rm cos}^2 \frac{\theta}{2}\Bigg[
R_{00} +\frac{\omega^2}{ q^2}\, R_{zz} 
\nonumber \\
&- \frac{\omega}{q} R_{0z} +
\left( {\rm tan}^2\frac{\theta}{2}+\frac{Q^2}{2\, q^2}\right) R_{xx} \nonumber \\
&\mp {\rm tan}\frac{\theta}{2}\, \sqrt{  {\rm tan}^2\frac{\theta}{2}+\frac{Q^2}{ q^2} } \, R_{xy}
\Bigg] \ ,
\label{eq:xsw}
\end{align}
where $G$=$G_F$ for the NC processes and $G$=$G_F \, {\rm cos}\, \theta_C$ for the
CC processes, and the $-$ ($+$) sign in the last term is relative to the $\nu$ ($\overline{\nu}$)
initiated reactions.  The value of
$G_F$ is $1.1803\times 10^{-5}$ GeV$^{-2}$ as obtained from an analysis of
super-allowed $0^+ \rightarrow 0^+$ $\beta$-decays by \citet{Towner:1999}---this
value includes radiative corrections---while ${\rm cos}\, \theta_C$ is taken as 0.97425
from \citet{PDG}.  The initial neutrino four-momentum is $k^\mu=(\epsilon, {\bf k})$, the final
lepton four momentum is $k^{\mu \,\prime}=(\epsilon^\prime,{\bf k}^\prime)$, and the
lepton scattering angle is denoted by $\theta$. 
The lepton energy and momentum transfers are defined as $\omega=\epsilon-\epsilon^\prime$
and ${\bf q}={\bf k}-{\bf k}^\prime$, respectively, and the squared four-momentum
transfer as $Q^2=q^2-\omega^2 > 0$.  The Fermi function $F(Z,k^\prime)$
accounts for the Coulomb distortion of the final lepton wave function in the charge-raising
reaction,
\begin{equation}
F(Z,k^\prime) = 2\, (1+\gamma)\, (2\, k^\prime\, r_A)^{2\,\gamma-2}\, {\rm exp} \left(\pi\, y\right)\,
\Bigg| \frac{\Gamma(\gamma+i\, y)}{\Gamma(1+2\,\gamma)} \Bigg|^2 \ , \\
\end{equation}
with
\begin{equation}
\gamma=\sqrt{1-\left(Z\,\alpha\right)^2} \ ;
\end{equation}
otherwise it is set to one.  Here $y = Z\, \alpha \, \epsilon^\prime/k^\prime$,
$\Gamma(z)$ is the gamma function, $r_A$ is the nuclear radius,
and $\alpha$ is the fine structure constant.  There are in principle
radiative corrections for the CC and NC processes due to bremsstrahlung and
virtual photon- and $Z$-exchanges.  These corrections have been evaluated in the deuteron
by \citet{Towner:1998}, and \citet{Kurylov:2002} at the low energies
($\sim 10$ MeV) relevant for the SNO experiment, which measured the neutrino
flux from the $^8$B decay in the sun.  They are not considered
further below, since our focus here is primarily on scattering of neutrinos
with energies larger than 100 MeV, and we are not concerned with
discussing cross section calculations with \% accuracy in this regime.
The nuclear response functions are defined as
\begin{eqnarray}
\label{eq:r1}
R_{00}(q,\omega) &=\sum_f \delta( \omega+E_0-E_f)\,
\nonumber \\
&\times\langle f\!\mid j^0({\bf q},\omega) \mid \!0\rangle\langle f\!\mid j^0({\bf q},\omega) \mid\! 0 \rangle^*  \ ,\\
\label{eq:r2}
R_{zz}(q,\omega) &=\sum_f \delta( \omega+E_0-E_f)\,
\nonumber \\
&\times\langle f\!\mid j^z({\bf q},\omega) \mid \! 0\rangle\langle f\!\mid j^z({\bf q},\omega) \mid\! 0 \rangle^*\ ,  \\
\label{eq:r3}
R_{0z}(q,\omega) &=2 \sum_f \delta( \omega+E_0-E_f)\, 
\nonumber \\
&\times{\rm Re}\Big[
\langle f\! \mid j^0({\bf q},\omega) \mid \! 0 \rangle\langle f\! \mid j^z({\bf q},\omega) \mid\! 0 \rangle^*\Big]  \ ,\\
\label{eq:r4}
R_{xx}(q,\omega) &= \sum_f \delta( \omega+E_0-E_f)\,
\nonumber \\
&\times\Big[
\langle f\!\mid j^x({\bf q},\omega) \mid \! 0 \rangle\langle f\!\mid j^x({\bf q},\omega) \mid\! 0\rangle^* 
\nonumber \\
&+\,\langle f\!\mid j^y({\bf q},\omega) \mid \! 0\rangle\langle f\! \mid j^y({\bf q},\omega) \mid\! 0 \rangle^*  \Big] \ ,\\
R_{xy}(q,\omega) &=2 \sum_f \delta( \omega+E_0-E_f)\, 
\nonumber \\
&\times{\rm Im}\Big[
\langle f\!\mid j^x({\bf q},\omega) \mid\! 0 \rangle\langle f\!\mid j^y({\bf q},\omega) \mid\! 0 \rangle^* \Big] \ ,
\label{eq:r5}
\end{eqnarray}
where $\mid \! 0\rangle$ represents the initial
ground state of the nucleus of energy $E_0$,
$\mid \!\! f\rangle$ its final state of energy $E_f$, and
an average over the initial spin projections is understood.
The three-momentum
transfer ${\bf q}$ is taken along the $z$-axis (i.e., the spin-quantization axis), and
 $j^\mu({\bf q},\omega)$ is the time component (for $\mu=0$) or space
component (for $\mu=x,y,z$) of the NC or CC.  Note that in the model
of electroweak currents adopted here, their $\omega$-dependence
enters through the dependence on $Q^2$ of the electroweak form
factors of the nucleon and $N$-to-$\Delta$ transition.  Below, when
discussing QMC calculations of
$R_{\alpha\beta}(q,\omega)$, the four-momentum $Q^2$
transfer is fixed at the top of the quasi-elastic peak, and the form factors
are evaluated at $Q_{\rm qe}^2=q^2-\omega^2_{\rm qe}$ with
$\omega_{\rm qe}=\sqrt{q^2+m^2}-m$, so that the only $\omega$-dependence
left in $R_{\alpha\beta}(q,\omega)$ is that from the energy-conserving $\delta$-function.

The expression above for the CC cross section is valid in the limit $\epsilon^\prime \simeq
k^\prime$, in which the lepton rest mass is neglected.   At small incident neutrino energy,
this approximation is not correct.  Inclusion of the lepton rest mass leads to
changes in the kinematical factors multiplying the various response functions.  The resulting
cross section can be found in \citet{Shen:2012}. 

The cross section for inclusive electron scattering follows from Eq.~(\ref{eq:xsw})
by using current conservation to relate the longitudinal component of the
current to the charge operator via $j^z_\gamma(q\hat{\bf z})=(\omega/q) 
j^0_\gamma(q\hat{\bf z})$ and by noting that the interference response $R_{xy}$
vanishes, since it involves matrix elements of the vector and axial
parts of the current ${\bf j}_{NC}$ or ${\bf j}_{CC}$ of the type
${\rm Im}\left(\langle j^x\rangle \langle j_5^y \rangle^*
+\langle j_5^x\rangle \langle j^y \rangle^*\right)$. One finds
\begin{align}
\left(\frac{ {\rm d}\sigma}{ {\rm d}\epsilon^\prime 
{\rm d}\Omega}\right)_{\!\! e}&=& \sigma_{M}
\Bigg[\frac{Q^4}{ q^4}\, R_L +
\left( {\rm tan}^2\frac{\theta}{2}+\frac{Q^2}{2\, q^2}\right) R_T \Bigg] \ ,
\label{eq:xsem}
\end{align}
where $\sigma_M$ is the Mott cross section, and the longitudinal ($L$) and
transverse ($T$) response functions are
defined as in Eqs.~(\ref{eq:r1}) and~(\ref{eq:r4}) with $j^\mu$
replaced by $j^\mu_\gamma$.

The accurate calculation of the inclusive response  at low and intermediate energy
and momentum transfers (say, $q \lesssim 0.5$ GeV/c and $\omega$ in the
quasi-elastic region) is a challenging quantum many-body problem, since
it requires knowledge of the whole excitation spectrum of the nucleus and
inclusion in the electroweak currents of one- and two-body terms.  In the
specific case of inclusive weak scattering, its difficulty is compounded by the
fact that the energy of the incoming neutrinos is not known (in contrast
to inclusive $(e,e^\prime)$ scattering where the initial and final electron energies are
precisely known).  The observed cross section for a given energy and angle of the
final lepton results from a folding with the energy distribution of the incoming neutrino
flux and, consequently, may include contributions from energy- and momentum-transfer
regions of the nuclear response where different mechanisms are at play: the threshold region,
where the structure of the low-lying energy spectrum and collective effects are important; the
quasi-elastic region, which is dominated by scattering off individual nucleons and nucleon pairs;
and the $\Delta$ resonance region, where one or more pions are produced in the final
state.

The simplest model of nuclear response is based on the plane-wave
impulse approximation (PWIA).  The response is assumed to be given
by an incoherent sum of scattering processes
off single nucleons that propagate freely in the final state.  In PWIA
the struck nucleon with initial momentum ${\bf p}$ absorbs the
momentum ${\bf q}$ of the external field and transitions to
a free particle state of momentum ${\bf p}+{\bf q}$ without
suffering any interactions with the residual $(A-1)$ system.
In the most naive formulation of PWIA, the response is obtained from the single-nucleon
momentum distribution in the ground-state of the nucleus and the nucleon electroweak
form factors, 
\begin{align}
R^{\rm PWIA}_{\alpha \beta} (q,\omega)
&=\int {\rm d}{\bf p}\, N({\bf p}) \, x_{\alpha\beta}({\bf q},{\bf p})\,
\nonumber \\
\delta\!\Bigg[ \omega-\overline{E}
&-\frac{\left({\bf p}+{\bf q}\right)^2}{2\,m} - 
\frac{p^2}{2\,(A-1) m}\Bigg] \ ,
\end{align}
where $x_{\alpha\beta}$ describes the coupling to the external electroweak
field, $N({\bf p})$ is the nucleon momentum distribution, and the effects
of nuclear interactions are subsumed in the single parameter $\overline{E}$, which
can be interpreted as
an average binding energy.  The remaining terms in the $\delta$-function
are the final energies of the struck nucleon and recoiling $(A-1)$
system, respectively.  In cases where the momentum transfer ${\bf q}$ is
large, it may be more appropriate to use relativistic expressions for the
coupling $x_{\alpha\beta}$ and final nucleon kinetic energy.  

More sophisticated formulations of PWIA are based on the spectral function,
thus removing the need for including the parameter $\overline{E}$.  To this
end, it is useful to first express the response in terms of the real-time
propagation of the final state as
\begin{align}
R_{\alpha \beta} (q,\omega)
&=\frac{1}{2\pi} \int^\infty_{-\infty} {\rm d}t\, {\rm e}^{i(\omega+E_0)t}\, 
\langle0\!\mid O^\dagger_\beta({\bf q}) \, {\rm e}^{-iHt} \,O_\alpha({\bf q}) \mid\!0\rangle \nonumber \\
&\equiv\frac{1}{2\pi} \int^\infty_{-\infty} {\rm d}t\, {\rm e}^{i(\omega+E_0)t}\, \widetilde{R}_{\alpha\beta}(q,t)
\end{align}
where the $O_\alpha$'s denote the relevant components of the
electroweak current of interest.  Since the interactions of
the struck nucleon with the remaining nucleons are neglected,
the $A$-body Hamiltonian reduces to $H\simeq K(A)+H(1,\dots,A-1)$,
where $K(A)$ is the kinetic energy operator of nucleon $A$ (the struck
nucleon) and $H(1,\dots,A-1)$ is the Hamiltonian for the
remaining  (and fully interacting) $A-1$ nucleons.  

Ignoring the energy dependence in the spectral function
reproduces the naive PWIA response, since integrating the spectral 
function $S({\bf p}, E)$
recovers the momentum distribution.  
At large values of the momentum transfer
($q \sim 1$ GeV/c), one would expect the spectral function approach
to be reasonably accurate.  There will be significant corrections,
however, arising from the fact that in some instances the struck
nucleon is not only in a mean field, but is
strongly interacting with one or more other nucleons.
More sophisticated treatments are required to get a complete picture.

PWIA calculations of the longitudinal
response measured in $(e,e^\prime)$ scattering, for example,
grossly overestimate the data
in the quasi-elastic peak region~\cite{carlson1998}.  They also lead to an incorrect strength
distribution, since they underestimate energy-weighted sum rules of the
longitudinal (and transverse) response functions.  
Much of this overestimate can be attributed to the fact the charge
can propagate through the interaction, not only through the movement of nucleons.

It is possible to compute sum rules of the electroweak response as 
ground state expectation values that are much more accurate than
approximations to the full response.  One can also calculate integral
transforms of the response, which can be directly compared to experimental
data and provide a great deal of information about the full response.
Here we review results for sum rules and Euclidean response.

\subsection{Sum rules of electroweak response functions}
Sum rules provide a powerful tool for studying integral
properties of the response of a nuclear many-body system
to an external probe.  Of particular interest
are those at constant three-momentum transfer,
since they can be expressed as ground-state expectation
values of appropriate combinations of the electroweak current operators
(and commutators of these combinations with the Hamiltonian in the
energy-weighted case), thus avoiding the need for
computing the nuclear excitation spectrum.  

In the electromagnetic case, the (non-energy-weighted) sum
rules are defined as~\cite{Carlson:2002}
\begin{equation}
\label{eq:sr}
S_\alpha (q)= C_\alpha \int_{\omega^+_{\rm th}}^\infty 
{\rm d}\omega\, \frac{R_\alpha(q,\omega)} {G_E^{p\, 2}(Q^2)}\ ,
\end{equation}
where $R_\alpha(q,\omega)$ is the longitudinal ($\alpha=L$) 
or transverse ($\alpha=T$) response function, $\omega_{\rm th}$ is the
energy transfer corresponding to the inelastic threshold, $G_E^p(Q^2)$
is the proton electric form factor (evaluated at four-momentum transfer
$Q^2=q^2-\omega^2$), and the $C_\alpha$'s are appropriate normalization
factors, given by
\begin{equation}
C_L=\frac{1}{Z} \ , 
\qquad C_T=\frac{2}{\left(Z\, \mu_p^2+N\, \mu_n^2\right)}\, \frac{m^2}{q^2} \ .
\end{equation}
Here $Z$ ($N$) and $\mu_p$ ($\mu_n$) are
the proton (neutron) number and magnetic moment, respectively.  These
factors have been introduced so that $S_\alpha( q\rightarrow \infty )\simeq 1$
under the approximation that the nuclear electromagnetic charge and current
operators originate solely from the charge and spin magnetization of individual
protons and neutrons and that relativistic corrections to these one-body
operators---such as the Darwin-Foldy and spin-orbit terms in the charge
operator---are ignored.  The sum rules above can be expressed~\cite{McVoy:1962}
as ground-state expectation values of the type
\begin{equation}
\label{eq:sri}
S_\alpha(q)\!=\!C_\alpha  \Big[ \langle 0|
O_\alpha^\dagger({\bf q}) \, O_\alpha({\bf q}) |0\rangle -|\langle 0;{\bf q}|
 O_\alpha({\bf q}) |0 \rangle|^2 \Big]  \ ,
\end{equation}
where $O_\alpha({\bf q})$ is either the charge $j^0_\gamma({\bf q})$ ($\alpha=L$)
or transverse current ${\bf j}_{\gamma,\perp}({\bf q})$ ($\alpha=T$) operator divided
by $G_E^p(Q^2)$, $|0;{\bf q}\rangle$ denotes the ground state of the nucleus
recoiling with total momentum ${\bf q}$, and an average over the spin projections
is understood.
The $S_\alpha(q)$ as defined in
Eq.~(\ref{eq:sr}) only includes the inelastic contribution to $R_\alpha(q,\omega)$,
i.e., the elastic contribution represented by the second term on the r.h.s.~of
Eq.~(\ref{eq:sri}) has been removed.  It is proportional to the square of the
longitudinal
$F_L$ or transverse $F_T$ elastic form factor.  For $J^\pi=0^+$ states
like $^4$He or $^{12}$C, $F_T$
vanishes, while $F_L(q)$, discussed in Sec.\ref{sec:form} is given by
$F_L(q)\!=\!G_E^p(Q^2_{\rm el}) \,\langle 0;{\bf q}|  O_L({\bf q}) |0 \rangle/Z$,
with the four-momentum transfer $Q^2_{\rm el}=q^2-\omega_{\rm el}^2$
and $\omega_{\rm el}$ corresponding to elastic scattering,
$\omega_{\rm el}=\sqrt{q^2+m_A^2}-m_A$ ($m_A$ is the rest mass
of the nucleus).

In the case of NC and CC weak response functions, the (non-energy-weighted)
sum rules are generally defined as~\cite{Lovato:2014}
\begin{equation}
S_{\alpha\beta} (q) = C_{\alpha\beta} \int_{\omega_{\rm el}}^\infty 
{\rm d}\omega\, \, R_{\alpha\beta}(q,\omega) \ .
\end{equation}
and can be expressed as
\begin{align}
S_{\alpha\beta}(q)&=C_{\alpha\beta} \,
 \langle 0 | j^{\alpha \dagger}({\bf q}) \,j^\beta({\bf q})
 + (1- \delta_{\alpha\beta})\,  j^{\beta \dagger}({\bf q})\, j^\alpha({\bf q}) | 0\rangle \\
S_{xy}(q)&=C_{xy}\,  {\rm Im}\, 
 \langle 0 | j^{x \dagger}({\bf q}) \, j^y({\bf q})
 -  j^{y \dagger}({\bf q})\, j^x({\bf q}) |0\rangle
\end{align}
where
the $C_{\alpha\beta}$'s are convenient
normalization factors (see below), $\alpha \beta=00$, $zz$, $0z$, and $xx$, and
for $\alpha\beta=xx$ the expectation value of
$j^{x \dagger} j^x+j^{y \dagger} j^y$ is computed.  Note that
the electroweak nucleon and $N$-to-$\Delta$ form factors
in $j^\mu_{NC/CC}$ are taken to be functions of $q$ only
by evaluating them at $Q^2_{\rm qe}$, at the top of the quasi-elastic peak.
In contrast to the electromagnetic sum rules above, the $S_{\alpha\beta}(q)$
include the elastic and inelastic contributions; the former
are proportional to the square of electroweak form factors of the nucleus.
In the large $q$ limit, these nuclear form factors decrease rapidly with $q$, and the
sum rules reduce to the incoherent sum of single-nucleon contributions.  The
normalization factors $C_{\alpha\beta}$ are chosen such that
$S_{\alpha\beta}(q\rightarrow \infty) \simeq 1$, for example
\begin{equation}
C^{-1}_{xy}=-\frac{q}{m}\, G_A(Q^2_{\rm qe})\left[ Z\, 
\widetilde{G}_M^p(Q^2_{\rm qe})-N\, \widetilde{G}_M^n(Q^2_{\rm qe})\right] \ ,
\end{equation}
where $Z$ ($N$) is the proton (neutron) number, $G_A$ is the weak axial form factor
of the nucleon normalized as $G_A(0)=g_A$, and
$\widetilde{G}_M^p=\left(1-4\, {\rm sin}^2 \theta_W\right)
G_M^p/2-G_M^n/2$ and $\widetilde{G}_M^n=\left(1-4\, {\rm sin}^2 \theta_W\right)
G_M^n/2-G_M^p/2$ are its weak vector form factors.  The $G_M^p$ and $G_M^n$
are the ordinary proton and neutron magnetic form factors, normalized
to the proton and neutron magnetic moments: $G_M^p(0)=\mu_p$ and
$G_M^n(0)=\mu_n$.  Thus the $S_{\alpha\beta}(q)$ give sum
rules of response functions corresponding to approximately point-like electroweak
couplings.  

Obviously, sum rules of weak response functions cannot be 
compared to experimental data.  Even in the electromagnetic
case, a direct comparison between the calculated and experimentally
extracted sum rules cannot be made unambiguously for two reasons.
First, the experimental determination of $S_\alpha$ requires measuring
the associated $R_\alpha$ in the whole energy-transfer region, from
threshold up to $\infty$.  Inclusive electron scattering experiments only
allow access to the space-like region of the four-momentum transfer
($ \omega < q$).  While the response in the time-like region ($\omega > q$)
could, in principle, be measured via $e^+ e^-$ annihilation, no
such experiments have been carried out to date.  Therefore, for a
meaningful comparison between theory and experiment, one needs 
to estimate the strength outside the region covered by the experiment. 
In the past this has been accomplished in the case of $S_L(q)$ either
by extrapolating the data~\cite{Jourdan:1996} or, in the few-nucleon
systems, by parametrizing the high-energy tail and using energy-weighted
sum rules to constrain it~\cite{Schiavilla:1989,Schiavilla:1993}.

The second reason that direct comparison of theoretical and
``experimental'' sum rules is difficult lies in the inherent
inadequacy of the dynamical framework adopted in this review
to account for explicit pion production mechanisms.  The
latter mostly affect the transverse response and
make its $\Delta$-peak region outside the range of applicability of this
approach.  At low and intermediate momentum transfers ($q\lesssim 500$ MeV/c),
the quasi-elastic and $\Delta$-peak are well separated, and 
it is therefore reasonable to study sum rules of the
electromagnetic transverse response.  In the quasi-elastic
region, where nucleon and (virtual) pion degrees of freedom
are expected to be dominant, the dynamical framework adopted in the present review
should provide a realistic and quantitative description of electromagnetic (and weak)
response functions.

\begin{figure}
\includegraphics[width=0.45\textwidth]{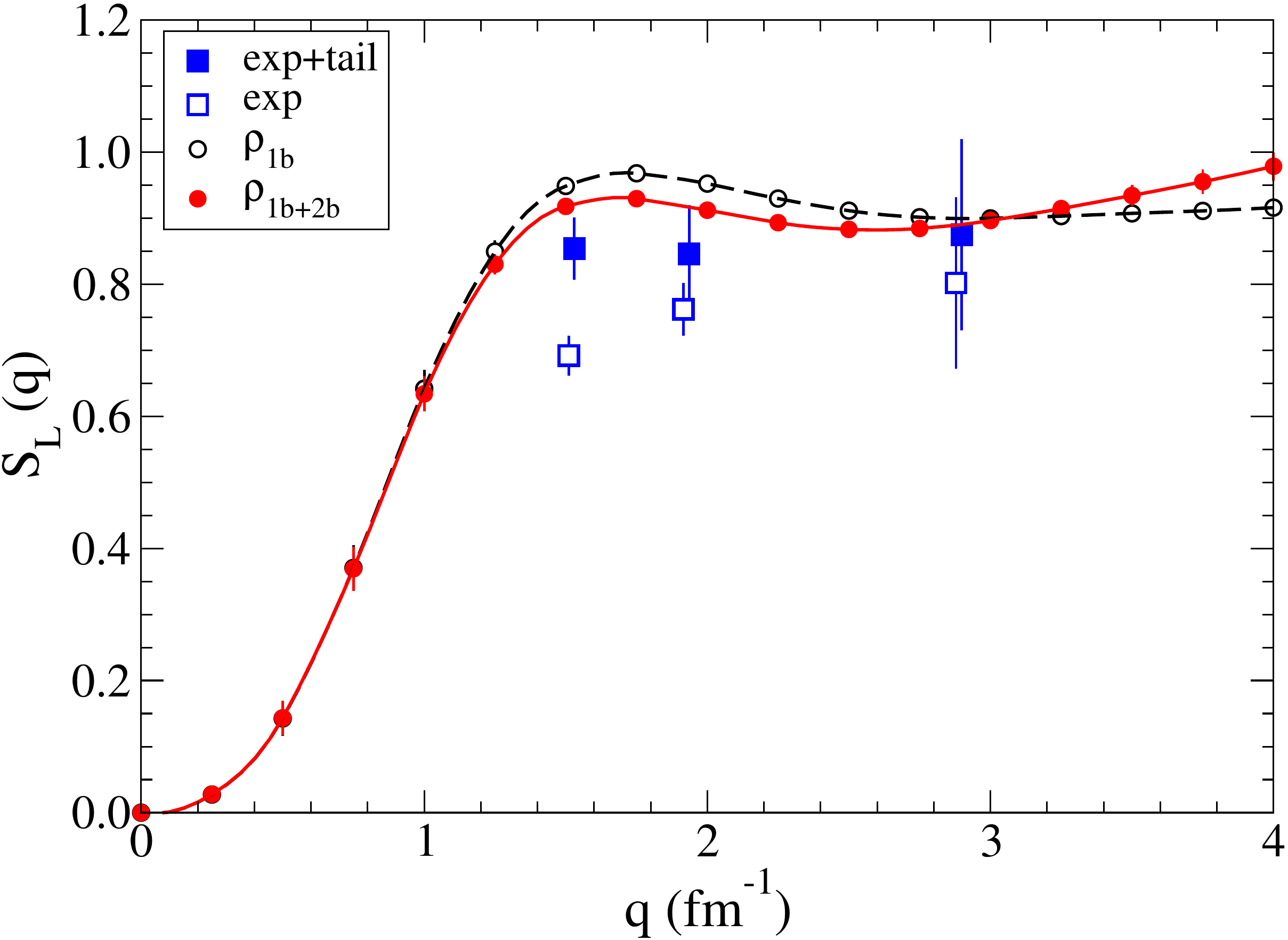}
\caption{The longitudinal sum rule of $^{12}$C
obtained with GFMC from the AV18+IL7 Hamiltonian with one-body only (empty circles, dashed line) and
one- and two-body (solid circles, solid line) terms in the charge operator is compared to experimental
data without (empty squares), and with (solid squares),
the tail contribution \cite{lovato2013}.}
\label{fig:f2a}
\end{figure}

\begin{figure}
\includegraphics[width=0.45\textwidth]{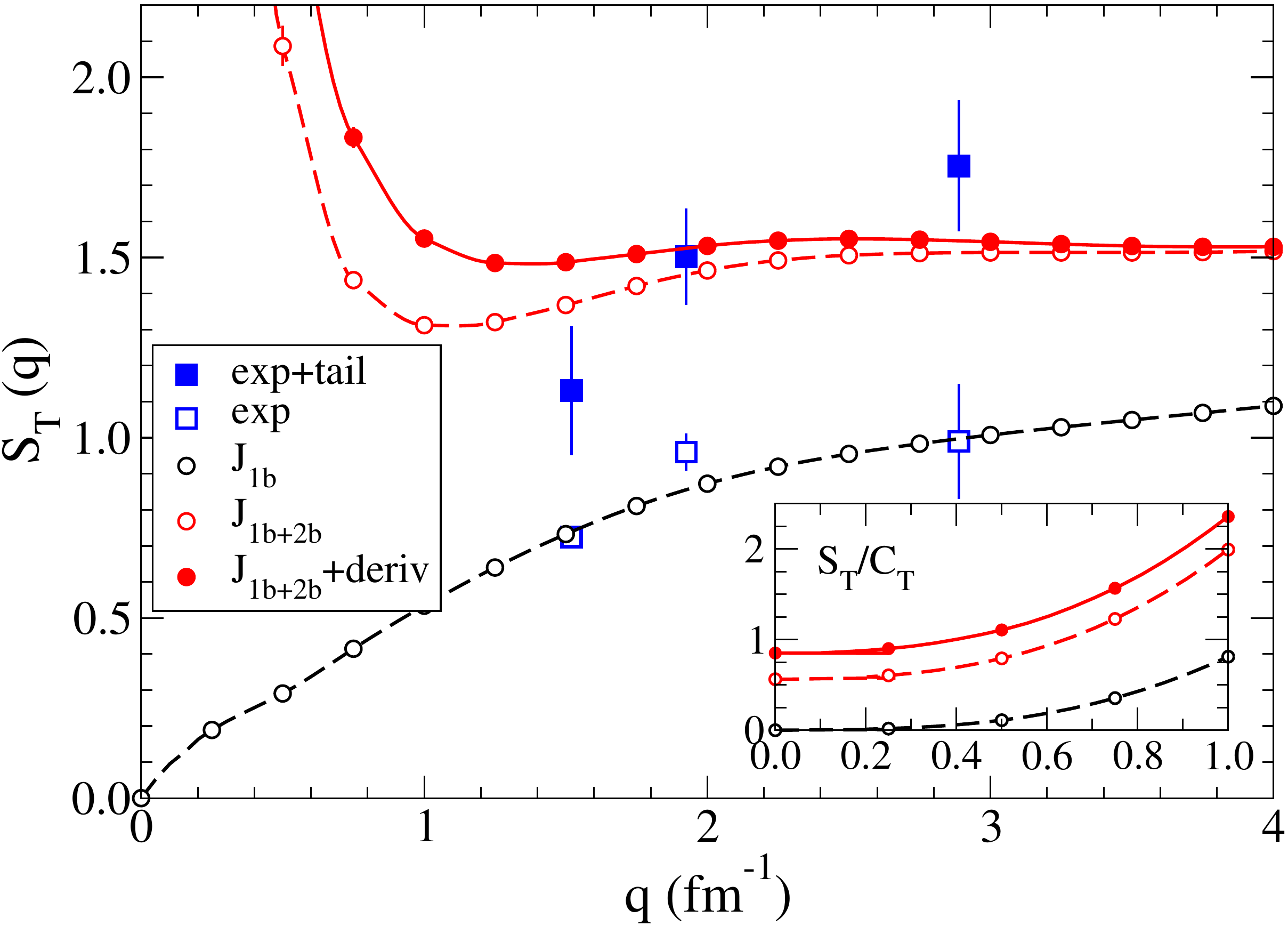}
\caption{Same as in Fig.~\ref{fig:f2}, but for the
transverse sum rule~\cite{lovato2013}.  The open symbols do not contain derivative terms
while a VMC evaluation of the derivative terms is included for
the solid red dots.
The inset shows $S_T(q)/C_T$ in the
small $q$-region.}
\label{fig:f3a}
\end{figure}

In Figs.~\ref{fig:f2a} and~\ref{fig:f3a}, we show recent results obtained
for the electromagnetic longitudinal and transverse sum rules in $^{12}$C.
The open squares give the experimental sum rules 
$S_L(q)$ and $S_T(q)$ obtained by integrating up to $\omega_{\rm max}$
(in the region where measurements are available) the
longitudinal and transverse response functions (divided by the square
of $G^p_E$) extracted from world data on inclusive $(e,e^\prime)$ scattering
off $^{12}$C~\cite{Jourdan:1996}; see \citet{lovato2013} for additional details.
We also show by the solid squares the experimental sum rules
obtained by estimating the contribution of strength in the region
$\omega > \omega_{\rm max}$.  This estimate $\Delta S_\alpha (q)$
is made by assuming that for $\omega > \omega_{\rm max}$,
i.e., well beyond the quasi-elastic peak, the longitudinal or transverse response in a
nucleus like $^{12}$C ($R_\alpha^A$) is proportional to that in the deuteron
($R_\alpha^{\, d}$), which can be accurately calculated~\cite{Shen:2012}.  
This scaling assumes that the high-energy
part of the response is dominated by $N\!N$ physics, and that the
most important contribution is from deuteron-like $np$ pairs. It is consistent
with the notion that at short times the
full propagator is governed by the product of pair propagators (assuming
$3N$ interactions are weak), discussed earlier in Sec.\ref{sec:eucl}.
Thus, one sets
$R^A_\alpha(q,\omega > \omega_{\rm max}) = \lambda(q)\, R^{\, d}_\alpha(q,\omega)$, and
determines $\lambda(q)$ by matching the experimental $^{12}$C response to the calculated
deuteron one.  It is worthwhile emphasizing that, for the transverse
case, this estimate is particularly uncertain for the reasons explained earlier;
the data on $R_T$~\cite{Jourdan:1996} indicate that at the higher $q$ values
for $\omega \sim \omega_{\rm max}$ there might be already
significant strength that has leaked in from the $\Delta$-peak region.

The sum rules computed with the AV18+IL7 Hamiltonian
and one-body only or one- and two-body terms in the electromagnetic
charge $S_L$ and current $S_T$ operators are shown,
respectively, by the dashed and solid lines in Figs.~\ref{fig:f2a}--\ref{fig:f3a}.
In the small $q$ limit, $S_L(q)$ vanishes quadratically, while
the divergent behavior in $S_T(q)$ is due to the $1/q^2$ present
in the normalization factor $C_T$.  In this limit,
$O_T({\bf q}\!=\!0)=i\left[\, H\, , \, \sum_i {\bf r}_i \, P_i \, \right]$~\cite{carlson1998,Marcucci:2005},
where $H$ is the Hamiltonian and $P_i$ is the proton projector,
and therefore $S_T(q)/C_T$ is finite; the associated
strength is due to collective excitations of electric-dipole
type in the nucleus.  In the large-$q$ limit, the one-body
sum rules differ from unity because of relativistic corrections
in $O_L({\bf q})$, primarily the Darwin-Foldy term which gives
a contribution $-\eta/(1+\eta)$ to $S_L^{\rm 1b}(q)$, where
$\eta \simeq q^2/(4\, m^2)$, and because of the convection
term in $O_T({\bf q})$, which gives a contribution $\simeq (4/3)\, C_T\, T_p/m$ to
$S^{\rm 1b}_T(q)$, where $T_p$ is the proton kinetic energy in the nucleus.

In contrast to $S_L$, the transverse sum rule has large two-body
contributions.  This is consistent with studies of Euclidean transverse response
functions in the few-nucleon systems~\cite{Carlson:2002}, which suggest
that a significant portion of this excess transverse strength is
in the quasi-elastic region.  Overall, the calculated $S_L(q)$
and $S_T(q)$ are in reasonable agreement with data.  However,
a direct calculation of the response functions is clearly needed for
a more meaningful comparison between theory and experiment.

\begin{figure}
\includegraphics[width=0.45\textwidth]{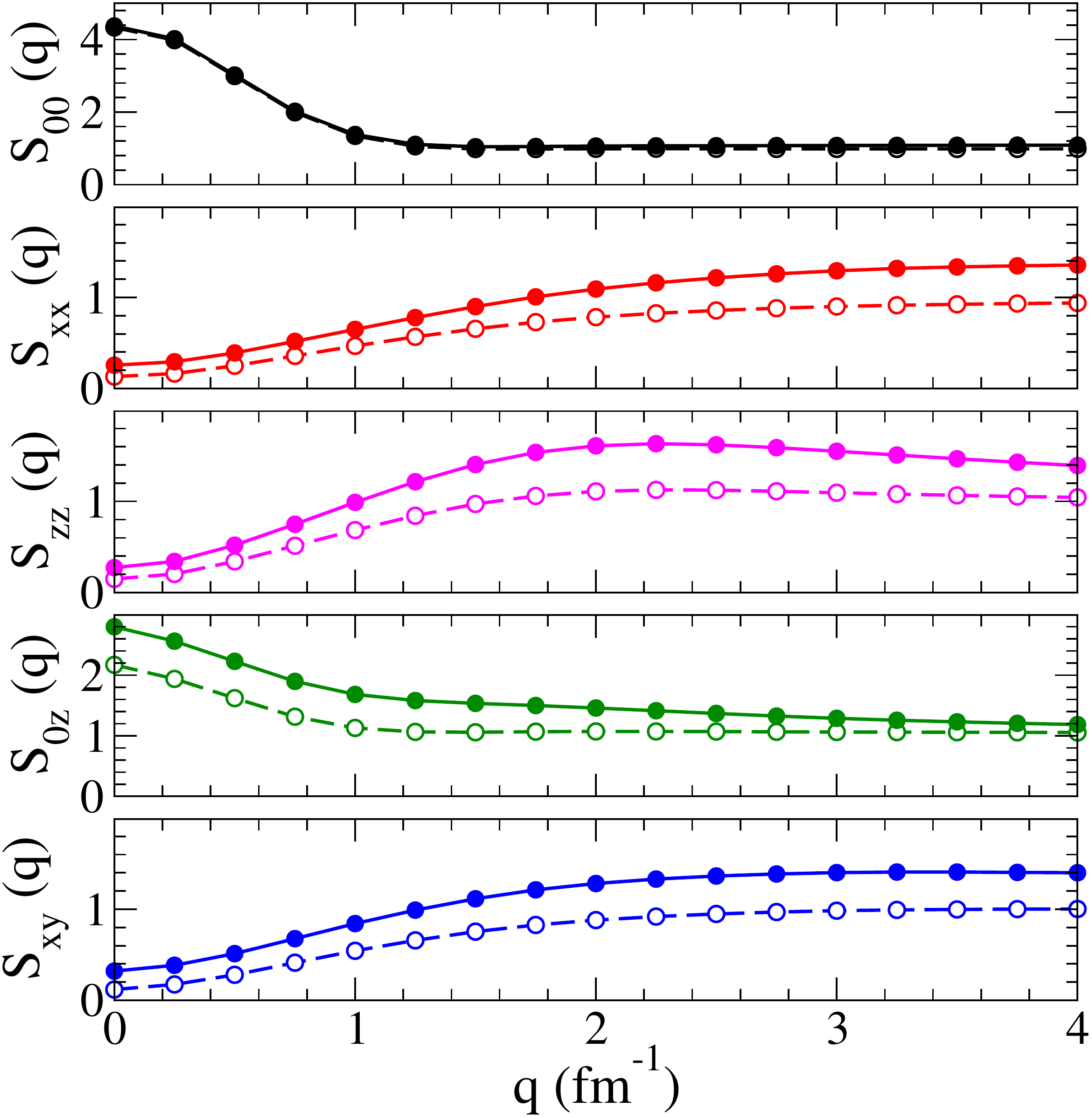}
\caption{The GFMC sum rules $S_{\alpha\beta}$ in $^{12}$C,
corresponding to the AV18+IL7 Hamiltonian and obtained with one-body
only (dashed lines) and one- and two-body (solid lines) terms in the NC~\cite{Lovato:2014}.}
\label{fig:f2}
\end{figure}

While sum rules of NC or CC weak sum rules are of a
more theoretical interest, they nevertheless
provide useful insights into the nature of the strength seen in the
quasi-elastic region of the response and, in particular, into the role of
two-body terms in the electroweak current.
Those corresponding to weak NC response functions and
relative to $^{12}$C are shown in Fig.~\ref{fig:f2}:
results $S^{\rm 1b}$  ($S^{\rm 2b}$) corresponding to one-body (one- and two-body)
terms in the NC are indicated by the dashed (solid) lines. 
Note that both $S_{\alpha\beta}^{\rm 1b}$ and $S_{\alpha\beta}^{\rm 2b}$
are normalized by the same factor $C_{\alpha\beta}$, which makes
$S^{\rm 1b}_{\alpha\beta}(q) \rightarrow 1$ in the large $q$ limit.  In the
small $q$ limit, $S^{\rm 1b}_{00}(q)$ and $S^{\rm 1b}_{0z}(q)$ are much
larger than $S^{\rm 1b}_{\alpha\beta}$ for $\alpha\beta \neq 00,0z$.

Except for $S_{00}^{\rm 2b} (q)$, the $S_{\alpha\beta}^{\rm 2b} (q)$ sum rules are considerably
larger than the $S_{\alpha\beta}^{\rm 1b} (q)$, by as much as 30-40\%. This 
enhancement is not seen in calculations of neutrino-deuteron scattering~\cite{Shen:2012}.
The increase due to two-body currents is quite 
substantial even down to small momentum transfers. 
At $q \simeq 1\ {\rm fm}^{-1}$, the  enhancement is about 50\% relative to
the one-body values.  In general, the additional
contributions of the two-body currents ($j_{\rm 2b}$)
to the sum rules are given by a combination of interference with 
one-body currents ($j_{\rm 1b}$), matrix elements of the type 
$\langle 0 \! \mid \! j_{\rm 1b}^\dagger\, j_{\rm 2b} \! \mid \! 0 \rangle 
+ \langle 0\! \mid\!  j_{\rm 2b}^\dagger\, j_{\rm 1b}\! \mid\! 0 \rangle$,
and contributions of the type 
$\langle 0 \! \mid\! j_{\rm 2b}^\dagger\, j_{\rm 2b} \! \mid \! 0 \rangle$. 
At low momentum transfers the dominant contributions 
are found to be of the latter $\langle 0\!\mid j_{\rm 2b}^\dagger\, j_{\rm 2b} \mid\! 0\rangle$ type, 
where the same pair is contributing in both left and right operators. 
Enhancements of the response due to two-body currents 
could be important in astrophysical settings, where the neutrino energies typically 
range up to 50 MeV. A direct calculation of the $^{12}$C response 
functions is required to determine whether the strength of the 
response at low $q$ extends to the low energies kinematically 
accessible to astrophysical neutrinos.

At higher momentum transfers the interference between 
one- and two-body currents plays a more important role.
The larger momentum transfer in the single-nucleon current connects the 
low-momentum components of the ground-state wave function directly with the 
high-momentum ones through the two-body current.
For nearly the same Hamiltonian as is used here, there is a 10\% probability
that the nucleons have momenta greater than 2~fm$^{-1}$
implying that $\approx 30\%$ of the wave function amplitude 
is in these high-momentum components~\cite{wiringa2014}.  The contribution of
$np$ pairs remains dominant at high momentum transfers, and matrix
elements of the type $\langle 0 \!\mid [\, j_{\rm 1b} (l) 
+ j_{\rm 1b} (m) ]^\dagger j_{\rm 2b} (lm) \mid\! 0 \rangle$ + c.c.
at short distances between nucleons $l$ and $m$ are critical.
\begin{figure}
\includegraphics[width=0.45\textwidth]{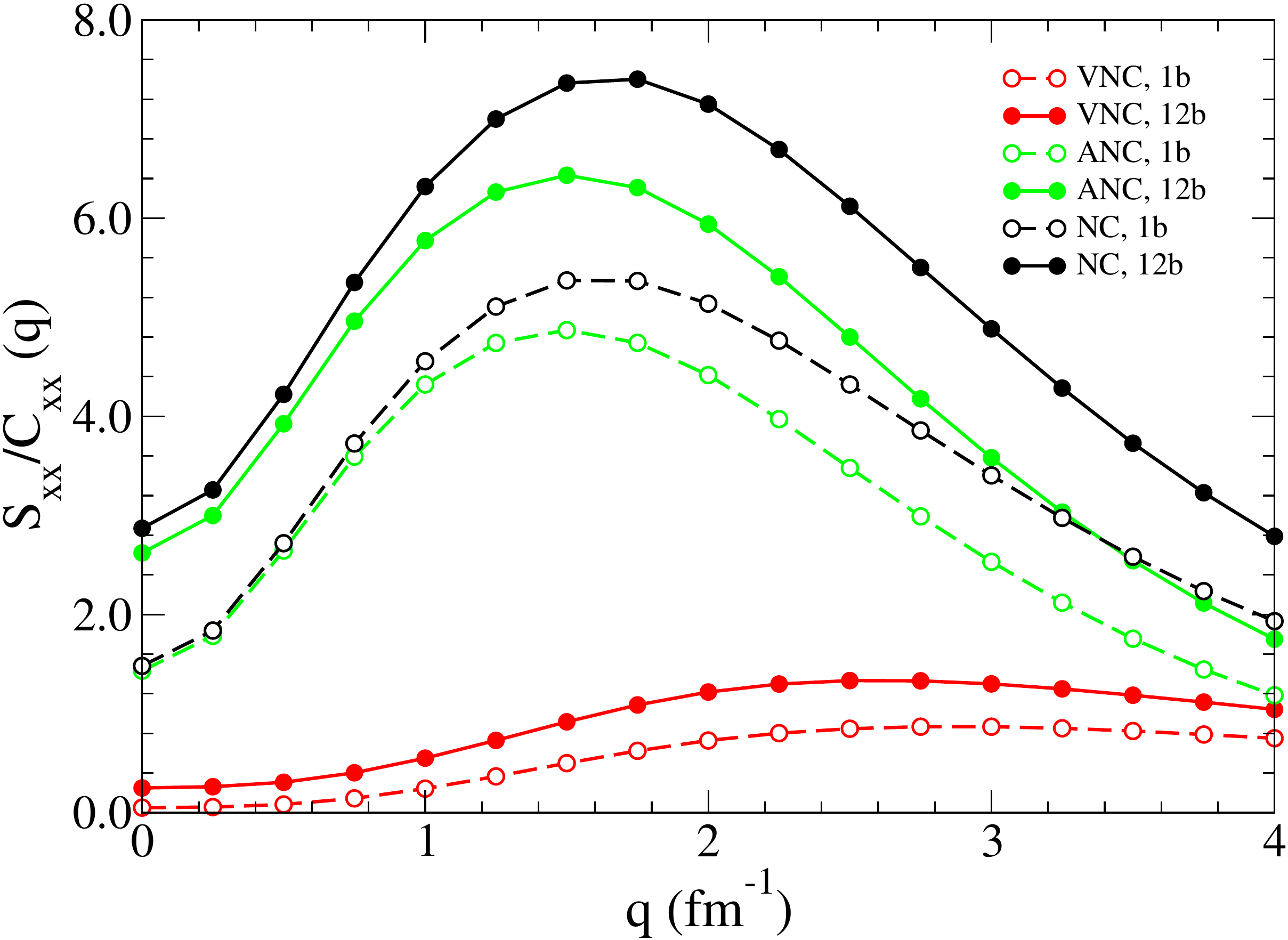}
\caption{The GFMC $S_{xx}/C_{xx}$ sum rules obtained with the NC
(curves labeled NC) and either its vector (curves labeled VNC) or axial-vector (curves labeled ANC)
parts only~\cite{Lovato:2014}. The corresponding one-body (one- and two-body) contributions are indicated
by dashed (solid) lines.  Note that the normalization factor $C_{xx}$ is not included.}
\label{fig:f3}
\end{figure}

Figure~\ref{fig:f3} shows the separate contributions associated with the 
vector (VNC) and axial-vector (ANC) parts of the $S_{xx} / C_{xx}$ sum rule. 
The ANC piece of the $S_{xx}$ sum rule is found to have large two-body 
contributions of the order of 30\% relative to the one-body part. 
Similar results are found for the $0z$ and $zz$ sum rules; the $xy$ sum rule is nonzero because of 
interference between the VNC and ANC and vanishes in the limit in which only one or the other is considered.
The ANC two-body contributions in the sum rules are much larger
than the contributions associated with axial two-body currents in weak charge-changing
transitions to specific states at low-momentum transfers, such as $\beta$-decays and
electron- and muon-capture processes involving nuclei with mass numbers
$A$=3--7~\cite{Marcucci:2011,Schiavilla:2002b}, where they amount to a few \%
(but are nevertheless necessary to reproduce the empirical data).

In summary, two-body currents
generate a significant enhancement of the single-nucleon neutral weak current response,
even at quasi-elastic kinematics.  This enhancement is driven by
strongly correlated $np$ pairs in nuclei.  The presence of these
correlated pairs also leads to important interference effects
between the amplitudes associated with one- and two-body
currents: the single-nucleon current can knock out two particles
from a correlated ground state, and the resulting amplitude
interferes with the amplitude induced by the action of the two-body
current on this correlated ground state. 

\subsection{Euclidean response functions}

Direct calculations of $R_{\alpha\beta}$ are difficult
in systems with $A> 2$, and at the moment one has to rely on techniques based
on integral transforms relative to the energy transfer, which eliminate
the need for summing explicitly over the final states.  Two such approaches
have been developed: one based on the Lorentz-integral transform (LIT) has
been used extensively in the few-nucleon systems, albeit so far by including
only one-body electroweak current operators.  It has been reviewed recently
\cite{Leidemann:2013}, and will not be discussed here.  
The other approach is based on the Laplace transform~\cite{Carlson:1994,Carlson:1994a}
and leads to Euclidean (or imaginary time) response functions, defined as
\begin{align}
E_{\alpha\beta}(q,\tau)&=
\int_0^\infty {\rm d}\omega\, {\rm e}^{-\tau\, \omega}\, R_{\alpha\beta}(q,\omega) \nonumber \\
&=\langle0\!\mid\! O^\dagger_\beta({\bf q})\, {\rm e}^{-\tau\left(H-E_0\right)}\,
O_\alpha({\bf q})\!\mid \! 0\rangle \ .
\end{align}
The Euclidean response is essentially a statistical mechanical formulation,
and hence can be evaluated with QMC methods similar to those
discussed earlier.
Electromagnetic Euclidean response functions
have been calculated for the few-nucleon systems 
($A$=3 and 4)~\cite{Carlson:1994,Carlson:1994a,Carlson:2002}, 
and very recently for $^{12}$C~\cite{Lovato:2015}.  It should be realized that
in a nucleus like $^{12}$C these are 
very computationally intensive calculations, requiring tens of millions of core
hours on modern machines.

In the case of $(e,e^\prime)$ scattering the electromagnetic Euclidean response functions
can be compared directly with experimental data, by simply evaluating the Laplace
transforms of the measured response functions, at least for values of $\tau$ large
enough so as to make $E_{L/T}(q,\tau)$ mostly sensitive to strength in the quasi-elastic and low-energy
regions of $R_{L/T}(q,\omega)$. 


The response at $\tau = 0$ is identical to the sum rule, and its 
slope at $\tau = 0$ is equivalent to the energy-weighted sum rule.
The simulation
proceeds by calculating the ground-state wave function using GFMC,
and then evaluating the imaginary-time dependent correlation functions
over a range of separations $\tau$ using the same paths sampled
in the original ground-state calculation.  Since the current operators
couple to states of different spin and isospin, the calculations
require recomputing the path integral for different current operators
$O_\alpha ({\bf q})$.

To more easily compare the Euclidean response to data for larger $\tau$, 
we multiply by a scaling
factor ${\tilde E}_{\alpha \beta} (q, \tau) = \exp [ q^2 \tau / ( 2 m ) ] E_{\alpha \beta} (q, \tau)$.  For a free nucleon initially at rest, this scaled
response is a constant independent of $\tau$, since the response is
a delta function in energy for each momentum transfer $q$. 
The slope and curvature 
of the calculated Euclidean response at low $\tau$ indicates the
strength at high energy, and the response at large $\tau$ is related to
the low-energy part of the nuclear response.  The calculated responses
have a higher average energy than simple PWIA-like approaches, and also
have greater strength at high energy (from $N\!N$ processes) and
at low energy (from low-lying nuclear states).

The difference between the full response and the simple PWIA is most
easily understood for the longitudinal response, which is dominated
by one-nucleon currents. The PWIA is sensitive to the momentum distribution
of the protons, as it assumes that the struck nucleon does not interact
with other nucleons.  The full calculation is also sensitive to the
propagation of charge through the NN interaction, since the struck
proton can charge exchange with other nucleons. This rapid propagation
of charge leads to an enhanced strength at high energy.



\begin{figure}
\includegraphics[width=0.45\textwidth]{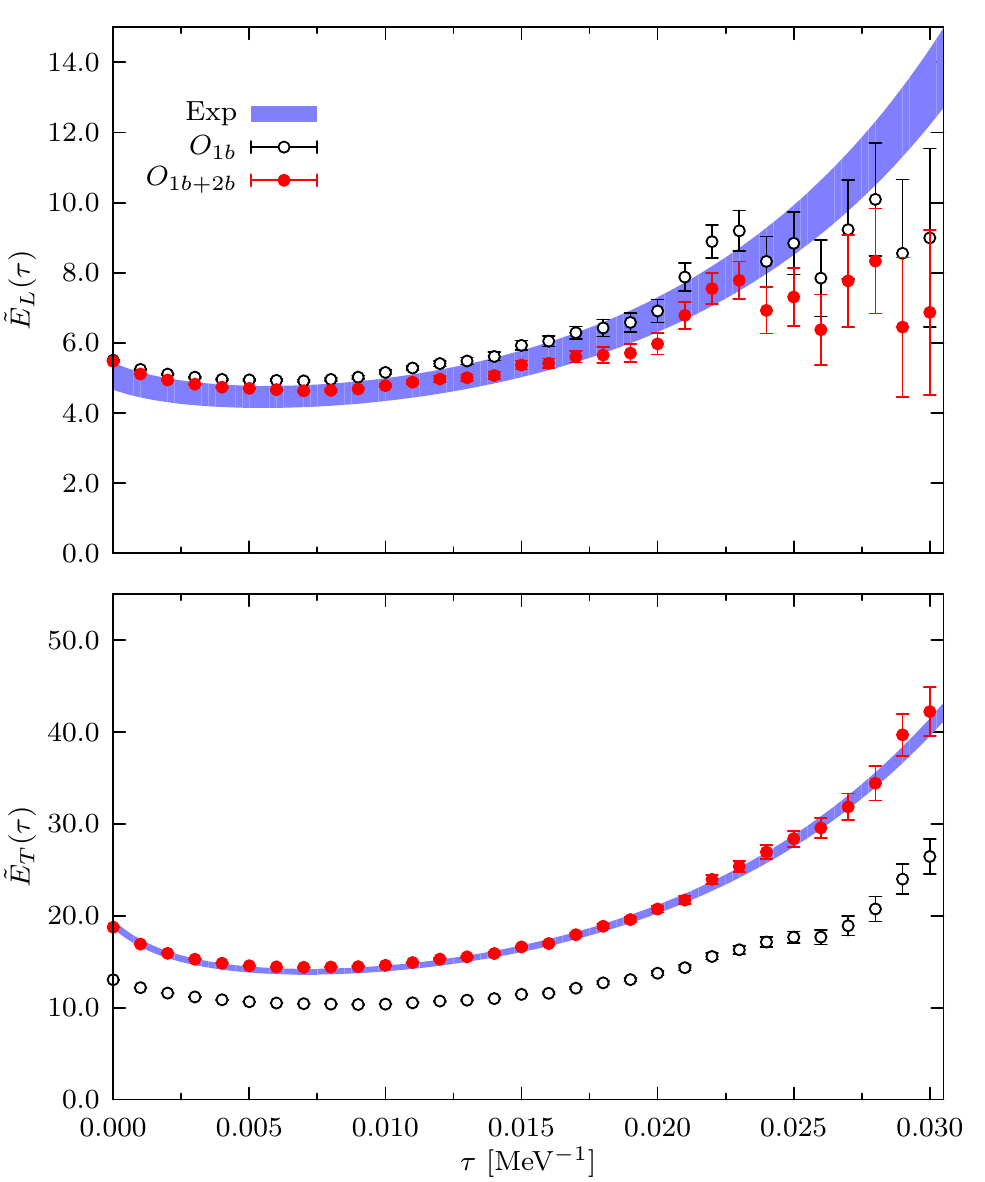}
\caption{The longitudinal (upper panel) and transverse (lower panel) 
electromagnetic Euclidean responses for $^{12}$C at q = 570 MeV/c. 
The bands represent the transform of the experimental data, and the
calculations with single-nucleon and two-nucleon currents are shown
as open and filled symbols, respectively.}
\label{fig:c12em}
\end{figure}

In Fig.~\ref{fig:c12em} we show recent calculations~\cite{Lovato:2015} 
of the $^{12}$C Euclidean electromagnetic longitudinal and transverse
response compared with experimental data.
The overall agreement with experiment in the longitudinal channel is excellent.
Here the calculation with the full currents is
very similar to that with one-nucleon currents alone.  The error bars
are higher at large $\tau$ (lower energy) because of the required subtraction
of the elastic contribution.

The transverse response is shown in the lower panel of Fig.~\ref{fig:c12em}.  
The difference
between single-nucleon currents and one- plus two-nucleon currents is quite
substantial and extends over the full range of $\tau$. This implies a 
substantial enhancement of the cross section in the full energy region,
including both the quasi-elastic peak and the low-energy regime.  
The full calculation
is in good agreement with experiment.
The enhancement can
in some cases be as large as 40\%, somewhat larger than typical effects
of two-nucleon currents on the squared matrix elements 
of low energy transitions, but
not dramatically so. The larger momentum transfers in these inclusive
experiments can be expected to lead to larger contributions from pion-
and $\Delta$ currents, and these are found to be the dominant two-nucleon
current contributions.

Ideally one would like to invert the Laplace transform to obtain a more
direct reconstruction of the response as a function of 
momentum and energy transfer. This has been accomplished already for
A=4, where the calculations are much faster and hence the simulations
can be carried out with high accuracy.  Recent calculations \cite{Lovato:2015} 
agree with earlier calculations of the EM response of $^4$He \cite{Carlson:2002}, 
but the statistical accuracy is at least an order of magnitude better.

\begin{figure}
\includegraphics[width=0.45\textwidth]{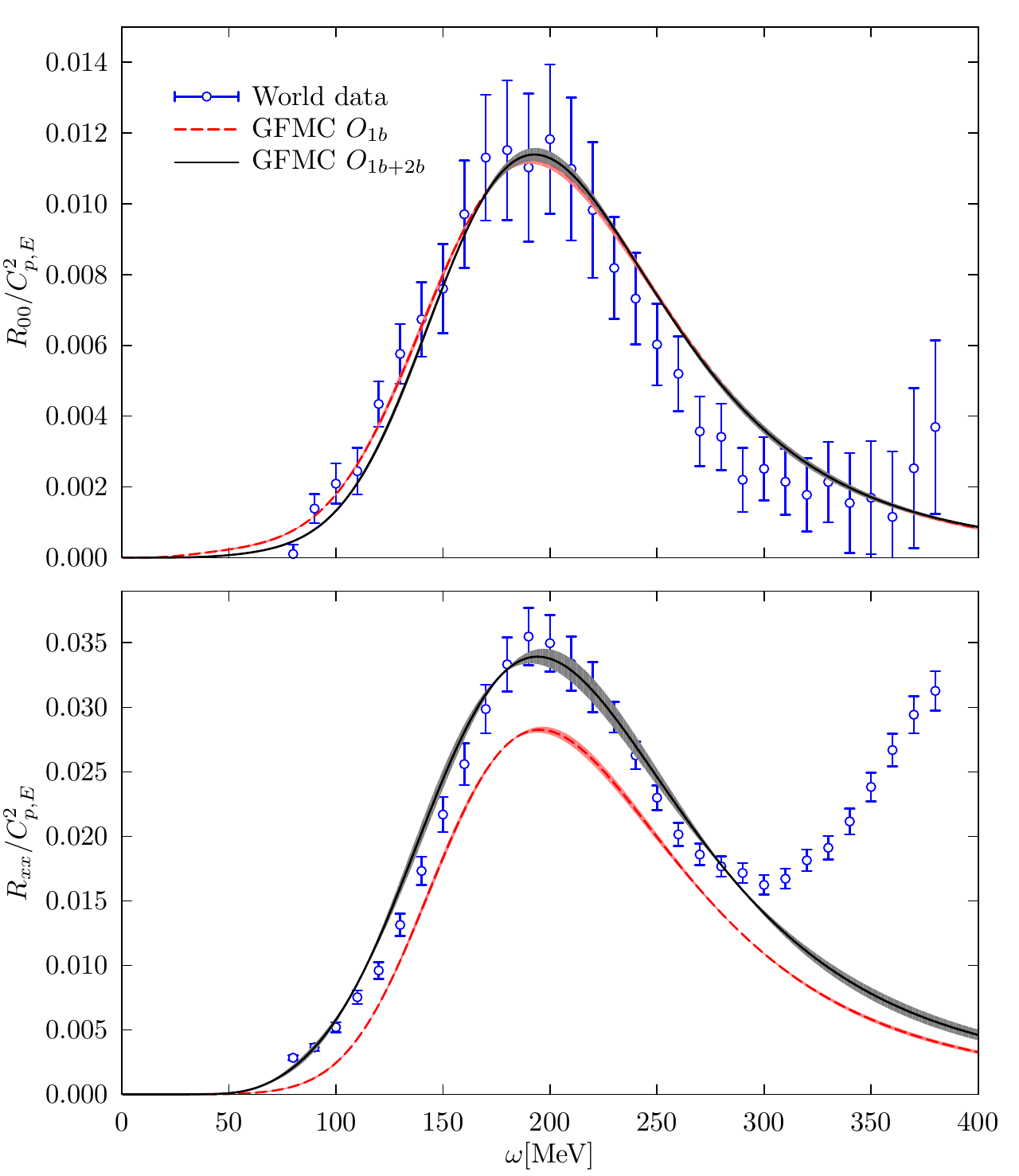}
\caption{The longitudinal (upper panel) and transverse (lower panel) EM response of $^4$He
at q=600 MeV/c reconstructed from the Euclidean response compared
to experimental data~\cite{Lovato:2015}.
The experimental results are shown as
symbols with error bars, and the bands show the reconstructed responses
and errors associated with the maximum entropy reconstruction.}
\label{fig:he4euc}
\end{figure}

For such accurate data the maximum entropy 
method~\cite{Bryan:1990,Jarrell:1996} can be used 
to reconstruct the response.
Results for $^4$He at q=600 MeV/c are shown in Fig.~\ref{fig:he4euc},
similar accuracy is obtained over a wide range of momentum transfers.
Again it is seen that the enhancement from two nucleon currents
is substantial and extends over the whole quasielastic regime.  
At higher energies
the calculated response does not include pion production and hence
fails to reproduce the strength associated with $\Delta$ production.

\begin{figure}
\includegraphics[width=0.45\textwidth]{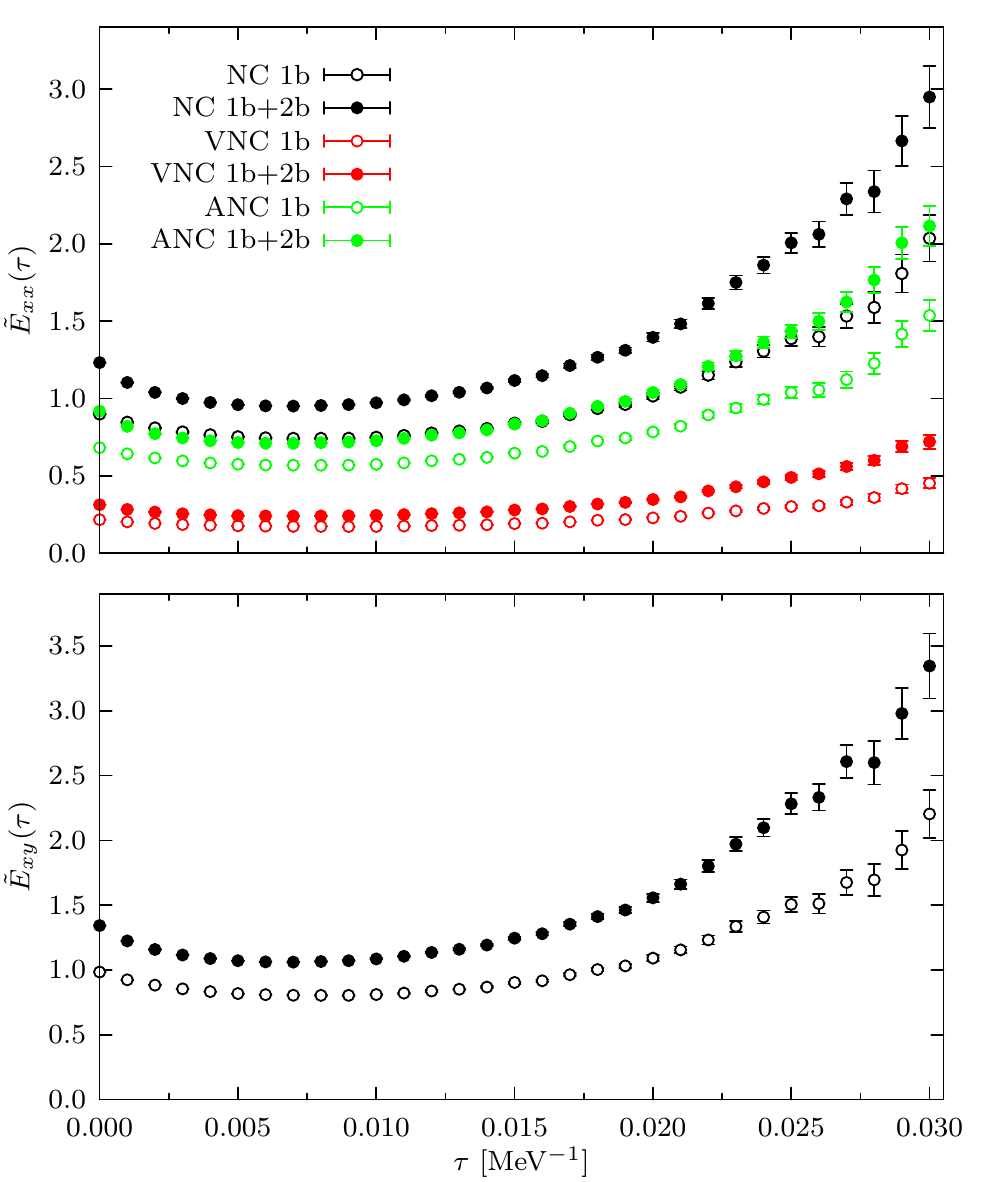}
\caption{The neutral current weak response of $^{12}C$ at q=570 MeV/c.
Calculations with single-nucleon currents are shown as open symbols,
and with the full currents as filled symbols. The upper panel shows
the transverse response and its vector-vector and axial-axial contributions,
while the lower panel shows the interference vector-axial vector response.}
\label{fig:c12nc}
\end{figure}

Imaginary time response functions for the neutral current response
of $^{12}$C have also been performed~\cite{Lovato:2015} and are shown
in Fig.~\ref{fig:c12nc}.
At present the statistical accuracy is not sufficient to
invert the response, but the Euclidean response already gives
important results. These calculations demonstrate an enhancement of the
axial currents in addition to the expected enhancement
in the vector channels. In particular, the vector-axial interference
response (lower panel) is significantly enhanced by the two-nucleon
currents.  It is this response that gives the difference
between neutrino and anti-neutrino cross sections. This is a very important
quantity in attempts to isolate the CP-violating phase in the neutrino
sector or the mass hierarchy in long-baseline experiments; see
for example~\cite{LBNE:2013}.  Future work on charge current responses
and inversions to the real-time response have many important applications
including accelerator neutrinos and neutrinos in astrophysical environments.

\section{The equation of state of neutron matter}

\subsection{\Pnm: Homogeneous phase}
\label{sec:afdmc-pnm}
The equation of state (EoS) of neutron matter is a key ingredient in understanding
the static and dynamic properties of neutron stars.
In the region between the inner crust and the outer core, neutron stars are
primarily neutrons, in equilibrium with a small fraction of protons,
electrons and muons in $\beta$-decay equilibrium.
It has been argued that when the chemical potential is large enough,
heavier particles containing strange quarks may appear. This is expected
to happen at densities $\gtrsim 3\rho_0$~\cite{Lonardoni:2014b}.  
However, while the determination of the maximum mass of neutron stars requires 
knowledge of the EoS up to several times nuclear densities,
the EoS around nuclear density and up to
about $2\rho_0$ largely determines their radii~\cite{Lattimer:2001}.
Astrophysical applications are not the only relevant ones.
The EoS of neutron matter is used
to constrain effective forces in the presence of large isospin-asymmetry.
For example, the bulk term of Skyrme models is sometimes fitted with input from
a neutron matter EoS.

Neutron matter is not directly accessible in terrestrial experiments, 
and all the indirect experimental evidence related to it is based on 
extrapolations of measurements on heavy nuclei, and on astrophysical
observations (see, e.g., \citet{Danielewicz:2002}). The role of ab-initio techniques becomes therefore
crucial as a tool for testing the model Hamiltonians that can be directly
fitted on experimental data for light nuclei against the constraints 
deriving from indirect measurements.

At low densities $\rho\leq 0.003$ fm$^{-3}$ properties of neutron matter are
very similar to ultra-cold Fermi gases that have been extensively studied 
in experiments. In this regime, the interaction is mainly s-wave, and the
system strongly paired. The nuclear interaction can be simplified, 
the standard DMC method for central potentials can be used, and very accurate 
results for the energy and the pairing gap obtained~\cite{Gezerlis:2008,
Gezerlis:2010,Carlson:2012}. Other results obtained using AFDMC with the full
nuclear Hamiltonian are qualitatively similar~\cite{Gandolfi:2008b,Gandolfi:2009b}.
At higher densities, the contribution of higher partial waves becomes important,
and the complete nuclear Hamiltonian has to be used to calculate the EoS.

Argonne and other modern interactions are very well suited to study dense matter.
The $N\!N$ scattering data are described well with AV18 in a very
wide range of laboratory energies, and this gives an idea to their validity to
study dense matter. A laboratory energy of 350 MeV (600 MeV) corresponds to a Fermi momentum
k$_F\approx$400 MeV (530 MeV) and to a neutron density $2\,\rho_0$ ($4\,\rho_0$).
This is not the case of softer potentials fitted to very low energy scattering data. 
The AV18 and AV8$^\prime$ two-body
interactions combined with the UIX three-body force 
have been extensively employed to calculate the
properties of neutron matter and its consequences for neutron star
structure (see, e.g., \citet{Akmal:1998}).  In the past, several attempts to use 
Illinois three-body forces were made,
but they provided unexpected overbinding of neutron matter at large
densities~\cite{Sarsa:2003}, and will not be discussed any further.
It has been recently shown that even the IL7 three-body interaction gives an
EoS too soft~\cite{Maris:2013}. It would be very interesting to calculate the 
EoS of symmetric nuclear matter using IL7, but unfortunately there are no such
calculations.

The first AFDMC calculations of the EoS of neutron matter including three-body
forces has been produced by~\citet{Sarsa:2003}.     Later 
using a different implementation of the constrained-path and with more
statistics, better agreement was obtained with GFMC where the comparison 
is available~\cite{Gandolfi:2009}.
To date, only the equation of state of pure neutron matter has been calculated with QMC
using realistic Hamiltonians, while nuclear matter can be studied by including only
two-body forces~\cite{Gandolfi:2014b}.

By imposing periodic boundary conditions it is possible 
to simulate an infinite system using a finite number of particles.
However, the energy and other physical quantities are affected by the 
spatial cut-offs that are required to make the wave function compatible
with periodic boundary conditions. The effect of cutting the potential energy at the edge
of the simulation box is made milder by summing the contributions
due to periodic images of the nucleons included in a given number of shells of
neighboring image simulation cells.
Finite size corrections to the kinetic energy already appear for the Fermi gas. 
In order to have a wave function that describes a system with zero total 
momentum and zero angular momentum, it is necessary to fill up a shell
characterized by the modulus of the single particle momentum. This fact
determines a set of magic numbers, which are commonly employed in
simulations of periodic systems. The kinetic energy corresponding to each
magic number is a non-regular and non-monotonic function of the
number of Fermions~\cite{Ceperley:1977}.
This fact suggests that for an interacting system it is necessary to proceed
with an accurate determination of the closed shell energies in order to
minimize the discrepancy with the infinite system limit.

To this end, the effect of using different number of neutrons was carefully studied by means
of the Periodic Box Fermi hypernetted chain (FHNC) method~\cite{Fantoni:2001c}. This study showed that the particular choice of 
33 Fermions (for each spin state) is the closest to the thermodynamic limit.
Another strategy for allowing an accurate extrapolation consists of using 
the Twisted Averaged Boundary Conditions. The method,
described in~\citet{Lin:2001} is based on randomly drifting the center of the Fermi sphere, which
adds a phase to the plane waves used in the Slater determinant, in order
to add contributions from wave vectors other than those strictly compatible
with the simulation box. This procedure smooths the behavior of the energy
as a function of  $N$, giving the possibility of better determining the
$N\rightarrow\infty$ limit~\cite{Gandolfi:2009}.

\begin{figure}
\includegraphics[scale=0.3]{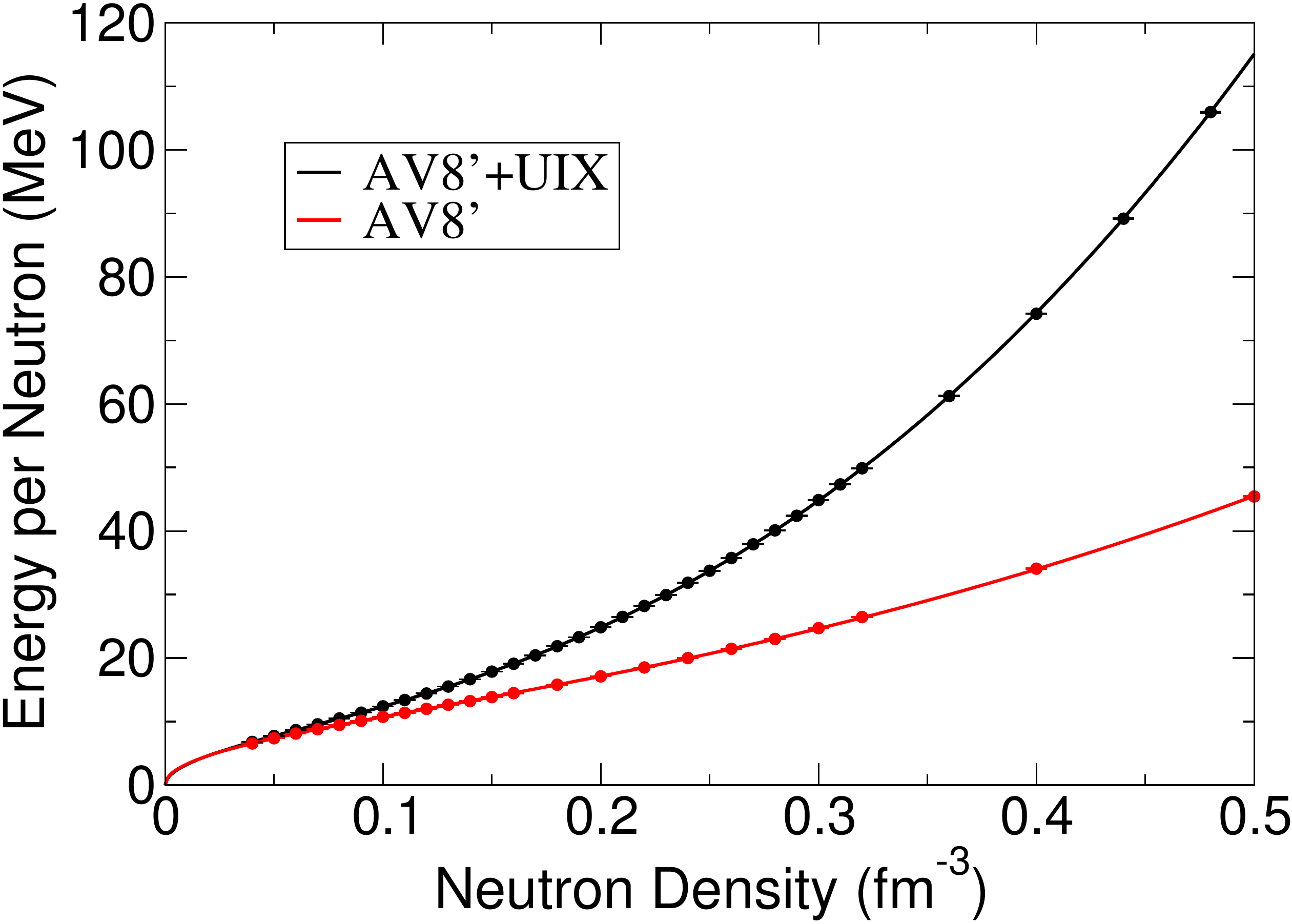}
\caption{The EoS of neutron matter as a function of the density,
obtained using the AV8$^\prime$ $N\!N$ interaction alone (lower red symbols/line),
and combined with the UIX $3N$ force~\cite{Gandolfi:2014}.}
\label{fig:pnmav8uix}
\end{figure}

In Fig.~\ref{fig:pnmav8uix} the EoS of neutron matter
computed by a simulation with $N=66$ is presented.
In order to check the consistency of the results given by AFDMC, a 
simulation was performed using only $N=14$ neutrons and by imposing the same 
boundary conditions to the interaction as in the GFMC calculation~\cite{Carlson:2003,Carlson:2003b}.
The comparison shows that the two methods are in good agreement~\cite{Gandolfi:2009}.
Particular care was taken in studying the effect of finite-size effects by repeating
each simulation using a different number of neutrons and using Twisted Averaged Boundary 
Conditions. The repulsive nature of the three-neutron
interaction is clear from the figure, where the EoS obtained with and without
UIX is shown.

The AFDMC results are conveniently fitted using the functional form
\begin{equation}
\label{eq:fit}
E(\rho_n)=a\left(\frac{\rho_n}{\rho_0}\right)^\alpha+
b\left(\frac{\rho_n}{\rho_0}\right)^\beta \,,
\end{equation}
where $E$ is the energy per neutron (in MeV) as a function of the density $\rho_n$
(in fm$^{-3}$).  The parameters of the fit for both
AV8$^\prime$ and the full AV8$^\prime$+UIX
Hamiltonian are reported in Table~\ref{tab:fitpnmav8}.

\begin{table}
\centering
\begin{tabular}{@{} lcccccc @{}}
\hline
$3N$ force       & $E_{\rm sym}$ &$L$ & $a$    &  $\alpha$ & $b$ & $\beta$ \\
               & (MeV)  & (MeV) & (MeV)    &        & (MeV)  \\
\hline
none                        & 30.5 & 31.3 & 12.7 & 0.49  & 1.78 & 2.26 \\
$V_{2\pi}^{PW}+V^R_{\mu=150}$&32.1 & 40.8 & 12.7 & 0.48  & 3.45 & 2.12 \\
$V_{2\pi}^{PW}+V^R_{\mu=300}$&32.0 & 40.6 & 12.8 & 0.488 & 3.19 & 2.20 \\
$V_{3\pi}+V_R$              & 32.0 & 44.0 & 13.0 & 0.49  & 3.21 & 2.47 \\
$V_{2\pi}^{PW}+V^R_{\mu=150}$&33.7 & 51.5 & 12.6 & 0.475 & 5.16 & 2.12  \\
$V_{3\pi}+V_R$              & 33.8 & 56.2 & 13.0 & 0.50  & 4.71 & 2.49  \\
UIX                         & 35.1 & 63.6 & 13.4 & 0.514 & 5.62 & 2.436 \\
\hline
\end{tabular}
\label{tab:fit}
\caption{The parameters of Eq.~(\ref{eq:fit}) fitting the equation of
state computed with the full AV8$^\prime$+UIX Hamiltonian and with the 
$N\!N$ interaction only (AV8$^\prime)$. The parametrization of 
selected EoSs shown in Fig.~\ref{fig:eospnmv3} are also
included. For each EoS, the corresponding $E_{\rm sym}$ and slope
$L$ are indicated.
}
\label{tab:fitpnmav8}
\end{table}

\begin{figure}
\includegraphics[scale=0.3]{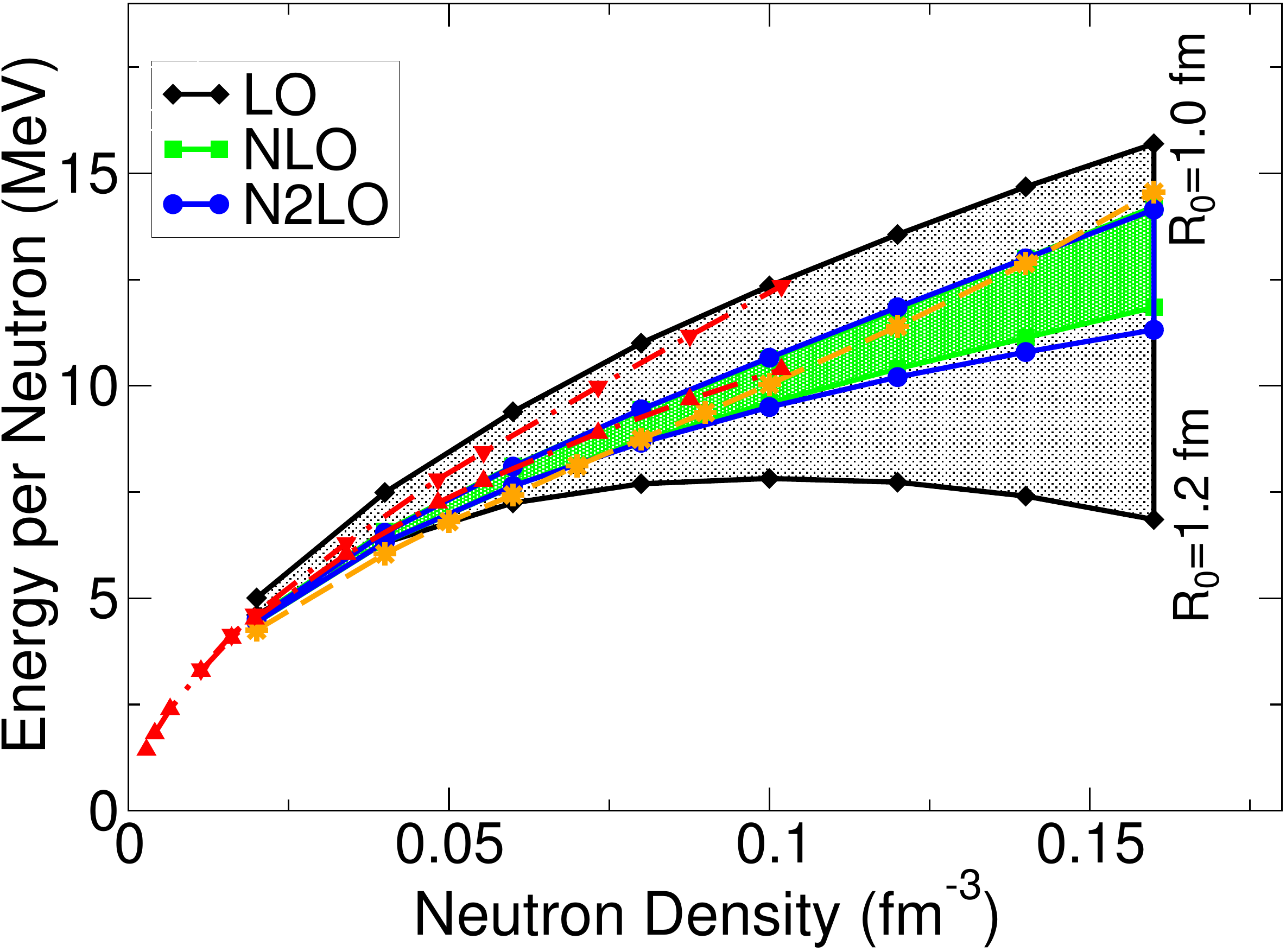}
\caption{The EoS of neutron matter as a function of the density,
calculated by~\citet{Gezerlis:2014} using AFDMC with chiral $N\!N$ interactions at 
LO, NLO and N$^2$LO for the two different cutoff indicated in the figure 
(three-body forces have not been included at N$^2$LO).
Also shown are the results obtained by~\citet{Wlazlowski:2014} using lattice QMC 
at N$^2$LO, by including the $3N$ interaction (upper red dot-dashed line) and 
without (lower red dot-dashed line), and the results of~\citet{Roggero:2014} using the N$^2$LO$_{\rm opt}$ 
without 3N
(orange dashed line).
}
\label{fig:pnmeft}
\end{figure}

The EoS of neutron matter up to $\rho_0$ has been recently
calculated by \citet{Gezerlis:2013,Gezerlis:2014} with nuclear two-body local interactions derived within the 
chiral effective field theory.
The AFDMC calculations for the $\chi$EFT interaction at LO, NLO, and N$^2$LO 
orders are shown in Fig.~\ref{fig:pnmeft}. (Note that three-body forces have not been included at N$^2$LO).
At each order in the 
chiral expansion, it is important to address the systematic uncertainties
entering through the regulators used to renormalize short-range correlations;
see~\citet{Gezerlis:2014} for more comprehensive details.
In the figure, the EoS obtained using cutoffs of R$_0$=1.0 fm and
1.2 fm are indicated. The figure shows that the results are converging in the chiral 
expansion, i.e. the energy per neutron at N$^2$LO is quite similar to NLO.
The three-neutron interaction entering at N$^2$LO has not been
included in the calculation but its contribution is expected to be 
small~\cite{Tews:inprep}.
Other approaches based on lattice-based QMC methods have been explored recently 
by~\citet{Wlazlowski:2014} and~\citet{Roggero:2014}, with very similar results
also included in Fig.~\ref{fig:pnmeft}.

\subsubsection{Three-neutron force and Symmetry energy}
\label{sec:esym}
As described in Sec.~\ref{sec:nn} the $N\!N$ force is obtained by accurately
fitting scattering data, but a $3N$ force is essential to have a
good description of the ground states of light nuclei.  The effect of the
$3N$ force on the nuclear matter EoS is particularly important,
as it is needed to correctly reproduce the saturation density $\rho_0$ 
and the energy.  The neutron matter EoS
is also sensitive to the particular choice of the $3N$ force,
and consequently the corresponding neutron star structure.

By assuming that the $N\!N$ Hamiltonian is well constrained by
scattering data, the effect of using different three-neutron forces to
compute the EoS of neutron matter has been studied carefully. As 
described in Sec.~\ref{sec:TNI} the $3N$ force can be split into different
parts: a long-range term given by $2\pi$-exchange, an intermediate
part described by $3\pi$-rings and a phenomenological short-range
repulsion. The role of the latter term is the least understood, although
in part it is probably mocking up a relativistic boost correction to the $N\!N$
interaction~\cite{Akmal:1998,pieper2001}. It is
important to address the effect of all these terms in the calculation of
neutron matter.  These parts have been tuned and the effective
range of the repulsive term changed to explore how these terms change the
many-body correlations in neutron matter.  The main part that has been
explored is the short-range term. This term is purely phenomenological and
it is mainly responsible for providing the correlations at high densities.
The expectation value of the $2\pi$-exchange Fujita-Miyazawa operator 
in neutron matter is small compared to $V_R$, and this limits
almost the whole effect of UIX to the short-range term \cite{Gandolfi:2012}.

From the experimental side, the EoS of neutron matter cannot be 
measured, but strong efforts have been made to measure the
isospin-symmetry energy, see the review by~\citet{Tsang:2012}. By assuming a quadratic dependence of the 
isospin-asymmetry $\delta=(\rho_n-\rho_p)/(\rho_n+\rho_p)$, the symmetry 
energy can be interpreted as the difference
between pure neutron matter ($\delta=1$) and symmetric nuclear matter ($\delta=0$)
\begin{equation}
\label{eq:esym}
E_{\rm sym}(\rho)=E_{\rm PNM}(\rho)-E_{\rm SNM}(\rho) \,,
\end{equation}
where $E_{\rm PNM}$ is the energy per neutron of pure neutron matter,
and $E_{\rm SNM}$ is the energy per nucleon of symmetric nuclear matter.
The total energy of nuclear matter will take the form
\begin{equation}
E(\rho,\delta)=E_{\rm SNM}+E_{\rm sym}(\rho)\delta^2 \,.
\end{equation}
Several experiments aim to measure the symmetry energy $E_{\rm sym}$ at the empirical
saturation density $\rho_0$=0.16 fm$^{-3}$, and the parameter $L$ related to 
its first derivative.
Around $\rho_0$ the symmetry energy can be expanded as
\begin{equation}
\label{eq:esym2}
E_{\rm sym}(\rho)=E_{\rm sym}+\frac{L}{3}\frac{\rho-\rho_0}{\rho_0}+\cdots \,.
\end{equation}

The present experimental constraints to $E_{\rm sym}$ have been used
to study the sensitivity of the EoS of neutron matter to the particular
choice of the $3N$ force. The
assumptions are that the empirical energy of nuclear matter at saturation is
$E_{\rm SNM}(\rho_0)=-16$ MeV, and through Eq.~(\ref{eq:esym}) there
is a consequent range of the energy of neutron matter at saturation,
$E_{\rm PNM}(\rho_0)$.  By following~\citet{Tsang:2009} 
the symmetry energy is expected to be in the
range $32\pm2$ MeV, corresponding to the neutron matter energy
$E_{\rm PNM}(\rho_0)=16\pm2$ MeV. Other papers report a wider range of
values of $E_{\rm sym}$; see for example~\citet{Chen:2010}.

\begin{figure}
\includegraphics[scale=0.3]{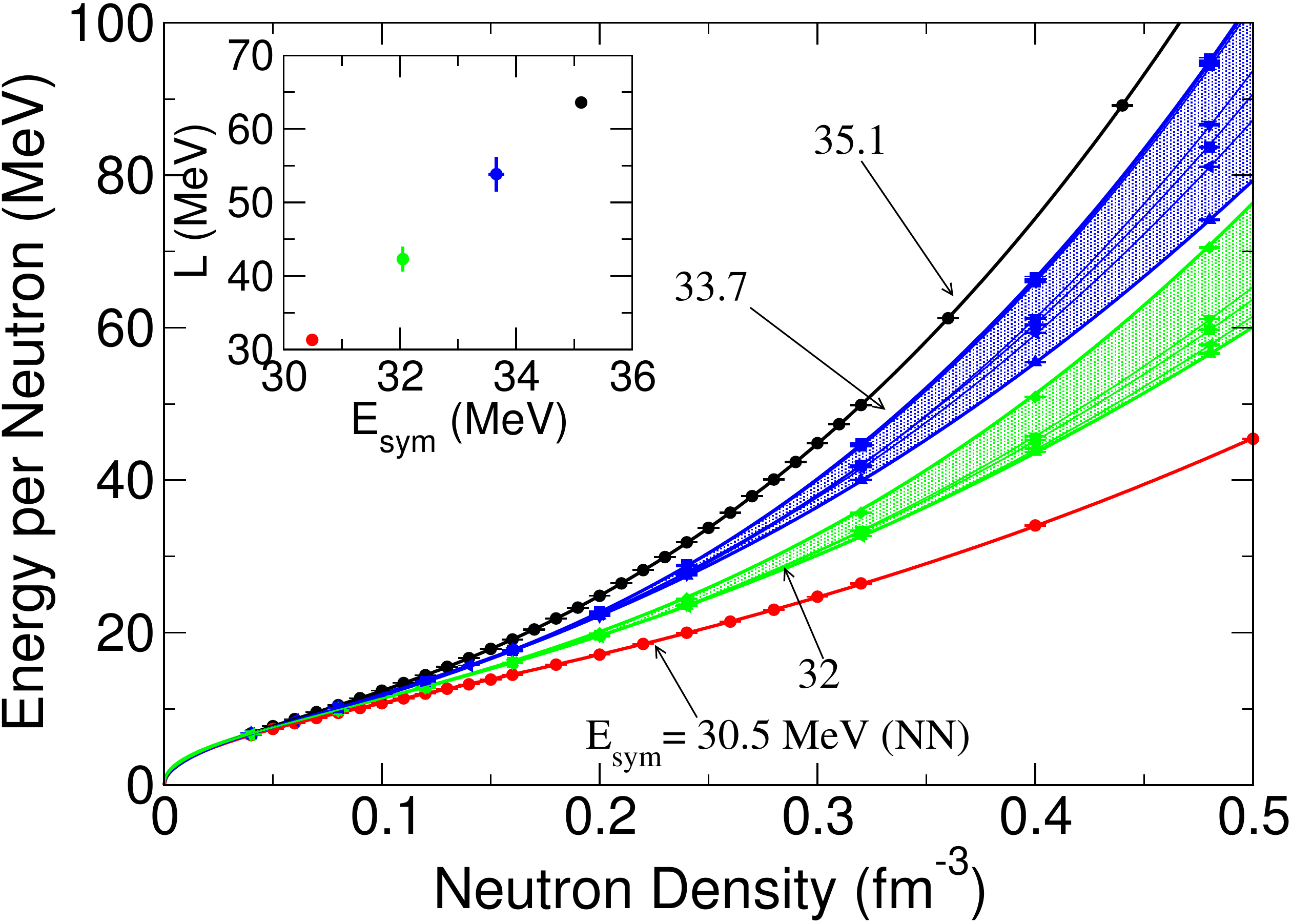}
\caption{The energy per particle of neutron matter for different values
of the nuclear symmetry energy ($E_{\rm sym}$)~\cite{Gandolfi:2012}.
For each value of $E_{\rm sym}$ the corresponding band shows the effect of
different spatial and spin structures of the three-neutron interaction.
The red and black line show the same result of Fig.~\ref{fig:pnmav8uix}
where just the two-body alone and with the original UIX three-body forces 
has been used.
The inset shows the linear correlation between $E_{\rm sym}$ and its
density derivative $L$.}
\label{fig:eospnmv3}
\end{figure}

Following~\citet{pieper2001}, different parametrizations of $A^{PW}_{2\pi}$, $A^{SW}_{2\pi}$
and $A_{3\pi}$ have been considered. 
Starting with the original strengths of these
parameters, the constant $A_R$ has been adjusted in order to reproduce
a particular value of $E_{\rm PNM}(\rho_0)$ and give a corresponding symmetry
energy. We show the various EoS
computed using different models of $3N$ interactions in Fig.~\ref{fig:eospnmv3},
compared to the AV8$^\prime$ $N\!N$ force alone and with the original UIX $3N$ 
force.
The blue and green bands in the figure show the EoS with a 
symmetry energy corresponding to 33.7 and 32 MeV, respectively.  
Each band covers the various results
obtained using different three-neutron forces adjusted to have the same
$E_{\rm sym}$. 
The parameters fitting the higher and the lower EoS 
for each band are reported in Table~\ref{tab:fitpnmav8}.
It is interesting to note that the bands are tiny
around $\rho_0$, and the uncertainty grows at larger densities. The
two bands show the sensitivity of the EoS to the three-neutron force.

The EoS are used to determine the value of $L$ as a function of $E_{\rm sym}$
in Eq.~(\ref{eq:esym2}), and the result is shown in the inset of
Fig.~\ref{fig:eospnmv3}.  As expected, the uncertainty in $L$ is very small, producing
a very accurate prediction of $L$ as a function of $E_{\rm sym}$~\cite{Gandolfi:2012}. 
These results generally agree with experimental constraints 
(see~\citet{Tsang:2012} and~\citet{Lattimer:2013}), and with constraints
from neutron stars~\cite{Steiner:2012}, as discussed in the next section.
Future
experiments with the aim to measure simultaneously $E_{\rm sym}$ and $L$
will provide a strong test of the assumed model.  Two important
aspects could be missing in this model: the relativistic effects
and the contribution of higher-order many-body forces.  However in the regime of
densities considered, these effects can probably be neglected.  First,
the relativistic effects have been previously studied in \citet{Akmal:1998},
where it has been shown that the density dependence of such effects
has roughly the same behavior as the short-range part of the three-body
force, i.e., that they can be incorporated in its short-range part. 
Second, the four-body force contributions should be suppressed relative to the
three-body force for densities up to 2-3$\rho_0$. Within $\chi$EFT this assumption can
be justified at nuclear density by the high precision fits to light-nuclei
obtained with only three-body forces~\cite{Epelbaum:2009}.
For phenomenological interactions, the contribution of the two-body potential 
energy is much larger than that of the three-body, and the four-body is then expected 
to be much smaller than the three-body in dense matter~\cite{Akmal:1998}.

\subsubsection{Neutron star structure}

\begin{figure}
\begin{center}
\includegraphics[width=8cm]{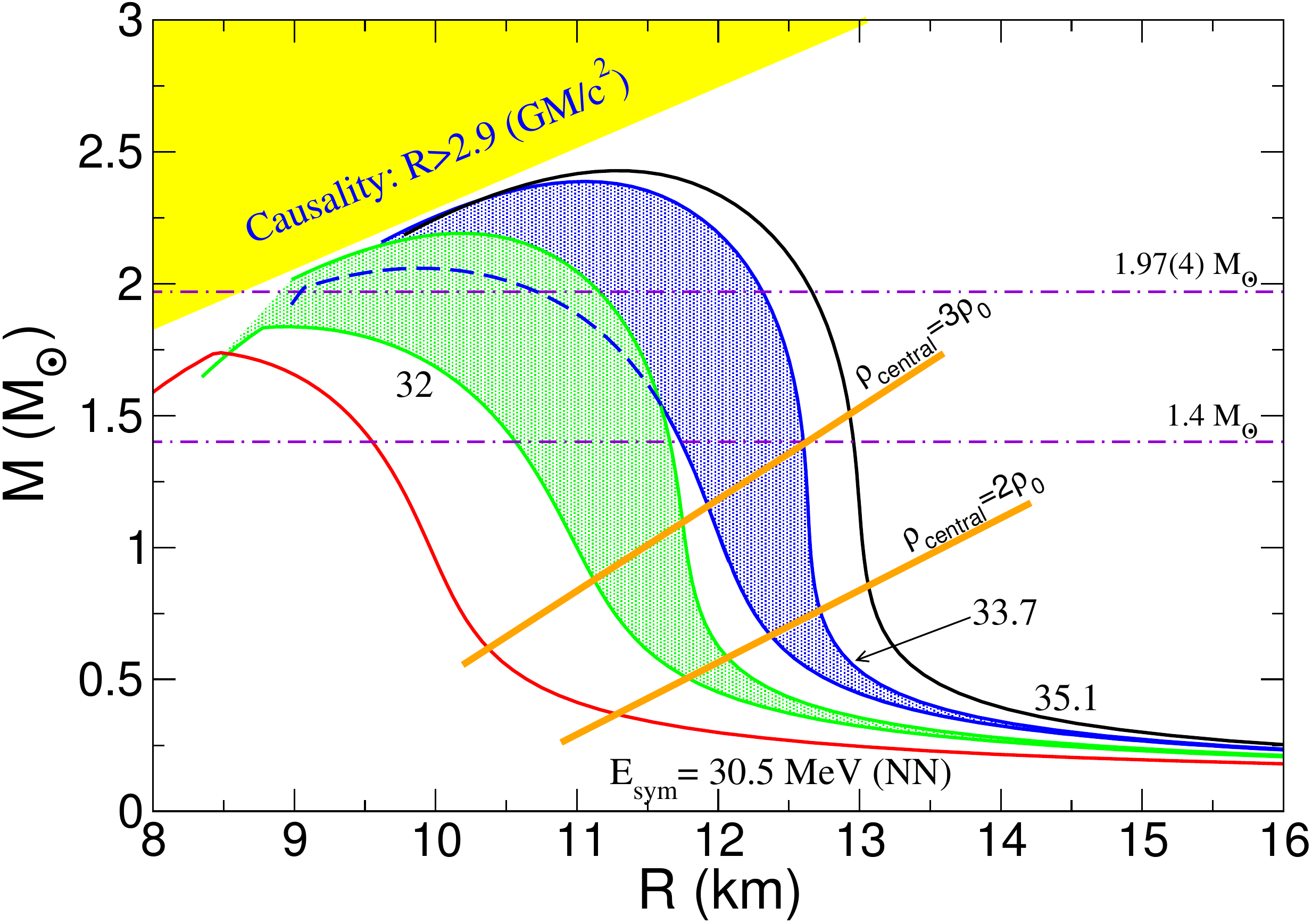}
\caption{Predicted neutron-star masses plotted as a function of stellar radius
~\cite{Gandolfi:2012,Gandolfi:2014}. Different
EoS are considered: those obtained with the AV8$^\prime$ $N\!N$ (red curve) 
and with the UIX $3N$ interactions (black solid line) presented in Sec.~\ref{sec:afdmc-pnm}.
The two green and 
blue bands show the results obtained using different $3N$ forces constrained
to have a particular value of the symmetry energy (indicate by numbers near the bands and curves).
}
\label{fig:mr} 
\end{center} 
\end{figure}

While real neutron stars are very complicated objects, their main global 
properties can usually be well-approximated by considering simple 
idealized models consisting of a perfect fluid in hydrostatic equilibrium. 
If rotation can be neglected to a first approximation (as is the case for 
the spin rates of most currently-known pulsars) then the model can be 
taken to be spherical and its structure obtained by solving the 
Tolman-Oppenheimer-Volkoff (TOV) equations \cite{TOV1939}, enabling one to calculate, for 
example, the stellar mass as a function of radius or of central density. 
Using the energy density $\epsilon(\rho)$ defined as
\begin{equation}
\epsilon(\rho)=\rho [E(\rho)+m_n c^2] \,,
\end{equation}
where $m_n$ is the mass of neutron,
and the pressure $P(\rho)$ at zero temperature is given by
\begin{equation}
P(\rho)=\rho^2\frac{\partial E(\rho)}{\partial\rho} \,,
\end{equation}
as inputs, the neutron star model is evaluated
by integrating the TOV equations:
\begin{equation}
\frac{dP}{dr}=-\frac{G[m(r)+4\pi r^3P/c^2][\epsilon+P/c^2]}{r[r-2Gm(r)/c^2]} \,,
\label{eq:tov1}
\end{equation}
\begin{equation}
\frac{dm(r)}{dr}=4\pi\epsilon r^2 \,.
\label{eq:tov2}
\end{equation}
Here $m(r)$ is the gravitational mass enclosed within a radius $r$,
and $G$ is the gravitational constant. The solution of the TOV equations
for a given central density gives the profiles of $\rho$, $\epsilon$ and
$P$ as functions of radius $r$, and also the total radius $R$ and mass
$M=m(R)$, with $R$ defined as the distance where the pressure $P$ drops to zero.
A sequence of models can be generated by specifying a succession of
values for the central density. In Fig.~\ref{fig:mr} the mass $M$
(measured in solar masses M$_{\odot}$) as a function of the radius $R$
(measured in km) is showed, as obtained from AFDMC calculations using different
prescriptions for the EoS presented in the previous sections.

It is interesting to make a comparison between these results so as to see
the changes caused by introduction of the various different features
in the Hamiltonian.  An objective of this type of work is to attempt
to constrain microphysical models for neutron-star matter by making
comparison with astronomical observations.  This has become
possible in the last few years, as discussed for example in \citet{Steiner:2010}, 
\citet{Ozel:2010}, \citet{Steiner:2012} and \citet{Steiner:2014}. 
Further progress is anticipated within the next few years if gravitational
waves from neutron star mergers can be detected.  The most recently observed
maximum neutron star masses are 1.97(2) M$_{\odot}$
\cite{Demorest:2010} and 2.01(4) M$_\odot$ \cite{Antoniadis:2013}.
These observations put the most severe constraints on the EoS,
although the precise hadronic composition is still undetermined.

\subsection{Inhomogeneous Neutron Matter}

\begin{figure}
\includegraphics[width=3.2in]{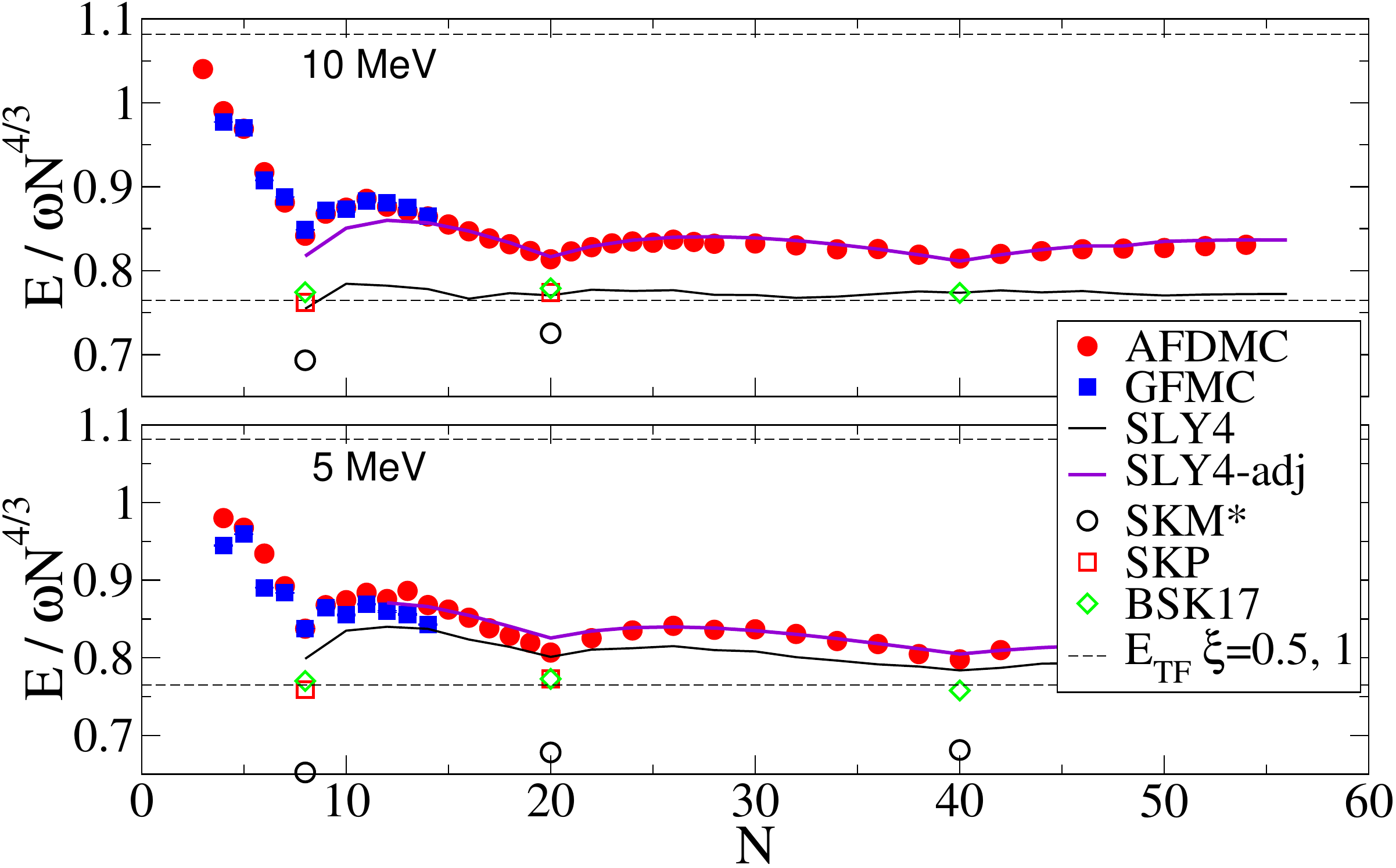}
\caption{Energies divided by $\hbar\omega N^{4/3}$ for neutrons in 
HO fields with $\hbar\omega = 10$~MeV (top) and 5~MeV
(bottom) from~\cite{Gandolfi:2011}.  Filled symbols indicate ab initio calculations;
the dashed lines are Thomas-Fermi results. Other results have been obtained 
using a variety of Skyrme forces indicated in the legend.
}
\label{fig:hospectra}
\end{figure}

While the mass and radius of a neutron star depend primarily on the equation
of state of neutron matter, the inner crust of the star contains inhomogeneous
neutron matter immersed between very neutron-rich nuclei
\cite{Ravenhall:1983,Shternin:2007,Brown:2009}.
Similarly, the exterior of very neutron-rich nuclei is believed to have a 
significant excess of neutrons.  This neutron distribution can be probed, for example, in parity-violating electron scattering.

Mean-field models including Skyrme and related models are typically fit
to bulk properties of known nuclei, which are much nearer to
isospin symmetry.  They have sometimes also included results
from ab-initio calculations of neutron matter directly in their fits,
e.g., \citet{Chabanat:1997,Chabanat:1998}.  Historically, this is
the only information used to constrain density functionals in the 
pure neutron matter limit.

\begin{figure}
\includegraphics[width=3.2in]{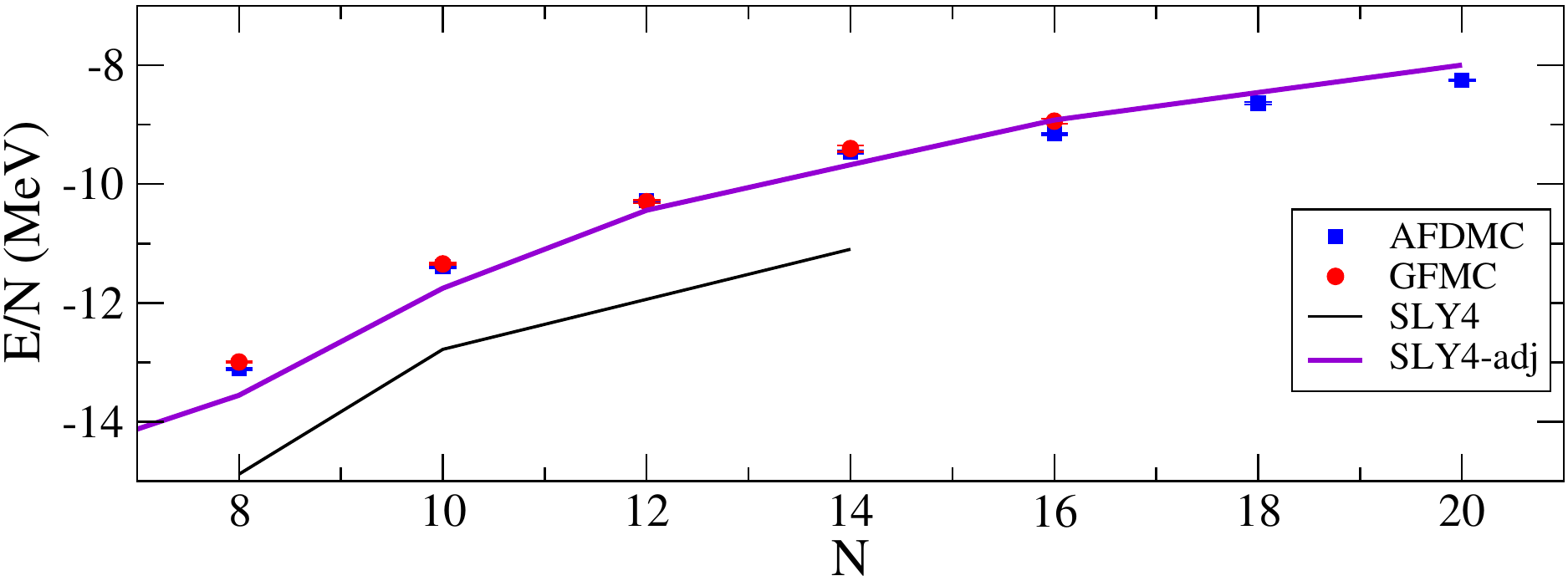}
\caption{Energies per particle for neutrons in the Woods-Saxon field~\cite{Gandolfi:2011}, 
symbols as in Fig.~\ref{fig:hospectra}.
}
\label{fig:wsspectra}
\end{figure}

Therefore it is useful to perform ab-initio studies
of inhomogeneous neutron matter at low and moderate densities.
A study of neutron drops can provide constraints on
density-functional studies of neutron-rich inhomogeneous matter, as
well as the properties of neutron-rich nuclei
that can be measured in terrestrial experiments.
\cite{Gandolfi:2011,Maris:2013}.
It is also possible to study neutron-rich nuclei with an
inert core of neutrons and protons, including realistic NN and 3N
interactions between the neutrons. This approach has been
used to study the binding energies of oxygen 
\cite{Gandolfi:2006,Chang:2004} and calcium isotopes~\cite{Gandolfi:2008}.

Calculations of neutron drops provide information about a variety
of quantities that enter in the energy-density functional.
Clearly the gradient term in pure neutron matter is important in neutron
drops, this term has a large uncertainty in fits of known nuclei.
The gradient term is important even in closed-shell arrangements of
neutrons in an external well. Studying drops between the closed
shell limits provides a variety of additional information.
One can study the superfluid pairing of pure neutron drops, a very
different environment from nuclei.  The pairing is expected to play
a more important role in dilute neutron matter, and may affect
the shell closure.  Similarly one can look at the purely 
isovector spin-orbit splitting by varying the number of 
neutrons around closed shells and possible sub-shell closures.

Early QMC calculations of very small neutron drops ($N$ = 6,7,8)
already indicated a substantial difference from traditional Skyrme
models, which overbind the drops and yield a too-large spin-orbit 
splitting~\cite{pudliner1996,Smerzi:1997,Pederiva:2004}. 
However these calculations did not systematically cover a wide range of
neutron numbers and confinement potentials.

Both GFMC and AFDMC have been used to provide ab-initio results for
neutron drops. 
The AV8$^\prime$+UIX Hamiltonian, which
produces an EoS consistent with known neutron star
masses (see the previous section), has been used 
to constrain several modern Skyrme models \cite{Gandolfi:2011,Maris:2013}.
Several forms of the external well were considered,
including harmonic oscillators (HO) of various frequency, as well
as Wood-Saxon wells. The former produce a wider range of densities,
particularly higher densities near the center of the trap, while in
the latter the density saturates as in nuclei.

The results of these calculations are shown in Figs. \ref{fig:hospectra}
and \ref{fig:wsspectra}. For the harmonic traps the energy is divided
by the frequency of the trap times N$^{4/3}$; this would be a constant
for a free Fermi Gas in the Thomas-Fermi or local density approximation.
The QMC results are shown as solid points.
For a given  Hamiltonian, the agreement between GFMC and AFDMC is
very satisfactory.  Results agree very well for the 10 MeV HO interaction,
while for $\hbar\omega = 5$~MeV, the AFDMC results are slightly higher
than the GFMC ones; the maximum difference is $3\%$, and more typically
results are within $1\%$. The bigger difference for the low-density
drops produced by the 5-MeV well
presumably arises because the importance function
used in AFDMC does not include pairing, in contrast to the more complete
treatment used in GFMC.

\begin{figure}
\includegraphics[width=3.2in]{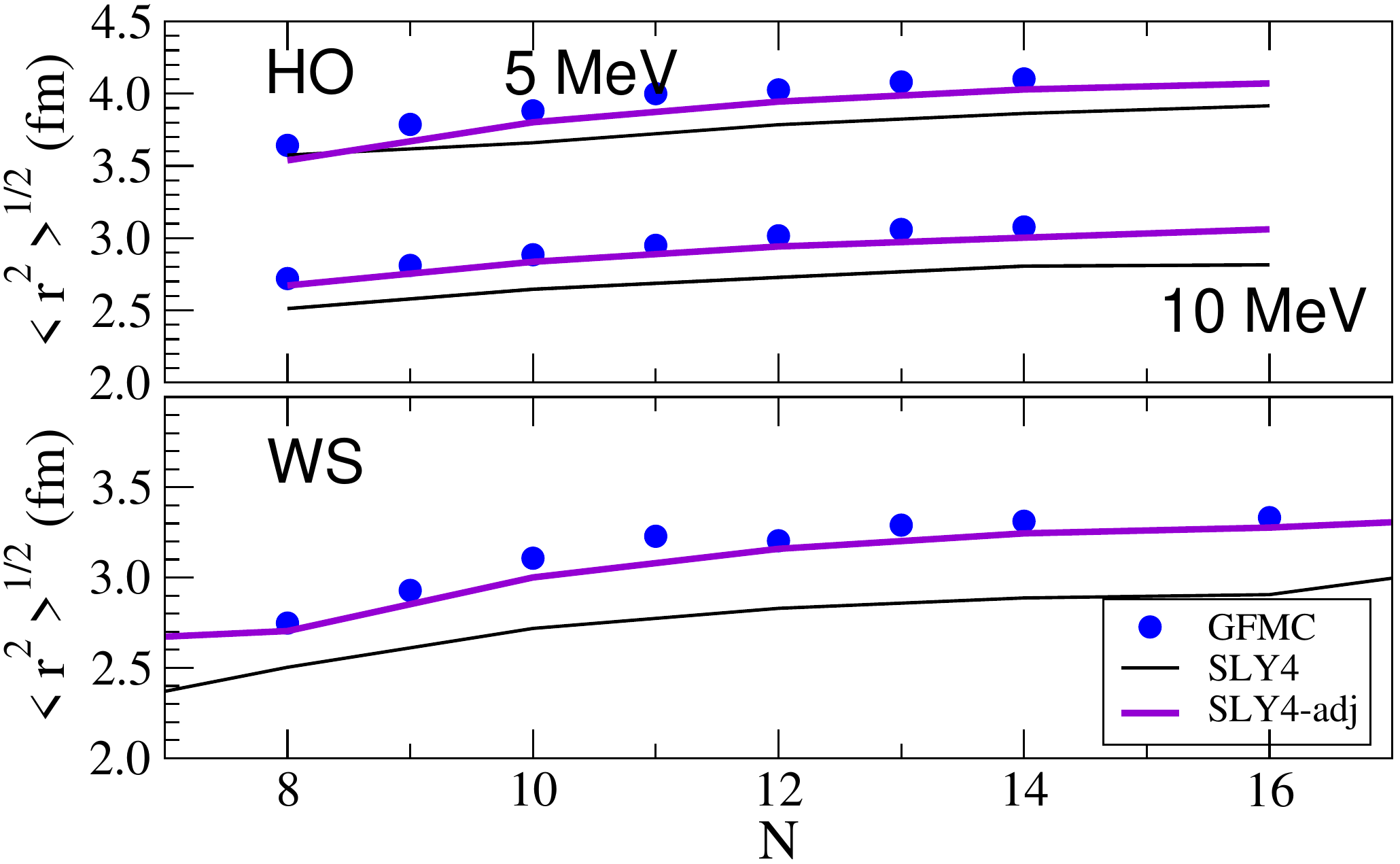}
\caption{Calculated radii of neutrons confined in HO
(upper) and WS (lower) fields compared to original and adjusted
Skyrme models (see text) from~\cite{Gandolfi:2011}.}
\label{fig:radii}
\end{figure}

In both cases conventional Skyrme models overbind the drops.
Since some of the Skyrme functionals have been fitted to the neutron matter EoS,
the overbinding might be explained by the contributions given by
the gradient term.  As is evident from
Fig.~\ref{fig:hospectra},
closed shells are still found at $N$ = 8, 20, and 40 neutrons in the HO wells.
These closed-shell states are almost exclusively sensitive to the neutron
matter EoS and the isovector gradient terms, while the contributions from
pairing and spin-orbit terms are very small.  
Instead, by examining drops with neutron numbers that differ slightly from
closed shells, one  can constrain the spin-orbit interaction. 
It has been found that a smaller isovector coupling, approximately 1/6
of the isoscalar coupling, reproduces rather accurately the ab-initio calculations for these drops.
Results for half-filled-shell drops (e.g. $N$ = 14 or 30) and odd-even staggerings are
sensitive to the pairing interactions as well as the spin-orbit force.
Fixing the spin-orbit strength from near closed-shell drops, the
pairing strength can be adjusted to fit the calculated spectra.

Adjusting these three parameters in the density
functional to better describe energies for selected number 
of neutrons in the HO as described in~\citet{Gandolfi:2011}
improves the agreement for all external fields and 
particle numbers considered.  This is shown by the upper solid curves (SLY4-adj)
in Fig.~\ref{fig:hospectra} and in Fig.~\ref{fig:wsspectra}.

\begin{figure}
\label{fig:dist}
\includegraphics[width=3.2in]{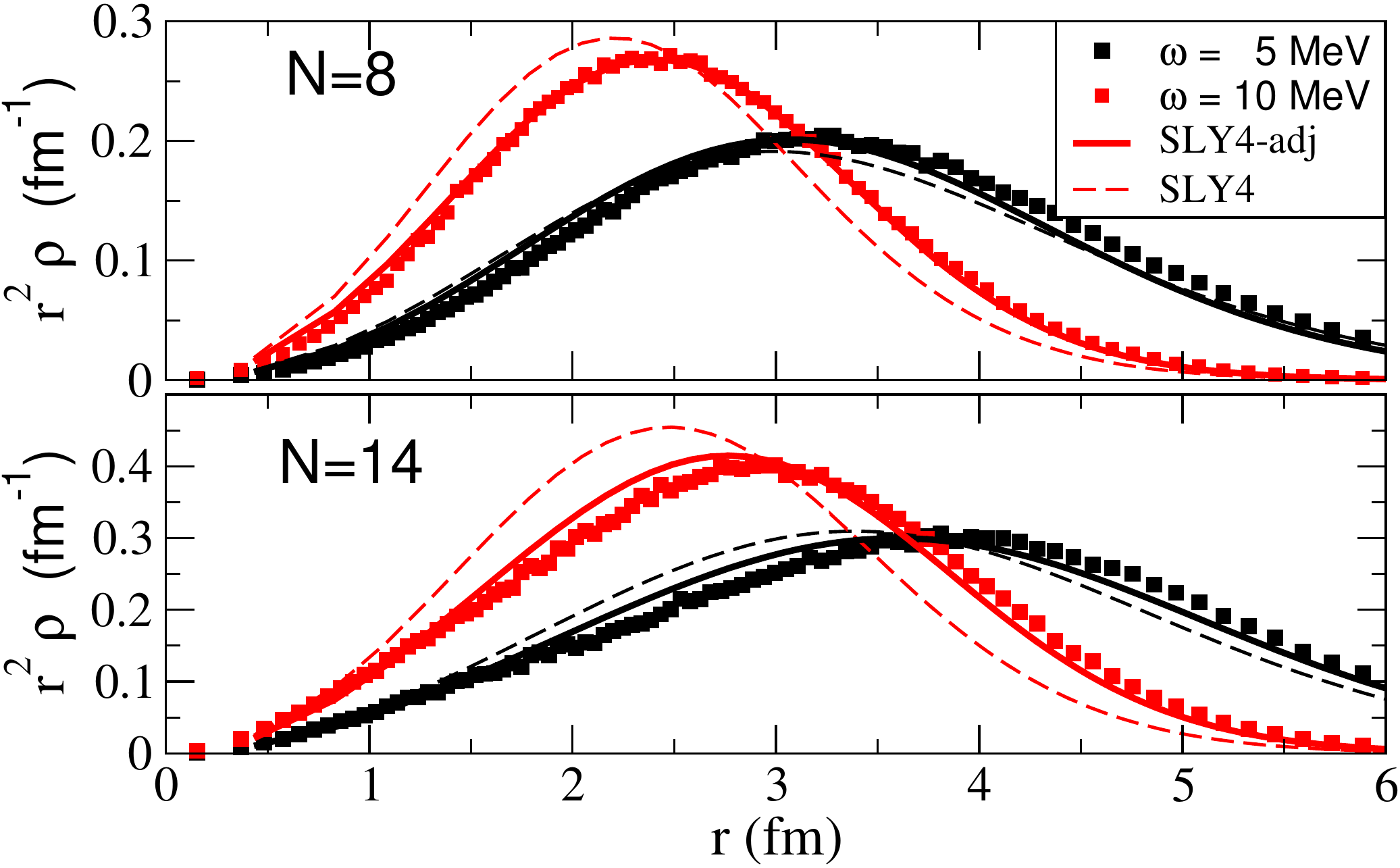}
\caption{Calculated densities of neutrons in HO
potentials compared to Skyrme models (see text) from~\cite{Gandolfi:2011}.}
\label{fig:hodist}
\end{figure}

The rms radii and density distributions of neutron drops
are also useful checks of the density functionals.
GFMC accurately computes these quantities. 
In Fig.~\ref{fig:radii} the radii computed using GFMC for different
drops are compared to those computed using the original SLY4 Skyrme and
the adjusted SLY4-adj for the two HO wells considered.
Comparisons of the densities for $N$ = 8 and 14 in the 
HO wells are shown in Fig.~\ref{fig:hodist}. These two systems
provide benchmarks of a closed-shell drop and of a half-filled-shell respectively.
The adjusted-SLY4 gives much better evaluations of these observables
than those obtained using the original SLY4 functional.

\begin{figure}
\includegraphics[width=0.45\textwidth]{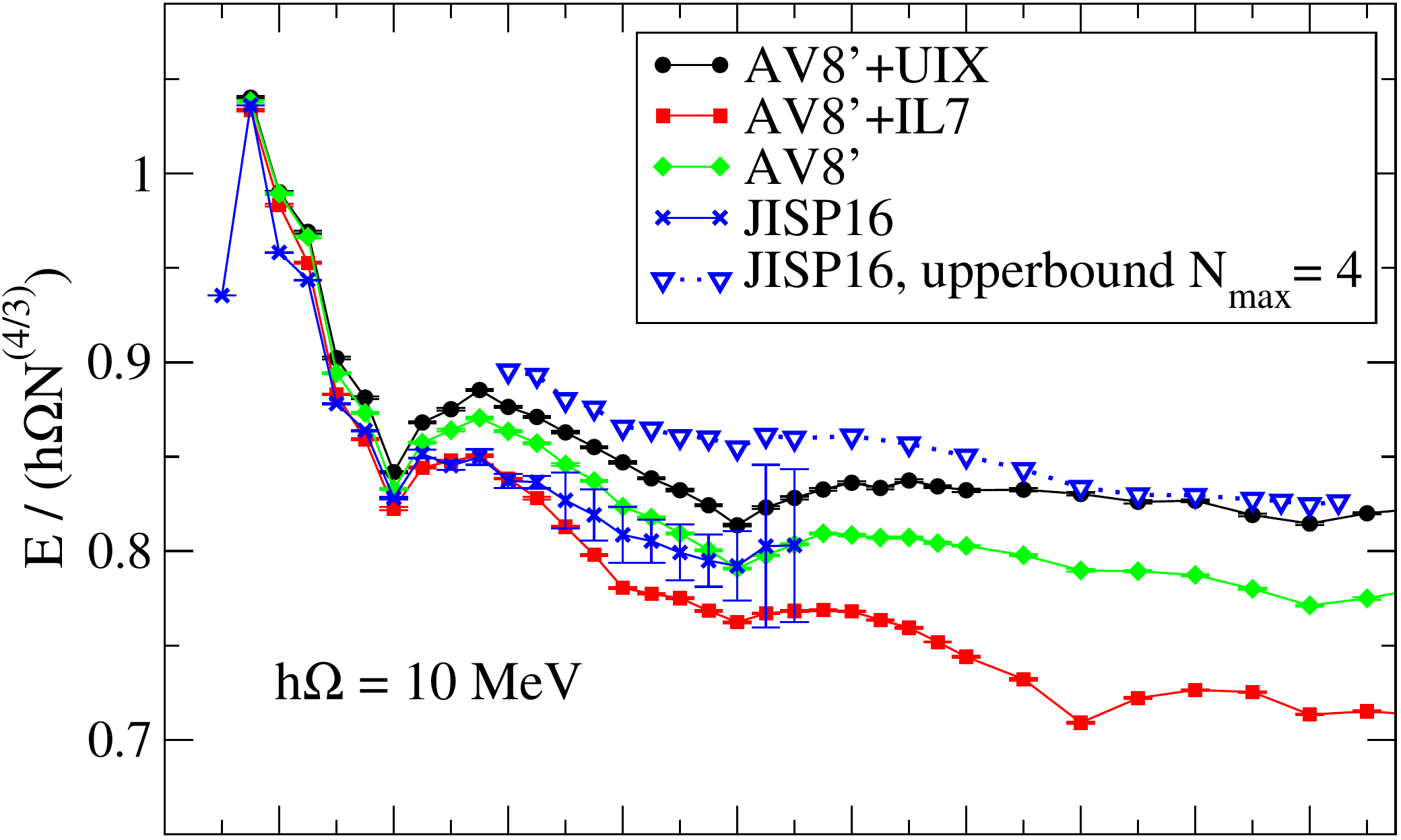}
\caption{Energies divided by $\hbar\omega N^{4/3}$ for neutrons in 
HO fields with $\hbar\omega = 10$~MeV obtained using AV8$'$ with and without 
three-neutron forces with AFDMC, and using JISP16 with the
NCFC method~\cite{Maris:2013}.
}
\label{fig:hocomp1}
\end{figure}

The QMC calculations can also be compared to predictions given by other methods.
For example, in Fig.~\ref{fig:hocomp1} the AFDMC results obtained using different
Hamiltonians (indicated in the legend) are compared to the no core full configuration
results obtained using the JISP16 interaction in no-core full configuration
(NCFC)
calculations \cite{Maris:2013}.

\begin{figure}
\includegraphics[width=0.45\textwidth]{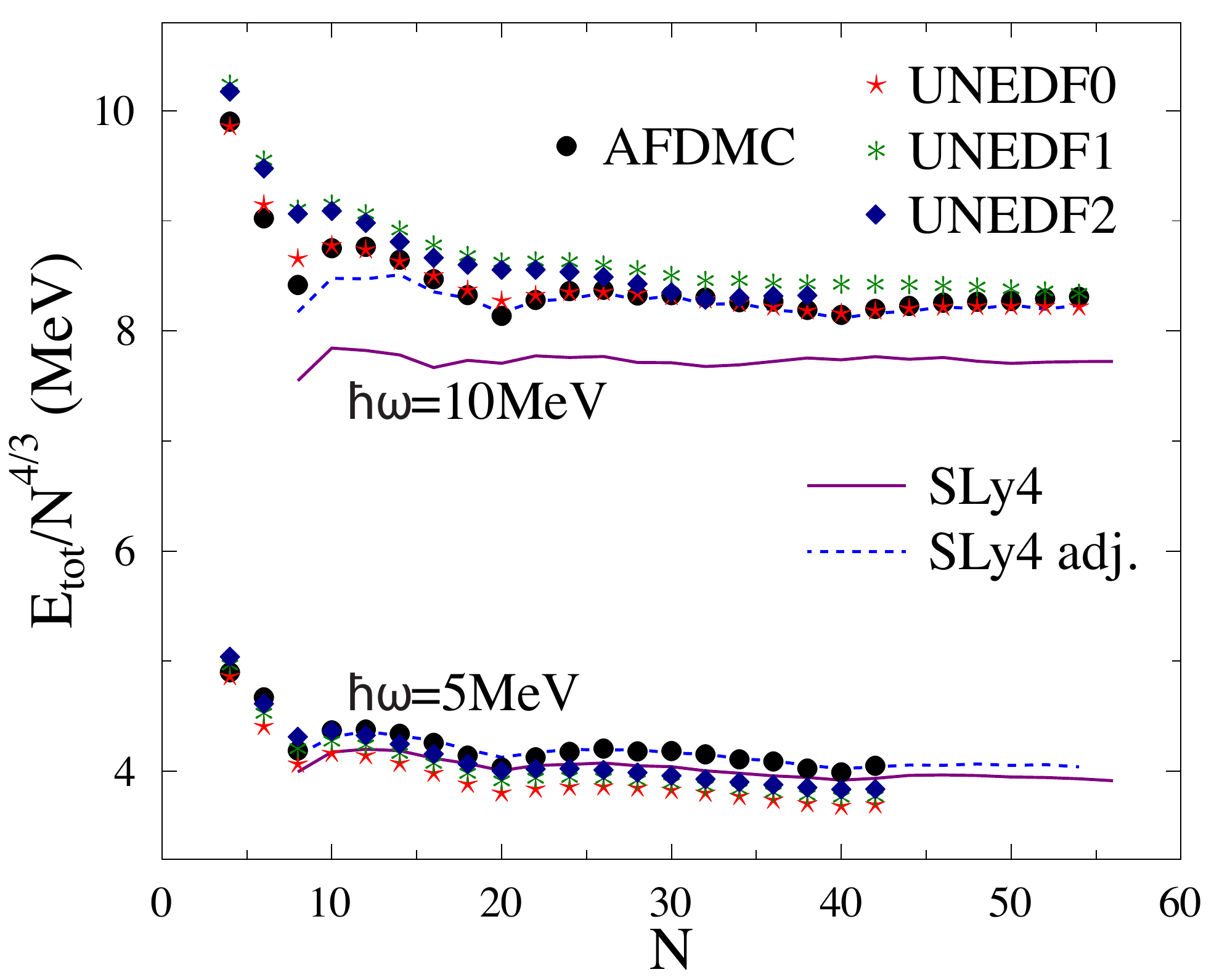}
\caption{Energies of neutron drops predicted using the UNEDF0, UNEDF1, and UNEDF2
Skyrme energy density functionals, compared to the AFDMC results~\cite{Kortelainen:2014}.
}
\label{fig:hocomp2}
\end{figure}

Recent density functionals successfully reproduce both the properties of
nuclei and neutrons drops.  The new Skyrme parametrizations 
UNEDF0, UNEDF1 and UNEDF2~\cite{Kortelainen:2014} are  compared
to QMC calculations in Fig.~\ref{fig:hocomp2}.  These new parametrizations
provide a much better fit to neutron drops.

\section{Conclusions and Future Directions}

  Quantum Monte Carlo methods have proved to be extremely 
valuable for studying the structure and reactions of nuclei and 
nucleonic matter with realistic nuclear interactions and currents.
As illustrated in this review, QMC methods can simultaneously 
treat diverse phenomena across a range of momentum scales 
including strong tensor correlations 
at short distances and the associated electroweak responses,
spectra and clustering and low-energy
EM transitions in light nuclei,
and superfluidity and the dense neutron matter equation of state.
Across this range from the lightest nuclei to neutron matter
the same nuclear models of interactions and currents are applicable. These
models have been directly obtained from nucleon-nucleon scattering
data and properties of the very lightest nuclei.

  QMC methods and accurate interaction and current models provide quantitative 
predictions for spectra, electromagnetic moments, transition rates, form factors,
asymptotic normalization constants, and other low-momentum properties 
of nuclei up to $A$=12. The recent results on electromagnetic transitions
in light nuclei is particularly encouraging, demonstrating conclusively
the importance of realistic models of two-nucleon currents even at
very low momentum transfer.  The wide range
of energies (up to $\sim$ 350 MeV lab) covered by these interactions also allow
one to study the electroweak response at rather large momentum transfers,
and to study the neutron matter equation of state up to the regime
where the Fermi momentum is $\sim 2.5$ fm$^{-1}$, a regime that controls the radius
and much of the structure of neutron stars.
Realistic models of the nuclear interaction predicted a stiff equation
of state at high densities from the two- and three-nucleon repulsion.
The recent observation of two-solar mass neutron stars confirms this
behavior.

   Progress has been due to a concerted effort of physicists studying
nuclear interactions and currents, novel quantum Monte Carlo methods,
and computer scientists and applied mathematicians enabling efficient 
computations on the largest available computers~\cite{Bogner:2013}. 
The dramatic advances in computer architecture, and the fairly 
wide availability of these machines, have also been key.

Many important challenges will be addressed in the near future,
in both light and heavy nuclei and nucleonic matter.  In light nuclei the study of
more complicated nuclear reactions will be important.  These can
address problems where it is difficult to conduct experiments,
including reactions at very low energies where the Coulomb barrier
suppresses the reaction rate, or reactions on unstable nuclei.
In addition tests of fundamental symmetries, including electric
dipole moments in light nuclei, can be addressed.
Many of these problems require only moderate advances in theory
and computation and it should be possible to address a significant
number in the next few years.

Neutrino scattering and nuclear response is of fundamental interest
in both fairly light nuclei like carbon and oxygen, and also in heavier nuclei
like argon.  Calculations of the charged-current carbon response
will be very illuminating, in particular regarding the difference of neutrino
to anti-neutrino cross sections.  This plays a key role in future
attempts to measure the neutrino mass hierarchy and the CP-violating
phase using accelerator neutrinos. Calculations in heavier nuclei
will allow us to explore the nuclear dependence of the quasielastic
scattering, which is expected to be fairly small 
as in electron scattering.

The properties of heavy neutron-rich nuclei are also very important,
particularly in light of the upcoming FRIB facility.  The extreme
neutron-rich nuclei play an important role in the r-process, and it
will be very interesting to explore questions including pairing in
neutron-rich nuclei and their weak response starting from fundamental
interactions. Of course larger nuclei also provide important tests
of fundamental symmetries, including electric dipole moments and
neutrinoless double beta decay. It will be an important challenge
to use quantum Monte Carlo techniques to study these problems.

The reliability and
dynamic range of these models are extremely important in extrapolating to
new regimes, particularly the neutron-rich matter found in supernovae
and neutron stars.  Questions to be addressed there include the equation of state
and weak response of beta-stable matter, relevant to the cooling of neutron
stars, and the response in hot low-density regimes characteristic of the
surface where the neutrinos decouple in the core-collapse supernovae.
Studies of the equation of state and its relevance to neutron star mergers
are also important.  Gravitational wave observations should be able to 
give much more precise information on the mass-radius relation in neutron stars.

We look forward to dramatic advances in theory
and computation, including a more refined understanding of nuclear interactions
and currents.  Combined with exciting prospects in experiments 
and observation, we believe there is a 
bright future for nuclear physics and its connections to quantum few-
and many-body theory, astrophysics, neutrino physics, and physics beyond
the Standard Model.

\begin{acknowledgments}
We are particularly indebted with A. Lovato for the careful reading of the 
manuscript and for the various critical comments.
We also would like to thank A. Bulgac, A. Gezerlis, D. Lonardoni, A. Lovato, 
J. Lynn, K. M. Nollett, S. Pastore, S. Reddy, and A. Roggero
for the useful discussions, and/or for sharing results.
The work of J.C., S.G., S.C.P., and R.B.W. has been supported
by the NUCLEI and previous UNEDF SciDAC projects.
This research is also supported by the U.S.~Department of Energy, Office of
Nuclear Physics, under contracts DE-AC02-05CH11231 (S.G.~and J.C.), 
DE-AC02-06CH11357 (S.C.P. and R.B.W.), and
DE-AC05-06OR23177 (R.S.), by
LISC, the Interdisciplinary Laboratory for
Computational Science, a joint venture of the University of Trento and
the Bruno Kessler Foundation (F.P.), and by National Science foundation grant PHY-1404405 (K.S.).
Under an award of computer time provided by the INCITE program,
this research used resources of the Argonne Leadership Computing
Facility at Argonne National Laboratory, which is supported by the
Office of Science of the U.S. Department of Energy under contract
DE-AC02-06CH11357.
This research also used resources provided by Los Alamos Open Supercomputing, by the National Energy
Research Scientific Computing Center (NERSC), and by Argonne's Laboratory
Computing Resource Center.
\end{acknowledgments}


\begin{thebibliography}{285}%
\makeatletter
\providecommand \@ifxundefined [1]{%
 \@ifx{#1\undefined}
}%
\providecommand \@ifnum [1]{%
 \ifnum #1\expandafter \@firstoftwo
 \else \expandafter \@secondoftwo
 \fi
}%
\providecommand \@ifx [1]{%
 \ifx #1\expandafter \@firstoftwo
 \else \expandafter \@secondoftwo
 \fi
}%
\providecommand \natexlab [1]{#1}%
\providecommand \enquote  [1]{``#1''}%
\providecommand \bibnamefont  [1]{#1}%
\providecommand \bibfnamefont [1]{#1}%
\providecommand \citenamefont [1]{#1}%
\providecommand \href@noop [0]{\@secondoftwo}%
\providecommand \href [0]{\begingroup \@sanitize@url \@href}%
\providecommand \@href[1]{\@@startlink{#1}\@@href}%
\providecommand \@@href[1]{\endgroup#1\@@endlink}%
\providecommand \@sanitize@url [0]{\catcode `\\12\catcode `\$12\catcode
  `\&12\catcode `\#12\catcode `\^12\catcode `\_12\catcode `\%12\relax}%
\providecommand \@@startlink[1]{}%
\providecommand \@@endlink[0]{}%
\providecommand \url  [0]{\begingroup\@sanitize@url \@url }%
\providecommand \@url [1]{\endgroup\@href {#1}{\urlprefix }}%
\providecommand \urlprefix  [0]{URL }%
\providecommand \Eprint [0]{\href }%
\providecommand \doibase [0]{http://dx.doi.org/}%
\providecommand \selectlanguage [0]{\@gobble}%
\providecommand \bibinfo  [0]{\@secondoftwo}%
\providecommand \bibfield  [0]{\@secondoftwo}%
\providecommand \translation [1]{[#1]}%
\providecommand \BibitemOpen [0]{}%
\providecommand \bibitemStop [0]{}%
\providecommand \bibitemNoStop [0]{.\EOS\space}%
\providecommand \EOS [0]{\spacefactor3000\relax}%
\providecommand \BibitemShut  [1]{\csname bibitem#1\endcsname}%
\let\auto@bib@innerbib\@empty
\bibitem [{\citenamefont {Abe}\ \emph {et~al.}(2012)\citenamefont {Abe},
  \citenamefont {Maris}, \citenamefont {Otsuka}, \citenamefont {Shimizu},
  \citenamefont {Utsuno},\ and\ \citenamefont {Vary}}]{Abe:2012}%
  \BibitemOpen
  \bibfield  {author} {\bibinfo {author} {\bibnamefont {Abe}, \bibfnamefont
  {T.}}, \bibinfo {author} {\bibfnamefont {P.}~\bibnamefont {Maris}}, \bibinfo
  {author} {\bibfnamefont {T.}~\bibnamefont {Otsuka}}, \bibinfo {author}
  {\bibfnamefont {N.}~\bibnamefont {Shimizu}}, \bibinfo {author} {\bibfnamefont
  {Y.}~\bibnamefont {Utsuno}}, \ and\ \bibinfo {author} {\bibfnamefont {J.~P.}\
  \bibnamefont {Vary}}} (\bibinfo {year} {2012}),\ \href {\doibase
  10.1103/PhysRevC.86.054301} {\bibfield  {journal} {\bibinfo  {journal} {Phys.
  Rev. C}\ }\textbf {\bibinfo {volume} {86}},\ \bibinfo {pages}
  {054301}}\BibitemShut {NoStop}%
\bibitem [{\citenamefont {Abe}\ and\ \citenamefont {Seki}(2009)}]{Abe:2009}%
  \BibitemOpen
  \bibfield  {author} {\bibinfo {author} {\bibnamefont {Abe}, \bibfnamefont
  {T.}}, \ and\ \bibinfo {author} {\bibfnamefont {R.}~\bibnamefont {Seki}}}
  (\bibinfo {year} {2009}),\ \href {\doibase 10.1103/PhysRevC.79.054003}
  {\bibfield  {journal} {\bibinfo  {journal} {Phys. Rev.}\ }\textbf {\bibinfo
  {volume} {C79}},\ \bibinfo {pages} {054003}}\BibitemShut {NoStop}%
\bibitem [{\citenamefont {Acha}\ \emph {et~al.}(2007)\citenamefont {Acha} \emph
  {et~al.}}]{Ach07}%
  \BibitemOpen
  \bibfield  {author} {\bibinfo {author} {\bibnamefont {Acha}, \bibfnamefont
  {A.}},  \emph {et~al.} (\bibinfo {collaboration} {HAPPEX Collaboration})}
  (\bibinfo {year} {2007}),\ \href {\doibase 10.1103/PhysRevLett.98.032301}
  {\bibfield  {journal} {\bibinfo  {journal} {Phys. Rev. Lett.}\ }\textbf
  {\bibinfo {volume} {98}},\ \bibinfo {pages} {032301}}\BibitemShut {NoStop}%
\bibitem [{\citenamefont {Ackerbauer}\ \emph {et~al.}(1998)\citenamefont
  {Ackerbauer} \emph {et~al.}}]{Ack98}%
  \BibitemOpen
  \bibfield  {author} {\bibinfo {author} {\bibnamefont {Ackerbauer},
  \bibfnamefont {P.}},  \emph {et~al.}} (\bibinfo {year} {1998}),\ \href
  {\doibase http://dx.doi.org/10.1016/S0370-2693(97)01382-8} {\bibfield
  {journal} {\bibinfo  {journal} {Phys. Lett. B}\ }\textbf {\bibinfo {volume}
  {417}},\ \bibinfo {pages} {224 }}\BibitemShut {NoStop}%
\bibitem [{\citenamefont {Adelberger}\ \emph {et~al.}(2011)\citenamefont
  {Adelberger} \emph {et~al.}}]{adelberger2011}%
  \BibitemOpen
  \bibfield  {author} {\bibinfo {author} {\bibnamefont {Adelberger},
  \bibfnamefont {E.~G.}},  \emph {et~al.}} (\bibinfo {year} {2011}),\ \href
  {\doibase 10.1103/RevModPhys.83.195} {\bibfield  {journal} {\bibinfo
  {journal} {Rev. Mod. Phys.}\ }\textbf {\bibinfo {volume} {83}},\ \bibinfo
  {pages} {195}}\BibitemShut {NoStop}%
\bibitem [{\citenamefont {Aguilar-Areval}(2008)}]{Aguilar:2008}%
  \BibitemOpen
  \bibfield  {author} {\bibinfo {author} {\bibnamefont {Aguilar-Areval},
  \bibfnamefont {A.~A.~{\it et al.}.}} (\bibinfo {collaboration} {MiniBooNE
  Collaboration})} (\bibinfo {year} {2008}),\ \href {\doibase
  10.1103/PhysRevLett.100.032301} {\bibfield  {journal} {\bibinfo  {journal}
  {Phys. Rev. Lett.}\ }\textbf {\bibinfo {volume} {100}},\ \bibinfo {pages}
  {032301}}\BibitemShut {NoStop}%
\bibitem [{\citenamefont {Ahmed}\ \emph {et~al.}(2012)\citenamefont {Ahmed}
  \emph {et~al.}}]{Ahm12}%
  \BibitemOpen
  \bibfield  {author} {\bibinfo {author} {\bibnamefont {Ahmed}, \bibfnamefont
  {Z.}},  \emph {et~al.} (\bibinfo {collaboration} {HAPPEX Collaboration})}
  (\bibinfo {year} {2012}),\ \href {\doibase 10.1103/PhysRevLett.108.102001}
  {\bibfield  {journal} {\bibinfo  {journal} {Phys. Rev. Lett.}\ }\textbf
  {\bibinfo {volume} {108}},\ \bibinfo {pages} {102001}}\BibitemShut {NoStop}%
\bibitem [{\citenamefont {Ahrens}\ \emph {et~al.}(1987)\citenamefont {Ahrens}
  \emph {et~al.}}]{Ahr87}%
  \BibitemOpen
  \bibfield  {author} {\bibinfo {author} {\bibnamefont {Ahrens}, \bibfnamefont
  {L.~A.}},  \emph {et~al.}} (\bibinfo {year} {1987}),\ \href {\doibase
  10.1103/PhysRevD.35.785} {\bibfield  {journal} {\bibinfo  {journal} {Phys.
  Rev. D}\ }\textbf {\bibinfo {volume} {35}},\ \bibinfo {pages}
  {785}}\BibitemShut {NoStop}%
\bibitem [{\citenamefont {Akmal}\ \emph {et~al.}(1998)\citenamefont {Akmal},
  \citenamefont {Pandharipande},\ and\ \citenamefont {Ravenhall}}]{Akmal:1998}%
  \BibitemOpen
  \bibfield  {author} {\bibinfo {author} {\bibnamefont {Akmal}, \bibfnamefont
  {A.}}, \bibinfo {author} {\bibfnamefont {V.~R.}\ \bibnamefont
  {Pandharipande}}, \ and\ \bibinfo {author} {\bibfnamefont {D.~G.}\
  \bibnamefont {Ravenhall}}} (\bibinfo {year} {1998}),\ \href {\doibase
  10.1103/PhysRevC.58.1804} {\bibfield  {journal} {\bibinfo  {journal} {Phys.
  Rev. C}\ }\textbf {\bibinfo {volume} {58}}~(\bibinfo {number} {3}),\ \bibinfo
  {pages} {1804}}\BibitemShut {NoStop}%
\bibitem [{\citenamefont {{Amaldi}}\ \emph {et~al.}(1979)\citenamefont
  {{Amaldi}}, \citenamefont {{Fubini}},\ and\ \citenamefont
  {{Furlan}}}]{Amaldi:1979}%
  \BibitemOpen
  \bibfield  {author} {\bibinfo {author} {\bibnamefont {{Amaldi}},
  \bibfnamefont {E.}}, \bibinfo {author} {\bibfnamefont {S.}~\bibnamefont
  {{Fubini}}}, \ and\ \bibinfo {author} {\bibfnamefont {G.}~\bibnamefont
  {{Furlan}}}} (\bibinfo {year} {1979}),\ \href {\doibase 10.1007/BFb0048208}
  {\bibfield  {journal} {\bibinfo  {journal} {Springer Tracts in Modern
  Physics}\ }\textbf {\bibinfo {volume} {93}},\ 10.1007/BFb0048208}\BibitemShut
  {NoStop}%
\bibitem [{\citenamefont {Amroun}\ \emph {et~al.}(1994)\citenamefont {Amroun},
  \citenamefont {Breton}, \citenamefont {Cavedon}, \citenamefont {Frois},
  \citenamefont {Goutte}, \citenamefont {Juster}, \citenamefont {Leconte},
  \citenamefont {Martino}, \citenamefont {Mizuno}, \citenamefont {Phan},
  \citenamefont {Platchkov}, \citenamefont {Sick},\ and\ \citenamefont
  {Williamson}}]{Amroun94}%
  \BibitemOpen
  \bibfield  {author} {\bibinfo {author} {\bibnamefont {Amroun}, \bibfnamefont
  {A.}}, \bibinfo {author} {\bibfnamefont {V.}~\bibnamefont {Breton}}, \bibinfo
  {author} {\bibfnamefont {J.-M.}\ \bibnamefont {Cavedon}}, \bibinfo {author}
  {\bibfnamefont {B.}~\bibnamefont {Frois}}, \bibinfo {author} {\bibfnamefont
  {D.}~\bibnamefont {Goutte}}, \bibinfo {author} {\bibfnamefont
  {F.}~\bibnamefont {Juster}}, \bibinfo {author} {\bibfnamefont
  {P.}~\bibnamefont {Leconte}}, \bibinfo {author} {\bibfnamefont
  {J.}~\bibnamefont {Martino}}, \bibinfo {author} {\bibfnamefont
  {Y.}~\bibnamefont {Mizuno}}, \bibinfo {author} {\bibfnamefont {X.-H.}\
  \bibnamefont {Phan}}, \bibinfo {author} {\bibfnamefont {S.}~\bibnamefont
  {Platchkov}}, \bibinfo {author} {\bibfnamefont {I.}~\bibnamefont {Sick}}, \
  and\ \bibinfo {author} {\bibfnamefont {S.}~\bibnamefont {Williamson}}}
  (\bibinfo {year} {1994}),\ \href@noop {} {\bibfield  {journal} {\bibinfo
  {journal} {Nucl. Phys. A}\ }\textbf {\bibinfo {volume} {579}},\ \bibinfo
  {pages} {596}}\BibitemShut {NoStop}%
\bibitem [{\citenamefont {Anderson}(1976)}]{anderson1976}%
  \BibitemOpen
  \bibfield  {author} {\bibinfo {author} {\bibnamefont {Anderson},
  \bibfnamefont {J.~B.}}} (\bibinfo {year} {1976}),\ \href@noop {} {\bibfield
  {journal} {\bibinfo  {journal} {J. Chem. Phys.}\ }\textbf {\bibinfo {volume}
  {65}}~(\bibinfo {number} {10}),\ \bibinfo {pages} {4121}}\BibitemShut
  {NoStop}%
\bibitem [{\citenamefont {Andreev}\ \emph {et~al.}(2007)\citenamefont {Andreev}
  \emph {et~al.}}]{And07}%
  \BibitemOpen
  \bibfield  {author} {\bibinfo {author} {\bibnamefont {Andreev}, \bibfnamefont
  {V.~A.}},  \emph {et~al.} (\bibinfo {collaboration} {MuCap Collaboration})}
  (\bibinfo {year} {2007}),\ \href {\doibase 10.1103/PhysRevLett.99.032002}
  {\bibfield  {journal} {\bibinfo  {journal} {Phys. Rev. Lett.}\ }\textbf
  {\bibinfo {volume} {99}},\ \bibinfo {pages} {032002}}\BibitemShut {NoStop}%
\bibitem [{\citenamefont {Aniol}\ \emph {et~al.}(2004)\citenamefont {Aniol}
  \emph {et~al.}}]{Ani04}%
  \BibitemOpen
  \bibfield  {author} {\bibinfo {author} {\bibnamefont {Aniol}, \bibfnamefont
  {K.~A.}},  \emph {et~al.} (\bibinfo {collaboration} {HAPPEX Collaboration})}
  (\bibinfo {year} {2004}),\ \href {\doibase 10.1103/PhysRevC.69.065501}
  {\bibfield  {journal} {\bibinfo  {journal} {Phys. Rev. C}\ }\textbf {\bibinfo
  {volume} {69}},\ \bibinfo {pages} {065501}}\BibitemShut {NoStop}%
\bibitem [{\citenamefont {Antoniadis}\ \emph {et~al.}(2013)\citenamefont
  {Antoniadis} \emph {et~al.}}]{Antoniadis:2013}%
  \BibitemOpen
  \bibfield  {author} {\bibinfo {author} {\bibnamefont {Antoniadis},
  \bibfnamefont {J.}},  \emph {et~al.}} (\bibinfo {year} {2013}),\ \href@noop
  {} {\bibfield  {journal} {\bibinfo  {journal} {Science}\ }\textbf {\bibinfo
  {volume} {340}}~(\bibinfo {number} {6131})}\BibitemShut {NoStop}%
\bibitem [{\citenamefont {Bacca}\ and\ \citenamefont
  {Pastore}(2014)}]{bacca2014}%
  \BibitemOpen
  \bibfield  {author} {\bibinfo {author} {\bibnamefont {Bacca}, \bibfnamefont
  {S.}}, \ and\ \bibinfo {author} {\bibfnamefont {S.}~\bibnamefont {Pastore}}}
  (\bibinfo {year} {2014}),\ \href@noop {} {\bibfield  {journal} {\bibinfo
  {journal} {J. Phys. G; Nucl. Part. Phys.}\ }\textbf {\bibinfo {volume}
  {41}},\ \bibinfo {pages} {123002}}\BibitemShut {NoStop}%
\bibitem [{\citenamefont {Baker}\ \emph {et~al.}(1981)\citenamefont {Baker},
  \citenamefont {Cnops}, \citenamefont {Connolly}, \citenamefont {Kahn},
  \citenamefont {Kirk}, \citenamefont {Murtagh}, \citenamefont {Palmer},
  \citenamefont {Samios},\ and\ \citenamefont {Tanaka}}]{Bak81}%
  \BibitemOpen
  \bibfield  {author} {\bibinfo {author} {\bibnamefont {Baker}, \bibfnamefont
  {N.~J.}}, \bibinfo {author} {\bibfnamefont {A.~M.}\ \bibnamefont {Cnops}},
  \bibinfo {author} {\bibfnamefont {P.~L.}\ \bibnamefont {Connolly}}, \bibinfo
  {author} {\bibfnamefont {S.~A.}\ \bibnamefont {Kahn}}, \bibinfo {author}
  {\bibfnamefont {H.~G.}\ \bibnamefont {Kirk}}, \bibinfo {author}
  {\bibfnamefont {M.~J.}\ \bibnamefont {Murtagh}}, \bibinfo {author}
  {\bibfnamefont {R.~B.}\ \bibnamefont {Palmer}}, \bibinfo {author}
  {\bibfnamefont {N.~P.}\ \bibnamefont {Samios}}, \ and\ \bibinfo {author}
  {\bibfnamefont {M.}~\bibnamefont {Tanaka}}} (\bibinfo {year} {1981}),\ \href
  {\doibase 10.1103/PhysRevD.23.2499} {\bibfield  {journal} {\bibinfo
  {journal} {Phys. Rev. D}\ }\textbf {\bibinfo {volume} {23}},\ \bibinfo
  {pages} {2499}}\BibitemShut {NoStop}%
\bibitem [{\citenamefont {Barletta}\ \emph {et~al.}(2009)\citenamefont
  {Barletta}, \citenamefont {Romero-Redondo}, \citenamefont {Kievsky},
  \citenamefont {Viviani},\ and\ \citenamefont {Garrido}}]{Barletta:2009}%
  \BibitemOpen
  \bibfield  {author} {\bibinfo {author} {\bibnamefont {Barletta},
  \bibfnamefont {P.}}, \bibinfo {author} {\bibfnamefont {C.}~\bibnamefont
  {Romero-Redondo}}, \bibinfo {author} {\bibfnamefont {A.}~\bibnamefont
  {Kievsky}}, \bibinfo {author} {\bibfnamefont {M.}~\bibnamefont {Viviani}}, \
  and\ \bibinfo {author} {\bibfnamefont {E.}~\bibnamefont {Garrido}}} (\bibinfo
  {year} {2009}),\ \href {\doibase 10.1103/PhysRevLett.103.090402} {\bibfield
  {journal} {\bibinfo  {journal} {Phys. Rev. Lett.}\ }\textbf {\bibinfo
  {volume} {103}},\ \bibinfo {pages} {090402}}\BibitemShut {NoStop}%
\bibitem [{\citenamefont {Barrett}\ \emph {et~al.}(2013)\citenamefont
  {Barrett}, \citenamefont {Navrátil},\ and\ \citenamefont
  {Vary}}]{Barrett:2013}%
  \BibitemOpen
  \bibfield  {author} {\bibinfo {author} {\bibnamefont {Barrett}, \bibfnamefont
  {B.~R.}}, \bibinfo {author} {\bibfnamefont {P.}~\bibnamefont {Navrátil}}, \
  and\ \bibinfo {author} {\bibfnamefont {J.~P.}\ \bibnamefont {Vary}}}
  (\bibinfo {year} {2013}),\ \href {\doibase
  http://dx.doi.org/10.1016/j.ppnp.2012.10.003} {\bibfield  {journal} {\bibinfo
   {journal} {Progress in Particle and Nuclear Physics}\ }\textbf {\bibinfo
  {volume} {69}}~(\bibinfo {number} {0}),\ \bibinfo {pages} {131 }}\BibitemShut
  {NoStop}%
\bibitem [{\citenamefont {Beane}\ \emph {et~al.}(2013)\citenamefont {Beane},
  \citenamefont {Chang}, \citenamefont {Cohen}, \citenamefont {Detmold},
  \citenamefont {Lin}, \citenamefont {Luu}, \citenamefont {Orginos},
  \citenamefont {Parre\~no}, \citenamefont {Savage},\ and\ \citenamefont
  {Walker-Loud}}]{Beane:2013}%
  \BibitemOpen
  \bibfield  {author} {\bibinfo {author} {\bibnamefont {Beane}, \bibfnamefont
  {S.~R.}}, \bibinfo {author} {\bibfnamefont {E.}~\bibnamefont {Chang}},
  \bibinfo {author} {\bibfnamefont {S.~D.}\ \bibnamefont {Cohen}}, \bibinfo
  {author} {\bibfnamefont {W.}~\bibnamefont {Detmold}}, \bibinfo {author}
  {\bibfnamefont {H.~W.}\ \bibnamefont {Lin}}, \bibinfo {author} {\bibfnamefont
  {T.~C.}\ \bibnamefont {Luu}}, \bibinfo {author} {\bibfnamefont
  {K.}~\bibnamefont {Orginos}}, \bibinfo {author} {\bibfnamefont
  {A.}~\bibnamefont {Parre\~no}}, \bibinfo {author} {\bibfnamefont {M.~J.}\
  \bibnamefont {Savage}}, \ and\ \bibinfo {author} {\bibfnamefont
  {A.}~\bibnamefont {Walker-Loud}}} (\bibinfo {year} {2013}),\ \href {\doibase
  10.1103/PhysRevD.87.034506} {\bibfield  {journal} {\bibinfo  {journal} {Phys.
  Rev. D}\ }\textbf {\bibinfo {volume} {87}},\ \bibinfo {pages}
  {034506}}\BibitemShut {NoStop}%
\bibitem [{\citenamefont {Bedaque}\ and\ \citenamefont {van
  Kolck}(2002)}]{Bed02}%
  \BibitemOpen
  \bibfield  {author} {\bibinfo {author} {\bibnamefont {Bedaque}, \bibfnamefont
  {P.}}, \ and\ \bibinfo {author} {\bibfnamefont {U.}~\bibnamefont {van
  Kolck}}} (\bibinfo {year} {2002}),\ \href@noop {} {\bibfield  {journal}
  {\bibinfo  {journal} {Ann.\ Rev.\ Nucl.\ Part.\ Sci.}\ }\textbf {\bibinfo
  {volume} {52}},\ \bibinfo {pages} {339}}\BibitemShut {NoStop}%
\bibitem [{\citenamefont {Beise}\ \emph {et~al.}(2005)\citenamefont {Beise},
  \citenamefont {Pitt},\ and\ \citenamefont {Spayde}}]{Bei05}%
  \BibitemOpen
  \bibfield  {author} {\bibinfo {author} {\bibnamefont {Beise}, \bibfnamefont
  {E.}}, \bibinfo {author} {\bibfnamefont {M.}~\bibnamefont {Pitt}}, \ and\
  \bibinfo {author} {\bibfnamefont {D.}~\bibnamefont {Spayde}}} (\bibinfo
  {year} {2005}),\ \href {\doibase
  http://dx.doi.org/10.1016/j.ppnp.2004.07.002} {\bibfield  {journal} {\bibinfo
   {journal} {Progress in Particle and Nuclear Physics}\ }\textbf {\bibinfo
  {volume} {54}}~(\bibinfo {number} {1}),\ \bibinfo {pages} {289 }}\BibitemShut
  {NoStop}%
\bibitem [{\citenamefont {Benhar}\ \emph {et~al.}(2010)\citenamefont {Benhar},
  \citenamefont {Coletti},\ and\ \citenamefont {Meloni}}]{Benhar:2010}%
  \BibitemOpen
  \bibfield  {author} {\bibinfo {author} {\bibnamefont {Benhar}, \bibfnamefont
  {O.}}, \bibinfo {author} {\bibfnamefont {P.}~\bibnamefont {Coletti}}, \ and\
  \bibinfo {author} {\bibfnamefont {D.}~\bibnamefont {Meloni}}} (\bibinfo
  {year} {2010}),\ \href {\doibase 10.1103/PhysRevLett.105.132301} {\bibfield
  {journal} {\bibinfo  {journal} {Phys. Rev. Lett.}\ }\textbf {\bibinfo
  {volume} {105}},\ \bibinfo {pages} {132301}}\BibitemShut {NoStop}%
\bibitem [{\citenamefont {Bernard}\ \emph {et~al.}(2011)\citenamefont
  {Bernard}, \citenamefont {Epelbaum}, \citenamefont {Krebs},\ and\
  \citenamefont {Meissner}}]{Ber11}%
  \BibitemOpen
  \bibfield  {author} {\bibinfo {author} {\bibnamefont {Bernard}, \bibfnamefont
  {V.}}, \bibinfo {author} {\bibfnamefont {E.}~\bibnamefont {Epelbaum}},
  \bibinfo {author} {\bibfnamefont {H.}~\bibnamefont {Krebs}}, \ and\ \bibinfo
  {author} {\bibfnamefont {U.-G.}\ \bibnamefont {Meissner}}} (\bibinfo {year}
  {2011}),\ \href@noop {} {\bibfield  {journal} {\bibinfo  {journal} {Phys.
  Rev. C}\ }\textbf {\bibinfo {volume} {84}},\ \bibinfo {pages}
  {054001}}\BibitemShut {NoStop}%
\bibitem [{\citenamefont {Bernard}\ \emph {et~al.}(1994)\citenamefont
  {Bernard}, \citenamefont {Kaiser},\ and\ \citenamefont {Meissner}}]{Ber94}%
  \BibitemOpen
  \bibfield  {author} {\bibinfo {author} {\bibnamefont {Bernard}, \bibfnamefont
  {V.}}, \bibinfo {author} {\bibfnamefont {N.}~\bibnamefont {Kaiser}}, \ and\
  \bibinfo {author} {\bibfnamefont {U.-G.}\ \bibnamefont {Meissner}}} (\bibinfo
  {year} {1994}),\ \href {\doibase 10.1103/PhysRevD.50.6899} {\bibfield
  {journal} {\bibinfo  {journal} {Phys. Rev. D}\ }\textbf {\bibinfo {volume}
  {50}},\ \bibinfo {pages} {6899}}\BibitemShut {NoStop}%
\bibitem [{\citenamefont {B.J.~Hammond}(1994)}]{Hammond:1994}%
  \BibitemOpen
  \bibfield  {author} {\bibinfo {author} {\bibnamefont {B.J.~Hammond},
  \bibfnamefont {P.~R., W.A.~Lester}}} (\bibinfo {year} {1994}),\ \href@noop {}
  {\emph {\bibinfo {title} {Monte Carlo Methods in ab Initio Quantum
  Chemistry}}}\ (\bibinfo  {publisher} {World Scientific,
  Singapore})\BibitemShut {NoStop}%
\bibitem [{\citenamefont {Bogner}\ \emph {et~al.}(2013)\citenamefont {Bogner},
  \citenamefont {Bulgac}, \citenamefont {Carlson}, \citenamefont {Engel},
  \citenamefont {Fann} \emph {et~al.}}]{Bogner:2013}%
  \BibitemOpen
  \bibfield  {author} {\bibinfo {author} {\bibnamefont {Bogner}, \bibfnamefont
  {S.}}, \bibinfo {author} {\bibfnamefont {A.}~\bibnamefont {Bulgac}}, \bibinfo
  {author} {\bibfnamefont {J.~A.}\ \bibnamefont {Carlson}}, \bibinfo {author}
  {\bibfnamefont {J.}~\bibnamefont {Engel}}, \bibinfo {author} {\bibfnamefont
  {G.}~\bibnamefont {Fann}},  \emph {et~al.}} (\bibinfo {year} {2013}),\ \href
  {\doibase 10.1016/j.cpc.2013.05.020} {\bibfield  {journal} {\bibinfo
  {journal} {Comput. Phys. Commun.}\ }\textbf {\bibinfo {volume} {184}},\
  \bibinfo {pages} {2235}}\BibitemShut {NoStop}%
\bibitem [{\citenamefont {Bogner}\ \emph {et~al.}(2010)\citenamefont {Bogner},
  \citenamefont {Furnstahl},\ and\ \citenamefont {Schwenk}}]{Bogner:2010}%
  \BibitemOpen
  \bibfield  {author} {\bibinfo {author} {\bibnamefont {Bogner}, \bibfnamefont
  {S.}}, \bibinfo {author} {\bibfnamefont {R.}~\bibnamefont {Furnstahl}}, \
  and\ \bibinfo {author} {\bibfnamefont {A.}~\bibnamefont {Schwenk}}} (\bibinfo
  {year} {2010}),\ \href@noop {} {\bibfield  {journal} {\bibinfo  {journal}
  {Progress in Particle and Nuclear Physics}\ }\textbf {\bibinfo {volume}
  {65}}~(\bibinfo {number} {1}),\ \bibinfo {pages} {94}}\BibitemShut {NoStop}%
\bibitem [{\citenamefont {Bonett-Matiz}\ \emph {et~al.}(2013)\citenamefont
  {Bonett-Matiz}, \citenamefont {Mukherjee},\ and\ \citenamefont
  {Alhassid}}]{Bonett-Matiz:2013}%
  \BibitemOpen
  \bibfield  {author} {\bibinfo {author} {\bibnamefont {Bonett-Matiz},
  \bibfnamefont {M.}}, \bibinfo {author} {\bibfnamefont {A.}~\bibnamefont
  {Mukherjee}}, \ and\ \bibinfo {author} {\bibfnamefont {Y.}~\bibnamefont
  {Alhassid}}} (\bibinfo {year} {2013}),\ \href {\doibase
  10.1103/PhysRevC.88.011302} {\bibfield  {journal} {\bibinfo  {journal} {Phys.
  Rev. C}\ }\textbf {\bibinfo {volume} {88}},\ \bibinfo {pages}
  {011302}}\BibitemShut {NoStop}%
\bibitem [{\citenamefont {Bonnard}\ and\ \citenamefont
  {Juillet}(2013)}]{Bonnard:2013}%
  \BibitemOpen
  \bibfield  {author} {\bibinfo {author} {\bibnamefont {Bonnard}, \bibfnamefont
  {J.}}, \ and\ \bibinfo {author} {\bibfnamefont {O.}~\bibnamefont {Juillet}}}
  (\bibinfo {year} {2013}),\ \href {\doibase 10.1103/PhysRevLett.111.012502}
  {\bibfield  {journal} {\bibinfo  {journal} {Phys. Rev. Lett.}\ }\textbf
  {\bibinfo {volume} {111}},\ \bibinfo {pages} {012502}}\BibitemShut {NoStop}%
\bibitem [{\citenamefont {Brida}\ \emph {et~al.}(2011)\citenamefont {Brida},
  \citenamefont {Pieper},\ and\ \citenamefont {Wiringa}}]{brida2011}%
  \BibitemOpen
  \bibfield  {author} {\bibinfo {author} {\bibnamefont {Brida}, \bibfnamefont
  {I.}}, \bibinfo {author} {\bibfnamefont {S.~C.}\ \bibnamefont {Pieper}}, \
  and\ \bibinfo {author} {\bibfnamefont {R.~B.}\ \bibnamefont {Wiringa}}}
  (\bibinfo {year} {2011}),\ \href@noop {} {\bibfield  {journal} {\bibinfo
  {journal} {Phys. Rev. C}\ }\textbf {\bibinfo {volume} {84}},\ \bibinfo
  {pages} {024319}}\BibitemShut {NoStop}%
\bibitem [{\citenamefont {{Brown}}\ and\ \citenamefont
  {{Cumming}}(2009)}]{Brown:2009}%
  \BibitemOpen
  \bibfield  {author} {\bibinfo {author} {\bibnamefont {{Brown}}, \bibfnamefont
  {E.~F.}}, \ and\ \bibinfo {author} {\bibfnamefont {A.}~\bibnamefont
  {{Cumming}}}} (\bibinfo {year} {2009}),\ \href {\doibase
  10.1088/0004-637X/698/2/1020} {\bibfield  {journal} {\bibinfo  {journal}
  {\apj}\ }\textbf {\bibinfo {volume} {698}},\ \bibinfo {pages}
  {1020}}\BibitemShut {NoStop}%
\bibitem [{\citenamefont {Bryan}(1990)}]{Bryan:1990}%
  \BibitemOpen
  \bibfield  {author} {\bibinfo {author} {\bibnamefont {Bryan}, \bibfnamefont
  {R.}}} (\bibinfo {year} {1990}),\ \href@noop {} {\bibfield  {journal}
  {\bibinfo  {journal} {European Biophysics Journal}\ }\textbf {\bibinfo
  {volume} {18}}~(\bibinfo {number} {3}),\ \bibinfo {pages} {165}}\BibitemShut
  {NoStop}%
\bibitem [{\citenamefont {Butkevich}(2010)}]{Butkevich10}%
  \BibitemOpen
  \bibfield  {author} {\bibinfo {author} {\bibnamefont {Butkevich},
  \bibfnamefont {A.~V.}}} (\bibinfo {year} {2010}),\ \href {\doibase
  10.1103/PhysRevC.82.055501} {\bibfield  {journal} {\bibinfo  {journal} {Phys.
  Rev. C}\ }\textbf {\bibinfo {volume} {82}},\ \bibinfo {pages}
  {055501}}\BibitemShut {NoStop}%
\bibitem [{\citenamefont {Carlson}(1986)}]{Car86}%
  \BibitemOpen
  \bibfield  {author} {\bibinfo {author} {\bibnamefont {Carlson}, \bibfnamefont
  {C.~E.}}} (\bibinfo {year} {1986}),\ \href {\doibase
  10.1103/PhysRevD.34.2704} {\bibfield  {journal} {\bibinfo  {journal} {Phys.
  Rev. D}\ }\textbf {\bibinfo {volume} {34}},\ \bibinfo {pages}
  {2704}}\BibitemShut {NoStop}%
\bibitem [{\citenamefont {Carlson}(1987)}]{carlson1987}%
  \BibitemOpen
  \bibfield  {author} {\bibinfo {author} {\bibnamefont {Carlson}, \bibfnamefont
  {J.}}} (\bibinfo {year} {1987}),\ \href {\doibase 10.1103/PhysRevC.36.2026}
  {\bibfield  {journal} {\bibinfo  {journal} {Phys. Rev. C}\ }\textbf {\bibinfo
  {volume} {36}}~(\bibinfo {number} {5}),\ \bibinfo {pages} {2026}}\BibitemShut
  {NoStop}%
\bibitem [{\citenamefont {Carlson}(1988)}]{carlson1988}%
  \BibitemOpen
  \bibfield  {author} {\bibinfo {author} {\bibnamefont {Carlson}, \bibfnamefont
  {J.}}} (\bibinfo {year} {1988}),\ \href {\doibase 10.1103/PhysRevC.38.1879}
  {\bibfield  {journal} {\bibinfo  {journal} {Phys. Rev. C}\ }\textbf {\bibinfo
  {volume} {38}}~(\bibinfo {number} {4}),\ \bibinfo {pages} {1879}}\BibitemShut
  {NoStop}%
\bibitem [{\citenamefont {Carlson}(2003)}]{Carlson:2003b}%
  \BibitemOpen
  \bibfield  {author} {\bibinfo {author} {\bibnamefont {Carlson}, \bibfnamefont
  {J.}}} (\bibinfo {year} {2003}),\ \href@noop {} {\bibfield  {journal}
  {\bibinfo  {journal} {Eur. Phys. J. A}\ }\textbf {\bibinfo {volume} {17}},\
  \bibinfo {pages} {463}}\BibitemShut {NoStop}%
\bibitem [{\citenamefont {Carlson}\ \emph
  {et~al.}(2003{\natexlab{a}})\citenamefont {Carlson}, \citenamefont {Chang},
  \citenamefont {Pandharipande},\ and\ \citenamefont
  {Schmidt}}]{Carlson:2003c}%
  \BibitemOpen
  \bibfield  {author} {\bibinfo {author} {\bibnamefont {Carlson}, \bibfnamefont
  {J.}}, \bibinfo {author} {\bibfnamefont {S.-Y.}\ \bibnamefont {Chang}},
  \bibinfo {author} {\bibfnamefont {V.~R.}\ \bibnamefont {Pandharipande}}, \
  and\ \bibinfo {author} {\bibfnamefont {K.~E.}\ \bibnamefont {Schmidt}}}
  (\bibinfo {year} {2003}{\natexlab{a}}),\ \href {\doibase
  10.1103/PhysRevLett.91.050401} {\bibfield  {journal} {\bibinfo  {journal}
  {Phys. Rev. Lett.}\ }\textbf {\bibinfo {volume} {91}}~(\bibinfo {number}
  {5}),\ \bibinfo {pages} {050401}}\BibitemShut {NoStop}%
\bibitem [{\citenamefont {Carlson}\ \emph {et~al.}(2012)\citenamefont
  {Carlson}, \citenamefont {Gandolfi},\ and\ \citenamefont
  {Gezerlis}}]{Carlson:2012}%
  \BibitemOpen
  \bibfield  {author} {\bibinfo {author} {\bibnamefont {Carlson}, \bibfnamefont
  {J.}}, \bibinfo {author} {\bibfnamefont {S.}~\bibnamefont {Gandolfi}}, \ and\
  \bibinfo {author} {\bibfnamefont {A.}~\bibnamefont {Gezerlis}}} (\bibinfo
  {year} {2012}),\ \href {\doibase 10.1093/ptep/pts031} {\bibfield  {journal}
  {\bibinfo  {journal} {PTEP}\ }\textbf {\bibinfo {volume} {2012}},\ \bibinfo
  {pages} {01A209}}\BibitemShut {NoStop}%
\bibitem [{\citenamefont {Carlson}\ \emph {et~al.}(2002)\citenamefont
  {Carlson}, \citenamefont {Jourdan}, \citenamefont {Schiavilla},\ and\
  \citenamefont {Sick}}]{Carlson:2002}%
  \BibitemOpen
  \bibfield  {author} {\bibinfo {author} {\bibnamefont {Carlson}, \bibfnamefont
  {J.}}, \bibinfo {author} {\bibfnamefont {J.}~\bibnamefont {Jourdan}},
  \bibinfo {author} {\bibfnamefont {R.}~\bibnamefont {Schiavilla}}, \ and\
  \bibinfo {author} {\bibfnamefont {I.}~\bibnamefont {Sick}}} (\bibinfo {year}
  {2002}),\ \href {\doibase 10.1103/PhysRevC.65.024002} {\bibfield  {journal}
  {\bibinfo  {journal} {Phys. Rev. C}\ }\textbf {\bibinfo {volume} {65}},\
  \bibinfo {pages} {024002}}\BibitemShut {NoStop}%
\bibitem [{\citenamefont {Carlson}\ \emph
  {et~al.}(2003{\natexlab{b}})\citenamefont {Carlson}, \citenamefont {Morales},
  \citenamefont {Pandharipande},\ and\ \citenamefont
  {Ravenhall}}]{Carlson:2003}%
  \BibitemOpen
  \bibfield  {author} {\bibinfo {author} {\bibnamefont {Carlson}, \bibfnamefont
  {J.}}, \bibinfo {author} {\bibfnamefont {J.}~\bibnamefont {Morales}},
  \bibinfo {author} {\bibfnamefont {V.~R.}\ \bibnamefont {Pandharipande}}, \
  and\ \bibinfo {author} {\bibfnamefont {D.~G.}\ \bibnamefont {Ravenhall}}}
  (\bibinfo {year} {2003}{\natexlab{b}}),\ \href {\doibase
  10.1103/PhysRevC.68.025802} {\bibfield  {journal} {\bibinfo  {journal} {Phys.
  Rev. C}\ }\textbf {\bibinfo {volume} {68}}~(\bibinfo {number} {2}),\ \bibinfo
  {pages} {025802}}\BibitemShut {NoStop}%
\bibitem [{\citenamefont {Carlson}\ \emph {et~al.}(1993)\citenamefont
  {Carlson}, \citenamefont {Pandharipande},\ and\ \citenamefont
  {Schiavilla}}]{carlson1993}%
  \BibitemOpen
  \bibfield  {author} {\bibinfo {author} {\bibnamefont {Carlson}, \bibfnamefont
  {J.}}, \bibinfo {author} {\bibfnamefont {V.~R.}\ \bibnamefont
  {Pandharipande}}, \ and\ \bibinfo {author} {\bibfnamefont {R.}~\bibnamefont
  {Schiavilla}}} (\bibinfo {year} {1993}),\ \href {\doibase
  10.1103/PhysRevC.47.484} {\bibfield  {journal} {\bibinfo  {journal} {Phys.
  Rev. C}\ }\textbf {\bibinfo {volume} {47}}~(\bibinfo {number} {2}),\ \bibinfo
  {pages} {484}}\BibitemShut {NoStop}%
\bibitem [{\citenamefont {Carlson}\ \emph {et~al.}(1983)\citenamefont
  {Carlson}, \citenamefont {Pandharipande},\ and\ \citenamefont
  {Wiringa}}]{carlson1983}%
  \BibitemOpen
  \bibfield  {author} {\bibinfo {author} {\bibnamefont {Carlson}, \bibfnamefont
  {J.}}, \bibinfo {author} {\bibfnamefont {V.~R.}\ \bibnamefont
  {Pandharipande}}, \ and\ \bibinfo {author} {\bibfnamefont {R.~B.}\
  \bibnamefont {Wiringa}}} (\bibinfo {year} {1983}),\ \href@noop {} {\bibfield
  {journal} {\bibinfo  {journal} {Nucl. Phys. A}\ }\textbf {\bibinfo {volume}
  {401}},\ \bibinfo {pages} {59}}\BibitemShut {NoStop}%
\bibitem [{\citenamefont {Carlson}\ \emph {et~al.}(1991)\citenamefont
  {Carlson}, \citenamefont {Riska}, \citenamefont {Schiavilla},\ and\
  \citenamefont {Wiringa}}]{Car91}%
  \BibitemOpen
  \bibfield  {author} {\bibinfo {author} {\bibnamefont {Carlson}, \bibfnamefont
  {J.}}, \bibinfo {author} {\bibfnamefont {D.~O.}\ \bibnamefont {Riska}},
  \bibinfo {author} {\bibfnamefont {R.}~\bibnamefont {Schiavilla}}, \ and\
  \bibinfo {author} {\bibfnamefont {R.~B.}\ \bibnamefont {Wiringa}}} (\bibinfo
  {year} {1991}),\ \href {\doibase 10.1103/PhysRevC.44.619} {\bibfield
  {journal} {\bibinfo  {journal} {Phys. Rev. C}\ }\textbf {\bibinfo {volume}
  {44}},\ \bibinfo {pages} {619}}\BibitemShut {NoStop}%
\bibitem [{\citenamefont {Carlson}\ and\ \citenamefont
  {Schiavilla}(1992)}]{Carlson:1994}%
  \BibitemOpen
  \bibfield  {author} {\bibinfo {author} {\bibnamefont {Carlson}, \bibfnamefont
  {J.}}, \ and\ \bibinfo {author} {\bibfnamefont {R.}~\bibnamefont
  {Schiavilla}}} (\bibinfo {year} {1992}),\ \href {\doibase
  10.1103/PhysRevLett.68.3682} {\bibfield  {journal} {\bibinfo  {journal}
  {Phys. Rev. Lett.}\ }\textbf {\bibinfo {volume} {68}},\ \bibinfo {pages}
  {3682}}\BibitemShut {NoStop}%
\bibitem [{\citenamefont {Carlson}\ and\ \citenamefont
  {Schiavilla}(1994)}]{Carlson:1994a}%
  \BibitemOpen
  \bibfield  {author} {\bibinfo {author} {\bibnamefont {Carlson}, \bibfnamefont
  {J.}}, \ and\ \bibinfo {author} {\bibfnamefont {R.}~\bibnamefont
  {Schiavilla}}} (\bibinfo {year} {1994}),\ \href {\doibase
  10.1103/PhysRevC.49.R2880} {\bibfield  {journal} {\bibinfo  {journal} {Phys.
  Rev. C}\ }\textbf {\bibinfo {volume} {49}},\ \bibinfo {pages}
  {R2880}}\BibitemShut {NoStop}%
\bibitem [{\citenamefont {Carlson}\ and\ \citenamefont
  {Schiavilla}(1998)}]{carlson1998}%
  \BibitemOpen
  \bibfield  {author} {\bibinfo {author} {\bibnamefont {Carlson}, \bibfnamefont
  {J.}}, \ and\ \bibinfo {author} {\bibfnamefont {R.}~\bibnamefont
  {Schiavilla}}} (\bibinfo {year} {1998}),\ \href {\doibase
  10.1103/RevModPhys.70.743} {\bibfield  {journal} {\bibinfo  {journal} {Rev.
  Mod. Phys.}\ }\textbf {\bibinfo {volume} {70}}~(\bibinfo {number} {3}),\
  \bibinfo {pages} {743}}\BibitemShut {NoStop}%
\bibitem [{\citenamefont {Carlson}\ \emph {et~al.}(1987)\citenamefont
  {Carlson}, \citenamefont {Schmidt},\ and\ \citenamefont
  {Kalos}}]{Carlson:1987}%
  \BibitemOpen
  \bibfield  {author} {\bibinfo {author} {\bibnamefont {Carlson}, \bibfnamefont
  {J.}}, \bibinfo {author} {\bibfnamefont {K.~E.}\ \bibnamefont {Schmidt}}, \
  and\ \bibinfo {author} {\bibfnamefont {M.~H.}\ \bibnamefont {Kalos}}}
  (\bibinfo {year} {1987}),\ \href {\doibase 10.1103/PhysRevC.36.27} {\bibfield
   {journal} {\bibinfo  {journal} {Phys. Rev. C}\ }\textbf {\bibinfo {volume}
  {36}},\ \bibinfo {pages} {27}}\BibitemShut {NoStop}%
\bibitem [{\citenamefont {Ceperley}\ \emph {et~al.}(1977)\citenamefont
  {Ceperley}, \citenamefont {Chester},\ and\ \citenamefont
  {Kalos}}]{Ceperley:1977}%
  \BibitemOpen
  \bibfield  {author} {\bibinfo {author} {\bibnamefont {Ceperley},
  \bibfnamefont {D.}}, \bibinfo {author} {\bibfnamefont {G.~V.}\ \bibnamefont
  {Chester}}, \ and\ \bibinfo {author} {\bibfnamefont {M.~H.}\ \bibnamefont
  {Kalos}}} (\bibinfo {year} {1977}),\ \href {\doibase
  10.1103/PhysRevB.16.3081} {\bibfield  {journal} {\bibinfo  {journal} {Phys.
  Rev. B}\ }\textbf {\bibinfo {volume} {16}}~(\bibinfo {number} {7}),\ \bibinfo
  {pages} {3081}}\BibitemShut {NoStop}%
\bibitem [{\citenamefont {Ceperley}(1995)}]{ceperley1995}%
  \BibitemOpen
  \bibfield  {author} {\bibinfo {author} {\bibnamefont {Ceperley},
  \bibfnamefont {D.~M.}}} (\bibinfo {year} {1995}),\ \href {\doibase
  10.1103/RevModPhys.67.279} {\bibfield  {journal} {\bibinfo  {journal} {Rev.
  Mod. Phys.}\ }\textbf {\bibinfo {volume} {67}},\ \bibinfo {pages}
  {279}}\BibitemShut {NoStop}%
\bibitem [{\citenamefont {{Chabanat}}\ \emph {et~al.}(1997)\citenamefont
  {{Chabanat}}, \citenamefont {{Bonche}}, \citenamefont {{Haensel}},
  \citenamefont {{Meyer}},\ and\ \citenamefont {{Schaeffer}}}]{Chabanat:1997}%
  \BibitemOpen
  \bibfield  {author} {\bibinfo {author} {\bibnamefont {{Chabanat}},
  \bibfnamefont {E.}}, \bibinfo {author} {\bibfnamefont {P.}~\bibnamefont
  {{Bonche}}}, \bibinfo {author} {\bibfnamefont {P.}~\bibnamefont {{Haensel}}},
  \bibinfo {author} {\bibfnamefont {J.}~\bibnamefont {{Meyer}}}, \ and\
  \bibinfo {author} {\bibfnamefont {R.}~\bibnamefont {{Schaeffer}}}} (\bibinfo
  {year} {1997}),\ \href {\doibase 10.1016/S0375-9474(97)00596-4} {\bibfield
  {journal} {\bibinfo  {journal} {Nucl. Phys. A}\ }\textbf {\bibinfo {volume}
  {627}},\ \bibinfo {pages} {710}}\BibitemShut {NoStop}%
\bibitem [{\citenamefont {{Chabanat}}\ \emph {et~al.}(1998)\citenamefont
  {{Chabanat}}, \citenamefont {{Bonche}}, \citenamefont {{Haensel}},
  \citenamefont {{Meyer}},\ and\ \citenamefont {{Schaeffer}}}]{Chabanat:1998}%
  \BibitemOpen
  \bibfield  {author} {\bibinfo {author} {\bibnamefont {{Chabanat}},
  \bibfnamefont {E.}}, \bibinfo {author} {\bibfnamefont {P.}~\bibnamefont
  {{Bonche}}}, \bibinfo {author} {\bibfnamefont {P.}~\bibnamefont {{Haensel}}},
  \bibinfo {author} {\bibfnamefont {J.}~\bibnamefont {{Meyer}}}, \ and\
  \bibinfo {author} {\bibfnamefont {R.}~\bibnamefont {{Schaeffer}}}} (\bibinfo
  {year} {1998}),\ \href {\doibase 10.1016/S0375-9474(98)00180-8} {\bibfield
  {journal} {\bibinfo  {journal} {Nucl. Phys. A}\ }\textbf {\bibinfo {volume}
  {635}},\ \bibinfo {pages} {231}}\BibitemShut {NoStop}%
\bibitem [{\citenamefont {Chang}\ \emph {et~al.}(2004)\citenamefont {Chang},
  \citenamefont {Morales}, \citenamefont {Jr.}, \citenamefont {Pandharipande},
  \citenamefont {Ravenhall}, \citenamefont {Carlson}, \citenamefont {Pieper},
  \citenamefont {Wiringa},\ and\ \citenamefont {Schmidt}}]{Chang:2004}%
  \BibitemOpen
  \bibfield  {author} {\bibinfo {author} {\bibnamefont {Chang}, \bibfnamefont
  {S.~Y.}}, \bibinfo {author} {\bibfnamefont {J.}~\bibnamefont {Morales}},
  \bibinfo {author} {\bibnamefont {Jr.}}, \bibinfo {author} {\bibfnamefont
  {V.~R.}\ \bibnamefont {Pandharipande}}, \bibinfo {author} {\bibfnamefont
  {D.~G.}\ \bibnamefont {Ravenhall}}, \bibinfo {author} {\bibfnamefont
  {J.}~\bibnamefont {Carlson}}, \bibinfo {author} {\bibfnamefont {S.~C.}\
  \bibnamefont {Pieper}}, \bibinfo {author} {\bibfnamefont {R.~B.}\
  \bibnamefont {Wiringa}}, \ and\ \bibinfo {author} {\bibfnamefont {K.~E.}\
  \bibnamefont {Schmidt}}} (\bibinfo {year} {2004}),\ \href@noop {} {\bibfield
  {journal} {\bibinfo  {journal} {Nucl. Phys. A}\ }\textbf {\bibinfo {volume}
  {746}},\ \bibinfo {pages} {215}}\BibitemShut {NoStop}%
\bibitem [{\citenamefont {Chemtob}\ and\ \citenamefont {Rho}(1971)}]{Che71}%
  \BibitemOpen
  \bibfield  {author} {\bibinfo {author} {\bibnamefont {Chemtob}, \bibfnamefont
  {M.}}, \ and\ \bibinfo {author} {\bibfnamefont {M.}~\bibnamefont {Rho}}}
  (\bibinfo {year} {1971}),\ \href@noop {} {\bibfield  {journal} {\bibinfo
  {journal} {Nucl. Phys. A}\ }\textbf {\bibinfo {volume} {163}},\ \bibinfo
  {pages} {1}}\BibitemShut {NoStop}%
\bibitem [{\citenamefont {Chen}\ \emph {et~al.}(2010)\citenamefont {Chen},
  \citenamefont {Ko}, \citenamefont {Li},\ and\ \citenamefont
  {Xu}}]{Chen:2010}%
  \BibitemOpen
  \bibfield  {author} {\bibinfo {author} {\bibnamefont {Chen}, \bibfnamefont
  {L.-W.}}, \bibinfo {author} {\bibfnamefont {C.~M.}\ \bibnamefont {Ko}},
  \bibinfo {author} {\bibfnamefont {B.-A.}\ \bibnamefont {Li}}, \ and\ \bibinfo
  {author} {\bibfnamefont {J.}~\bibnamefont {Xu}}} (\bibinfo {year} {2010}),\
  \href {\doibase 10.1103/PhysRevC.82.024321} {\bibfield  {journal} {\bibinfo
  {journal} {Phys. Rev. C}\ }\textbf {\bibinfo {volume} {82}}~(\bibinfo
  {number} {2}),\ \bibinfo {pages} {024321}}\BibitemShut {NoStop}%
\bibitem [{\citenamefont {Chernykh}\ \emph {et~al.}(2010)\citenamefont
  {Chernykh}, \citenamefont {Feldmeier}, \citenamefont {Neff}, \citenamefont
  {von Neumann-Cosel},\ and\ \citenamefont {Richter}}]{Chernylch:2010}%
  \BibitemOpen
  \bibfield  {author} {\bibinfo {author} {\bibnamefont {Chernykh},
  \bibfnamefont {M.}}, \bibinfo {author} {\bibfnamefont {H.}~\bibnamefont
  {Feldmeier}}, \bibinfo {author} {\bibfnamefont {T.}~\bibnamefont {Neff}},
  \bibinfo {author} {\bibfnamefont {P.}~\bibnamefont {von Neumann-Cosel}}, \
  and\ \bibinfo {author} {\bibfnamefont {A.}~\bibnamefont {Richter}}} (\bibinfo
  {year} {2010}),\ \href {\doibase 10.1103/PhysRevLett.105.022501} {\bibfield
  {journal} {\bibinfo  {journal} {Phys. Rev. Lett.}\ }\textbf {\bibinfo
  {volume} {105}},\ \bibinfo {pages} {022501}}\BibitemShut {NoStop}%
\bibitem [{\citenamefont {Coelho}\ \emph {et~al.}(1983)\citenamefont {Coelho},
  \citenamefont {Das},\ and\ \citenamefont {Robilotta}}]{coelho1983}%
  \BibitemOpen
  \bibfield  {author} {\bibinfo {author} {\bibnamefont {Coelho}, \bibfnamefont
  {H.~T.}}, \bibinfo {author} {\bibfnamefont {T.~K.}\ \bibnamefont {Das}}, \
  and\ \bibinfo {author} {\bibfnamefont {M.~R.}\ \bibnamefont {Robilotta}}}
  (\bibinfo {year} {1983}),\ \href@noop {} {\bibfield  {journal} {\bibinfo
  {journal} {Phys. Rev. C}\ }\textbf {\bibinfo {volume} {28}},\ \bibinfo
  {pages} {1812}}\BibitemShut {NoStop}%
\bibitem [{\citenamefont {Cohen}\ and\ \citenamefont
  {Kurath}(1967)}]{cohen1967}%
  \BibitemOpen
  \bibfield  {author} {\bibinfo {author} {\bibnamefont {Cohen}, \bibfnamefont
  {S.}}, \ and\ \bibinfo {author} {\bibfnamefont {D.}~\bibnamefont {Kurath}}}
  (\bibinfo {year} {1967}),\ \href@noop {} {\bibfield  {journal} {\bibinfo
  {journal} {Nucl. Phys. A}\ }\textbf {\bibinfo {volume} {101}},\ \bibinfo
  {pages} {1}}\BibitemShut {NoStop}%
\bibitem [{\citenamefont {Coon}\ \emph {et~al.}(1979)\citenamefont {Coon},
  \citenamefont {Scadron}, \citenamefont {McNamee}, \citenamefont {Barrett},
  \citenamefont {Blatt},\ and\ \citenamefont {McKellar}}]{coon1979}%
  \BibitemOpen
  \bibfield  {author} {\bibinfo {author} {\bibnamefont {Coon}, \bibfnamefont
  {S.~A.}}, \bibinfo {author} {\bibfnamefont {M.~D.}\ \bibnamefont {Scadron}},
  \bibinfo {author} {\bibfnamefont {P.~C.}\ \bibnamefont {McNamee}}, \bibinfo
  {author} {\bibfnamefont {B.~R.}\ \bibnamefont {Barrett}}, \bibinfo {author}
  {\bibfnamefont {D.~W.~E.}\ \bibnamefont {Blatt}}, \ and\ \bibinfo {author}
  {\bibfnamefont {B.~H.~J.}\ \bibnamefont {McKellar}}} (\bibinfo {year}
  {1979}),\ \href@noop {} {\bibfield  {journal} {\bibinfo  {journal} {Nucl.
  Phys. A}\ }\textbf {\bibinfo {volume} {317}},\ \bibinfo {pages}
  {242}}\BibitemShut {NoStop}%
\bibitem [{\citenamefont {Czarnecki}\ \emph {et~al.}(2007)\citenamefont
  {Czarnecki}, \citenamefont {Marciano},\ and\ \citenamefont {Sirlin}}]{Cza07}%
  \BibitemOpen
  \bibfield  {author} {\bibinfo {author} {\bibnamefont {Czarnecki},
  \bibfnamefont {A.}}, \bibinfo {author} {\bibfnamefont {W.~J.}\ \bibnamefont
  {Marciano}}, \ and\ \bibinfo {author} {\bibfnamefont {A.}~\bibnamefont
  {Sirlin}}} (\bibinfo {year} {2007}),\ \href {\doibase
  10.1103/PhysRevLett.99.032003} {\bibfield  {journal} {\bibinfo  {journal}
  {Phys. Rev. Lett.}\ }\textbf {\bibinfo {volume} {99}},\ \bibinfo {pages}
  {032003}}\BibitemShut {NoStop}%
\bibitem [{\citenamefont {Danielewicz}\ \emph {et~al.}(2002)\citenamefont
  {Danielewicz}, \citenamefont {Lacey},\ and\ \citenamefont
  {Lynch}}]{Danielewicz:2002}%
  \BibitemOpen
  \bibfield  {author} {\bibinfo {author} {\bibnamefont {Danielewicz},
  \bibfnamefont {P.}}, \bibinfo {author} {\bibfnamefont {R.}~\bibnamefont
  {Lacey}}, \ and\ \bibinfo {author} {\bibfnamefont {W.~G.}\ \bibnamefont
  {Lynch}}} (\bibinfo {year} {2002}),\ \href@noop {} {\bibfield  {journal}
  {\bibinfo  {journal} {Science}\ }\textbf {\bibinfo {volume} {298}},\ \bibinfo
  {pages} {1592}}\BibitemShut {NoStop}%
\bibitem [{\citenamefont {De~Vries}\ \emph {et~al.}(1987)\citenamefont
  {De~Vries}, \citenamefont {De~Jager},\ and\ \citenamefont
  {De~Vries}}]{DeVries:1987}%
  \BibitemOpen
  \bibfield  {author} {\bibinfo {author} {\bibnamefont {De~Vries},
  \bibfnamefont {H.}}, \bibinfo {author} {\bibfnamefont {C.~W.}\ \bibnamefont
  {De~Jager}}, \ and\ \bibinfo {author} {\bibfnamefont {C.}~\bibnamefont
  {De~Vries}}} (\bibinfo {year} {1987}),\ \href@noop {} {\bibfield  {journal}
  {\bibinfo  {journal} {Atomic Data and Nuclear Data Tables}\ }\textbf
  {\bibinfo {volume} {36}},\ \bibinfo {pages} {495}}\BibitemShut {NoStop}%
\bibitem [{\citenamefont {Demorest}\ \emph {et~al.}(2010)\citenamefont
  {Demorest}, \citenamefont {Pennucci}, \citenamefont {Ransom}, \citenamefont
  {Roberts},\ and\ \citenamefont {Hessels}}]{Demorest:2010}%
  \BibitemOpen
  \bibfield  {author} {\bibinfo {author} {\bibnamefont {Demorest},
  \bibfnamefont {P.}}, \bibinfo {author} {\bibfnamefont {T.}~\bibnamefont
  {Pennucci}}, \bibinfo {author} {\bibfnamefont {S.}~\bibnamefont {Ransom}},
  \bibinfo {author} {\bibfnamefont {M.}~\bibnamefont {Roberts}}, \ and\
  \bibinfo {author} {\bibfnamefont {J.}~\bibnamefont {Hessels}}} (\bibinfo
  {year} {2010}),\ \href@noop {} {\bibfield  {journal} {\bibinfo  {journal}
  {Nature}\ }\textbf {\bibinfo {volume} {467}},\ \bibinfo {pages}
  {1081}}\BibitemShut {NoStop}%
\bibitem [{\citenamefont {Dickhoff}\ and\ \citenamefont
  {Barbieri}(2004)}]{Dickhoff:2004}%
  \BibitemOpen
  \bibfield  {author} {\bibinfo {author} {\bibnamefont {Dickhoff},
  \bibfnamefont {W.}}, \ and\ \bibinfo {author} {\bibfnamefont
  {C.}~\bibnamefont {Barbieri}}} (\bibinfo {year} {2004}),\ \href {\doibase
  http://dx.doi.org/10.1016/j.ppnp.2004.02.038} {\bibfield  {journal} {\bibinfo
   {journal} {Progress in Particle and Nuclear Physics}\ }\textbf {\bibinfo
  {volume} {52}}~(\bibinfo {number} {2}),\ \bibinfo {pages} {377 }}\BibitemShut
  {NoStop}%
\bibitem [{\citenamefont {Donnelly}\ and\ \citenamefont {Sick}(1984)}]{Don84}%
  \BibitemOpen
  \bibfield  {author} {\bibinfo {author} {\bibnamefont {Donnelly},
  \bibfnamefont {T.~W.}}, \ and\ \bibinfo {author} {\bibfnamefont
  {I.}~\bibnamefont {Sick}}} (\bibinfo {year} {1984}),\ \href {\doibase
  10.1103/RevModPhys.56.461} {\bibfield  {journal} {\bibinfo  {journal} {Rev.
  Mod. Phys.}\ }\textbf {\bibinfo {volume} {56}},\ \bibinfo {pages}
  {461}}\BibitemShut {NoStop}%
\bibitem [{\citenamefont {Edmonds}(1957)}]{Edm57}%
  \BibitemOpen
  \bibfield  {author} {\bibinfo {author} {\bibnamefont {Edmonds}, \bibfnamefont
  {A.}}} (\bibinfo {year} {1957}),\ \href@noop {} {\bibfield  {journal}
  {\bibinfo  {journal} {Princeton University Press}\ }\textbf {\bibinfo
  {volume} {16}},\ \bibinfo {pages} {1}}\BibitemShut {NoStop}%
\bibitem [{\citenamefont {Entem}\ and\ \citenamefont
  {Machleidt}(2002)}]{entem2002}%
  \BibitemOpen
  \bibfield  {author} {\bibinfo {author} {\bibnamefont {Entem}, \bibfnamefont
  {D.}}, \ and\ \bibinfo {author} {\bibfnamefont {R.}~\bibnamefont
  {Machleidt}}} (\bibinfo {year} {2002}),\ \href@noop {} {\bibfield  {journal}
  {\bibinfo  {journal} {Phys. Lett. B}\ }\textbf {\bibinfo {volume} {524}},\
  \bibinfo {pages} {93}}\BibitemShut {NoStop}%
\bibitem [{\citenamefont {Entem}\ and\ \citenamefont
  {Machleidt}(2003)}]{Ent03}%
  \BibitemOpen
  \bibfield  {author} {\bibinfo {author} {\bibnamefont {Entem}, \bibfnamefont
  {D.}}, \ and\ \bibinfo {author} {\bibfnamefont {R.}~\bibnamefont
  {Machleidt}}} (\bibinfo {year} {2003}),\ \href@noop {} {\bibfield  {journal}
  {\bibinfo  {journal} {Phys. Rev. C}\ }\textbf {\bibinfo {volume} {68}},\
  \bibinfo {pages} {041001}}\BibitemShut {NoStop}%
\bibitem [{\citenamefont {Epelbaum}\ \emph {et~al.}(1998)\citenamefont
  {Epelbaum}, \citenamefont {Gloeckle},\ and\ \citenamefont
  {Meissner}}]{Epe98}%
  \BibitemOpen
  \bibfield  {author} {\bibinfo {author} {\bibnamefont {Epelbaum},
  \bibfnamefont {E.}}, \bibinfo {author} {\bibfnamefont {W.}~\bibnamefont
  {Gloeckle}}, \ and\ \bibinfo {author} {\bibfnamefont {U.-G.}\ \bibnamefont
  {Meissner}}} (\bibinfo {year} {1998}),\ \href@noop {} {\bibfield  {journal}
  {\bibinfo  {journal} {Nuc. Phys. A}\ }\textbf {\bibinfo {volume} {637}},\
  \bibinfo {pages} {107}}\BibitemShut {NoStop}%
\bibitem [{\citenamefont {Epelbaum}\ \emph {et~al.}(2009)\citenamefont
  {Epelbaum}, \citenamefont {Hammer},\ and\ \citenamefont
  {Mei\ss{}ner}}]{Epelbaum:2009}%
  \BibitemOpen
  \bibfield  {author} {\bibinfo {author} {\bibnamefont {Epelbaum},
  \bibfnamefont {E.}}, \bibinfo {author} {\bibfnamefont {H.-W.}\ \bibnamefont
  {Hammer}}, \ and\ \bibinfo {author} {\bibfnamefont {U.-G.}\ \bibnamefont
  {Mei\ss{}ner}}} (\bibinfo {year} {2009}),\ \href {\doibase
  10.1103/RevModPhys.81.1773} {\bibfield  {journal} {\bibinfo  {journal} {Rev.
  Mod. Phys.}\ }\textbf {\bibinfo {volume} {81}}~(\bibinfo {number} {4}),\
  \bibinfo {pages} {1773}}\BibitemShut {NoStop}%
\bibitem [{\citenamefont {Epelbaum}\ \emph {et~al.}(2012)\citenamefont
  {Epelbaum}, \citenamefont {Krebs}, \citenamefont {Lahde}, \citenamefont
  {Lee},\ and\ \citenamefont {Meissner}}]{Epelbaum:2012}%
  \BibitemOpen
  \bibfield  {author} {\bibinfo {author} {\bibnamefont {Epelbaum},
  \bibfnamefont {E.}}, \bibinfo {author} {\bibfnamefont {H.}~\bibnamefont
  {Krebs}}, \bibinfo {author} {\bibfnamefont {T.~A.}\ \bibnamefont {Lahde}},
  \bibinfo {author} {\bibfnamefont {D.}~\bibnamefont {Lee}}, \ and\ \bibinfo
  {author} {\bibfnamefont {U.-G.}\ \bibnamefont {Meissner}}} (\bibinfo {year}
  {2012}),\ \href {\doibase 10.1103/PhysRevLett.109.252501} {\bibfield
  {journal} {\bibinfo  {journal} {Phys. Rev. Lett.}\ }\textbf {\bibinfo
  {volume} {109}},\ \bibinfo {pages} {252501}}\BibitemShut {NoStop}%
\bibitem [{\citenamefont {Epelbaum}\ \emph {et~al.}(2014)\citenamefont
  {Epelbaum}, \citenamefont {Krebs}, \citenamefont {L\"ahde}, \citenamefont
  {Lee}, \citenamefont {Mei\ss{}ner},\ and\ \citenamefont
  {Rupak}}]{Epelbaum:2014}%
  \BibitemOpen
  \bibfield  {author} {\bibinfo {author} {\bibnamefont {Epelbaum},
  \bibfnamefont {E.}}, \bibinfo {author} {\bibfnamefont {H.}~\bibnamefont
  {Krebs}}, \bibinfo {author} {\bibfnamefont {T.~A.}\ \bibnamefont {L\"ahde}},
  \bibinfo {author} {\bibfnamefont {D.}~\bibnamefont {Lee}}, \bibinfo {author}
  {\bibfnamefont {U.-G.}\ \bibnamefont {Mei\ss{}ner}}, \ and\ \bibinfo {author}
  {\bibfnamefont {G.}~\bibnamefont {Rupak}}} (\bibinfo {year} {2014}),\ \href
  {\doibase 10.1103/PhysRevLett.112.102501} {\bibfield  {journal} {\bibinfo
  {journal} {Phys. Rev. Lett.}\ }\textbf {\bibinfo {volume} {112}},\ \bibinfo
  {pages} {102501}}\BibitemShut {NoStop}%
\bibitem [{\citenamefont {Epelbaum}\ \emph {et~al.}(2011)\citenamefont
  {Epelbaum}, \citenamefont {Krebs}, \citenamefont {Lee},\ and\ \citenamefont
  {Meissner}}]{Epelbaum:2011}%
  \BibitemOpen
  \bibfield  {author} {\bibinfo {author} {\bibnamefont {Epelbaum},
  \bibfnamefont {E.}}, \bibinfo {author} {\bibfnamefont {H.}~\bibnamefont
  {Krebs}}, \bibinfo {author} {\bibfnamefont {D.}~\bibnamefont {Lee}}, \ and\
  \bibinfo {author} {\bibfnamefont {U.-G.}\ \bibnamefont {Meissner}}} (\bibinfo
  {year} {2011}),\ \href {\doibase 10.1103/PhysRevLett.106.192501} {\bibfield
  {journal} {\bibinfo  {journal} {Phys. Rev. Lett.}\ }\textbf {\bibinfo
  {volume} {106}},\ \bibinfo {pages} {192501}}\BibitemShut {NoStop}%
\bibitem [{\citenamefont {Epelbaum}\ and\ \citenamefont
  {Meissner}(1999)}]{Epe99}%
  \BibitemOpen
  \bibfield  {author} {\bibinfo {author} {\bibnamefont {Epelbaum},
  \bibfnamefont {E.}}, \ and\ \bibinfo {author} {\bibfnamefont {U.-G.}\
  \bibnamefont {Meissner}}} (\bibinfo {year} {1999}),\ \href@noop {} {\bibfield
   {journal} {\bibinfo  {journal} {Phys. Lett. B}\ }\textbf {\bibinfo {volume}
  {461}},\ \bibinfo {pages} {287}}\BibitemShut {NoStop}%
\bibitem [{\citenamefont {Epelbaum}\ \emph {et~al.}(2002)\citenamefont
  {Epelbaum}, \citenamefont {Nogga}, \citenamefont {Gloeckle}, \citenamefont
  {Kamada}, \citenamefont {Meissner} \emph {et~al.}}]{Epe02}%
  \BibitemOpen
  \bibfield  {author} {\bibinfo {author} {\bibnamefont {Epelbaum},
  \bibfnamefont {E.}}, \bibinfo {author} {\bibfnamefont {A.}~\bibnamefont
  {Nogga}}, \bibinfo {author} {\bibfnamefont {W.}~\bibnamefont {Gloeckle}},
  \bibinfo {author} {\bibfnamefont {H.}~\bibnamefont {Kamada}}, \bibinfo
  {author} {\bibfnamefont {U.~G.}\ \bibnamefont {Meissner}},  \emph {et~al.}}
  (\bibinfo {year} {2002}),\ \href@noop {} {\bibfield  {journal} {\bibinfo
  {journal} {Phys. Rev. C}\ }\textbf {\bibinfo {volume} {66}},\ \bibinfo
  {pages} {064001}}\BibitemShut {NoStop}%
\bibitem [{\citenamefont {Esbensen}(2004)}]{esbensen2004}%
  \BibitemOpen
  \bibfield  {author} {\bibinfo {author} {\bibnamefont {Esbensen},
  \bibfnamefont {H.}}} (\bibinfo {year} {2004}),\ \href {\doibase
  10.1103/PhysRevC.70.047603} {\bibfield  {journal} {\bibinfo  {journal} {Phys.
  Rev. C}\ }\textbf {\bibinfo {volume} {70}},\ \bibinfo {pages}
  {047603}}\BibitemShut {NoStop}%
\bibitem [{\citenamefont {Fantoni}\ and\ \citenamefont
  {Schmidt}(2001)}]{Fantoni:2001c}%
  \BibitemOpen
  \bibfield  {author} {\bibinfo {author} {\bibnamefont {Fantoni}, \bibfnamefont
  {S.}}, \ and\ \bibinfo {author} {\bibfnamefont {K.~E.}\ \bibnamefont
  {Schmidt}}} (\bibinfo {year} {2001}),\ \href@noop {} {\bibfield  {journal}
  {\bibinfo  {journal} {Nucl. Phys. A}\ }\textbf {\bibinfo {volume} {690}},\
  \bibinfo {pages} {456}}\BibitemShut {NoStop}%
\bibitem [{\citenamefont {Feenberg}\ and\ \citenamefont
  {Wigner}(1937)}]{feenberg1937}%
  \BibitemOpen
  \bibfield  {author} {\bibinfo {author} {\bibnamefont {Feenberg},
  \bibfnamefont {E.}}, \ and\ \bibinfo {author} {\bibfnamefont
  {E.}~\bibnamefont {Wigner}}} (\bibinfo {year} {1937}),\ \href {\doibase
  10.1103/PhysRev.51.95} {\bibfield  {journal} {\bibinfo  {journal} {Phys.
  Rev.}\ }\textbf {\bibinfo {volume} {51}},\ \bibinfo {pages} {95}}\BibitemShut
  {NoStop}%
\bibitem [{\citenamefont {Foulkes}\ \emph {et~al.}(2001)\citenamefont
  {Foulkes}, \citenamefont {Mitas}, \citenamefont {Needs},\ and\ \citenamefont
  {Rajagopal}}]{Foulkes:2001}%
  \BibitemOpen
  \bibfield  {author} {\bibinfo {author} {\bibnamefont {Foulkes}, \bibfnamefont
  {W.~M.~C.}}, \bibinfo {author} {\bibfnamefont {L.}~\bibnamefont {Mitas}},
  \bibinfo {author} {\bibfnamefont {R.~J.}\ \bibnamefont {Needs}}, \ and\
  \bibinfo {author} {\bibfnamefont {G.}~\bibnamefont {Rajagopal}}} (\bibinfo
  {year} {2001}),\ \href {\doibase 10.1103/RevModPhys.73.33} {\bibfield
  {journal} {\bibinfo  {journal} {Rev. Mod. Phys.}\ }\textbf {\bibinfo {volume}
  {73}},\ \bibinfo {pages} {33}}\BibitemShut {NoStop}%
\bibitem [{\citenamefont {Friar}(1977)}]{Fri77}%
  \BibitemOpen
  \bibfield  {author} {\bibinfo {author} {\bibnamefont {Friar}, \bibfnamefont
  {J.}}} (\bibinfo {year} {1977}),\ \href@noop {} {\bibfield  {journal}
  {\bibinfo  {journal} {Ann. Phys. (N.Y.)}\ }\textbf {\bibinfo {volume}
  {104}},\ \bibinfo {pages} {380}}\BibitemShut {NoStop}%
\bibitem [{\citenamefont {Friar}\ \emph {et~al.}(1999)\citenamefont {Friar},
  \citenamefont {H\"uber},\ and\ \citenamefont {van Kolck}}]{friar1999}%
  \BibitemOpen
  \bibfield  {author} {\bibinfo {author} {\bibnamefont {Friar}, \bibfnamefont
  {J.~L.}}, \bibinfo {author} {\bibfnamefont {D.}~\bibnamefont {H\"uber}}, \
  and\ \bibinfo {author} {\bibfnamefont {U.}~\bibnamefont {van Kolck}}}
  (\bibinfo {year} {1999}),\ \href {\doibase 10.1103/PhysRevC.59.53} {\bibfield
   {journal} {\bibinfo  {journal} {Phys. Rev. C}\ }\textbf {\bibinfo {volume}
  {59}},\ \bibinfo {pages} {53}}\BibitemShut {NoStop}%
\bibitem [{\citenamefont {Friar}\ and\ \citenamefont {van
  Kolck}(1999)}]{Fri99}%
  \BibitemOpen
  \bibfield  {author} {\bibinfo {author} {\bibnamefont {Friar}, \bibfnamefont
  {J.~L.}}, \ and\ \bibinfo {author} {\bibfnamefont {U.}~\bibnamefont {van
  Kolck}}} (\bibinfo {year} {1999}),\ \href@noop {} {\bibfield  {journal}
  {\bibinfo  {journal} {Phys. Rev. C}\ }\textbf {\bibinfo {volume} {60}},\
  \bibinfo {pages} {034006}}\BibitemShut {NoStop}%
\bibitem [{\citenamefont {Friar}\ \emph {et~al.}(2004)\citenamefont {Friar},
  \citenamefont {van Kolck}, \citenamefont {Rentmeester},\ and\ \citenamefont
  {Timmermans}}]{Fri04}%
  \BibitemOpen
  \bibfield  {author} {\bibinfo {author} {\bibnamefont {Friar}, \bibfnamefont
  {J.~L.}}, \bibinfo {author} {\bibfnamefont {U.}~\bibnamefont {van Kolck}},
  \bibinfo {author} {\bibfnamefont {M.~C.~M.}\ \bibnamefont {Rentmeester}}, \
  and\ \bibinfo {author} {\bibfnamefont {R.~G.~E.}\ \bibnamefont {Timmermans}}}
  (\bibinfo {year} {2004}),\ \href@noop {} {\bibfield  {journal} {\bibinfo
  {journal} {Phys. Rev. C}\ }\textbf {\bibinfo {volume} {70}},\ \bibinfo
  {pages} {044001}}\BibitemShut {NoStop}%
\bibitem [{\citenamefont {Friar}\ \emph {et~al.}(2005)\citenamefont {Friar},
  \citenamefont {Payne},\ and\ \citenamefont {van Kolck}}]{Fri05}%
  \BibitemOpen
  \bibfield  {author} {\bibinfo {author} {\bibnamefont {Friar}, \bibfnamefont
  {J.~L.}}, \bibinfo {author} {\bibfnamefont {G.~L.}\ \bibnamefont {Payne}}, \
  and\ \bibinfo {author} {\bibfnamefont {U.}~\bibnamefont {van Kolck}}}
  (\bibinfo {year} {2005}),\ \href@noop {} {\bibfield  {journal} {\bibinfo
  {journal} {Phys. Rev. C}\ }\textbf {\bibinfo {volume} {71}},\ \bibinfo
  {pages} {024003}}\BibitemShut {NoStop}%
\bibitem [{\citenamefont {Fujita}\ and\ \citenamefont
  {Miyazawa}(1957)}]{Fujita1957}%
  \BibitemOpen
  \bibfield  {author} {\bibinfo {author} {\bibnamefont {Fujita}, \bibfnamefont
  {J.}}, \ and\ \bibinfo {author} {\bibfnamefont {H.}~\bibnamefont {Miyazawa}}}
  (\bibinfo {year} {1957}),\ \href@noop {} {\bibfield  {journal} {\bibinfo
  {journal} {Prog. Theor. Phys.}\ }\textbf {\bibinfo {volume} {17}},\ \bibinfo
  {pages} {360}}\BibitemShut {NoStop}%
\bibitem [{\citenamefont {Gandolfi}(2007)}]{Gandolfi:2007c}%
  \BibitemOpen
  \bibfield  {author} {\bibinfo {author} {\bibnamefont {Gandolfi},
  \bibfnamefont {S.}}} (\bibinfo {year} {2007}),\ \href@noop {} {\bibfield
  {journal} {\bibinfo  {journal} {arXiv:0712.1364 [nucl-th]}\ }}\bibinfo {note}
  {{P}h.{D}. thesis, {U}niversity of {T}rento, {I}taly}\BibitemShut {NoStop}%
\bibitem [{\citenamefont {Gandolfi}\ \emph {et~al.}(2011)\citenamefont
  {Gandolfi}, \citenamefont {Carlson},\ and\ \citenamefont
  {Pieper}}]{Gandolfi:2011}%
  \BibitemOpen
  \bibfield  {author} {\bibinfo {author} {\bibnamefont {Gandolfi},
  \bibfnamefont {S.}}, \bibinfo {author} {\bibfnamefont {J.}~\bibnamefont
  {Carlson}}, \ and\ \bibinfo {author} {\bibfnamefont {S.~C.}\ \bibnamefont
  {Pieper}}} (\bibinfo {year} {2011}),\ \href {\doibase
  10.1103/PhysRevLett.106.012501} {\bibfield  {journal} {\bibinfo  {journal}
  {Phys. Rev. Lett.}\ }\textbf {\bibinfo {volume} {106}}~(\bibinfo {number}
  {1}),\ \bibinfo {pages} {012501}}\BibitemShut {NoStop}%
\bibitem [{\citenamefont {{Gandolfi}}\ \emph {et~al.}(2012)\citenamefont
  {{Gandolfi}}, \citenamefont {{Carlson}},\ and\ \citenamefont
  {{Reddy}}}]{Gandolfi:2012}%
  \BibitemOpen
  \bibfield  {author} {\bibinfo {author} {\bibnamefont {{Gandolfi}},
  \bibfnamefont {S.}}, \bibinfo {author} {\bibfnamefont {J.}~\bibnamefont
  {{Carlson}}}, \ and\ \bibinfo {author} {\bibfnamefont {S.}~\bibnamefont
  {{Reddy}}}} (\bibinfo {year} {2012}),\ \href {\doibase
  10.1103/PhysRevC.85.032801} {\bibfield  {journal} {\bibinfo  {journal} {Phys.
  Rev. C}\ }\textbf {\bibinfo {volume} {85}}~(\bibinfo {number} {3}),\ \bibinfo
  {eid} {032801}}\BibitemShut {NoStop}%
\bibitem [{\citenamefont {{Gandolfi}}\ \emph {et~al.}(2014)\citenamefont
  {{Gandolfi}}, \citenamefont {{Carlson}}, \citenamefont {{Reddy}},
  \citenamefont {{Steiner}},\ and\ \citenamefont {{Wiringa}}}]{Gandolfi:2014}%
  \BibitemOpen
  \bibfield  {author} {\bibinfo {author} {\bibnamefont {{Gandolfi}},
  \bibfnamefont {S.}}, \bibinfo {author} {\bibfnamefont {J.}~\bibnamefont
  {{Carlson}}}, \bibinfo {author} {\bibfnamefont {S.}~\bibnamefont {{Reddy}}},
  \bibinfo {author} {\bibfnamefont {A.~W.}\ \bibnamefont {{Steiner}}}, \ and\
  \bibinfo {author} {\bibfnamefont {R.~B.}\ \bibnamefont {{Wiringa}}}}
  (\bibinfo {year} {2014}),\ \href {\doibase 10.1140/epja/i2014-14010-5}
  {\bibfield  {journal} {\bibinfo  {journal} {Eur. Phys. J. A}\ }\textbf
  {\bibinfo {volume} {50}},\ \bibinfo {pages} {10}}\BibitemShut {NoStop}%
\bibitem [{\citenamefont {Gandolfi}\ \emph
  {et~al.}(2008{\natexlab{a}})\citenamefont {Gandolfi}, \citenamefont
  {Illarionov}, \citenamefont {Fantoni}, \citenamefont {Pederiva},\ and\
  \citenamefont {Schmidt}}]{Gandolfi:2008b}%
  \BibitemOpen
  \bibfield  {author} {\bibinfo {author} {\bibnamefont {Gandolfi},
  \bibfnamefont {S.}}, \bibinfo {author} {\bibfnamefont {A.~Y.}\ \bibnamefont
  {Illarionov}}, \bibinfo {author} {\bibfnamefont {S.}~\bibnamefont {Fantoni}},
  \bibinfo {author} {\bibfnamefont {F.}~\bibnamefont {Pederiva}}, \ and\
  \bibinfo {author} {\bibfnamefont {K.~E.}\ \bibnamefont {Schmidt}}} (\bibinfo
  {year} {2008}{\natexlab{a}}),\ \href@noop {} {\bibfield  {journal} {\bibinfo
  {journal} {Phys. Rev. Lett.}\ }\textbf {\bibinfo {volume} {101}},\ \bibinfo
  {pages} {132501}}\BibitemShut {NoStop}%
\bibitem [{\citenamefont {Gandolfi}\ \emph
  {et~al.}(2009{\natexlab{a}})\citenamefont {Gandolfi}, \citenamefont
  {Illarionov}, \citenamefont {Pederiva}, \citenamefont {Schmidt},\ and\
  \citenamefont {Fantoni}}]{Gandolfi:2009b}%
  \BibitemOpen
  \bibfield  {author} {\bibinfo {author} {\bibnamefont {Gandolfi},
  \bibfnamefont {S.}}, \bibinfo {author} {\bibfnamefont {A.~Y.}\ \bibnamefont
  {Illarionov}}, \bibinfo {author} {\bibfnamefont {F.}~\bibnamefont
  {Pederiva}}, \bibinfo {author} {\bibfnamefont {K.~E.}\ \bibnamefont
  {Schmidt}}, \ and\ \bibinfo {author} {\bibfnamefont {S.}~\bibnamefont
  {Fantoni}}} (\bibinfo {year} {2009}{\natexlab{a}}),\ \href {\doibase
  10.1103/PhysRevC.80.045802} {\bibfield  {journal} {\bibinfo  {journal} {Phys.
  Rev. C}\ }\textbf {\bibinfo {volume} {80}},\ \bibinfo {pages}
  {045802}}\BibitemShut {NoStop}%
\bibitem [{\citenamefont {Gandolfi}\ \emph
  {et~al.}(2009{\natexlab{b}})\citenamefont {Gandolfi}, \citenamefont
  {Illarionov}, \citenamefont {Schmidt}, \citenamefont {Pederiva},\ and\
  \citenamefont {Fantoni}}]{Gandolfi:2009}%
  \BibitemOpen
  \bibfield  {author} {\bibinfo {author} {\bibnamefont {Gandolfi},
  \bibfnamefont {S.}}, \bibinfo {author} {\bibfnamefont {A.~Y.}\ \bibnamefont
  {Illarionov}}, \bibinfo {author} {\bibfnamefont {K.~E.}\ \bibnamefont
  {Schmidt}}, \bibinfo {author} {\bibfnamefont {F.}~\bibnamefont {Pederiva}}, \
  and\ \bibinfo {author} {\bibfnamefont {S.}~\bibnamefont {Fantoni}}} (\bibinfo
  {year} {2009}{\natexlab{b}}),\ \href@noop {} {\bibfield  {journal} {\bibinfo
  {journal} {Phys. Rev. C}\ }\textbf {\bibinfo {volume} {79}},\ \bibinfo
  {pages} {054005}}\BibitemShut {NoStop}%
\bibitem [{\citenamefont {Gandolfi}\ \emph {et~al.}(2014)\citenamefont
  {Gandolfi}, \citenamefont {Lovato}, \citenamefont {Carlson},\ and\
  \citenamefont {Schmidt}}]{Gandolfi:2014b}%
  \BibitemOpen
  \bibfield  {author} {\bibinfo {author} {\bibnamefont {Gandolfi},
  \bibfnamefont {S.}}, \bibinfo {author} {\bibfnamefont {A.}~\bibnamefont
  {Lovato}}, \bibinfo {author} {\bibfnamefont {J.}~\bibnamefont {Carlson}}, \
  and\ \bibinfo {author} {\bibfnamefont {K.~E.}\ \bibnamefont {Schmidt}}}
  (\bibinfo {year} {2014}),\ \href {\doibase 10.1103/PhysRevC.90.061306}
  {\bibfield  {journal} {\bibinfo  {journal} {Phys. Rev. C}\ }\textbf {\bibinfo
  {volume} {90}},\ \bibinfo {pages} {061306}}\BibitemShut {NoStop}%
\bibitem [{\citenamefont {Gandolfi}\ \emph
  {et~al.}(2008{\natexlab{b}})\citenamefont {Gandolfi}, \citenamefont
  {Pederiva},\ and\ \citenamefont {a~Beccara}}]{Gandolfi:2008}%
  \BibitemOpen
  \bibfield  {author} {\bibinfo {author} {\bibnamefont {Gandolfi},
  \bibfnamefont {S.}}, \bibinfo {author} {\bibfnamefont {F.}~\bibnamefont
  {Pederiva}}, \ and\ \bibinfo {author} {\bibfnamefont {S.}~\bibnamefont
  {a~Beccara}}} (\bibinfo {year} {2008}{\natexlab{b}}),\ \href@noop {}
  {\bibfield  {journal} {\bibinfo  {journal} {Eur. Phys. J. A}\ }\textbf
  {\bibinfo {volume} {35}},\ \bibinfo {pages} {207}}\BibitemShut {NoStop}%
\bibitem [{\citenamefont {Gandolfi}\ \emph {et~al.}(2006)\citenamefont
  {Gandolfi}, \citenamefont {Pederiva}, \citenamefont {Fantoni},\ and\
  \citenamefont {Schmidt}}]{Gandolfi:2006}%
  \BibitemOpen
  \bibfield  {author} {\bibinfo {author} {\bibnamefont {Gandolfi},
  \bibfnamefont {S.}}, \bibinfo {author} {\bibfnamefont {F.}~\bibnamefont
  {Pederiva}}, \bibinfo {author} {\bibfnamefont {S.}~\bibnamefont {Fantoni}}, \
  and\ \bibinfo {author} {\bibfnamefont {K.~E.}\ \bibnamefont {Schmidt}}}
  (\bibinfo {year} {2006}),\ \href@noop {} {\bibfield  {journal} {\bibinfo
  {journal} {Phys. Rev. C}\ }\textbf {\bibinfo {volume} {73}},\ \bibinfo
  {pages} {044304}}\BibitemShut {NoStop}%
\bibitem [{\citenamefont {Gezerlis}\ and\ \citenamefont
  {Carlson}(2008)}]{Gezerlis:2008}%
  \BibitemOpen
  \bibfield  {author} {\bibinfo {author} {\bibnamefont {Gezerlis},
  \bibfnamefont {A.}}, \ and\ \bibinfo {author} {\bibfnamefont
  {J.}~\bibnamefont {Carlson}}} (\bibinfo {year} {2008}),\ \href@noop {}
  {\bibfield  {journal} {\bibinfo  {journal} {Phys. Rev. C}\ }\textbf {\bibinfo
  {volume} {77}},\ \bibinfo {pages} {032801(R)}}\BibitemShut {NoStop}%
\bibitem [{\citenamefont {Gezerlis}\ and\ \citenamefont
  {Carlson}(2010)}]{Gezerlis:2010}%
  \BibitemOpen
  \bibfield  {author} {\bibinfo {author} {\bibnamefont {Gezerlis},
  \bibfnamefont {A.}}, \ and\ \bibinfo {author} {\bibfnamefont
  {J.}~\bibnamefont {Carlson}}} (\bibinfo {year} {2010}),\ \href {\doibase
  10.1103/PhysRevC.81.025803} {\bibfield  {journal} {\bibinfo  {journal} {Phys.
  Rev. C}\ }\textbf {\bibinfo {volume} {81}},\ \bibinfo {pages}
  {025803}}\BibitemShut {NoStop}%
\bibitem [{\citenamefont {Gezerlis}\ \emph {et~al.}(2014)\citenamefont
  {Gezerlis}, \citenamefont {Tews}, \citenamefont {Epelbaum}, \citenamefont
  {Freunek}, \citenamefont {Gandolfi}, \citenamefont {Hebeler}, \citenamefont
  {Nogga},\ and\ \citenamefont {Schwenk}}]{Gezerlis:2014}%
  \BibitemOpen
  \bibfield  {author} {\bibinfo {author} {\bibnamefont {Gezerlis},
  \bibfnamefont {A.}}, \bibinfo {author} {\bibfnamefont {I.}~\bibnamefont
  {Tews}}, \bibinfo {author} {\bibfnamefont {E.}~\bibnamefont {Epelbaum}},
  \bibinfo {author} {\bibfnamefont {M.}~\bibnamefont {Freunek}}, \bibinfo
  {author} {\bibfnamefont {S.}~\bibnamefont {Gandolfi}}, \bibinfo {author}
  {\bibfnamefont {K.}~\bibnamefont {Hebeler}}, \bibinfo {author} {\bibfnamefont
  {A.}~\bibnamefont {Nogga}}, \ and\ \bibinfo {author} {\bibfnamefont
  {A.}~\bibnamefont {Schwenk}}} (\bibinfo {year} {2014}),\ \href {\doibase
  10.1103/PhysRevC.90.054323} {\bibfield  {journal} {\bibinfo  {journal} {Phys.
  Rev. C}\ }\textbf {\bibinfo {volume} {90}},\ \bibinfo {pages}
  {054323}}\BibitemShut {NoStop}%
\bibitem [{\citenamefont {Gezerlis}\ \emph {et~al.}(2013)\citenamefont
  {Gezerlis}, \citenamefont {Tews}, \citenamefont {Epelbaum}, \citenamefont
  {Gandolfi}, \citenamefont {Hebeler} \emph {et~al.}}]{Gezerlis:2013}%
  \BibitemOpen
  \bibfield  {author} {\bibinfo {author} {\bibnamefont {Gezerlis},
  \bibfnamefont {A.}}, \bibinfo {author} {\bibfnamefont {I.}~\bibnamefont
  {Tews}}, \bibinfo {author} {\bibfnamefont {E.}~\bibnamefont {Epelbaum}},
  \bibinfo {author} {\bibfnamefont {S.}~\bibnamefont {Gandolfi}}, \bibinfo
  {author} {\bibfnamefont {K.}~\bibnamefont {Hebeler}},  \emph {et~al.}}
  (\bibinfo {year} {2013}),\ \href {\doibase 10.1103/PhysRevLett.111.032501}
  {\bibfield  {journal} {\bibinfo  {journal} {Phys. Rev. Lett.}\ }\textbf
  {\bibinfo {volume} {111}}~(\bibinfo {number} {3}),\ \bibinfo {pages}
  {032501}}\BibitemShut {NoStop}%
\bibitem [{\citenamefont {Giorgini}\ \emph {et~al.}(2008)\citenamefont
  {Giorgini}, \citenamefont {Pitaevskii},\ and\ \citenamefont
  {Stringari}}]{Giorgini:2008}%
  \BibitemOpen
  \bibfield  {author} {\bibinfo {author} {\bibnamefont {Giorgini},
  \bibfnamefont {S.}}, \bibinfo {author} {\bibfnamefont {L.~P.}\ \bibnamefont
  {Pitaevskii}}, \ and\ \bibinfo {author} {\bibfnamefont {S.}~\bibnamefont
  {Stringari}}} (\bibinfo {year} {2008}),\ \href {\doibase
  10.1103/RevModPhys.80.1215} {\bibfield  {journal} {\bibinfo  {journal} {Rev.
  Mod. Phys.}\ }\textbf {\bibinfo {volume} {80}},\ \bibinfo {pages}
  {1215}}\BibitemShut {NoStop}%
\bibitem [{\citenamefont {Girlanda}(2008)}]{Gir08}%
  \BibitemOpen
  \bibfield  {author} {\bibinfo {author} {\bibnamefont {Girlanda},
  \bibfnamefont {L.}}} (\bibinfo {year} {2008}),\ \href@noop {} {\bibfield
  {journal} {\bibinfo  {journal} {Phys. Rev. C}\ }\textbf {\bibinfo {volume}
  {77}},\ \bibinfo {pages} {067001}}\BibitemShut {NoStop}%
\bibitem [{\citenamefont {Girlanda}\ \emph {et~al.}(2010)\citenamefont
  {Girlanda}, \citenamefont {Kievsky}, \citenamefont {Marcucci}, \citenamefont
  {Pastore}, \citenamefont {Schiavilla} \emph {et~al.}}]{Gir10a}%
  \BibitemOpen
  \bibfield  {author} {\bibinfo {author} {\bibnamefont {Girlanda},
  \bibfnamefont {L.}}, \bibinfo {author} {\bibfnamefont {A.}~\bibnamefont
  {Kievsky}}, \bibinfo {author} {\bibfnamefont {L.}~\bibnamefont {Marcucci}},
  \bibinfo {author} {\bibfnamefont {S.}~\bibnamefont {Pastore}}, \bibinfo
  {author} {\bibfnamefont {R.}~\bibnamefont {Schiavilla}},  \emph {et~al.}}
  (\bibinfo {year} {2010}),\ \href@noop {} {\bibfield  {journal} {\bibinfo
  {journal} {Phys. Rev. Lett.}\ }\textbf {\bibinfo {volume} {105}},\ \bibinfo
  {pages} {232502}}\BibitemShut {NoStop}%
\bibitem [{\citenamefont {Girlanda}\ \emph {et~al.}(2011)\citenamefont
  {Girlanda}, \citenamefont {Kievsky},\ and\ \citenamefont {Viviani}}]{Gir11}%
  \BibitemOpen
  \bibfield  {author} {\bibinfo {author} {\bibnamefont {Girlanda},
  \bibfnamefont {L.}}, \bibinfo {author} {\bibfnamefont {A.}~\bibnamefont
  {Kievsky}}, \ and\ \bibinfo {author} {\bibfnamefont {M.}~\bibnamefont
  {Viviani}}} (\bibinfo {year} {2011}),\ \href@noop {} {\bibfield  {journal}
  {\bibinfo  {journal} {Phys. Rev. C}\ }\textbf {\bibinfo {volume} {84}},\
  \bibinfo {pages} {014001}}\BibitemShut {NoStop}%
\bibitem [{\citenamefont {Gorringe}\ and\ \citenamefont
  {Fearing}(2003)}]{Gor03}%
  \BibitemOpen
  \bibfield  {author} {\bibinfo {author} {\bibnamefont {Gorringe},
  \bibfnamefont {T.}}, \ and\ \bibinfo {author} {\bibfnamefont {H.~W.}\
  \bibnamefont {Fearing}}} (\bibinfo {year} {2003}),\ \href {\doibase
  10.1103/RevModPhys.76.31} {\bibfield  {journal} {\bibinfo  {journal} {Rev.
  Mod. Phys.}\ }\textbf {\bibinfo {volume} {76}},\ \bibinfo {pages}
  {31}}\BibitemShut {NoStop}%
\bibitem [{\citenamefont {Grinyer}\ \emph {et~al.}(2011)\citenamefont {Grinyer}
  \emph {et~al.}}]{grinyer2011}%
  \BibitemOpen
  \bibfield  {author} {\bibinfo {author} {\bibnamefont {Grinyer}, \bibfnamefont
  {G.~F.}},  \emph {et~al.}} (\bibinfo {year} {2011}),\ \href {\doibase
  10.1103/PhysRevLett.106.162502} {\bibfield  {journal} {\bibinfo  {journal}
  {Phys. Rev. Lett.}\ }\textbf {\bibinfo {volume} {106}},\ \bibinfo {pages}
  {162502}}\BibitemShut {NoStop}%
\bibitem [{\citenamefont {Grinyer}\ \emph {et~al.}(2012)\citenamefont {Grinyer}
  \emph {et~al.}}]{grinyer2012}%
  \BibitemOpen
  \bibfield  {author} {\bibinfo {author} {\bibnamefont {Grinyer}, \bibfnamefont
  {G.~F.}},  \emph {et~al.}} (\bibinfo {year} {2012}),\ \href {\doibase
  10.1103/PhysRevC.86.024315} {\bibfield  {journal} {\bibinfo  {journal} {Phys.
  Rev. C}\ }\textbf {\bibinfo {volume} {86}},\ \bibinfo {pages}
  {024315}}\BibitemShut {NoStop}%
\bibitem [{\citenamefont {Hadjimichael}\ \emph {et~al.}(1983)\citenamefont
  {Hadjimichael}, \citenamefont {Goulard},\ and\ \citenamefont
  {Bornais}}]{Had83}%
  \BibitemOpen
  \bibfield  {author} {\bibinfo {author} {\bibnamefont {Hadjimichael},
  \bibfnamefont {E.}}, \bibinfo {author} {\bibfnamefont {B.}~\bibnamefont
  {Goulard}}, \ and\ \bibinfo {author} {\bibfnamefont {R.}~\bibnamefont
  {Bornais}}} (\bibinfo {year} {1983}),\ \href {\doibase
  10.1103/PhysRevC.27.831} {\bibfield  {journal} {\bibinfo  {journal} {Phys.
  Rev. C}\ }\textbf {\bibinfo {volume} {27}},\ \bibinfo {pages}
  {831}}\BibitemShut {NoStop}%
\bibitem [{\citenamefont {Hagen}\ \emph {et~al.}(2014)\citenamefont {Hagen},
  \citenamefont {Papenbrock}, \citenamefont {Ekstr\"om}, \citenamefont {Wendt},
  \citenamefont {Baardsen}, \citenamefont {Gandolfi}, \citenamefont
  {Hjorth-Jensen},\ and\ \citenamefont {Horowitz}}]{Hagen:2014b}%
  \BibitemOpen
  \bibfield  {author} {\bibinfo {author} {\bibnamefont {Hagen}, \bibfnamefont
  {G.}}, \bibinfo {author} {\bibfnamefont {T.}~\bibnamefont {Papenbrock}},
  \bibinfo {author} {\bibfnamefont {A.}~\bibnamefont {Ekstr\"om}}, \bibinfo
  {author} {\bibfnamefont {K.~A.}\ \bibnamefont {Wendt}}, \bibinfo {author}
  {\bibfnamefont {G.}~\bibnamefont {Baardsen}}, \bibinfo {author}
  {\bibfnamefont {S.}~\bibnamefont {Gandolfi}}, \bibinfo {author}
  {\bibfnamefont {M.}~\bibnamefont {Hjorth-Jensen}}, \ and\ \bibinfo {author}
  {\bibfnamefont {C.~J.}\ \bibnamefont {Horowitz}}} (\bibinfo {year} {2014}),\
  \href {\doibase 10.1103/PhysRevC.89.014319} {\bibfield  {journal} {\bibinfo
  {journal} {Phys. Rev. C}\ }\textbf {\bibinfo {volume} {89}},\ \bibinfo
  {pages} {014319}}\BibitemShut {NoStop}%
\bibitem [{\citenamefont {{Hagen}}\ \emph {et~al.}(2014)\citenamefont
  {{Hagen}}, \citenamefont {{Papenbrock}}, \citenamefont {{Hjorth-Jensen}},\
  and\ \citenamefont {{Dean}}}]{Hagen:2014}%
  \BibitemOpen
  \bibfield  {author} {\bibinfo {author} {\bibnamefont {{Hagen}}, \bibfnamefont
  {G.}}, \bibinfo {author} {\bibfnamefont {T.}~\bibnamefont {{Papenbrock}}},
  \bibinfo {author} {\bibfnamefont {M.}~\bibnamefont {{Hjorth-Jensen}}}, \ and\
  \bibinfo {author} {\bibfnamefont {D.~J.}\ \bibnamefont {{Dean}}}} (\bibinfo
  {year} {2014}),\ \href {\doibase 10.1088/0034-4885/77/9/096302} {\bibfield
  {journal} {\bibinfo  {journal} {Reports on Progress in Physics}\ }\textbf
  {\bibinfo {volume} {77}}~(\bibinfo {number} {9}),\ \bibinfo {eid}
  {096302}}\BibitemShut {NoStop}%
\bibitem [{\citenamefont {Haxton}\ and\ \citenamefont
  {Holstein}(2013)}]{Hax13}%
  \BibitemOpen
  \bibfield  {author} {\bibinfo {author} {\bibnamefont {Haxton}, \bibfnamefont
  {W.~C.}}, \ and\ \bibinfo {author} {\bibfnamefont {B.~R.}\ \bibnamefont
  {Holstein}}} (\bibinfo {year} {2013}),\ \href@noop {} {\bibfield  {journal}
  {\bibinfo  {journal} {Prog. Part. Nucl. Phys.}\ }\textbf {\bibinfo {volume}
  {71}},\ \bibinfo {pages} {185}}\BibitemShut {NoStop}%
\bibitem [{\citenamefont {Henley}\ \emph {et~al.}(1979)\citenamefont {Henley},
  \citenamefont {Miller}, \citenamefont {Rho},\ and\ \citenamefont
  {Wilkinson}}]{henley1979}%
  \BibitemOpen
  \bibfield  {author} {\bibinfo {author} {\bibnamefont {Henley}, \bibfnamefont
  {E.~M.}}, \bibinfo {author} {\bibfnamefont {G.~A.}\ \bibnamefont {Miller}},
  \bibinfo {author} {\bibfnamefont {M.}~\bibnamefont {Rho}}, \ and\ \bibinfo
  {author} {\bibfnamefont {D.}~\bibnamefont {Wilkinson}}} (\bibinfo {year}
  {1979}),\ \href@noop {} {\bibinfo  {journal} {{\it Mesons in Nuclei}, Vol. 1,
  (North-Holland, Amsterdam)}\ ,\ \bibinfo {pages} {405}}\BibitemShut {NoStop}%
\bibitem [{\citenamefont {Hyde-Wright}\ and\ \citenamefont
  {de~Jager}(2004)}]{Hyd04}%
  \BibitemOpen
\bibfield  {journal} {  }\bibfield  {author} {\bibinfo {author} {\bibnamefont
  {Hyde-Wright}, \bibfnamefont {C.~E.}}, \ and\ \bibinfo {author}
  {\bibfnamefont {K.}~\bibnamefont {de~Jager}}} (\bibinfo {year} {2004}),\
  \href {\doibase 10.1146/annurev.nucl.53.041002.110443} {\bibfield  {journal}
  {\bibinfo  {journal} {Annual Review of Nuclear and Particle Science}\
  }\textbf {\bibinfo {volume} {54}}~(\bibinfo {number} {1}),\ \bibinfo {pages}
  {217}}\BibitemShut {NoStop}%
\bibitem [{\citenamefont {Ishii}\ \emph {et~al.}(2007)\citenamefont {Ishii},
  \citenamefont {Aoki},\ and\ \citenamefont {Hatsuda}}]{Ishii:2007}%
  \BibitemOpen
  \bibfield  {author} {\bibinfo {author} {\bibnamefont {Ishii}, \bibfnamefont
  {N.}}, \bibinfo {author} {\bibfnamefont {S.}~\bibnamefont {Aoki}}, \ and\
  \bibinfo {author} {\bibfnamefont {T.}~\bibnamefont {Hatsuda}}} (\bibinfo
  {year} {2007}),\ \href {\doibase 10.1103/PhysRevLett.99.022001} {\bibfield
  {journal} {\bibinfo  {journal} {Phys. Rev. Lett.}\ }\textbf {\bibinfo
  {volume} {99}},\ \bibinfo {pages} {022001}}\BibitemShut {NoStop}%
\bibitem [{\citenamefont {Jarrell}\ and\ \citenamefont
  {Gubernatis}(1996)}]{Jarrell:1996}%
  \BibitemOpen
  \bibfield  {author} {\bibinfo {author} {\bibnamefont {Jarrell}, \bibfnamefont
  {M.}}, \ and\ \bibinfo {author} {\bibfnamefont {J.~E.}\ \bibnamefont
  {Gubernatis}}} (\bibinfo {year} {1996}),\ \href@noop {} {\bibfield  {journal}
  {\bibinfo  {journal} {Physics Reports}\ }\textbf {\bibinfo {volume}
  {269}}~(\bibinfo {number} {3}),\ \bibinfo {pages} {133}}\BibitemShut
  {NoStop}%
\bibitem [{\citenamefont {{Jourdan}}(1996)}]{Jourdan:1996}%
  \BibitemOpen
  \bibfield  {author} {\bibinfo {author} {\bibnamefont {{Jourdan}},
  \bibfnamefont {J.}}} (\bibinfo {year} {1996}),\ \href {\doibase
  10.1016/0375-9474(96)00143-1} {\bibfield  {journal} {\bibinfo  {journal}
  {Nucl. Phys. A}\ }\textbf {\bibinfo {volume} {603}},\ \bibinfo {pages}
  {117}}\BibitemShut {NoStop}%
\bibitem [{\citenamefont {Kalos}(1962)}]{Kalos:1962}%
  \BibitemOpen
  \bibfield  {author} {\bibinfo {author} {\bibnamefont {Kalos}, \bibfnamefont
  {M.~H.}}} (\bibinfo {year} {1962}),\ \href {\doibase
  10.1103/PhysRev.128.1791} {\bibfield  {journal} {\bibinfo  {journal} {Phys.
  Rev.}\ }\textbf {\bibinfo {volume} {128}},\ \bibinfo {pages}
  {1791}}\BibitemShut {NoStop}%
\bibitem [{\citenamefont {Kamada}\ \emph {et~al.}(2001)\citenamefont {Kamada},
  \citenamefont {Nogga}, \citenamefont {Gl\"ockle}, \citenamefont {Hiyama},
  \citenamefont {Kamimura}, \citenamefont {Varga}, \citenamefont {Suzuki},
  \citenamefont {Viviani}, \citenamefont {Kievsky}, \citenamefont {Rosati},
  \citenamefont {Carlson}, \citenamefont {Pieper}, \citenamefont {Wiringa},
  \citenamefont {Navr\'atil}, \citenamefont {Barrett}, \citenamefont {Barnea},
  \citenamefont {Leidemann},\ and\ \citenamefont {Orlandini}}]{kamada2001}%
  \BibitemOpen
  \bibfield  {author} {\bibinfo {author} {\bibnamefont {Kamada}, \bibfnamefont
  {H.}}, \bibinfo {author} {\bibfnamefont {A.}~\bibnamefont {Nogga}}, \bibinfo
  {author} {\bibfnamefont {W.}~\bibnamefont {Gl\"ockle}}, \bibinfo {author}
  {\bibfnamefont {E.}~\bibnamefont {Hiyama}}, \bibinfo {author} {\bibfnamefont
  {M.}~\bibnamefont {Kamimura}}, \bibinfo {author} {\bibfnamefont
  {K.}~\bibnamefont {Varga}}, \bibinfo {author} {\bibfnamefont
  {Y.}~\bibnamefont {Suzuki}}, \bibinfo {author} {\bibfnamefont
  {M.}~\bibnamefont {Viviani}}, \bibinfo {author} {\bibfnamefont
  {A.}~\bibnamefont {Kievsky}}, \bibinfo {author} {\bibfnamefont
  {S.}~\bibnamefont {Rosati}}, \bibinfo {author} {\bibfnamefont
  {J.}~\bibnamefont {Carlson}}, \bibinfo {author} {\bibfnamefont {S.~C.}\
  \bibnamefont {Pieper}}, \bibinfo {author} {\bibfnamefont {R.~B.}\
  \bibnamefont {Wiringa}}, \bibinfo {author} {\bibfnamefont {P.}~\bibnamefont
  {Navr\'atil}}, \bibinfo {author} {\bibfnamefont {B.~R.}\ \bibnamefont
  {Barrett}}, \bibinfo {author} {\bibfnamefont {N.}~\bibnamefont {Barnea}},
  \bibinfo {author} {\bibfnamefont {W.}~\bibnamefont {Leidemann}}, \ and\
  \bibinfo {author} {\bibfnamefont {G.}~\bibnamefont {Orlandini}}} (\bibinfo
  {year} {2001}),\ \href {\doibase 10.1103/PhysRevC.64.044001} {\bibfield
  {journal} {\bibinfo  {journal} {Phys. Rev. C}\ }\textbf {\bibinfo {volume}
  {64}}~(\bibinfo {number} {4}),\ \bibinfo {pages} {044001}}\BibitemShut
  {NoStop}%
\bibitem [{\citenamefont {Kammel}\ and\ \citenamefont
  {Kubodera}(2010)}]{Kam10}%
  \BibitemOpen
  \bibfield  {author} {\bibinfo {author} {\bibnamefont {Kammel}, \bibfnamefont
  {P.}}, \ and\ \bibinfo {author} {\bibfnamefont {K.}~\bibnamefont {Kubodera}}}
  (\bibinfo {year} {2010}),\ \href {\doibase
  10.1146/annurev-nucl-100809-131946} {\bibfield  {journal} {\bibinfo
  {journal} {Annual Review of Nuclear and Particle Science}\ }\textbf {\bibinfo
  {volume} {60}}~(\bibinfo {number} {1}),\ \bibinfo {pages} {327}}\BibitemShut
  {NoStop}%
\bibitem [{\citenamefont {Kay}\ \emph {et~al.}(2013)\citenamefont {Kay},
  \citenamefont {Schiffer},\ and\ \citenamefont {Freeman}}]{kay2013}%
  \BibitemOpen
  \bibfield  {author} {\bibinfo {author} {\bibnamefont {Kay}, \bibfnamefont
  {B.~P.}}, \bibinfo {author} {\bibfnamefont {J.~P.}\ \bibnamefont {Schiffer}},
  \ and\ \bibinfo {author} {\bibfnamefont {S.~J.}\ \bibnamefont {Freeman}}}
  (\bibinfo {year} {2013}),\ \href {\doibase 10.1103/PhysRevLett.111.042502}
  {\bibfield  {journal} {\bibinfo  {journal} {Phys. Rev. Lett.}\ }\textbf
  {\bibinfo {volume} {111}},\ \bibinfo {pages} {042502}}\BibitemShut {NoStop}%
\bibitem [{\citenamefont {Kitagaki}\ \emph {et~al.}(1983)\citenamefont
  {Kitagaki} \emph {et~al.}}]{Kit83}%
  \BibitemOpen
  \bibfield  {author} {\bibinfo {author} {\bibnamefont {Kitagaki},
  \bibfnamefont {T.}},  \emph {et~al.}} (\bibinfo {year} {1983}),\ \href
  {\doibase 10.1103/PhysRevD.28.436} {\bibfield  {journal} {\bibinfo  {journal}
  {Phys. Rev. D}\ }\textbf {\bibinfo {volume} {28}},\ \bibinfo {pages}
  {436}}\BibitemShut {NoStop}%
\bibitem [{\citenamefont {van Kolck}(1994)}]{van94}%
  \BibitemOpen
  \bibfield  {author} {\bibinfo {author} {\bibnamefont {van Kolck},
  \bibfnamefont {U.}}} (\bibinfo {year} {1994}),\ \href@noop {} {\bibfield
  {journal} {\bibinfo  {journal} {Phys. Rev. C}\ }\textbf {\bibinfo {volume}
  {49}},\ \bibinfo {pages} {2932}}\BibitemShut {NoStop}%
\bibitem [{\citenamefont {K\"olling}\ \emph {et~al.}(2009)\citenamefont
  {K\"olling}, \citenamefont {Epelbaum}, \citenamefont {Krebs},\ and\
  \citenamefont {Meissner}}]{Koe09}%
  \BibitemOpen
  \bibfield  {author} {\bibinfo {author} {\bibnamefont {K\"olling},
  \bibfnamefont {S.}}, \bibinfo {author} {\bibfnamefont {E.}~\bibnamefont
  {Epelbaum}}, \bibinfo {author} {\bibfnamefont {H.}~\bibnamefont {Krebs}}, \
  and\ \bibinfo {author} {\bibfnamefont {U.-G.}\ \bibnamefont {Meissner}}}
  (\bibinfo {year} {2009}),\ \href@noop {} {\bibfield  {journal} {\bibinfo
  {journal} {Phys.\ Rev.\ C}\ }\textbf {\bibinfo {volume} {80}},\ \bibinfo
  {pages} {045502}}\BibitemShut {NoStop}%
\bibitem [{\citenamefont {K\"olling}\ \emph {et~al.}(2011)\citenamefont
  {K\"olling}, \citenamefont {Epelbaum}, \citenamefont {Krebs},\ and\
  \citenamefont {Meissner}}]{Koe11}%
  \BibitemOpen
  \bibfield  {author} {\bibinfo {author} {\bibnamefont {K\"olling},
  \bibfnamefont {S.}}, \bibinfo {author} {\bibfnamefont {E.}~\bibnamefont
  {Epelbaum}}, \bibinfo {author} {\bibfnamefont {H.}~\bibnamefont {Krebs}}, \
  and\ \bibinfo {author} {\bibfnamefont {U.-G.}\ \bibnamefont {Meissner}}}
  (\bibinfo {year} {2011}),\ \href@noop {} {\bibfield  {journal} {\bibinfo
  {journal} {Phys.\ Rev.\ C}\ }\textbf {\bibinfo {volume} {84}},\ \bibinfo
  {pages} {054008}}\BibitemShut {NoStop}%
\bibitem [{\citenamefont {Koonin}\ \emph {et~al.}(1997)\citenamefont {Koonin},
  \citenamefont {Dean},\ and\ \citenamefont {Langanke}}]{koonin1997}%
  \BibitemOpen
  \bibfield  {author} {\bibinfo {author} {\bibnamefont {Koonin}, \bibfnamefont
  {S.~E.}}, \bibinfo {author} {\bibfnamefont {D.~J.}\ \bibnamefont {Dean}}, \
  and\ \bibinfo {author} {\bibfnamefont {K.}~\bibnamefont {Langanke}}}
  (\bibinfo {year} {1997}),\ \href@noop {} {\bibfield  {journal} {\bibinfo
  {journal} {Phys. Rep.}\ }\textbf {\bibinfo {volume} {278}},\ \bibinfo {pages}
  {1}}\BibitemShut {NoStop}%
\bibitem [{\citenamefont {Korover}\ \emph {et~al.}(2014)\citenamefont {Korover}
  \emph {et~al.}}]{korover2014}%
  \BibitemOpen
  \bibfield  {author} {\bibinfo {author} {\bibnamefont {Korover}, \bibfnamefont
  {I.}},  \emph {et~al.} (\bibinfo {collaboration} {(Jefferson Lab Hall A
  Collaboration)})} (\bibinfo {year} {2014}),\ \href {\doibase
  10.1103/PhysRevLett.113.022501} {\bibfield  {journal} {\bibinfo  {journal}
  {Phys. Rev. Lett.}\ }\textbf {\bibinfo {volume} {113}},\ \bibinfo {pages}
  {022501}}\BibitemShut {NoStop}%
\bibitem [{\citenamefont {Kortelainen}\ \emph {et~al.}(2014)\citenamefont
  {Kortelainen}, \citenamefont {McDonnell}, \citenamefont {Nazarewicz},
  \citenamefont {Olsen}, \citenamefont {Reinhard}, \citenamefont {Sarich},
  \citenamefont {Schunck}, \citenamefont {Wild}, \citenamefont {Davesne},
  \citenamefont {Erler},\ and\ \citenamefont {Pastore}}]{Kortelainen:2014}%
  \BibitemOpen
  \bibfield  {author} {\bibinfo {author} {\bibnamefont {Kortelainen},
  \bibfnamefont {M.}}, \bibinfo {author} {\bibfnamefont {J.}~\bibnamefont
  {McDonnell}}, \bibinfo {author} {\bibfnamefont {W.}~\bibnamefont
  {Nazarewicz}}, \bibinfo {author} {\bibfnamefont {E.}~\bibnamefont {Olsen}},
  \bibinfo {author} {\bibfnamefont {P.-G.}\ \bibnamefont {Reinhard}}, \bibinfo
  {author} {\bibfnamefont {J.}~\bibnamefont {Sarich}}, \bibinfo {author}
  {\bibfnamefont {N.}~\bibnamefont {Schunck}}, \bibinfo {author} {\bibfnamefont
  {S.~M.}\ \bibnamefont {Wild}}, \bibinfo {author} {\bibfnamefont
  {D.}~\bibnamefont {Davesne}}, \bibinfo {author} {\bibfnamefont
  {J.}~\bibnamefont {Erler}}, \ and\ \bibinfo {author} {\bibfnamefont
  {A.}~\bibnamefont {Pastore}}} (\bibinfo {year} {2014}),\ \href {\doibase
  10.1103/PhysRevC.89.054314} {\bibfield  {journal} {\bibinfo  {journal} {Phys.
  Rev. C}\ }\textbf {\bibinfo {volume} {89}},\ \bibinfo {pages}
  {054314}}\BibitemShut {NoStop}%
\bibitem [{\citenamefont {Kubodera}\ \emph {et~al.}(1978)\citenamefont
  {Kubodera}, \citenamefont {Delorme},\ and\ \citenamefont {Rho}}]{Kub78}%
  \BibitemOpen
  \bibfield  {author} {\bibinfo {author} {\bibnamefont {Kubodera},
  \bibfnamefont {K.}}, \bibinfo {author} {\bibfnamefont {J.}~\bibnamefont
  {Delorme}}, \ and\ \bibinfo {author} {\bibfnamefont {M.}~\bibnamefont {Rho}}}
  (\bibinfo {year} {1978}),\ \href {\doibase 10.1103/PhysRevLett.40.755}
  {\bibfield  {journal} {\bibinfo  {journal} {Phys. Rev. Lett.}\ }\textbf
  {\bibinfo {volume} {40}},\ \bibinfo {pages} {755}}\BibitemShut {NoStop}%
\bibitem [{\citenamefont {Kumar}(1974)}]{kumar1974}%
  \BibitemOpen
  \bibfield  {author} {\bibinfo {author} {\bibnamefont {Kumar}, \bibfnamefont
  {N.}}} (\bibinfo {year} {1974}),\ \href {\doibase
  http://dx.doi.org/10.1016/0375-9474(74)90540-5} {\bibfield  {journal}
  {\bibinfo  {journal} {Nuclear Physics A}\ }\textbf {\bibinfo {volume}
  {225}}~(\bibinfo {number} {2}),\ \bibinfo {pages} {221 }}\BibitemShut
  {NoStop}%
\bibitem [{\citenamefont {Kurath}(1979)}]{kurath1979}%
  \BibitemOpen
  \bibfield  {author} {\bibinfo {author} {\bibnamefont {Kurath}, \bibfnamefont
  {D.}}} (\bibinfo {year} {1979}),\ \href {\doibase
  http://dx.doi.org/10.1016/0375-9474(79)90458-5} {\bibfield  {journal}
  {\bibinfo  {journal} {Nuclear Physics A}\ }\textbf {\bibinfo {volume}
  {317}}~(\bibinfo {number} {1}),\ \bibinfo {pages} {175 }}\BibitemShut
  {NoStop}%
\bibitem [{\citenamefont {Kurylov}\ \emph {et~al.}(2002)\citenamefont
  {Kurylov}, \citenamefont {Ramsey-Musolf},\ and\ \citenamefont
  {Vogel}}]{Kurylov:2002}%
  \BibitemOpen
  \bibfield  {author} {\bibinfo {author} {\bibnamefont {Kurylov}, \bibfnamefont
  {A.}}, \bibinfo {author} {\bibfnamefont {M.~J.}\ \bibnamefont
  {Ramsey-Musolf}}, \ and\ \bibinfo {author} {\bibfnamefont {P.}~\bibnamefont
  {Vogel}}} (\bibinfo {year} {2002}),\ \href {\doibase
  10.1103/PhysRevC.65.055501} {\bibfield  {journal} {\bibinfo  {journal} {Phys.
  Rev. C}\ }\textbf {\bibinfo {volume} {65}},\ \bibinfo {pages}
  {055501}}\BibitemShut {NoStop}%
\bibitem [{\citenamefont {Lapik\'as}\ \emph {et~al.}(1999)\citenamefont
  {Lapik\'as}, \citenamefont {Wesseling},\ and\ \citenamefont
  {Wiringa}}]{lapikas1999}%
  \BibitemOpen
  \bibfield  {author} {\bibinfo {author} {\bibnamefont {Lapik\'as},
  \bibfnamefont {L.}}, \bibinfo {author} {\bibfnamefont {J.}~\bibnamefont
  {Wesseling}}, \ and\ \bibinfo {author} {\bibfnamefont {R.~B.}\ \bibnamefont
  {Wiringa}}} (\bibinfo {year} {1999}),\ \href {\doibase
  10.1103/PhysRevLett.82.4404} {\bibfield  {journal} {\bibinfo  {journal}
  {Phys. Rev. Lett.}\ }\textbf {\bibinfo {volume} {82}},\ \bibinfo {pages}
  {4404}}\BibitemShut {NoStop}%
\bibitem [{\citenamefont {{Lattimer}}\ and\ \citenamefont
  {{Lim}}(2013)}]{Lattimer:2013}%
  \BibitemOpen
  \bibfield  {author} {\bibinfo {author} {\bibnamefont {{Lattimer}},
  \bibfnamefont {J.~M.}}, \ and\ \bibinfo {author} {\bibfnamefont
  {Y.}~\bibnamefont {{Lim}}}} (\bibinfo {year} {2013}),\ \href {\doibase
  10.1088/0004-637X/771/1/51} {\bibfield  {journal} {\bibinfo  {journal}
  {\apj}\ }\textbf {\bibinfo {volume} {771}},\ \bibinfo {eid} {51}}\BibitemShut
  {NoStop}%
\bibitem [{\citenamefont {{Lattimer}}\ and\ \citenamefont
  {{Prakash}}(2001)}]{Lattimer:2001}%
  \BibitemOpen
  \bibfield  {author} {\bibinfo {author} {\bibnamefont {{Lattimer}},
  \bibfnamefont {J.~M.}}, \ and\ \bibinfo {author} {\bibfnamefont
  {M.}~\bibnamefont {{Prakash}}}} (\bibinfo {year} {2001}),\ \href {\doibase
  10.1086/319702} {\bibfield  {journal} {\bibinfo  {journal} {\apj}\ }\textbf
  {\bibinfo {volume} {550}},\ \bibinfo {pages} {426}}\BibitemShut {NoStop}%
\bibitem [{\citenamefont {{LBNE Collaboration}}\ \emph
  {et~al.}(2013)\citenamefont {{LBNE Collaboration}}, \citenamefont {{Adams}},
  \citenamefont {{Adams}}, \citenamefont {{Akiri}}, \citenamefont {{Alion}},
  \citenamefont {{Anderson}}, \citenamefont {{Andreopoulos}}, \citenamefont
  {{Andrews}}, \citenamefont {{Anghel}}, \citenamefont {{Costa dos Anjos}},\
  and\ \citenamefont {et~al.}}]{LBNE:2013}%
  \BibitemOpen
  \bibfield  {author} {\bibinfo {author} {\bibnamefont {{LBNE
  Collaboration}},}, \bibinfo {author} {\bibfnamefont {C.}~\bibnamefont
  {{Adams}}}, \bibinfo {author} {\bibfnamefont {D.}~\bibnamefont {{Adams}}},
  \bibinfo {author} {\bibfnamefont {T.}~\bibnamefont {{Akiri}}}, \bibinfo
  {author} {\bibfnamefont {T.}~\bibnamefont {{Alion}}}, \bibinfo {author}
  {\bibfnamefont {K.}~\bibnamefont {{Anderson}}}, \bibinfo {author}
  {\bibfnamefont {C.}~\bibnamefont {{Andreopoulos}}}, \bibinfo {author}
  {\bibfnamefont {M.}~\bibnamefont {{Andrews}}}, \bibinfo {author}
  {\bibfnamefont {I.}~\bibnamefont {{Anghel}}}, \bibinfo {author}
  {\bibfnamefont {J.~C.}\ \bibnamefont {{Costa dos Anjos}}}, \ and\ \bibinfo
  {author} {\bibnamefont {et~al.}}} (\bibinfo {year} {2013}),\ \href@noop {}
  {\bibfield  {journal} {\bibinfo  {journal} {ArXiv e-prints}\ }}\Eprint
  {http://arxiv.org/abs/1307.7335} {arXiv:1307.7335 [hep-ex]} \BibitemShut
  {NoStop}%
\bibitem [{\citenamefont {{Lee}}(2009)}]{lee2009}%
  \BibitemOpen
  \bibfield  {author} {\bibinfo {author} {\bibnamefont {{Lee}}, \bibfnamefont
  {D.}}} (\bibinfo {year} {2009}),\ \href {\doibase 10.1016/j.ppnp.2008.12.001}
  {\bibfield  {journal} {\bibinfo  {journal} {Prog. in Part. and Nucl. Phys.}\
  }\textbf {\bibinfo {volume} {63}},\ \bibinfo {pages} {117}}\BibitemShut
  {NoStop}%
\bibitem [{\citenamefont {Lee}\ \emph {et~al.}(2004)\citenamefont {Lee},
  \citenamefont {Borasoy},\ and\ \citenamefont {Schaefer}}]{Lee:2004}%
  \BibitemOpen
  \bibfield  {author} {\bibinfo {author} {\bibnamefont {Lee}, \bibfnamefont
  {D.}}, \bibinfo {author} {\bibfnamefont {B.}~\bibnamefont {Borasoy}}, \ and\
  \bibinfo {author} {\bibfnamefont {T.}~\bibnamefont {Schaefer}}} (\bibinfo
  {year} {2004}),\ \href {\doibase 10.1103/PhysRevC.70.014007} {\bibfield
  {journal} {\bibinfo  {journal} {Phys. Rev. C}\ }\textbf {\bibinfo {volume}
  {70}},\ \bibinfo {pages} {014007}}\BibitemShut {NoStop}%
\bibitem [{\citenamefont {Lee}\ and\ \citenamefont
  {Sch{\"a}fer}(2006)}]{Lee:2005}%
  \BibitemOpen
  \bibfield  {author} {\bibinfo {author} {\bibnamefont {Lee}, \bibfnamefont
  {D.}}, \ and\ \bibinfo {author} {\bibfnamefont {T.}~\bibnamefont
  {Sch{\"a}fer}}} (\bibinfo {year} {2006}),\ \href {\doibase
  10.1103/PhysRevC.73.015201} {\bibfield  {journal} {\bibinfo  {journal} {Phys.
  Rev.}\ }\textbf {\bibinfo {volume} {C73}},\ \bibinfo {pages}
  {015201}}\BibitemShut {NoStop}%
\bibitem [{\citenamefont {Leidemann}\ and\ \citenamefont
  {Orlandini}(2013)}]{Leidemann:2013}%
  \BibitemOpen
  \bibfield  {author} {\bibinfo {author} {\bibnamefont {Leidemann},
  \bibfnamefont {W.}}, \ and\ \bibinfo {author} {\bibfnamefont
  {G.}~\bibnamefont {Orlandini}}} (\bibinfo {year} {2013}),\ \href {\doibase
  10.1016/j.ppnp.2012.09.001} {\bibfield  {journal} {\bibinfo  {journal} {Prog.
  Part. Nucl. Phys.}\ }\textbf {\bibinfo {volume} {68}},\ \bibinfo {pages}
  {158}}\BibitemShut {NoStop}%
\bibitem [{\citenamefont {Lin}\ \emph {et~al.}(2001)\citenamefont {Lin},
  \citenamefont {Zong},\ and\ \citenamefont {Ceperley}}]{Lin:2001}%
  \BibitemOpen
  \bibfield  {author} {\bibinfo {author} {\bibnamefont {Lin}, \bibfnamefont
  {C.}}, \bibinfo {author} {\bibfnamefont {F.~H.}\ \bibnamefont {Zong}}, \ and\
  \bibinfo {author} {\bibfnamefont {D.~M.}\ \bibnamefont {Ceperley}}} (\bibinfo
  {year} {2001}),\ \href {\doibase 10.1103/PhysRevE.64.016702} {\bibfield
  {journal} {\bibinfo  {journal} {Phys. Rev. E}\ }\textbf {\bibinfo {volume}
  {64}}~(\bibinfo {number} {1}),\ \bibinfo {pages} {016702}}\BibitemShut
  {NoStop}%
\bibitem [{\citenamefont {Lin}\ and\ \citenamefont {Liou}(1991)}]{Lin91}%
  \BibitemOpen
  \bibfield  {author} {\bibinfo {author} {\bibnamefont {Lin}, \bibfnamefont
  {D.}}, \ and\ \bibinfo {author} {\bibfnamefont {M.~K.}\ \bibnamefont {Liou}}}
  (\bibinfo {year} {1991}),\ \href {\doibase 10.1103/PhysRevC.43.R930}
  {\bibfield  {journal} {\bibinfo  {journal} {Phys. Rev. C}\ }\textbf {\bibinfo
  {volume} {43}},\ \bibinfo {pages} {R930}}\BibitemShut {NoStop}%
\bibitem [{\citenamefont {Liu}\ \emph {et~al.}(1974)\citenamefont {Liu},
  \citenamefont {Kalos},\ and\ \citenamefont {Chester}}]{Liu:1974}%
  \BibitemOpen
  \bibfield  {author} {\bibinfo {author} {\bibnamefont {Liu}, \bibfnamefont
  {K.~S.}}, \bibinfo {author} {\bibfnamefont {M.~H.}\ \bibnamefont {Kalos}}, \
  and\ \bibinfo {author} {\bibfnamefont {G.~V.}\ \bibnamefont {Chester}}}
  (\bibinfo {year} {1974}),\ \href {\doibase 10.1103/PhysRevA.10.303}
  {\bibfield  {journal} {\bibinfo  {journal} {Phys. Rev. A}\ }\textbf {\bibinfo
  {volume} {10}},\ \bibinfo {pages} {303}}\BibitemShut {NoStop}%
\bibitem [{\citenamefont {Lomnitz-Adler}\ \emph {et~al.}(1981)\citenamefont
  {Lomnitz-Adler}, \citenamefont {Pandharipande},\ and\ \citenamefont
  {Smith}}]{lomnitz1981monte}%
  \BibitemOpen
  \bibfield  {author} {\bibinfo {author} {\bibnamefont {Lomnitz-Adler},
  \bibfnamefont {J.}}, \bibinfo {author} {\bibfnamefont {V.}~\bibnamefont
  {Pandharipande}}, \ and\ \bibinfo {author} {\bibfnamefont {R.}~\bibnamefont
  {Smith}}} (\bibinfo {year} {1981}),\ \href@noop {} {\bibfield  {journal}
  {\bibinfo  {journal} {Nucl. Phys. A}\ }\textbf {\bibinfo {volume}
  {361}}~(\bibinfo {number} {2}),\ \bibinfo {pages} {399}}\BibitemShut
  {NoStop}%
\bibitem [{\citenamefont {Lonardoni}\ \emph {et~al.}(2013)\citenamefont
  {Lonardoni}, \citenamefont {Gandolfi},\ and\ \citenamefont
  {Pederiva}}]{Lonardoni:2013}%
  \BibitemOpen
  \bibfield  {author} {\bibinfo {author} {\bibnamefont {Lonardoni},
  \bibfnamefont {D.}}, \bibinfo {author} {\bibfnamefont {S.}~\bibnamefont
  {Gandolfi}}, \ and\ \bibinfo {author} {\bibfnamefont {F.}~\bibnamefont
  {Pederiva}}} (\bibinfo {year} {2013}),\ \href {\doibase
  10.1103/PhysRevC.87.041303} {\bibfield  {journal} {\bibinfo  {journal} {Phys.
  Rev. C}\ }\textbf {\bibinfo {volume} {87}},\ \bibinfo {pages}
  {041303}}\BibitemShut {NoStop}%
\bibitem [{\citenamefont {Lonardoni}\ \emph {et~al.}(2015)\citenamefont
  {Lonardoni}, \citenamefont {Lovato}, \citenamefont {Gandolfi},\ and\
  \citenamefont {Pederiva}}]{Lonardoni:2014b}%
  \BibitemOpen
  \bibfield  {author} {\bibinfo {author} {\bibnamefont {Lonardoni},
  \bibfnamefont {D.}}, \bibinfo {author} {\bibfnamefont {A.}~\bibnamefont
  {Lovato}}, \bibinfo {author} {\bibfnamefont {S.}~\bibnamefont {Gandolfi}}, \
  and\ \bibinfo {author} {\bibfnamefont {F.}~\bibnamefont {Pederiva}}}
  (\bibinfo {year} {2015}),\ \href {\doibase 10.1103/PhysRevLett.114.092301}
  {\bibfield  {journal} {\bibinfo  {journal} {Phys. Rev. Lett.}\ }\textbf
  {\bibinfo {volume} {114}},\ \bibinfo {pages} {092301}}\BibitemShut {NoStop}%
\bibitem [{\citenamefont {Lonardoni}\ \emph {et~al.}(2014)\citenamefont
  {Lonardoni}, \citenamefont {Pederiva},\ and\ \citenamefont
  {Gandolfi}}]{Lonardoni:2014}%
  \BibitemOpen
  \bibfield  {author} {\bibinfo {author} {\bibnamefont {Lonardoni},
  \bibfnamefont {D.}}, \bibinfo {author} {\bibfnamefont {F.}~\bibnamefont
  {Pederiva}}, \ and\ \bibinfo {author} {\bibfnamefont {S.}~\bibnamefont
  {Gandolfi}}} (\bibinfo {year} {2014}),\ \href {\doibase
  10.1103/PhysRevC.89.014314} {\bibfield  {journal} {\bibinfo  {journal} {Phys.
  Rev. C}\ }\textbf {\bibinfo {volume} {89}},\ \bibinfo {pages}
  {014314}}\BibitemShut {NoStop}%
\bibitem [{\citenamefont {Lovato}\ \emph {et~al.}(2013)\citenamefont {Lovato},
  \citenamefont {Gandolfi}, \citenamefont {Butler}, \citenamefont {Carlson},
  \citenamefont {Lusk}, \citenamefont {Pieper},\ and\ \citenamefont
  {Schiavilla}}]{lovato2013}%
  \BibitemOpen
  \bibfield  {author} {\bibinfo {author} {\bibnamefont {Lovato}, \bibfnamefont
  {A.}}, \bibinfo {author} {\bibfnamefont {S.}~\bibnamefont {Gandolfi}},
  \bibinfo {author} {\bibfnamefont {R.}~\bibnamefont {Butler}}, \bibinfo
  {author} {\bibfnamefont {J.}~\bibnamefont {Carlson}}, \bibinfo {author}
  {\bibfnamefont {E.}~\bibnamefont {Lusk}}, \bibinfo {author} {\bibfnamefont
  {S.~C.}\ \bibnamefont {Pieper}}, \ and\ \bibinfo {author} {\bibfnamefont
  {R.}~\bibnamefont {Schiavilla}}} (\bibinfo {year} {2013}),\ \href@noop {}
  {\bibfield  {journal} {\bibinfo  {journal} {Phys. Rev. Lett.}\ }\textbf
  {\bibinfo {volume} {111}},\ \bibinfo {pages} {092501}}\BibitemShut {NoStop}%
\bibitem [{\citenamefont {Lovato}\ \emph {et~al.}(2014)\citenamefont {Lovato},
  \citenamefont {Gandolfi}, \citenamefont {Carlson}, \citenamefont {Pieper},\
  and\ \citenamefont {Schiavilla}}]{Lovato:2014}%
  \BibitemOpen
  \bibfield  {author} {\bibinfo {author} {\bibnamefont {Lovato}, \bibfnamefont
  {A.}}, \bibinfo {author} {\bibfnamefont {S.}~\bibnamefont {Gandolfi}},
  \bibinfo {author} {\bibfnamefont {J.}~\bibnamefont {Carlson}}, \bibinfo
  {author} {\bibfnamefont {S.~C.}\ \bibnamefont {Pieper}}, \ and\ \bibinfo
  {author} {\bibfnamefont {R.}~\bibnamefont {Schiavilla}}} (\bibinfo {year}
  {2014}),\ \href {\doibase 10.1103/PhysRevLett.112.182502} {\bibfield
  {journal} {\bibinfo  {journal} {Phys. Rev. Lett.}\ }\textbf {\bibinfo
  {volume} {112}},\ \bibinfo {pages} {182502}}\BibitemShut {NoStop}%
\bibitem [{\citenamefont {Lovato}\ \emph {et~al.}(2015)\citenamefont {Lovato},
  \citenamefont {Gandolfi}, \citenamefont {Carlson}, \citenamefont {Pieper},\
  and\ \citenamefont {Schiavilla}}]{Lovato:2015}%
  \BibitemOpen
  \bibfield  {author} {\bibinfo {author} {\bibnamefont {Lovato}, \bibfnamefont
  {A.}}, \bibinfo {author} {\bibfnamefont {S.}~\bibnamefont {Gandolfi}},
  \bibinfo {author} {\bibfnamefont {J.}~\bibnamefont {Carlson}}, \bibinfo
  {author} {\bibfnamefont {S.~C.}\ \bibnamefont {Pieper}}, \ and\ \bibinfo
  {author} {\bibfnamefont {R.}~\bibnamefont {Schiavilla}}} (\bibinfo {year}
  {2015}),\ \href@noop {} {\ }\Eprint
  {http://arxiv.org/abs/arxiv.org/abs/1501.01981} {arxiv.org/abs/1501.01981}
  \BibitemShut {NoStop}%
\bibitem [{\citenamefont {Lu}\ \emph {et~al.}(2013)\citenamefont {Lu},
  \citenamefont {Mueller}, \citenamefont {Drake}, \citenamefont
  {N\"ortersh\"auser}, \citenamefont {Pieper},\ and\ \citenamefont
  {Yan}}]{lu2013}%
  \BibitemOpen
  \bibfield  {author} {\bibinfo {author} {\bibnamefont {Lu}, \bibfnamefont
  {Z.-T.}}, \bibinfo {author} {\bibfnamefont {P.}~\bibnamefont {Mueller}},
  \bibinfo {author} {\bibfnamefont {G.~W.~F.}\ \bibnamefont {Drake}}, \bibinfo
  {author} {\bibfnamefont {W.}~\bibnamefont {N\"ortersh\"auser}}, \bibinfo
  {author} {\bibfnamefont {S.~C.}\ \bibnamefont {Pieper}}, \ and\ \bibinfo
  {author} {\bibfnamefont {Z.-C.}\ \bibnamefont {Yan}}} (\bibinfo {year}
  {2013}),\ \href {\doibase 10.1103/RevModPhys.85.1383} {\bibfield  {journal}
  {\bibinfo  {journal} {Rev. Mod. Phys.}\ }\textbf {\bibinfo {volume} {85}},\
  \bibinfo {pages} {1383}}\BibitemShut {NoStop}%
\bibitem [{\citenamefont {Lusk}\ \emph {et~al.}(2010)\citenamefont {Lusk},
  \citenamefont {Pieper},\ and\ \citenamefont {Butler}}]{Lusk:2010}%
  \BibitemOpen
  \bibfield  {author} {\bibinfo {author} {\bibnamefont {Lusk}, \bibfnamefont
  {E.}}, \bibinfo {author} {\bibfnamefont {S.}~\bibnamefont {Pieper}}, \ and\
  \bibinfo {author} {\bibfnamefont {R.}~\bibnamefont {Butler}}} (\bibinfo
  {year} {2010}),\ \href {http://www.scidacreview.org/1002/html/adlb.html}
  {\bibfield  {journal} {\bibinfo  {journal} {SciDAC Review}\ }\textbf
  {\bibinfo {volume} {17}},\ \bibinfo {pages} {30}}\BibitemShut {NoStop}%
\bibitem [{\citenamefont {Lynn}\ \emph {et~al.}(2014)\citenamefont {Lynn},
  \citenamefont {Carlson}, \citenamefont {Epelbaum}, \citenamefont {Gandolfi},
  \citenamefont {Gezerlis} \emph {et~al.}}]{Lynn:2014}%
  \BibitemOpen
  \bibfield  {author} {\bibinfo {author} {\bibnamefont {Lynn}, \bibfnamefont
  {J.}}, \bibinfo {author} {\bibfnamefont {J.}~\bibnamefont {Carlson}},
  \bibinfo {author} {\bibfnamefont {E.}~\bibnamefont {Epelbaum}}, \bibinfo
  {author} {\bibfnamefont {S.}~\bibnamefont {Gandolfi}}, \bibinfo {author}
  {\bibfnamefont {A.}~\bibnamefont {Gezerlis}},  \emph {et~al.}} (\bibinfo
  {year} {2014}),\ \href {\doibase 10.1103/PhysRevLett.113.192501} {\bibfield
  {journal} {\bibinfo  {journal} {Phys. Rev. Lett.}\ }\textbf {\bibinfo
  {volume} {113}}~(\bibinfo {number} {19}),\ \bibinfo {pages}
  {192501}}\BibitemShut {NoStop}%
\bibitem [{\citenamefont {{Lynn}}\ and\ \citenamefont
  {{Schmidt}}(2012)}]{Lynn2012}%
  \BibitemOpen
  \bibfield  {author} {\bibinfo {author} {\bibnamefont {{Lynn}}, \bibfnamefont
  {J.~E.}}, \ and\ \bibinfo {author} {\bibfnamefont {K.~E.}\ \bibnamefont
  {{Schmidt}}}} (\bibinfo {year} {2012}),\ \href {\doibase
  10.1103/PhysRevC.86.014324} {\bibfield  {journal} {\bibinfo  {journal}
  {\prc}\ }\textbf {\bibinfo {volume} {86}}~(\bibinfo {number} {1}),\ \bibinfo
  {eid} {014324}}\BibitemShut {NoStop}%
\bibitem [{\citenamefont {Macfarlane}\ and\ \citenamefont
  {French}(1960)}]{macfarlane1960}%
  \BibitemOpen
  \bibfield  {author} {\bibinfo {author} {\bibnamefont {Macfarlane},
  \bibfnamefont {M.~H.}}, \ and\ \bibinfo {author} {\bibfnamefont {J.~B.}\
  \bibnamefont {French}}} (\bibinfo {year} {1960}),\ \href {\doibase
  10.1103/RevModPhys.32.567} {\bibfield  {journal} {\bibinfo  {journal} {Rev.
  Mod. Phys.}\ }\textbf {\bibinfo {volume} {32}},\ \bibinfo {pages}
  {567}}\BibitemShut {NoStop}%
\bibitem [{\citenamefont {Machleidt}(2001)}]{Mac01}%
  \BibitemOpen
  \bibfield  {author} {\bibinfo {author} {\bibnamefont {Machleidt},
  \bibfnamefont {R.}}} (\bibinfo {year} {2001}),\ \href@noop {} {\bibfield
  {journal} {\bibinfo  {journal} {Phys. Rev. C}\ }\textbf {\bibinfo {volume}
  {63}},\ \bibinfo {pages} {024001}}\BibitemShut {NoStop}%
\bibitem [{\citenamefont {Machleidt}\ and\ \citenamefont
  {Entem}(2011)}]{Mac11}%
  \BibitemOpen
  \bibfield  {author} {\bibinfo {author} {\bibnamefont {Machleidt},
  \bibfnamefont {R.}}, \ and\ \bibinfo {author} {\bibfnamefont
  {D.}~\bibnamefont {Entem}}} (\bibinfo {year} {2011}),\ \href@noop {}
  {\bibfield  {journal} {\bibinfo  {journal} {Phys.\ Rep.}\ }\textbf {\bibinfo
  {volume} {503}},\ \bibinfo {pages} {1}}\BibitemShut {NoStop}%
\bibitem [{\citenamefont {Machleidt}\ \emph {et~al.}(1996)\citenamefont
  {Machleidt}, \citenamefont {Sammarruca},\ and\ \citenamefont
  {Song}}]{machleidt1996}%
  \BibitemOpen
  \bibfield  {author} {\bibinfo {author} {\bibnamefont {Machleidt},
  \bibfnamefont {R.}}, \bibinfo {author} {\bibfnamefont {F.}~\bibnamefont
  {Sammarruca}}, \ and\ \bibinfo {author} {\bibfnamefont {Y.}~\bibnamefont
  {Song}}} (\bibinfo {year} {1996}),\ \href {\doibase
  10.1103/PhysRevC.53.R1483} {\bibfield  {journal} {\bibinfo  {journal} {Phys.
  Rev. C}\ }\textbf {\bibinfo {volume} {53}}~(\bibinfo {number} {4}),\ \bibinfo
  {pages} {R1483}}\BibitemShut {NoStop}%
\bibitem [{\citenamefont {Marcucci}\ \emph {et~al.}(2012)\citenamefont
  {Marcucci}, \citenamefont {Kievsky}, \citenamefont {Rosati}, \citenamefont
  {Schiavilla},\ and\ \citenamefont {Viviani}}]{Mar12}%
  \BibitemOpen
  \bibfield  {author} {\bibinfo {author} {\bibnamefont {Marcucci},
  \bibfnamefont {L.~E.}}, \bibinfo {author} {\bibfnamefont {A.}~\bibnamefont
  {Kievsky}}, \bibinfo {author} {\bibfnamefont {S.}~\bibnamefont {Rosati}},
  \bibinfo {author} {\bibfnamefont {R.}~\bibnamefont {Schiavilla}}, \ and\
  \bibinfo {author} {\bibfnamefont {M.}~\bibnamefont {Viviani}}} (\bibinfo
  {year} {2012}),\ \href {\doibase 10.1103/PhysRevLett.108.052502} {\bibfield
  {journal} {\bibinfo  {journal} {Phys. Rev. Lett.}\ }\textbf {\bibinfo
  {volume} {108}},\ \bibinfo {pages} {052502}}\BibitemShut {NoStop}%
\bibitem [{\citenamefont {Marcucci}\ \emph {et~al.}(2008)\citenamefont
  {Marcucci}, \citenamefont {Pervin}, \citenamefont {Pieper}, \citenamefont
  {Schiavilla},\ and\ \citenamefont {Wiringa}}]{marcucci2008}%
  \BibitemOpen
  \bibfield  {author} {\bibinfo {author} {\bibnamefont {Marcucci},
  \bibfnamefont {L.~E.}}, \bibinfo {author} {\bibfnamefont {M.}~\bibnamefont
  {Pervin}}, \bibinfo {author} {\bibfnamefont {S.~C.}\ \bibnamefont {Pieper}},
  \bibinfo {author} {\bibfnamefont {R.}~\bibnamefont {Schiavilla}}, \ and\
  \bibinfo {author} {\bibfnamefont {R.~B.}\ \bibnamefont {Wiringa}}} (\bibinfo
  {year} {2008}),\ \href {\doibase 10.1103/PhysRevC.78.065501} {\bibfield
  {journal} {\bibinfo  {journal} {Phys. Rev. C}\ }\textbf {\bibinfo {volume}
  {78}},\ \bibinfo {pages} {065501}}\BibitemShut {NoStop}%
\bibitem [{\citenamefont {Marcucci}\ \emph {et~al.}(2011)\citenamefont
  {Marcucci}, \citenamefont {Piarulli}, \citenamefont {Viviani}, \citenamefont
  {Girlanda}, \citenamefont {Kievsky}, \citenamefont {Rosati},\ and\
  \citenamefont {Schiavilla}}]{Marcucci:2011}%
  \BibitemOpen
  \bibfield  {author} {\bibinfo {author} {\bibnamefont {Marcucci},
  \bibfnamefont {L.~E.}}, \bibinfo {author} {\bibfnamefont {M.}~\bibnamefont
  {Piarulli}}, \bibinfo {author} {\bibfnamefont {M.}~\bibnamefont {Viviani}},
  \bibinfo {author} {\bibfnamefont {L.}~\bibnamefont {Girlanda}}, \bibinfo
  {author} {\bibfnamefont {A.}~\bibnamefont {Kievsky}}, \bibinfo {author}
  {\bibfnamefont {S.}~\bibnamefont {Rosati}}, \ and\ \bibinfo {author}
  {\bibfnamefont {R.}~\bibnamefont {Schiavilla}}} (\bibinfo {year} {2011}),\
  \href {\doibase 10.1103/PhysRevC.83.014002} {\bibfield  {journal} {\bibinfo
  {journal} {Phys. Rev. C}\ }\textbf {\bibinfo {volume} {83}},\ \bibinfo
  {pages} {014002}}\BibitemShut {NoStop}%
\bibitem [{\citenamefont {Marcucci}\ \emph {et~al.}(2000)\citenamefont
  {Marcucci}, \citenamefont {Schiavilla}, \citenamefont {Viviani},
  \citenamefont {Kievsky}, \citenamefont {Rosati},\ and\ \citenamefont
  {Beacom}}]{Mar00a}%
  \BibitemOpen
  \bibfield  {author} {\bibinfo {author} {\bibnamefont {Marcucci},
  \bibfnamefont {L.~E.}}, \bibinfo {author} {\bibfnamefont {R.}~\bibnamefont
  {Schiavilla}}, \bibinfo {author} {\bibfnamefont {M.}~\bibnamefont {Viviani}},
  \bibinfo {author} {\bibfnamefont {A.}~\bibnamefont {Kievsky}}, \bibinfo
  {author} {\bibfnamefont {S.}~\bibnamefont {Rosati}}, \ and\ \bibinfo {author}
  {\bibfnamefont {J.~F.}\ \bibnamefont {Beacom}}} (\bibinfo {year} {2000}),\
  \href {\doibase 10.1103/PhysRevC.63.015801} {\bibfield  {journal} {\bibinfo
  {journal} {Phys. Rev. C}\ }\textbf {\bibinfo {volume} {63}},\ \bibinfo
  {pages} {015801}}\BibitemShut {NoStop}%
\bibitem [{\citenamefont {Marcucci}\ \emph {et~al.}(2005)\citenamefont
  {Marcucci}, \citenamefont {Viviani}, \citenamefont {Schiavilla},
  \citenamefont {Kievsky},\ and\ \citenamefont {Rosati}}]{Marcucci:2005}%
  \BibitemOpen
  \bibfield  {author} {\bibinfo {author} {\bibnamefont {Marcucci},
  \bibfnamefont {L.~E.}}, \bibinfo {author} {\bibfnamefont {M.}~\bibnamefont
  {Viviani}}, \bibinfo {author} {\bibfnamefont {R.}~\bibnamefont {Schiavilla}},
  \bibinfo {author} {\bibfnamefont {A.}~\bibnamefont {Kievsky}}, \ and\
  \bibinfo {author} {\bibfnamefont {S.}~\bibnamefont {Rosati}}} (\bibinfo
  {year} {2005}),\ \href@noop {} {\bibfield  {journal} {\bibinfo  {journal}
  {Phys. Rev. C}\ }\textbf {\bibinfo {volume} {72}},\ \bibinfo {pages}
  {014001}}\BibitemShut {NoStop}%
\bibitem [{\citenamefont {Maris}\ \emph {et~al.}(2013)\citenamefont {Maris},
  \citenamefont {Vary}, \citenamefont {Gandolfi}, \citenamefont {Carlson},\
  and\ \citenamefont {Pieper}}]{Maris:2013}%
  \BibitemOpen
  \bibfield  {author} {\bibinfo {author} {\bibnamefont {Maris}, \bibfnamefont
  {P.}}, \bibinfo {author} {\bibfnamefont {J.~P.}\ \bibnamefont {Vary}},
  \bibinfo {author} {\bibfnamefont {S.}~\bibnamefont {Gandolfi}}, \bibinfo
  {author} {\bibfnamefont {J.}~\bibnamefont {Carlson}}, \ and\ \bibinfo
  {author} {\bibfnamefont {S.~C.}\ \bibnamefont {Pieper}}} (\bibinfo {year}
  {2013}),\ \href {\doibase 10.1103/PhysRevC.87.054318} {\bibfield  {journal}
  {\bibinfo  {journal} {Phys. Rev. C}\ }\textbf {\bibinfo {volume} {87}},\
  \bibinfo {pages} {054318}}\BibitemShut {NoStop}%
\bibitem [{\citenamefont {Mathiot}(1989)}]{Mat89}%
  \BibitemOpen
  \bibfield  {author} {\bibinfo {author} {\bibnamefont {Mathiot}, \bibfnamefont
  {J.-F.}}} (\bibinfo {year} {1989}),\ \href {\doibase
  http://dx.doi.org/10.1016/0370-1573(89)90079-3} {\bibfield  {journal}
  {\bibinfo  {journal} {Physics Reports}\ }\textbf {\bibinfo {volume} {173}},\
  \bibinfo {pages} {63 }}\BibitemShut {NoStop}%
\bibitem [{\citenamefont {McCutchan}\ \emph {et~al.}(2012)\citenamefont
  {McCutchan}, \citenamefont {Lister}, \citenamefont {Pieper}, \citenamefont
  {Wiringa}, \citenamefont {Seweryniak}, \citenamefont {Greene}, \citenamefont
  {Bertone}, \citenamefont {Carpenter}, \citenamefont {Chiara}, \citenamefont
  {G\"urdal}, \citenamefont {Hoffman}, \citenamefont {Janssens}, \citenamefont
  {Khoo}, \citenamefont {Lauritsen},\ and\ \citenamefont
  {Zhu}}]{mccutchan2012}%
  \BibitemOpen
  \bibfield  {author} {\bibinfo {author} {\bibnamefont {McCutchan},
  \bibfnamefont {E.~A.}}, \bibinfo {author} {\bibfnamefont {C.~J.}\
  \bibnamefont {Lister}}, \bibinfo {author} {\bibfnamefont {S.~C.}\
  \bibnamefont {Pieper}}, \bibinfo {author} {\bibfnamefont {R.~B.}\
  \bibnamefont {Wiringa}}, \bibinfo {author} {\bibfnamefont {D.}~\bibnamefont
  {Seweryniak}}, \bibinfo {author} {\bibfnamefont {J.~P.}\ \bibnamefont
  {Greene}}, \bibinfo {author} {\bibfnamefont {P.~F.}\ \bibnamefont {Bertone}},
  \bibinfo {author} {\bibfnamefont {M.~P.}\ \bibnamefont {Carpenter}}, \bibinfo
  {author} {\bibfnamefont {C.~J.}\ \bibnamefont {Chiara}}, \bibinfo {author}
  {\bibfnamefont {G.}~\bibnamefont {G\"urdal}}, \bibinfo {author}
  {\bibfnamefont {C.~R.}\ \bibnamefont {Hoffman}}, \bibinfo {author}
  {\bibfnamefont {R.~V.~F.}\ \bibnamefont {Janssens}}, \bibinfo {author}
  {\bibfnamefont {T.~L.}\ \bibnamefont {Khoo}}, \bibinfo {author}
  {\bibfnamefont {T.}~\bibnamefont {Lauritsen}}, \ and\ \bibinfo {author}
  {\bibfnamefont {S.}~\bibnamefont {Zhu}}} (\bibinfo {year} {2012}),\ \href
  {\doibase 10.1103/PhysRevC.86.014312} {\bibfield  {journal} {\bibinfo
  {journal} {Phys. Rev. C}\ }\textbf {\bibinfo {volume} {86}},\ \bibinfo
  {pages} {014312}}\BibitemShut {NoStop}%
\bibitem [{\citenamefont {McVoy}\ and\ \citenamefont
  {Van~Hove}(1962)}]{McVoy:1962}%
  \BibitemOpen
  \bibfield  {author} {\bibinfo {author} {\bibnamefont {McVoy}, \bibfnamefont
  {K.~W.}}, \ and\ \bibinfo {author} {\bibfnamefont {L.}~\bibnamefont
  {Van~Hove}}} (\bibinfo {year} {1962}),\ \href {\doibase
  10.1103/PhysRev.125.1034} {\bibfield  {journal} {\bibinfo  {journal} {Phys.
  Rev.}\ }\textbf {\bibinfo {volume} {125}},\ \bibinfo {pages}
  {1034}}\BibitemShut {NoStop}%
\bibitem [{\citenamefont {Metropolis}\ \emph {et~al.}(1953)\citenamefont
  {Metropolis}, \citenamefont {Rosenbluth}, \citenamefont {Rosenbluth},
  \citenamefont {Teller},\ and\ \citenamefont {Teller}}]{metropolis1953}%
  \BibitemOpen
  \bibfield  {author} {\bibinfo {author} {\bibnamefont {Metropolis},
  \bibfnamefont {N.}}, \bibinfo {author} {\bibfnamefont {A.~W.}\ \bibnamefont
  {Rosenbluth}}, \bibinfo {author} {\bibfnamefont {M.~N.}\ \bibnamefont
  {Rosenbluth}}, \bibinfo {author} {\bibfnamefont {A.~H.}\ \bibnamefont
  {Teller}}, \ and\ \bibinfo {author} {\bibfnamefont {E.}~\bibnamefont
  {Teller}}} (\bibinfo {year} {1953}),\ \href {\doibase
  http://dx.doi.org/10.1063/1.1699114} {\bibfield  {journal} {\bibinfo
  {journal} {The Journal of Chemical Physics}\ }\textbf {\bibinfo {volume}
  {21}}~(\bibinfo {number} {6}),\ \bibinfo {pages} {1087}}\BibitemShut
  {NoStop}%
\bibitem [{\citenamefont {Miller}\ \emph {et~al.}(1982)\citenamefont {Miller}
  \emph {et~al.}}]{Mil82}%
  \BibitemOpen
  \bibfield  {author} {\bibinfo {author} {\bibnamefont {Miller}, \bibfnamefont
  {K.~L.}},  \emph {et~al.}} (\bibinfo {year} {1982}),\ \href {\doibase
  10.1103/PhysRevD.26.537} {\bibfield  {journal} {\bibinfo  {journal} {Phys.
  Rev. D}\ }\textbf {\bibinfo {volume} {26}},\ \bibinfo {pages}
  {537}}\BibitemShut {NoStop}%
\bibitem [{\citenamefont {Moskowitz}\ \emph {et~al.}(1982)\citenamefont
  {Moskowitz}, \citenamefont {Schmidt}, \citenamefont {Lee},\ and\
  \citenamefont {Kalos}}]{Moskowitz:1982}%
  \BibitemOpen
  \bibfield  {author} {\bibinfo {author} {\bibnamefont {Moskowitz},
  \bibfnamefont {J.~W.}}, \bibinfo {author} {\bibfnamefont {K.~E.}\
  \bibnamefont {Schmidt}}, \bibinfo {author} {\bibfnamefont {M.}~\bibnamefont
  {Lee}}, \ and\ \bibinfo {author} {\bibfnamefont {M.~H.}\ \bibnamefont
  {Kalos}}} (\bibinfo {year} {1982}),\ \href@noop {} {\bibfield  {journal}
  {\bibinfo  {journal} {J. Chem. Phys.}\ }\textbf {\bibinfo {volume} {77}},\
  \bibinfo {pages} {349}}\BibitemShut {NoStop}%
\bibitem [{\citenamefont {Muller}\ \emph {et~al.}(2000)\citenamefont {Muller},
  \citenamefont {Koonin}, \citenamefont {Seki},\ and\ \citenamefont {van
  Kolck}}]{Muller:1999}%
  \BibitemOpen
  \bibfield  {author} {\bibinfo {author} {\bibnamefont {Muller}, \bibfnamefont
  {H.}}, \bibinfo {author} {\bibfnamefont {S.}~\bibnamefont {Koonin}}, \bibinfo
  {author} {\bibfnamefont {R.}~\bibnamefont {Seki}}, \ and\ \bibinfo {author}
  {\bibfnamefont {U.}~\bibnamefont {van Kolck}}} (\bibinfo {year} {2000}),\
  \href {\doibase 10.1103/PhysRevC.61.044320} {\bibfield  {journal} {\bibinfo
  {journal} {Phys. Rev.}\ }\textbf {\bibinfo {volume} {C61}},\ \bibinfo {pages}
  {044320}}\BibitemShut {NoStop}%
\bibitem [{\citenamefont {Nakamura}(2010)}]{PDG}%
  \BibitemOpen
  \bibfield  {author} {\bibinfo {author} {\bibnamefont {Nakamura},
  \bibfnamefont {K.~{\it et al.}.}}} (\bibinfo {year} {2010}),\ \href
  {http://stacks.iop.org/0954-3899/37/i=7A/a=075021} {\bibfield  {journal}
  {\bibinfo  {journal} {Journal of Physics G: Nuclear and Particle Physics}\
  }\textbf {\bibinfo {volume} {37}}~(\bibinfo {number} {7A}),\ \bibinfo {pages}
  {075021}}\BibitemShut {NoStop}%
\bibitem [{\citenamefont {Navratil}(2007)}]{Nav07}%
  \BibitemOpen
  \bibfield  {author} {\bibinfo {author} {\bibnamefont {Navratil},
  \bibfnamefont {P.}}} (\bibinfo {year} {2007}),\ \href@noop {} {\bibfield
  {journal} {\bibinfo  {journal} {Few-Body Syst.}\ }\textbf {\bibinfo {volume}
  {41}},\ \bibinfo {pages} {117}}\BibitemShut {NoStop}%
\bibitem [{\citenamefont {Navr\'atil}\ \emph {et~al.}(2007)\citenamefont
  {Navr\'atil}, \citenamefont {Gueorguiev}, \citenamefont {Vary}, \citenamefont
  {Ormand},\ and\ \citenamefont {Nogga}}]{navratil2007}%
  \BibitemOpen
  \bibfield  {author} {\bibinfo {author} {\bibnamefont {Navr\'atil},
  \bibfnamefont {P.}}, \bibinfo {author} {\bibfnamefont {V.~G.}\ \bibnamefont
  {Gueorguiev}}, \bibinfo {author} {\bibfnamefont {J.~P.}\ \bibnamefont
  {Vary}}, \bibinfo {author} {\bibfnamefont {W.~E.}\ \bibnamefont {Ormand}}, \
  and\ \bibinfo {author} {\bibfnamefont {A.}~\bibnamefont {Nogga}}} (\bibinfo
  {year} {2007}),\ \href {\doibase 10.1103/PhysRevLett.99.042501} {\bibfield
  {journal} {\bibinfo  {journal} {Phys. Rev. Lett.}\ }\textbf {\bibinfo
  {volume} {99}},\ \bibinfo {pages} {042501}}\BibitemShut {NoStop}%
\bibitem [{\citenamefont {Nelder}\ and\ \citenamefont
  {Mead}(1965)}]{nelder1965}%
  \BibitemOpen
  \bibfield  {author} {\bibinfo {author} {\bibnamefont {Nelder}, \bibfnamefont
  {J.~A.}}, \ and\ \bibinfo {author} {\bibfnamefont {R.}~\bibnamefont {Mead}}}
  (\bibinfo {year} {1965}),\ \href@noop {} {\bibfield  {journal} {\bibinfo
  {journal} {The Computer Journal}\ }\textbf {\bibinfo {volume} {7}}~(\bibinfo
  {number} {4}),\ \bibinfo {pages} {308}}\BibitemShut {NoStop}%
\bibitem [{\citenamefont {Nightingale}\ and\ \citenamefont
  {Umrigar}(1999)}]{Nightingale:1999}%
  \BibitemOpen
  \bibfield  {author} {\bibinfo {author} {\bibnamefont {Nightingale},
  \bibfnamefont {M.}}, \ and\ \bibinfo {author} {\bibfnamefont
  {C.}~\bibnamefont {Umrigar}}} (\bibinfo {year} {1999}),\ \href@noop {} {\emph
  {\bibinfo {title} {Quantum Monte Carlo Methods in Physics and Chemistry}}}\
  (\bibinfo  {publisher} {Springer})\BibitemShut {NoStop}%
\bibitem [{\citenamefont {NNDC}(2014)}]{NNDC-data}%
  \BibitemOpen
  \bibfield  {author} {\bibinfo {author} {\bibnamefont {NNDC},}} (\bibinfo
  {year} {2014}),\ \href@noop {} {\enquote {\bibinfo {title} {Nudat 2},}\
  }\bibinfo {howpublished}
  {http://www.nndc.bnl.gov/nudat2/chartNuc.jsp}\BibitemShut {NoStop}%
\bibitem [{\citenamefont {Nollett}(2012)}]{nollett2012}%
  \BibitemOpen
  \bibfield  {author} {\bibinfo {author} {\bibnamefont {Nollett}, \bibfnamefont
  {K.~M.}}} (\bibinfo {year} {2012}),\ \href {\doibase
  10.1103/PhysRevC.86.044330} {\bibfield  {journal} {\bibinfo  {journal} {Phys.
  Rev. C}\ }\textbf {\bibinfo {volume} {86}},\ \bibinfo {pages}
  {044330}}\BibitemShut {NoStop}%
\bibitem [{\citenamefont {Nollett}\ \emph {et~al.}(2007)\citenamefont
  {Nollett}, \citenamefont {Pieper}, \citenamefont {Wiringa}, \citenamefont
  {Carlson},\ and\ \citenamefont {Hale}}]{Nollett:2007}%
  \BibitemOpen
  \bibfield  {author} {\bibinfo {author} {\bibnamefont {Nollett}, \bibfnamefont
  {K.~M.}}, \bibinfo {author} {\bibfnamefont {S.~C.}\ \bibnamefont {Pieper}},
  \bibinfo {author} {\bibfnamefont {R.~B.}\ \bibnamefont {Wiringa}}, \bibinfo
  {author} {\bibfnamefont {J.}~\bibnamefont {Carlson}}, \ and\ \bibinfo
  {author} {\bibfnamefont {G.}~\bibnamefont {Hale}}} (\bibinfo {year} {2007}),\
  \href {\doibase 10.1103/PhysRevLett.99.022502} {\bibfield  {journal}
  {\bibinfo  {journal} {Phys. Rev. Lett.}\ }\textbf {\bibinfo {volume} {99}},\
  \bibinfo {pages} {022502}}\BibitemShut {NoStop}%
\bibitem [{\citenamefont {Nollett}\ and\ \citenamefont
  {Wiringa}(2011)}]{nollett2011}%
  \BibitemOpen
  \bibfield  {author} {\bibinfo {author} {\bibnamefont {Nollett}, \bibfnamefont
  {K.~M.}}, \ and\ \bibinfo {author} {\bibfnamefont {R.~B.}\ \bibnamefont
  {Wiringa}}} (\bibinfo {year} {2011}),\ \href {\doibase
  10.1103/PhysRevC.83.041001} {\bibfield  {journal} {\bibinfo  {journal} {Phys.
  Rev. C}\ }\textbf {\bibinfo {volume} {83}},\ \bibinfo {pages}
  {041001}}\BibitemShut {NoStop}%
\bibitem [{\citenamefont {N\"{o}rtersh\"{a}user}\ and\ \citenamefont {{\it et
  al.}}(2009)}]{Nortershauser09}%
  \BibitemOpen
  \bibfield  {author} {\bibinfo {author} {\bibnamefont {N\"{o}rtersh\"{a}user},
  \bibfnamefont {N.}}, \ and\ \bibinfo {author} {\bibnamefont {{\it et al.}}}}
  (\bibinfo {year} {2009}),\ \href@noop {} {\bibfield  {journal} {\bibinfo
  {journal} {Phys. Rev. Lett.}\ }\textbf {\bibinfo {volume} {102}},\ \bibinfo
  {pages} {062503}}\BibitemShut {NoStop}%
\bibitem [{\citenamefont {N\"{o}rtersh\"{a}user}\ \emph
  {et~al.}(2011)\citenamefont {N\"{o}rtersh\"{a}user}, \citenamefont {Neff},
  \citenamefont {S\'{a}nchez},\ and\ \citenamefont {Sick}}]{Nortershauser11}%
  \BibitemOpen
  \bibfield  {author} {\bibinfo {author} {\bibnamefont {N\"{o}rtersh\"{a}user},
  \bibfnamefont {N.}}, \bibinfo {author} {\bibfnamefont {T.}~\bibnamefont
  {Neff}}, \bibinfo {author} {\bibfnamefont {R.}~\bibnamefont {S\'{a}nchez}}, \
  and\ \bibinfo {author} {\bibfnamefont {I.}~\bibnamefont {Sick}}} (\bibinfo
  {year} {2011}),\ \href@noop {} {\bibfield  {journal} {\bibinfo  {journal}
  {Phys. Rev. C}\ }\textbf {\bibinfo {volume} {84}},\ \bibinfo {pages}
  {024307}}\BibitemShut {NoStop}%
\bibitem [{\citenamefont {Oppenheimer}\ and\ \citenamefont
  {Volkoff}(1939)}]{TOV1939}%
  \BibitemOpen
  \bibfield  {author} {\bibinfo {author} {\bibnamefont {Oppenheimer},
  \bibfnamefont {J.~R.}}, \ and\ \bibinfo {author} {\bibfnamefont {G.~M.}\
  \bibnamefont {Volkoff}}} (\bibinfo {year} {1939}),\ \href {\doibase
  10.1103/PhysRev.55.374} {\bibfield  {journal} {\bibinfo  {journal} {Phys.
  Rev.}\ }\textbf {\bibinfo {volume} {55}},\ \bibinfo {pages}
  {374}}\BibitemShut {NoStop}%
\bibitem [{\citenamefont {Ordonez}\ \emph {et~al.}(1996)\citenamefont
  {Ordonez}, \citenamefont {Ray},\ and\ \citenamefont {van Kolck}}]{Ord95}%
  \BibitemOpen
  \bibfield  {author} {\bibinfo {author} {\bibnamefont {Ordonez}, \bibfnamefont
  {C.}}, \bibinfo {author} {\bibfnamefont {L.}~\bibnamefont {Ray}}, \ and\
  \bibinfo {author} {\bibfnamefont {U.}~\bibnamefont {van Kolck}}} (\bibinfo
  {year} {1996}),\ \href@noop {} {\bibfield  {journal} {\bibinfo  {journal}
  {Phys. Rev. C}\ }\textbf {\bibinfo {volume} {53}},\ \bibinfo {pages}
  {2086}}\BibitemShut {NoStop}%
\bibitem [{\citenamefont {Otsuka}\ \emph {et~al.}(2001)\citenamefont {Otsuka},
  \citenamefont {Honma}, \citenamefont {Mizusaki}, \citenamefont {Shimizu},\
  and\ \citenamefont {Utsuno}}]{Otsuka:2001}%
  \BibitemOpen
  \bibfield  {author} {\bibinfo {author} {\bibnamefont {Otsuka}, \bibfnamefont
  {T.}}, \bibinfo {author} {\bibfnamefont {M.}~\bibnamefont {Honma}}, \bibinfo
  {author} {\bibfnamefont {T.}~\bibnamefont {Mizusaki}}, \bibinfo {author}
  {\bibfnamefont {N.}~\bibnamefont {Shimizu}}, \ and\ \bibinfo {author}
  {\bibfnamefont {Y.}~\bibnamefont {Utsuno}}} (\bibinfo {year} {2001}),\ \href
  {\doibase http://dx.doi.org/10.1016/S0146-6410(01)00157-0} {\bibfield
  {journal} {\bibinfo  {journal} {Progress in Particle and Nuclear Physics}\
  }\textbf {\bibinfo {volume} {47}}~(\bibinfo {number} {1}),\ \bibinfo {pages}
  {319 }}\BibitemShut {NoStop}%
\bibitem [{\citenamefont {Ozel}\ \emph {et~al.}(2010)\citenamefont {Ozel},
  \citenamefont {Baym},\ and\ \citenamefont {Guver}}]{Ozel:2010}%
  \BibitemOpen
  \bibfield  {author} {\bibinfo {author} {\bibnamefont {Ozel}, \bibfnamefont
  {F.}}, \bibinfo {author} {\bibfnamefont {G.}~\bibnamefont {Baym}}, \ and\
  \bibinfo {author} {\bibfnamefont {T.}~\bibnamefont {Guver}}} (\bibinfo {year}
  {2010}),\ \href@noop {} {\bibfield  {journal} {\bibinfo  {journal} {Phys.
  Rev.}\ }\textbf {\bibinfo {volume} {D82}},\ \bibinfo {pages}
  {101301}}\BibitemShut {NoStop}%
\bibitem [{\citenamefont {Pandharipande}\ \emph {et~al.}(1997)\citenamefont
  {Pandharipande}, \citenamefont {Sick},\ and\ \citenamefont {deWitt
  Huberts}}]{pandharipande1997}%
  \BibitemOpen
  \bibfield  {author} {\bibinfo {author} {\bibnamefont {Pandharipande},
  \bibfnamefont {V.~R.}}, \bibinfo {author} {\bibfnamefont {I.}~\bibnamefont
  {Sick}}, \ and\ \bibinfo {author} {\bibfnamefont {P.~K.~A.}\ \bibnamefont
  {deWitt Huberts}}} (\bibinfo {year} {1997}),\ \href {\doibase
  10.1103/RevModPhys.51.821} {\bibfield  {journal} {\bibinfo  {journal} {Rev.
  Mod. Phys.}\ }\textbf {\bibinfo {volume} {69}}~(\bibinfo {number} {3}),\
  \bibinfo {pages} {961}}\BibitemShut {NoStop}%
\bibitem [{\citenamefont {Pandharipande}\ and\ \citenamefont
  {Wiringa}(1979)}]{pandharipande1979}%
  \BibitemOpen
  \bibfield  {author} {\bibinfo {author} {\bibnamefont {Pandharipande},
  \bibfnamefont {V.~R.}}, \ and\ \bibinfo {author} {\bibfnamefont {R.~B.}\
  \bibnamefont {Wiringa}}} (\bibinfo {year} {1979}),\ \href {\doibase
  10.1103/RevModPhys.51.821} {\bibfield  {journal} {\bibinfo  {journal} {Rev.
  Mod. Phys.}\ }\textbf {\bibinfo {volume} {51}}~(\bibinfo {number} {4}),\
  \bibinfo {pages} {821}}\BibitemShut {NoStop}%
\bibitem [{\citenamefont {Park}\ \emph {et~al.}(1993)\citenamefont {Park},
  \citenamefont {Min},\ and\ \citenamefont {Rho}}]{Par93}%
  \BibitemOpen
  \bibfield  {author} {\bibinfo {author} {\bibnamefont {Park}, \bibfnamefont
  {T.-S.}}, \bibinfo {author} {\bibfnamefont {D.-P.}\ \bibnamefont {Min}}, \
  and\ \bibinfo {author} {\bibfnamefont {M.}~\bibnamefont {Rho}}} (\bibinfo
  {year} {1993}),\ \href@noop {} {\bibfield  {journal} {\bibinfo  {journal}
  {Phys. Rep.}\ }\textbf {\bibinfo {volume} {233}},\ \bibinfo {pages}
  {341}}\BibitemShut {NoStop}%
\bibitem [{\citenamefont {Park}\ \emph {et~al.}(1996)\citenamefont {Park},
  \citenamefont {Min},\ and\ \citenamefont {Rho}}]{Par96}%
  \BibitemOpen
  \bibfield  {author} {\bibinfo {author} {\bibnamefont {Park}, \bibfnamefont
  {T.-S.}}, \bibinfo {author} {\bibfnamefont {D.-P.}\ \bibnamefont {Min}}, \
  and\ \bibinfo {author} {\bibfnamefont {M.}~\bibnamefont {Rho}}} (\bibinfo
  {year} {1996}),\ \href@noop {} {\bibfield  {journal} {\bibinfo  {journal}
  {Nucl.\ Phys.\ A}\ }\textbf {\bibinfo {volume} {596}},\ \bibinfo {pages}
  {515}}\BibitemShut {NoStop}%
\bibitem [{\citenamefont {Pastore}(2014)}]{pastore_pc}%
  \BibitemOpen
  \bibfield  {author} {\bibinfo {author} {\bibnamefont {Pastore}, \bibfnamefont
  {S.}}} (\bibinfo {year} {2014}),\ \href@noop {} {\bibinfo  {journal} {private
  communication}\ }\BibitemShut {NoStop}%
\bibitem [{\citenamefont {Pastore}\ \emph {et~al.}(2011)\citenamefont
  {Pastore}, \citenamefont {Girlanda}, \citenamefont {Schiavilla},\ and\
  \citenamefont {Viviani}}]{Pas11}%
  \BibitemOpen
\bibfield  {journal} {  }\bibfield  {author} {\bibinfo {author} {\bibnamefont
  {Pastore}, \bibfnamefont {S.}}, \bibinfo {author} {\bibfnamefont
  {L.}~\bibnamefont {Girlanda}}, \bibinfo {author} {\bibfnamefont
  {R.}~\bibnamefont {Schiavilla}}, \ and\ \bibinfo {author} {\bibfnamefont
  {M.}~\bibnamefont {Viviani}}} (\bibinfo {year} {2011}),\ \href@noop {}
  {\bibfield  {journal} {\bibinfo  {journal} {Phys.\ Rev.\ C}\ }\textbf
  {\bibinfo {volume} {84}},\ \bibinfo {pages} {024001}}\BibitemShut {NoStop}%
\bibitem [{\citenamefont {Pastore}\ \emph {et~al.}(2009)\citenamefont
  {Pastore}, \citenamefont {Girlanda}, \citenamefont {Schiavilla},
  \citenamefont {Viviani},\ and\ \citenamefont {Wiringa}}]{Pas09}%
  \BibitemOpen
  \bibfield  {author} {\bibinfo {author} {\bibnamefont {Pastore}, \bibfnamefont
  {S.}}, \bibinfo {author} {\bibfnamefont {L.}~\bibnamefont {Girlanda}},
  \bibinfo {author} {\bibfnamefont {R.}~\bibnamefont {Schiavilla}}, \bibinfo
  {author} {\bibfnamefont {M.}~\bibnamefont {Viviani}}, \ and\ \bibinfo
  {author} {\bibfnamefont {R.}~\bibnamefont {Wiringa}}} (\bibinfo {year}
  {2009}),\ \href@noop {} {\bibfield  {journal} {\bibinfo  {journal} {Phys.\
  Rev.\ C}\ }\textbf {\bibinfo {volume} {80}},\ \bibinfo {pages}
  {034004}}\BibitemShut {NoStop}%
\bibitem [{\citenamefont {Pastore}\ \emph {et~al.}(2013)\citenamefont
  {Pastore}, \citenamefont {Pieper}, \citenamefont {Schiavilla},\ and\
  \citenamefont {Wiringa}}]{pastore2013}%
  \BibitemOpen
  \bibfield  {author} {\bibinfo {author} {\bibnamefont {Pastore}, \bibfnamefont
  {S.}}, \bibinfo {author} {\bibfnamefont {S.~C.}\ \bibnamefont {Pieper}},
  \bibinfo {author} {\bibfnamefont {R.}~\bibnamefont {Schiavilla}}, \ and\
  \bibinfo {author} {\bibfnamefont {R.~B.}\ \bibnamefont {Wiringa}}} (\bibinfo
  {year} {2013}),\ \href@noop {} {\bibfield  {journal} {\bibinfo  {journal}
  {Phys. Rev. C}\ }\textbf {\bibinfo {volume} {87}},\ \bibinfo {pages}
  {035503}}\BibitemShut {NoStop}%
\bibitem [{\citenamefont {Pastore}\ \emph {et~al.}(2014)\citenamefont
  {Pastore}, \citenamefont {Wiringa}, \citenamefont {Pieper},\ and\
  \citenamefont {Schiavilla}}]{pastore2014}%
  \BibitemOpen
  \bibfield  {author} {\bibinfo {author} {\bibnamefont {Pastore}, \bibfnamefont
  {S.}}, \bibinfo {author} {\bibfnamefont {R.~B.}\ \bibnamefont {Wiringa}},
  \bibinfo {author} {\bibfnamefont {S.~C.}\ \bibnamefont {Pieper}}, \ and\
  \bibinfo {author} {\bibfnamefont {R.}~\bibnamefont {Schiavilla}}} (\bibinfo
  {year} {2014}),\ \href@noop {} {\bibfield  {journal} {\bibinfo  {journal}
  {Phys. Rev. C}\ }\textbf {\bibinfo {volume} {90}},\ \bibinfo {pages}
  {024321}}\BibitemShut {NoStop}%
\bibitem [{\citenamefont {Pederiva}\ \emph {et~al.}(2004)\citenamefont
  {Pederiva}, \citenamefont {Sarsa}, \citenamefont {Schmidt},\ and\
  \citenamefont {Fantoni}}]{Pederiva:2004}%
  \BibitemOpen
  \bibfield  {author} {\bibinfo {author} {\bibnamefont {Pederiva},
  \bibfnamefont {F.}}, \bibinfo {author} {\bibfnamefont {A.}~\bibnamefont
  {Sarsa}}, \bibinfo {author} {\bibfnamefont {K.~E.}\ \bibnamefont {Schmidt}},
  \ and\ \bibinfo {author} {\bibfnamefont {S.}~\bibnamefont {Fantoni}}}
  (\bibinfo {year} {2004}),\ \href@noop {} {\bibfield  {journal} {\bibinfo
  {journal} {Nucl. Phys. A}\ }\textbf {\bibinfo {volume} {742}},\ \bibinfo
  {pages} {255}}\BibitemShut {NoStop}%
\bibitem [{\citenamefont {Pervin}\ \emph {et~al.}(2007)\citenamefont {Pervin},
  \citenamefont {Pieper},\ and\ \citenamefont {Wiringa}}]{pervin2007}%
  \BibitemOpen
  \bibfield  {author} {\bibinfo {author} {\bibnamefont {Pervin}, \bibfnamefont
  {M.}}, \bibinfo {author} {\bibfnamefont {S.~C.}\ \bibnamefont {Pieper}}, \
  and\ \bibinfo {author} {\bibfnamefont {R.~B.}\ \bibnamefont {Wiringa}}}
  (\bibinfo {year} {2007}),\ \href {\doibase 10.1103/PhysRevC.76.064319}
  {\bibfield  {journal} {\bibinfo  {journal} {Phys. Rev. C}\ }\textbf {\bibinfo
  {volume} {76}},\ \bibinfo {pages} {064319}}\BibitemShut {NoStop}%
\bibitem [{\citenamefont {Peshkin}(1960)}]{peshkin1960}%
  \BibitemOpen
  \bibfield  {author} {\bibinfo {author} {\bibnamefont {Peshkin}, \bibfnamefont
  {M.}}} (\bibinfo {year} {1960}),\ \href@noop {} {\bibfield  {journal}
  {\bibinfo  {journal} {Phys. Rev.}\ }\textbf {\bibinfo {volume} {121}},\
  \bibinfo {pages} {636}}\BibitemShut {NoStop}%
\bibitem [{\citenamefont {Piarulli}\ \emph {et~al.}(2013)\citenamefont
  {Piarulli}, \citenamefont {Girlanda}, \citenamefont {Marcucci}, \citenamefont
  {Pastore}, \citenamefont {Schiavilla},\ and\ \citenamefont
  {Viviani}}]{Pia13}%
  \BibitemOpen
  \bibfield  {author} {\bibinfo {author} {\bibnamefont {Piarulli},
  \bibfnamefont {M.}}, \bibinfo {author} {\bibfnamefont {L.}~\bibnamefont
  {Girlanda}}, \bibinfo {author} {\bibfnamefont {L.}~\bibnamefont {Marcucci}},
  \bibinfo {author} {\bibfnamefont {S.}~\bibnamefont {Pastore}}, \bibinfo
  {author} {\bibfnamefont {R.}~\bibnamefont {Schiavilla}}, \ and\ \bibinfo
  {author} {\bibfnamefont {M.}~\bibnamefont {Viviani}}} (\bibinfo {year}
  {2013}),\ \href@noop {} {\bibfield  {journal} {\bibinfo  {journal} {Phys.\
  Rev.\ C}\ }\textbf {\bibinfo {volume} {87}},\ \bibinfo {pages}
  {014006}}\BibitemShut {NoStop}%
\bibitem [{\citenamefont {Piarulli}\ \emph {et~al.}(2015)\citenamefont
  {Piarulli}, \citenamefont {Girlanda}, \citenamefont {Schiavilla},
  \citenamefont {P\'erez}, \citenamefont {Amaro},\ and\ \citenamefont
  {Arriola}}]{Piarulli:2015}%
  \BibitemOpen
  \bibfield  {author} {\bibinfo {author} {\bibnamefont {Piarulli},
  \bibfnamefont {M.}}, \bibinfo {author} {\bibfnamefont {L.}~\bibnamefont
  {Girlanda}}, \bibinfo {author} {\bibfnamefont {R.}~\bibnamefont
  {Schiavilla}}, \bibinfo {author} {\bibfnamefont {R.~N.}\ \bibnamefont
  {P\'erez}}, \bibinfo {author} {\bibfnamefont {J.~E.}\ \bibnamefont {Amaro}},
  \ and\ \bibinfo {author} {\bibfnamefont {E.~R.}\ \bibnamefont {Arriola}}}
  (\bibinfo {year} {2015}),\ \href {\doibase 10.1103/PhysRevC.91.024003}
  {\bibfield  {journal} {\bibinfo  {journal} {Phys. Rev. C}\ }\textbf {\bibinfo
  {volume} {91}},\ \bibinfo {pages} {024003}}\BibitemShut {NoStop}%
\bibitem [{\citenamefont {Pieper}(2008{\natexlab{a}})}]{pieper2008}%
  \BibitemOpen
  \bibfield  {author} {\bibinfo {author} {\bibnamefont {Pieper}, \bibfnamefont
  {S.~C.}}} (\bibinfo {year} {2008}{\natexlab{a}}),\ \href@noop {} {\bibfield
  {journal} {\bibinfo  {journal} {AIP Conf. Proc.}\ }\textbf {\bibinfo {volume}
  {1011}},\ \bibinfo {pages} {143}}\BibitemShut {NoStop}%
\bibitem [{\citenamefont {Pieper}(2008{\natexlab{b}})}]{pieper2008b}%
  \BibitemOpen
  \bibfield  {author} {\bibinfo {author} {\bibnamefont {Pieper}, \bibfnamefont
  {S.~C.}}} (\bibinfo {year} {2008}{\natexlab{b}}),\ in\ \href@noop {} {\emph
  {\bibinfo {booktitle} {{Proceedings of the "Enrico Fermi" Summer School,
  Course CLXIX, {\it Nuclear Structure far from Stability: New Physics and new
  Technology}}}}},\ \bibinfo {editor} {edited by\ \bibinfo {editor}
  {\bibfnamefont {A.}~\bibnamefont {Covello}}, \bibinfo {editor} {\bibfnamefont
  {F.}~\bibnamefont {Iachello}}, \bibinfo {editor} {\bibfnamefont {R.~A.}\
  \bibnamefont {Ricci}}, \ and\ \bibinfo {editor} {\bibfnamefont
  {G.}~\bibnamefont {Maino}}}\ (\bibinfo  {publisher} {IOS Press},\ \bibinfo
  {address} {Amsterdam})\ p.\ \bibinfo {pages} {111},\ \bibinfo {note}
  {reprinted in La Rivista del Nuovo Cimento, {\bf 31}, 709,
  (2008)}\BibitemShut {NoStop}%
\bibitem [{\citenamefont {Pieper}(2015)}]{Pieper:unpub}%
  \BibitemOpen
  \bibfield  {author} {\bibinfo {author} {\bibnamefont {Pieper}, \bibfnamefont
  {S.~C.}}} (\bibinfo {year} {2015}),\ \href@noop {} {\bibinfo  {journal}
  {unpublished}\ }\BibitemShut {NoStop}%
\bibitem [{\citenamefont {Pieper}\ and\ \citenamefont
  {Carlson}(2015)}]{Pieper:2015}%
  \BibitemOpen
\bibfield  {journal} {  }\bibfield  {author} {\bibinfo {author} {\bibnamefont
  {Pieper}, \bibfnamefont {S.~C.}}, \ and\ \bibinfo {author} {\bibfnamefont
  {J.}~\bibnamefont {Carlson}}} (\bibinfo {year} {2015}),\ \href@noop {}
  {\bibinfo  {journal} {unpublished}\ }\BibitemShut {NoStop}%
\bibitem [{\citenamefont {Pieper}\ \emph {et~al.}(2001)\citenamefont {Pieper},
  \citenamefont {Pandharipande}, \citenamefont {Wiringa},\ and\ \citenamefont
  {Carlson}}]{pieper2001}%
  \BibitemOpen
\bibfield  {journal} {  }\bibfield  {author} {\bibinfo {author} {\bibnamefont
  {Pieper}, \bibfnamefont {S.~C.}}, \bibinfo {author} {\bibfnamefont {V.~R.}\
  \bibnamefont {Pandharipande}}, \bibinfo {author} {\bibfnamefont {R.~B.}\
  \bibnamefont {Wiringa}}, \ and\ \bibinfo {author} {\bibfnamefont
  {J.}~\bibnamefont {Carlson}}} (\bibinfo {year} {2001}),\ \href {\doibase
  10.1103/PhysRevC.64.014001} {\bibfield  {journal} {\bibinfo  {journal} {Phys.
  Rev. C}\ }\textbf {\bibinfo {volume} {64}}~(\bibinfo {number} {1}),\ \bibinfo
  {pages} {014001}}\BibitemShut {NoStop}%
\bibitem [{\citenamefont {Pieper}\ \emph {et~al.}(2002)\citenamefont {Pieper},
  \citenamefont {Varga},\ and\ \citenamefont {Wiringa}}]{pieper2002}%
  \BibitemOpen
  \bibfield  {author} {\bibinfo {author} {\bibnamefont {Pieper}, \bibfnamefont
  {S.~C.}}, \bibinfo {author} {\bibfnamefont {K.}~\bibnamefont {Varga}}, \ and\
  \bibinfo {author} {\bibfnamefont {R.~B.}\ \bibnamefont {Wiringa}}} (\bibinfo
  {year} {2002}),\ \href {\doibase 10.1103/PhysRevC.66.044310} {\bibfield
  {journal} {\bibinfo  {journal} {Phys. Rev. C}\ }\textbf {\bibinfo {volume}
  {66}}~(\bibinfo {number} {4}),\ \bibinfo {pages} {044310}}\BibitemShut
  {NoStop}%
\bibitem [{\citenamefont {Pieper}\ \emph {et~al.}(2004)\citenamefont {Pieper},
  \citenamefont {Wiringa},\ and\ \citenamefont {Carlson}}]{pieper2004}%
  \BibitemOpen
  \bibfield  {author} {\bibinfo {author} {\bibnamefont {Pieper}, \bibfnamefont
  {S.~C.}}, \bibinfo {author} {\bibfnamefont {R.~B.}\ \bibnamefont {Wiringa}},
  \ and\ \bibinfo {author} {\bibfnamefont {J.}~\bibnamefont {Carlson}}}
  (\bibinfo {year} {2004}),\ \href {\doibase 10.1103/PhysRevC.70.054325}
  {\bibfield  {journal} {\bibinfo  {journal} {Phys. Rev. C}\ }\textbf {\bibinfo
  {volume} {70}},\ \bibinfo {pages} {054325}}\BibitemShut {NoStop}%
\bibitem [{\citenamefont {Pieper}\ \emph {et~al.}(1992)\citenamefont {Pieper},
  \citenamefont {Wiringa},\ and\ \citenamefont {Pandharipande}}]{pieper1992}%
  \BibitemOpen
  \bibfield  {author} {\bibinfo {author} {\bibnamefont {Pieper}, \bibfnamefont
  {S.~C.}}, \bibinfo {author} {\bibfnamefont {R.~B.}\ \bibnamefont {Wiringa}},
  \ and\ \bibinfo {author} {\bibfnamefont {V.~R.}\ \bibnamefont
  {Pandharipande}}} (\bibinfo {year} {1992}),\ \href {\doibase
  10.1103/PhysRevC.46.1741} {\bibfield  {journal} {\bibinfo  {journal} {Phys.
  Rev. C}\ }\textbf {\bibinfo {volume} {46}}~(\bibinfo {number} {5}),\ \bibinfo
  {pages} {1741}}\BibitemShut {NoStop}%
\bibitem [{\citenamefont {{Pinkston}}\ and\ \citenamefont
  {{Satchler}}(1965)}]{Pinkston1965}%
  \BibitemOpen
  \bibfield  {author} {\bibinfo {author} {\bibnamefont {{Pinkston}},
  \bibfnamefont {W.~T.}}, \ and\ \bibinfo {author} {\bibfnamefont {G.~R.}\
  \bibnamefont {{Satchler}}}} (\bibinfo {year} {1965}),\ \href {\doibase
  10.1016/0029-5582(65)90417-7} {\bibfield  {journal} {\bibinfo  {journal}
  {Nucl. Phys.}\ }\textbf {\bibinfo {volume} {72}},\ \bibinfo {pages}
  {641}}\BibitemShut {NoStop}%
\bibitem [{\citenamefont {Preston}(1962)}]{preston1962}%
  \BibitemOpen
  \bibfield  {author} {\bibinfo {author} {\bibnamefont {Preston}, \bibfnamefont
  {M.~A.}}} (\bibinfo {year} {1962}),\ \href@noop {} {\bibinfo  {journal} {{\it
  Physics of the Nucleus}, (Addison-Wesley, Reading, Massachusetts)}\ ,\
  \bibinfo {pages} {299}}\BibitemShut {NoStop}%
\bibitem [{\citenamefont {Pudliner}\ \emph {et~al.}(1997)\citenamefont
  {Pudliner}, \citenamefont {Pandharipande}, \citenamefont {Carlson},
  \citenamefont {Pieper},\ and\ \citenamefont {Wiringa}}]{pudliner1997}%
  \BibitemOpen
\bibfield  {journal} {  }\bibfield  {author} {\bibinfo {author} {\bibnamefont
  {Pudliner}, \bibfnamefont {B.~S.}}, \bibinfo {author} {\bibfnamefont {V.~R.}\
  \bibnamefont {Pandharipande}}, \bibinfo {author} {\bibfnamefont
  {J.}~\bibnamefont {Carlson}}, \bibinfo {author} {\bibfnamefont {S.~C.}\
  \bibnamefont {Pieper}}, \ and\ \bibinfo {author} {\bibfnamefont {R.~B.}\
  \bibnamefont {Wiringa}}} (\bibinfo {year} {1997}),\ \href {\doibase
  10.1103/PhysRevC.56.1720} {\bibfield  {journal} {\bibinfo  {journal} {Phys.
  Rev. C}\ }\textbf {\bibinfo {volume} {56}}~(\bibinfo {number} {4}),\ \bibinfo
  {pages} {1720}}\BibitemShut {NoStop}%
\bibitem [{\citenamefont {Pudliner}\ \emph {et~al.}(1995)\citenamefont
  {Pudliner}, \citenamefont {Pandharipande}, \citenamefont {Carlson},\ and\
  \citenamefont {Wiringa}}]{pudliner1995}%
  \BibitemOpen
  \bibfield  {author} {\bibinfo {author} {\bibnamefont {Pudliner},
  \bibfnamefont {B.~S.}}, \bibinfo {author} {\bibfnamefont {V.~R.}\
  \bibnamefont {Pandharipande}}, \bibinfo {author} {\bibfnamefont
  {J.}~\bibnamefont {Carlson}}, \ and\ \bibinfo {author} {\bibfnamefont
  {R.~B.}\ \bibnamefont {Wiringa}}} (\bibinfo {year} {1995}),\ \href {\doibase
  10.1103/PhysRevLett.74.4396} {\bibfield  {journal} {\bibinfo  {journal}
  {Phys. Rev. Lett.}\ }\textbf {\bibinfo {volume} {74}}~(\bibinfo {number}
  {22}),\ \bibinfo {pages} {4396}}\BibitemShut {NoStop}%
\bibitem [{\citenamefont {Pudliner}\ \emph {et~al.}(1996)\citenamefont
  {Pudliner}, \citenamefont {Smerzi}, \citenamefont {Carlson}, \citenamefont
  {Pandharipande}, \citenamefont {Pieper},\ and\ \citenamefont
  {Ravenhall}}]{pudliner1996}%
  \BibitemOpen
  \bibfield  {author} {\bibinfo {author} {\bibnamefont {Pudliner},
  \bibfnamefont {B.~S.}}, \bibinfo {author} {\bibfnamefont {A.}~\bibnamefont
  {Smerzi}}, \bibinfo {author} {\bibfnamefont {J.}~\bibnamefont {Carlson}},
  \bibinfo {author} {\bibfnamefont {V.~R.}\ \bibnamefont {Pandharipande}},
  \bibinfo {author} {\bibfnamefont {S.~C.}\ \bibnamefont {Pieper}}, \ and\
  \bibinfo {author} {\bibfnamefont {D.~G.}\ \bibnamefont {Ravenhall}}}
  (\bibinfo {year} {1996}),\ \href {\doibase 10.1103/PhysRevLett.76.2416}
  {\bibfield  {journal} {\bibinfo  {journal} {Phys. Rev. Lett.}\ }\textbf
  {\bibinfo {volume} {76}}~(\bibinfo {number} {14}),\ \bibinfo {pages}
  {2416}}\BibitemShut {NoStop}%
\bibitem [{\citenamefont {Purcell}\ \emph {et~al.}(2010)\citenamefont
  {Purcell}, \citenamefont {Kelley}, \citenamefont {Kwan}, \citenamefont
  {Sheu},\ and\ \citenamefont {Weller}}]{Purcell10}%
  \BibitemOpen
  \bibfield  {author} {\bibinfo {author} {\bibnamefont {Purcell}, \bibfnamefont
  {J.~E.}}, \bibinfo {author} {\bibfnamefont {J.~H.}\ \bibnamefont {Kelley}},
  \bibinfo {author} {\bibfnamefont {E.}~\bibnamefont {Kwan}}, \bibinfo {author}
  {\bibfnamefont {C.~G.}\ \bibnamefont {Sheu}}, \ and\ \bibinfo {author}
  {\bibfnamefont {H.~R.}\ \bibnamefont {Weller}}} (\bibinfo {year} {2010}),\
  \href@noop {} {\bibfield  {journal} {\bibinfo  {journal} {Nucl. Phys. A}\
  }\textbf {\bibinfo {volume} {848}},\ \bibinfo {pages} {1}}\BibitemShut
  {NoStop}%
\bibitem [{\citenamefont {Ravenhall}\ \emph {et~al.}(1983)\citenamefont
  {Ravenhall}, \citenamefont {Pethick},\ and\ \citenamefont
  {Wilson}}]{Ravenhall:1983}%
  \BibitemOpen
  \bibfield  {author} {\bibinfo {author} {\bibnamefont {Ravenhall},
  \bibfnamefont {D.~G.}}, \bibinfo {author} {\bibfnamefont {C.~J.}\
  \bibnamefont {Pethick}}, \ and\ \bibinfo {author} {\bibfnamefont {J.~R.}\
  \bibnamefont {Wilson}}} (\bibinfo {year} {1983}),\ \href {\doibase
  10.1103/PhysRevLett.50.2066} {\bibfield  {journal} {\bibinfo  {journal}
  {Phys. Rev. Lett.}\ }\textbf {\bibinfo {volume} {50}}~(\bibinfo {number}
  {26}),\ \bibinfo {pages} {2066}}\BibitemShut {NoStop}%
\bibitem [{\citenamefont {Riska}\ and\ \citenamefont {Brown}(1972)}]{Ris72}%
  \BibitemOpen
  \bibfield  {author} {\bibinfo {author} {\bibnamefont {Riska}, \bibfnamefont
  {D.}}, \ and\ \bibinfo {author} {\bibfnamefont {G.}~\bibnamefont {Brown}}}
  (\bibinfo {year} {1972}),\ \href {\doibase
  http://dx.doi.org/10.1016/0370-2693(72)90376-0} {\bibfield  {journal}
  {\bibinfo  {journal} {Phys. Lett. B}\ }\textbf {\bibinfo {volume}
  {38}}~(\bibinfo {number} {4}),\ \bibinfo {pages} {193 }}\BibitemShut
  {NoStop}%
\bibitem [{\citenamefont {Riska}(1985{\natexlab{a}})}]{Ris85a}%
  \BibitemOpen
  \bibfield  {author} {\bibinfo {author} {\bibnamefont {Riska}, \bibfnamefont
  {D.~O.}}} (\bibinfo {year} {1985}{\natexlab{a}}),\ \href@noop {} {\bibfield
  {journal} {\bibinfo  {journal} {Phys. Scr.}\ }\textbf {\bibinfo {volume}
  {31}},\ \bibinfo {pages} {107}}\BibitemShut {NoStop}%
\bibitem [{\citenamefont {Riska}(1985{\natexlab{b}})}]{Ris85b}%
  \BibitemOpen
  \bibfield  {author} {\bibinfo {author} {\bibnamefont {Riska}, \bibfnamefont
  {D.~O.}}} (\bibinfo {year} {1985}{\natexlab{b}}),\ \href@noop {} {\bibfield
  {journal} {\bibinfo  {journal} {Phys. Scr.}\ }\textbf {\bibinfo {volume}
  {31}},\ \bibinfo {pages} {471}}\BibitemShut {NoStop}%
\bibitem [{\citenamefont {Riska}(1989)}]{Ris89}%
  \BibitemOpen
  \bibfield  {author} {\bibinfo {author} {\bibnamefont {Riska}, \bibfnamefont
  {D.~O.}}} (\bibinfo {year} {1989}),\ \href@noop {} {\bibfield  {journal}
  {\bibinfo  {journal} {Phys. Rep.}\ }\textbf {\bibinfo {volume} {181}},\
  \bibinfo {pages} {207}}\BibitemShut {NoStop}%
\bibitem [{\citenamefont {Riska}\ and\ \citenamefont {Poppius}(1985)}]{Ris85c}%
  \BibitemOpen
  \bibfield  {author} {\bibinfo {author} {\bibnamefont {Riska}, \bibfnamefont
  {D.~O.}}, \ and\ \bibinfo {author} {\bibfnamefont {M.}~\bibnamefont
  {Poppius}}} (\bibinfo {year} {1985}),\ \href@noop {} {\bibfield  {journal}
  {\bibinfo  {journal} {Phys. Scr.}\ }\textbf {\bibinfo {volume} {32}},\
  \bibinfo {pages} {581}}\BibitemShut {NoStop}%
\bibitem [{\citenamefont {Roggero}\ \emph {et~al.}(2014)\citenamefont
  {Roggero}, \citenamefont {Mukherjee},\ and\ \citenamefont
  {Pederiva}}]{Roggero:2014}%
  \BibitemOpen
  \bibfield  {author} {\bibinfo {author} {\bibnamefont {Roggero}, \bibfnamefont
  {A.}}, \bibinfo {author} {\bibfnamefont {A.}~\bibnamefont {Mukherjee}}, \
  and\ \bibinfo {author} {\bibfnamefont {F.}~\bibnamefont {Pederiva}}}
  (\bibinfo {year} {2014}),\ \href {\doibase 10.1103/PhysRevLett.112.221103}
  {\bibfield  {journal} {\bibinfo  {journal} {Phys. Rev. Lett.}\ }\textbf
  {\bibinfo {volume} {112}},\ \bibinfo {pages} {221103}}\BibitemShut {NoStop}%
\bibitem [{\citenamefont {Sachs}(1948)}]{Sac48}%
  \BibitemOpen
  \bibfield  {author} {\bibinfo {author} {\bibnamefont {Sachs}, \bibfnamefont
  {R.~G.}}} (\bibinfo {year} {1948}),\ \href@noop {} {\bibfield  {journal}
  {\bibinfo  {journal} {Phys. Rev.}\ }\textbf {\bibinfo {volume} {74}},\
  \bibinfo {pages} {433}}\BibitemShut {NoStop}%
\bibitem [{\citenamefont {Saito}\ \emph {et~al.}(1990)\citenamefont {Saito},
  \citenamefont {Wu}, \citenamefont {Ishikawa},\ and\ \citenamefont
  {Sasakawa}}]{Sai90}%
  \BibitemOpen
  \bibfield  {author} {\bibinfo {author} {\bibnamefont {Saito}, \bibfnamefont
  {T.-Y.}}, \bibinfo {author} {\bibfnamefont {Y.}~\bibnamefont {Wu}}, \bibinfo
  {author} {\bibfnamefont {S.}~\bibnamefont {Ishikawa}}, \ and\ \bibinfo
  {author} {\bibfnamefont {T.}~\bibnamefont {Sasakawa}}} (\bibinfo {year}
  {1990}),\ \href {\doibase http://dx.doi.org/10.1016/0370-2693(90)91586-Z}
  {\bibfield  {journal} {\bibinfo  {journal} {Phys. Lett. B}\ }\textbf
  {\bibinfo {volume} {242}}~(\bibinfo {number} {1}),\ \bibinfo {pages} {12
  }}\BibitemShut {NoStop}%
\bibitem [{\citenamefont {Sarsa}\ \emph {et~al.}(2003)\citenamefont {Sarsa},
  \citenamefont {Fantoni}, \citenamefont {Schmidt},\ and\ \citenamefont
  {Pederiva}}]{Sarsa:2003}%
  \BibitemOpen
  \bibfield  {author} {\bibinfo {author} {\bibnamefont {Sarsa}, \bibfnamefont
  {A.}}, \bibinfo {author} {\bibfnamefont {S.}~\bibnamefont {Fantoni}},
  \bibinfo {author} {\bibfnamefont {K.~E.}\ \bibnamefont {Schmidt}}, \ and\
  \bibinfo {author} {\bibfnamefont {F.}~\bibnamefont {Pederiva}}} (\bibinfo
  {year} {2003}),\ \href {\doibase 10.1103/PhysRevC.68.024308} {\bibfield
  {journal} {\bibinfo  {journal} {Phys. Rev. C}\ }\textbf {\bibinfo {volume}
  {68}}~(\bibinfo {number} {2}),\ \bibinfo {pages} {024308}}\BibitemShut
  {NoStop}%
\bibitem [{\citenamefont {Schiavilla}\ and\ \citenamefont
  {Pandharipande}(2002)}]{Schiavilla:2002}%
  \BibitemOpen
  \bibfield  {author} {\bibinfo {author} {\bibnamefont {Schiavilla},
  \bibfnamefont {R.}}, \ and\ \bibinfo {author} {\bibfnamefont {V.~R.}\
  \bibnamefont {Pandharipande}}} (\bibinfo {year} {2002}),\ \href {\doibase
  10.1103/PhysRevC.65.064009} {\bibfield  {journal} {\bibinfo  {journal} {Phys.
  Rev. C}\ }\textbf {\bibinfo {volume} {65}},\ \bibinfo {pages}
  {064009}}\BibitemShut {NoStop}%
\bibitem [{\citenamefont {Schiavilla}\ \emph {et~al.}(1989)\citenamefont
  {Schiavilla}, \citenamefont {Pandharipande},\ and\ \citenamefont
  {Riska}}]{Schiavilla:1989}%
  \BibitemOpen
  \bibfield  {author} {\bibinfo {author} {\bibnamefont {Schiavilla},
  \bibfnamefont {R.}}, \bibinfo {author} {\bibfnamefont {V.~R.}\ \bibnamefont
  {Pandharipande}}, \ and\ \bibinfo {author} {\bibfnamefont {D.~O.}\
  \bibnamefont {Riska}}} (\bibinfo {year} {1989}),\ \href {\doibase
  10.1103/PhysRevC.40.2294} {\bibfield  {journal} {\bibinfo  {journal} {Phys.
  Rev. C}\ }\textbf {\bibinfo {volume} {40}},\ \bibinfo {pages}
  {2294}}\BibitemShut {NoStop}%
\bibitem [{\citenamefont {Schiavilla}\ \emph {et~al.}(1990)\citenamefont
  {Schiavilla}, \citenamefont {Pandharipande},\ and\ \citenamefont
  {Riska}}]{Sch90}%
  \BibitemOpen
  \bibfield  {author} {\bibinfo {author} {\bibnamefont {Schiavilla},
  \bibfnamefont {R.}}, \bibinfo {author} {\bibfnamefont {V.~R.}\ \bibnamefont
  {Pandharipande}}, \ and\ \bibinfo {author} {\bibfnamefont {D.~O.}\
  \bibnamefont {Riska}}} (\bibinfo {year} {1990}),\ \href@noop {} {\bibfield
  {journal} {\bibinfo  {journal} {Phys. Rev. C}\ }\textbf {\bibinfo {volume}
  {41}},\ \bibinfo {pages} {309}}\BibitemShut {NoStop}%
\bibitem [{\citenamefont {Schiavilla}\ \emph {et~al.}(1986)\citenamefont
  {Schiavilla}, \citenamefont {Pandharipande},\ and\ \citenamefont
  {Wiringa}}]{schiavilla1986}%
  \BibitemOpen
  \bibfield  {author} {\bibinfo {author} {\bibnamefont {Schiavilla},
  \bibfnamefont {R.}}, \bibinfo {author} {\bibfnamefont {V.~R.}\ \bibnamefont
  {Pandharipande}}, \ and\ \bibinfo {author} {\bibfnamefont {R.~B.}\
  \bibnamefont {Wiringa}}} (\bibinfo {year} {1986}),\ \href@noop {} {\bibfield
  {journal} {\bibinfo  {journal} {Nucl. Phys.}\ }\textbf {\bibinfo {volume}
  {A449}},\ \bibinfo {pages} {219}}\BibitemShut {NoStop}%
\bibitem [{\citenamefont {Schiavilla}\ \emph {et~al.}(1998)\citenamefont
  {Schiavilla}, \citenamefont {Stoks}, \citenamefont {Gl\"ockle}, \citenamefont
  {Kamada}, \citenamefont {Nogga}, \citenamefont {Carlson}, \citenamefont
  {Machleidt}, \citenamefont {Pandharipande}, \citenamefont {Wiringa},
  \citenamefont {Kievsky}, \citenamefont {Rosati},\ and\ \citenamefont
  {Viviani}}]{Sch98}%
  \BibitemOpen
  \bibfield  {author} {\bibinfo {author} {\bibnamefont {Schiavilla},
  \bibfnamefont {R.}}, \bibinfo {author} {\bibfnamefont {V.~G.~J.}\
  \bibnamefont {Stoks}}, \bibinfo {author} {\bibfnamefont {W.}~\bibnamefont
  {Gl\"ockle}}, \bibinfo {author} {\bibfnamefont {H.}~\bibnamefont {Kamada}},
  \bibinfo {author} {\bibfnamefont {A.}~\bibnamefont {Nogga}}, \bibinfo
  {author} {\bibfnamefont {J.}~\bibnamefont {Carlson}}, \bibinfo {author}
  {\bibfnamefont {R.}~\bibnamefont {Machleidt}}, \bibinfo {author}
  {\bibfnamefont {V.~R.}\ \bibnamefont {Pandharipande}}, \bibinfo {author}
  {\bibfnamefont {R.~B.}\ \bibnamefont {Wiringa}}, \bibinfo {author}
  {\bibfnamefont {A.}~\bibnamefont {Kievsky}}, \bibinfo {author} {\bibfnamefont
  {S.}~\bibnamefont {Rosati}}, \ and\ \bibinfo {author} {\bibfnamefont
  {M.}~\bibnamefont {Viviani}}} (\bibinfo {year} {1998}),\ \href {\doibase
  10.1103/PhysRevC.58.1263} {\bibfield  {journal} {\bibinfo  {journal} {Phys.
  Rev. C}\ }\textbf {\bibinfo {volume} {58}},\ \bibinfo {pages}
  {1263}}\BibitemShut {NoStop}%
\bibitem [{\citenamefont {Schiavilla}\ and\ \citenamefont
  {Wiringa}(2002)}]{Schiavilla:2002b}%
  \BibitemOpen
  \bibfield  {author} {\bibinfo {author} {\bibnamefont {Schiavilla},
  \bibfnamefont {R.}}, \ and\ \bibinfo {author} {\bibfnamefont
  {R.}~\bibnamefont {Wiringa}}} (\bibinfo {year} {2002}),\ \href {\doibase
  10.1103/PhysRevC.65.054302} {\bibfield  {journal} {\bibinfo  {journal} {Phys.
  Rev. C}\ }\textbf {\bibinfo {volume} {65}},\ \bibinfo {pages}
  {054302}}\BibitemShut {NoStop}%
\bibitem [{\citenamefont {Schiavilla}\ \emph {et~al.}(1993)\citenamefont
  {Schiavilla}, \citenamefont {Wiringa},\ and\ \citenamefont
  {Carlson}}]{Schiavilla:1993}%
  \BibitemOpen
  \bibfield  {author} {\bibinfo {author} {\bibnamefont {Schiavilla},
  \bibfnamefont {R.}}, \bibinfo {author} {\bibfnamefont {R.~B.}\ \bibnamefont
  {Wiringa}}, \ and\ \bibinfo {author} {\bibfnamefont {J.}~\bibnamefont
  {Carlson}}} (\bibinfo {year} {1993}),\ \href {\doibase
  10.1103/PhysRevLett.70.3856} {\bibfield  {journal} {\bibinfo  {journal}
  {Phys. Rev. Lett.}\ }\textbf {\bibinfo {volume} {70}},\ \bibinfo {pages}
  {3856}}\BibitemShut {NoStop}%
\bibitem [{\citenamefont {Schiavilla}\ \emph {et~al.}(1992)\citenamefont
  {Schiavilla}, \citenamefont {Wiringa}, \citenamefont {Pandharipande},\ and\
  \citenamefont {Carlson}}]{Sch92}%
  \BibitemOpen
  \bibfield  {author} {\bibinfo {author} {\bibnamefont {Schiavilla},
  \bibfnamefont {R.}}, \bibinfo {author} {\bibfnamefont {R.~B.}\ \bibnamefont
  {Wiringa}}, \bibinfo {author} {\bibfnamefont {V.~R.}\ \bibnamefont
  {Pandharipande}}, \ and\ \bibinfo {author} {\bibfnamefont {J.}~\bibnamefont
  {Carlson}}} (\bibinfo {year} {1992}),\ \href@noop {} {\bibfield  {journal}
  {\bibinfo  {journal} {Phys. Rev. C}\ }\textbf {\bibinfo {volume} {45}},\
  \bibinfo {pages} {2628}}\BibitemShut {NoStop}%
\bibitem [{\citenamefont {Schiavilla}\ \emph {et~al.}(2007)\citenamefont
  {Schiavilla}, \citenamefont {Wiringa}, \citenamefont {Pieper},\ and\
  \citenamefont {Carlson}}]{schiavilla2007}%
  \BibitemOpen
  \bibfield  {author} {\bibinfo {author} {\bibnamefont {Schiavilla},
  \bibfnamefont {R.}}, \bibinfo {author} {\bibfnamefont {R.~B.}\ \bibnamefont
  {Wiringa}}, \bibinfo {author} {\bibfnamefont {S.~C.}\ \bibnamefont {Pieper}},
  \ and\ \bibinfo {author} {\bibfnamefont {J.}~\bibnamefont {Carlson}}}
  (\bibinfo {year} {2007}),\ \href {\doibase 10.1103/PhysRevLett.98.132501}
  {\bibfield  {journal} {\bibinfo  {journal} {Phys. Rev. Lett.}\ }\textbf
  {\bibinfo {volume} {98}},\ \bibinfo {pages} {132501}}\BibitemShut {NoStop}%
\bibitem [{\citenamefont {Schmidt}\ and\ \citenamefont
  {Ceperley}(1992)}]{Schmidt:1992}%
  \BibitemOpen
  \bibfield  {author} {\bibinfo {author} {\bibnamefont {Schmidt}, \bibfnamefont
  {K.}}, \ and\ \bibinfo {author} {\bibfnamefont {D.}~\bibnamefont {Ceperley}}}
  (\bibinfo {year} {1992}),\ \href@noop {} {\emph {\bibinfo {title} {The Monte
  Carlo Method in Condensed Matter Physics}}}\ (\bibinfo  {publisher} {ed by K.
  Binder Springer, Berlin})\BibitemShut {NoStop}%
\bibitem [{\citenamefont {Schmidt}\ and\ \citenamefont
  {Fantoni}(1999)}]{Schmidt:1999}%
  \BibitemOpen
  \bibfield  {author} {\bibinfo {author} {\bibnamefont {Schmidt}, \bibfnamefont
  {K.~E.}}, \ and\ \bibinfo {author} {\bibfnamefont {S.}~\bibnamefont
  {Fantoni}}} (\bibinfo {year} {1999}),\ \href@noop {} {\bibfield  {journal}
  {\bibinfo  {journal} {Phys. Lett. B}\ }\textbf {\bibinfo {volume} {446}},\
  \bibinfo {pages} {99}}\BibitemShut {NoStop}%
\bibitem [{\citenamefont {Schmidt}\ and\ \citenamefont
  {Lee}(1995)}]{Schmidt:1995}%
  \BibitemOpen
  \bibfield  {author} {\bibinfo {author} {\bibnamefont {Schmidt}, \bibfnamefont
  {K.~E.}}, \ and\ \bibinfo {author} {\bibfnamefont {M.~A.}\ \bibnamefont
  {Lee}}} (\bibinfo {year} {1995}),\ \href {\doibase 10.1103/PhysRevE.51.5495}
  {\bibfield  {journal} {\bibinfo  {journal} {Phys. Rev. E}\ }\textbf {\bibinfo
  {volume} {51}},\ \bibinfo {pages} {5495}}\BibitemShut {NoStop}%
\bibitem [{\citenamefont {Seki}\ and\ \citenamefont {van
  Kolck}(2006)}]{Seki:2006}%
  \BibitemOpen
  \bibfield  {author} {\bibinfo {author} {\bibnamefont {Seki}, \bibfnamefont
  {R.}}, \ and\ \bibinfo {author} {\bibfnamefont {U.}~\bibnamefont {van
  Kolck}}} (\bibinfo {year} {2006}),\ \href {\doibase
  10.1103/PhysRevC.73.044006} {\bibfield  {journal} {\bibinfo  {journal} {Phys.
  Rev.}\ }\textbf {\bibinfo {volume} {C73}},\ \bibinfo {pages}
  {044006}}\BibitemShut {NoStop}%
\bibitem [{\citenamefont {Shen}\ \emph {et~al.}(2012)\citenamefont {Shen},
  \citenamefont {Marcucci}, \citenamefont {Carlson}, \citenamefont {Gandolfi},\
  and\ \citenamefont {Schiavilla}}]{Shen:2012}%
  \BibitemOpen
  \bibfield  {author} {\bibinfo {author} {\bibnamefont {Shen}, \bibfnamefont
  {G.}}, \bibinfo {author} {\bibfnamefont {L.}~\bibnamefont {Marcucci}},
  \bibinfo {author} {\bibfnamefont {J.}~\bibnamefont {Carlson}}, \bibinfo
  {author} {\bibfnamefont {S.}~\bibnamefont {Gandolfi}}, \ and\ \bibinfo
  {author} {\bibfnamefont {R.}~\bibnamefont {Schiavilla}}} (\bibinfo {year}
  {2012}),\ \href {\doibase 10.1103/PhysRevC.86.035503} {\bibfield  {journal}
  {\bibinfo  {journal} {Phys. Rev.}\ }\textbf {\bibinfo {volume} {C86}},\
  \bibinfo {pages} {035503}}\BibitemShut {NoStop}%
\bibitem [{\citenamefont {Shiner}\ \emph {et~al.}(1994)\citenamefont {Shiner},
  \citenamefont {Dixson},\ and\ \citenamefont {Vedantham}}]{Shiner94}%
  \BibitemOpen
  \bibfield  {author} {\bibinfo {author} {\bibnamefont {Shiner}, \bibfnamefont
  {D.}}, \bibinfo {author} {\bibfnamefont {R.}~\bibnamefont {Dixson}}, \ and\
  \bibinfo {author} {\bibfnamefont {V.}~\bibnamefont {Vedantham}}} (\bibinfo
  {year} {1994}),\ \href@noop {} {\bibfield  {journal} {\bibinfo  {journal}
  {Phys. Rev. Lett.}\ }\textbf {\bibinfo {volume} {74}},\ \bibinfo {pages}
  {3553}}\BibitemShut {NoStop}%
\bibitem [{\citenamefont {{Shternin}}\ \emph {et~al.}(2007)\citenamefont
  {{Shternin}}, \citenamefont {{Yakovlev}}, \citenamefont {{Haensel}},\ and\
  \citenamefont {{Potekhin}}}]{Shternin:2007}%
  \BibitemOpen
  \bibfield  {author} {\bibinfo {author} {\bibnamefont {{Shternin}},
  \bibfnamefont {P.~S.}}, \bibinfo {author} {\bibfnamefont {D.~G.}\
  \bibnamefont {{Yakovlev}}}, \bibinfo {author} {\bibfnamefont
  {P.}~\bibnamefont {{Haensel}}}, \ and\ \bibinfo {author} {\bibfnamefont
  {A.~Y.}\ \bibnamefont {{Potekhin}}}} (\bibinfo {year} {2007}),\ \href
  {\doibase 10.1111/j.1745-3933.2007.00386.x} {\bibfield  {journal} {\bibinfo
  {journal} {Mon. Not. R. Astron. Soc.}\ }\textbf {\bibinfo {volume} {382}},\
  \bibinfo {pages} {L43}}\BibitemShut {NoStop}%
\bibitem [{\citenamefont {{Sick}}(1982)}]{Sick:1982}%
  \BibitemOpen
  \bibfield  {author} {\bibinfo {author} {\bibnamefont {{Sick}}, \bibfnamefont
  {I.}}} (\bibinfo {year} {1982}),\ \href {\doibase
  10.1016/0370-2693(82)90327-6} {\bibfield  {journal} {\bibinfo  {journal}
  {Phys. Lett. B}\ }\textbf {\bibinfo {volume} {116}},\ \bibinfo {pages}
  {212}}\BibitemShut {NoStop}%
\bibitem [{\citenamefont {{Sick}}(2013)}]{Sick:2013}%
  \BibitemOpen
  \bibfield  {author} {\bibinfo {author} {\bibnamefont {{Sick}}, \bibfnamefont
  {I.}}} (\bibinfo {year} {2013}),\ \href@noop {} {\ }\bibinfo {note} {Private
  communication}\BibitemShut {NoStop}%
\bibitem [{\citenamefont {Smerzi}\ \emph {et~al.}(1997)\citenamefont {Smerzi},
  \citenamefont {Ravenhall},\ and\ \citenamefont
  {Pandharipande}}]{Smerzi:1997}%
  \BibitemOpen
  \bibfield  {author} {\bibinfo {author} {\bibnamefont {Smerzi}, \bibfnamefont
  {A.}}, \bibinfo {author} {\bibfnamefont {D.~G.}\ \bibnamefont {Ravenhall}}, \
  and\ \bibinfo {author} {\bibfnamefont {V.~R.}\ \bibnamefont {Pandharipande}}}
  (\bibinfo {year} {1997}),\ \href {\doibase 10.1103/PhysRevC.56.2549}
  {\bibfield  {journal} {\bibinfo  {journal} {Phys. Rev. C}\ }\textbf {\bibinfo
  {volume} {56}}~(\bibinfo {number} {5}),\ \bibinfo {pages} {2549}}\BibitemShut
  {NoStop}%
\bibitem [{\citenamefont {Spayde}\ \emph {et~al.}(2000)\citenamefont {Spayde}
  \emph {et~al.}}]{Spa00}%
  \BibitemOpen
  \bibfield  {author} {\bibinfo {author} {\bibnamefont {Spayde}, \bibfnamefont
  {D.~T.}},  \emph {et~al.} (\bibinfo {collaboration} {(SAMPLE
  Collaboration)})} (\bibinfo {year} {2000}),\ \href {\doibase
  10.1103/PhysRevLett.84.1106} {\bibfield  {journal} {\bibinfo  {journal}
  {Phys. Rev. Lett.}\ }\textbf {\bibinfo {volume} {84}},\ \bibinfo {pages}
  {1106}}\BibitemShut {NoStop}%
\bibitem [{\citenamefont {Steiner}\ and\ \citenamefont
  {Gandolfi}(2012)}]{Steiner:2012}%
  \BibitemOpen
  \bibfield  {author} {\bibinfo {author} {\bibnamefont {Steiner}, \bibfnamefont
  {A.~W.}}, \ and\ \bibinfo {author} {\bibfnamefont {S.}~\bibnamefont
  {Gandolfi}}} (\bibinfo {year} {2012}),\ \href {\doibase
  10.1103/PhysRevLett.108.081102} {\bibfield  {journal} {\bibinfo  {journal}
  {Phys. Rev. Lett.}\ }\textbf {\bibinfo {volume} {108}},\ \bibinfo {pages}
  {081102}}\BibitemShut {NoStop}%
\bibitem [{\citenamefont {Steiner}\ \emph {et~al.}(2015)\citenamefont
  {Steiner}, \citenamefont {Gandolfi}, \citenamefont {Fattoyev},\ and\
  \citenamefont {Newton}}]{Steiner:2014}%
  \BibitemOpen
  \bibfield  {author} {\bibinfo {author} {\bibnamefont {Steiner}, \bibfnamefont
  {A.~W.}}, \bibinfo {author} {\bibfnamefont {S.}~\bibnamefont {Gandolfi}},
  \bibinfo {author} {\bibfnamefont {F.~J.}\ \bibnamefont {Fattoyev}}, \ and\
  \bibinfo {author} {\bibfnamefont {W.~G.}\ \bibnamefont {Newton}}} (\bibinfo
  {year} {2015}),\ \href {\doibase 10.1103/PhysRevC.91.015804} {\bibfield
  {journal} {\bibinfo  {journal} {Phys. Rev. C}\ }\textbf {\bibinfo {volume}
  {91}},\ \bibinfo {pages} {015804}}\BibitemShut {NoStop}%
\bibitem [{\citenamefont {Steiner}\ \emph {et~al.}(2010)\citenamefont
  {Steiner}, \citenamefont {Lattimer},\ and\ \citenamefont
  {Brown}}]{Steiner:2010}%
  \BibitemOpen
  \bibfield  {author} {\bibinfo {author} {\bibnamefont {Steiner}, \bibfnamefont
  {A.~W.}}, \bibinfo {author} {\bibfnamefont {J.~M.}\ \bibnamefont {Lattimer}},
  \ and\ \bibinfo {author} {\bibfnamefont {E.~F.}\ \bibnamefont {Brown}}}
  (\bibinfo {year} {2010}),\ \href@noop {} {\bibfield  {journal} {\bibinfo
  {journal} {Astrophys. J.}\ }\textbf {\bibinfo {volume} {722}},\ \bibinfo
  {pages} {33}}\BibitemShut {NoStop}%
\bibitem [{\citenamefont {Stoks}\ \emph
  {et~al.}(1993{\natexlab{a}})\citenamefont {Stoks}, \citenamefont
  {Timmermans},\ and\ \citenamefont {de~Swart}}]{stoks1993}%
  \BibitemOpen
  \bibfield  {author} {\bibinfo {author} {\bibnamefont {Stoks}, \bibfnamefont
  {V.}}, \bibinfo {author} {\bibfnamefont {R.}~\bibnamefont {Timmermans}}, \
  and\ \bibinfo {author} {\bibfnamefont {J.~J.}\ \bibnamefont {de~Swart}}}
  (\bibinfo {year} {1993}{\natexlab{a}}),\ \href {\doibase
  10.1103/PhysRevC.47.512} {\bibfield  {journal} {\bibinfo  {journal} {Phys.
  Rev. C}\ }\textbf {\bibinfo {volume} {47}}~(\bibinfo {number} {2}),\ \bibinfo
  {pages} {512}}\BibitemShut {NoStop}%
\bibitem [{\citenamefont {Stoks}\ \emph
  {et~al.}(1993{\natexlab{b}})\citenamefont {Stoks}, \citenamefont {Klomp},
  \citenamefont {Rentmeester},\ and\ \citenamefont {de~Swart}}]{stoks1993b}%
  \BibitemOpen
  \bibfield  {author} {\bibinfo {author} {\bibnamefont {Stoks}, \bibfnamefont
  {V.~G.~J.}}, \bibinfo {author} {\bibfnamefont {R.~A.~M.}\ \bibnamefont
  {Klomp}}, \bibinfo {author} {\bibfnamefont {M.~C.~M.}\ \bibnamefont
  {Rentmeester}}, \ and\ \bibinfo {author} {\bibfnamefont {J.~J.}\ \bibnamefont
  {de~Swart}}} (\bibinfo {year} {1993}{\natexlab{b}}),\ \href {\doibase
  10.1103/PhysRevC.48.792} {\bibfield  {journal} {\bibinfo  {journal} {Phys.
  Rev. C}\ }\textbf {\bibinfo {volume} {48}}~(\bibinfo {number} {2}),\ \bibinfo
  {pages} {792}}\BibitemShut {NoStop}%
\bibitem [{\citenamefont {Stoks}\ \emph {et~al.}(1994)\citenamefont {Stoks},
  \citenamefont {Klomp}, \citenamefont {Terheggen},\ and\ \citenamefont
  {de~Swart}}]{stoks1994}%
  \BibitemOpen
  \bibfield  {author} {\bibinfo {author} {\bibnamefont {Stoks}, \bibfnamefont
  {V.~G.~J.}}, \bibinfo {author} {\bibfnamefont {R.~A.~M.}\ \bibnamefont
  {Klomp}}, \bibinfo {author} {\bibfnamefont {C.~P.~F.}\ \bibnamefont
  {Terheggen}}, \ and\ \bibinfo {author} {\bibfnamefont {J.~J.}\ \bibnamefont
  {de~Swart}}} (\bibinfo {year} {1994}),\ \href {\doibase
  10.1103/PhysRevC.49.2950} {\bibfield  {journal} {\bibinfo  {journal} {Phys.
  Rev. C}\ }\textbf {\bibinfo {volume} {49}}~(\bibinfo {number} {6}),\ \bibinfo
  {pages} {2950}}\BibitemShut {NoStop}%
\bibitem [{\citenamefont {Strueve}\ \emph {et~al.}(1987)\citenamefont
  {Strueve}, \citenamefont {Hajduk}, \citenamefont {Sauer},\ and\ \citenamefont
  {Theis}}]{Str87}%
  \BibitemOpen
  \bibfield  {author} {\bibinfo {author} {\bibnamefont {Strueve}, \bibfnamefont
  {W.}}, \bibinfo {author} {\bibfnamefont {C.}~\bibnamefont {Hajduk}}, \bibinfo
  {author} {\bibfnamefont {P.}~\bibnamefont {Sauer}}, \ and\ \bibinfo {author}
  {\bibfnamefont {W.}~\bibnamefont {Theis}}} (\bibinfo {year} {1987}),\ \href
  {\doibase http://dx.doi.org/10.1016/0375-9474(87)90560-4} {\bibfield
  {journal} {\bibinfo  {journal} {Nucl. Phys. A}\ }\textbf {\bibinfo {volume}
  {465}}~(\bibinfo {number} {4}),\ \bibinfo {pages} {651 }}\BibitemShut
  {NoStop}%
\bibitem [{\citenamefont {Subedi}\ \emph {et~al.}(2008)\citenamefont {Subedi},
  \citenamefont {Shneor}, \citenamefont {Monaghan}, \citenamefont {Anderson},
  \citenamefont {Aniol} \emph {et~al.}}]{Subedi:2008}%
  \BibitemOpen
  \bibfield  {author} {\bibinfo {author} {\bibnamefont {Subedi}, \bibfnamefont
  {R.}}, \bibinfo {author} {\bibfnamefont {R.}~\bibnamefont {Shneor}}, \bibinfo
  {author} {\bibfnamefont {P.}~\bibnamefont {Monaghan}}, \bibinfo {author}
  {\bibfnamefont {B.}~\bibnamefont {Anderson}}, \bibinfo {author}
  {\bibfnamefont {K.}~\bibnamefont {Aniol}},  \emph {et~al.}} (\bibinfo {year}
  {2008}),\ \href {\doibase 10.1126/science.1156675} {\bibfield  {journal}
  {\bibinfo  {journal} {Science}\ }\textbf {\bibinfo {volume} {320}},\ \bibinfo
  {pages} {1476}},\ \Eprint {http://arxiv.org/abs/0908.1514} {0908.1514}
  \BibitemShut {NoStop}%
\bibitem [{\citenamefont {Tews}\ \emph {et~al.}(2015)\citenamefont {Tews},
  \citenamefont {Gandolfi}, \citenamefont {Gezerlis},\ and\ \citenamefont
  {Schwenk}}]{Tews:inprep}%
  \BibitemOpen
  \bibfield  {author} {\bibinfo {author} {\bibnamefont {Tews}, \bibfnamefont
  {I.}}, \bibinfo {author} {\bibfnamefont {S.}~\bibnamefont {Gandolfi}},
  \bibinfo {author} {\bibfnamefont {A.}~\bibnamefont {Gezerlis}}, \ and\
  \bibinfo {author} {\bibfnamefont {A.}~\bibnamefont {Schwenk}}} (\bibinfo
  {year} {2015}),\ \href@noop {} {\ }\bibinfo {note} {In
  preparation}\BibitemShut {NoStop}%
\bibitem [{\citenamefont {Tilley}\ \emph {et~al.}(2002)\citenamefont {Tilley},
  \citenamefont {Cheves}, \citenamefont {Godwin}, \citenamefont {Hale},
  \citenamefont {Hofmann}, \citenamefont {Kelley}, \citenamefont {Sheu},\ and\
  \citenamefont {Weller}}]{Tilley02}%
  \BibitemOpen
  \bibfield  {author} {\bibinfo {author} {\bibnamefont {Tilley}, \bibfnamefont
  {D.~R.}}, \bibinfo {author} {\bibfnamefont {C.~M.}\ \bibnamefont {Cheves}},
  \bibinfo {author} {\bibfnamefont {J.~L.}\ \bibnamefont {Godwin}}, \bibinfo
  {author} {\bibfnamefont {G.~M.}\ \bibnamefont {Hale}}, \bibinfo {author}
  {\bibfnamefont {H.~M.}\ \bibnamefont {Hofmann}}, \bibinfo {author}
  {\bibfnamefont {J.~H.}\ \bibnamefont {Kelley}}, \bibinfo {author}
  {\bibfnamefont {C.~G.}\ \bibnamefont {Sheu}}, \ and\ \bibinfo {author}
  {\bibfnamefont {H.~R.}\ \bibnamefont {Weller}}} (\bibinfo {year} {2002}),\
  \href@noop {} {\bibfield  {journal} {\bibinfo  {journal} {Nucl. Phys. A}\
  }\textbf {\bibinfo {volume} {708}},\ \bibinfo {pages} {3}}\BibitemShut
  {NoStop}%
\bibitem [{\citenamefont {Tilley}\ \emph {et~al.}(2004)\citenamefont {Tilley},
  \citenamefont {Kelley}, \citenamefont {Godwin}, \citenamefont {Millener},
  \citenamefont {Purcell}, \citenamefont {Sheu},\ and\ \citenamefont
  {Weller}}]{Tilley04}%
  \BibitemOpen
  \bibfield  {author} {\bibinfo {author} {\bibnamefont {Tilley}, \bibfnamefont
  {D.~R.}}, \bibinfo {author} {\bibfnamefont {J.~H.}\ \bibnamefont {Kelley}},
  \bibinfo {author} {\bibfnamefont {J.~L.}\ \bibnamefont {Godwin}}, \bibinfo
  {author} {\bibfnamefont {D.~J.}\ \bibnamefont {Millener}}, \bibinfo {author}
  {\bibfnamefont {J.~E.}\ \bibnamefont {Purcell}}, \bibinfo {author}
  {\bibfnamefont {C.~G.}\ \bibnamefont {Sheu}}, \ and\ \bibinfo {author}
  {\bibfnamefont {H.~R.}\ \bibnamefont {Weller}}} (\bibinfo {year} {2004}),\
  \href@noop {} {\bibfield  {journal} {\bibinfo  {journal} {Nucl. Phys. A}\
  }\textbf {\bibinfo {volume} {745}},\ \bibinfo {pages} {155}}\BibitemShut
  {NoStop}%
\bibitem [{\citenamefont {Towner}(1987)}]{Tow87}%
  \BibitemOpen
  \bibfield  {author} {\bibinfo {author} {\bibnamefont {Towner}, \bibfnamefont
  {I.~S.}}} (\bibinfo {year} {1987}),\ \href {\doibase
  http://dx.doi.org/10.1016/0370-1573(87)90138-4} {\bibfield  {journal}
  {\bibinfo  {journal} {Phys. Rep.}\ }\textbf {\bibinfo {volume}
  {155}}~(\bibinfo {number} {5}),\ \bibinfo {pages} {263 }}\BibitemShut
  {NoStop}%
\bibitem [{\citenamefont {{Towner}}\ and\ \citenamefont
  {{Hardy}}(1998)}]{Towner:1998}%
  \BibitemOpen
  \bibfield  {author} {\bibinfo {author} {\bibnamefont {{Towner}},
  \bibfnamefont {I.~S.}}, \ and\ \bibinfo {author} {\bibfnamefont {J.~C.}\
  \bibnamefont {{Hardy}}}} (\bibinfo {year} {1998}),\ \href@noop {} {\bibfield
  {journal} {\bibinfo  {journal} {ArXiv Nuclear Theory e-prints}\ }}\Eprint
  {http://arxiv.org/abs/nucl-th/9809087} {nucl-th/9809087} \BibitemShut
  {NoStop}%
\bibitem [{\citenamefont {Towner}\ and\ \citenamefont
  {Hardy}(1999)}]{Towner:1999}%
  \BibitemOpen
  \bibfield  {author} {\bibinfo {author} {\bibnamefont {Towner}, \bibfnamefont
  {I.~S.}}, \ and\ \bibinfo {author} {\bibfnamefont {J.~C.}\ \bibnamefont
  {Hardy}}} (\bibinfo {year} {1999}),\ \href@noop {} {\ ,\ \bibinfo {pages}
  {338}}\bibinfo {note} {Edited by P.\ Herczeg, C.M.\ Hoffman, and H.V.\
  Klapdor-Kleingrothaus (World Scientific, Singapore)}\BibitemShut {NoStop}%
\bibitem [{\citenamefont {Tsang}\ \emph {et~al.}(2012)\citenamefont {Tsang},
  \citenamefont {Stone}, \citenamefont {Camera}, \citenamefont {Danielewicz},
  \citenamefont {Gandolfi}, \citenamefont {Hebeler}, \citenamefont {Horowitz},
  \citenamefont {Lee}, \citenamefont {Lynch}, \citenamefont {Kohley},
  \citenamefont {Lemmon}, \citenamefont {M\"oller}, \citenamefont {Murakami},
  \citenamefont {Riordan}, \citenamefont {Roca-Maza}, \citenamefont
  {Sammarruca}, \citenamefont {Steiner}, \citenamefont {Vida\~na},\ and\
  \citenamefont {Yennello}}]{Tsang:2012}%
  \BibitemOpen
  \bibfield  {author} {\bibinfo {author} {\bibnamefont {Tsang}, \bibfnamefont
  {M.~B.}}, \bibinfo {author} {\bibfnamefont {J.~R.}\ \bibnamefont {Stone}},
  \bibinfo {author} {\bibfnamefont {F.}~\bibnamefont {Camera}}, \bibinfo
  {author} {\bibfnamefont {P.}~\bibnamefont {Danielewicz}}, \bibinfo {author}
  {\bibfnamefont {S.}~\bibnamefont {Gandolfi}}, \bibinfo {author}
  {\bibfnamefont {K.}~\bibnamefont {Hebeler}}, \bibinfo {author} {\bibfnamefont
  {C.~J.}\ \bibnamefont {Horowitz}}, \bibinfo {author} {\bibfnamefont
  {J.}~\bibnamefont {Lee}}, \bibinfo {author} {\bibfnamefont {W.~G.}\
  \bibnamefont {Lynch}}, \bibinfo {author} {\bibfnamefont {Z.}~\bibnamefont
  {Kohley}}, \bibinfo {author} {\bibfnamefont {R.}~\bibnamefont {Lemmon}},
  \bibinfo {author} {\bibfnamefont {P.}~\bibnamefont {M\"oller}}, \bibinfo
  {author} {\bibfnamefont {T.}~\bibnamefont {Murakami}}, \bibinfo {author}
  {\bibfnamefont {S.}~\bibnamefont {Riordan}}, \bibinfo {author} {\bibfnamefont
  {X.}~\bibnamefont {Roca-Maza}}, \bibinfo {author} {\bibfnamefont
  {F.}~\bibnamefont {Sammarruca}}, \bibinfo {author} {\bibfnamefont {A.~W.}\
  \bibnamefont {Steiner}}, \bibinfo {author} {\bibfnamefont {I.}~\bibnamefont
  {Vida\~na}}, \ and\ \bibinfo {author} {\bibfnamefont {S.~J.}\ \bibnamefont
  {Yennello}}} (\bibinfo {year} {2012}),\ \href {\doibase
  10.1103/PhysRevC.86.015803} {\bibfield  {journal} {\bibinfo  {journal} {Phys.
  Rev. C}\ }\textbf {\bibinfo {volume} {86}},\ \bibinfo {pages}
  {015803}}\BibitemShut {NoStop}%
\bibitem [{\citenamefont {Tsang}\ \emph {et~al.}(2009)\citenamefont {Tsang},
  \citenamefont {Zhang}, \citenamefont {Danielewicz}, \citenamefont {Famiano},
  \citenamefont {Li}, \citenamefont {Lynch},\ and\ \citenamefont
  {Steiner}}]{Tsang:2009}%
  \BibitemOpen
  \bibfield  {author} {\bibinfo {author} {\bibnamefont {Tsang}, \bibfnamefont
  {M.~B.}}, \bibinfo {author} {\bibfnamefont {Y.}~\bibnamefont {Zhang}},
  \bibinfo {author} {\bibfnamefont {P.}~\bibnamefont {Danielewicz}}, \bibinfo
  {author} {\bibfnamefont {M.}~\bibnamefont {Famiano}}, \bibinfo {author}
  {\bibfnamefont {Z.}~\bibnamefont {Li}}, \bibinfo {author} {\bibfnamefont
  {W.~G.}\ \bibnamefont {Lynch}}, \ and\ \bibinfo {author} {\bibfnamefont
  {A.~W.}\ \bibnamefont {Steiner}}} (\bibinfo {year} {2009}),\ \href {\doibase
  10.1103/PhysRevLett.102.122701} {\bibfield  {journal} {\bibinfo  {journal}
  {Phys. Rev. Lett.}\ }\textbf {\bibinfo {volume} {102}}~(\bibinfo {number}
  {12}),\ \bibinfo {pages} {122701}}\BibitemShut {NoStop}%
\bibitem [{\citenamefont {TUNL}(2014)}]{TUNL-data}%
  \BibitemOpen
  \bibfield  {author} {\bibinfo {author} {\bibnamefont {TUNL},}} (\bibinfo
  {year} {2014}),\ \href@noop {} {\enquote {\bibinfo {title} {Nuclear data
  evaluation project},}\ }\bibinfo {howpublished}
  {http://www.tunl.duke.edu/NuclData/}\BibitemShut {NoStop}%
\bibitem [{\citenamefont {Usmani}\ \emph {et~al.}(2012)\citenamefont {Usmani},
  \citenamefont {Anwar},\ and\ \citenamefont {Abdullah}}]{usmani2012}%
  \BibitemOpen
  \bibfield  {author} {\bibinfo {author} {\bibnamefont {Usmani}, \bibfnamefont
  {Q.~N.}}, \bibinfo {author} {\bibfnamefont {K.}~\bibnamefont {Anwar}}, \ and\
  \bibinfo {author} {\bibfnamefont {N.}~\bibnamefont {Abdullah}}} (\bibinfo
  {year} {2012}),\ \href {\doibase 10.1103/PhysRevC.86.034323} {\bibfield
  {journal} {\bibinfo  {journal} {Phys. Rev. C}\ }\textbf {\bibinfo {volume}
  {86}},\ \bibinfo {pages} {034323}}\BibitemShut {NoStop}%
\bibitem [{\citenamefont {Usmani}\ \emph {et~al.}(2009)\citenamefont {Usmani},
  \citenamefont {Singh}, \citenamefont {Anwar},\ and\ \citenamefont
  {Rawitscher}}]{usmani2009}%
  \BibitemOpen
  \bibfield  {author} {\bibinfo {author} {\bibnamefont {Usmani}, \bibfnamefont
  {Q.~N.}}, \bibinfo {author} {\bibfnamefont {A.}~\bibnamefont {Singh}},
  \bibinfo {author} {\bibfnamefont {K.}~\bibnamefont {Anwar}}, \ and\ \bibinfo
  {author} {\bibfnamefont {G.}~\bibnamefont {Rawitscher}}} (\bibinfo {year}
  {2009}),\ \href {\doibase 10.1103/PhysRevC.80.034309} {\bibfield  {journal}
  {\bibinfo  {journal} {Phys. Rev. C}\ }\textbf {\bibinfo {volume} {80}},\
  \bibinfo {pages} {034309}}\BibitemShut {NoStop}%
\bibitem [{\citenamefont {Viviani}\ \emph {et~al.}(2014)\citenamefont
  {Viviani}, \citenamefont {Baroni}, \citenamefont {Girlanda}, \citenamefont
  {Kievsky}, \citenamefont {Marcucci},\ and\ \citenamefont
  {Schiavilla}}]{Viv14}%
  \BibitemOpen
  \bibfield  {author} {\bibinfo {author} {\bibnamefont {Viviani}, \bibfnamefont
  {M.}}, \bibinfo {author} {\bibfnamefont {A.}~\bibnamefont {Baroni}}, \bibinfo
  {author} {\bibfnamefont {L.}~\bibnamefont {Girlanda}}, \bibinfo {author}
  {\bibfnamefont {A.}~\bibnamefont {Kievsky}}, \bibinfo {author} {\bibfnamefont
  {L.~E.}\ \bibnamefont {Marcucci}}, \ and\ \bibinfo {author} {\bibfnamefont
  {R.}~\bibnamefont {Schiavilla}}} (\bibinfo {year} {2014}),\ \href@noop {}
  {\bibfield  {journal} {\bibinfo  {journal} {Phys. Rev. C}\ }\textbf {\bibinfo
  {volume} {89}},\ \bibinfo {pages} {064004}}\BibitemShut {NoStop}%
\bibitem [{\citenamefont {Walecka}(1995)}]{Wal95}%
  \BibitemOpen
  \bibfield  {author} {\bibinfo {author} {\bibnamefont {Walecka}, \bibfnamefont
  {J.}}} (\bibinfo {year} {1995}),\ \href@noop {} {\bibfield  {journal}
  {\bibinfo  {journal} {Oxford Stud.Nucl.Phys.}\ }\textbf {\bibinfo {volume}
  {16}},\ \bibinfo {pages} {1}}\BibitemShut {NoStop}%
\bibitem [{\citenamefont {Weinberg}(1990)}]{Wei90a}%
  \BibitemOpen
  \bibfield  {author} {\bibinfo {author} {\bibnamefont {Weinberg},
  \bibfnamefont {S.}}} (\bibinfo {year} {1990}),\ \href@noop {} {\bibfield
  {journal} {\bibinfo  {journal} {Phys.\ Lett.\ B}\ }\textbf {\bibinfo {volume}
  {251}},\ \bibinfo {pages} {288}}\BibitemShut {NoStop}%
\bibitem [{\citenamefont {Weinberg}(1991)}]{Wei90b}%
  \BibitemOpen
  \bibfield  {author} {\bibinfo {author} {\bibnamefont {Weinberg},
  \bibfnamefont {S.}}} (\bibinfo {year} {1991}),\ \href@noop {} {\bibfield
  {journal} {\bibinfo  {journal} {Nucl. Phys.\ B}\ }\textbf {\bibinfo {volume}
  {363}},\ \bibinfo {pages} {3}}\BibitemShut {NoStop}%
\bibitem [{\citenamefont {Weinberg}(1992)}]{Wei90c}%
  \BibitemOpen
  \bibfield  {author} {\bibinfo {author} {\bibnamefont {Weinberg},
  \bibfnamefont {S.}}} (\bibinfo {year} {1992}),\ \href@noop {} {\bibfield
  {journal} {\bibinfo  {journal} {Phys.\ Lett.\ B}\ }\textbf {\bibinfo {volume}
  {295}},\ \bibinfo {pages} {114}}\BibitemShut {NoStop}%
\bibitem [{\citenamefont {Wiringa}(1991)}]{wiringa1991}%
  \BibitemOpen
  \bibfield  {author} {\bibinfo {author} {\bibnamefont {Wiringa}, \bibfnamefont
  {R.~B.}}} (\bibinfo {year} {1991}),\ \href {\doibase
  10.1103/PhysRevC.43.1585} {\bibfield  {journal} {\bibinfo  {journal} {Phys.
  Rev. C}\ }\textbf {\bibinfo {volume} {43}},\ \bibinfo {pages}
  {1585}}\BibitemShut {NoStop}%
\bibitem [{\citenamefont {Wiringa}(2006)}]{wiringa2006}%
  \BibitemOpen
  \bibfield  {author} {\bibinfo {author} {\bibnamefont {Wiringa}, \bibfnamefont
  {R.~B.}}} (\bibinfo {year} {2006}),\ \href@noop {} {\bibfield  {journal}
  {\bibinfo  {journal} {Phys. Rev. C}\ }\textbf {\bibinfo {volume} {73}},\
  \bibinfo {pages} {034317}}\BibitemShut {NoStop}%
\bibitem [{\citenamefont {Wiringa}(2014{\natexlab{a}})}]{webmomenta1}%
  \BibitemOpen
  \bibfield  {author} {\bibinfo {author} {\bibnamefont {Wiringa}, \bibfnamefont
  {R.~B.}}} (\bibinfo {year} {2014}{\natexlab{a}}),\ \href@noop {} {\emph
  {\bibinfo {title} {Single-Nucleon Momentum Distributions}}},\ \bibinfo {note}
  {\url{http://www.phy.anl.gov/theory/research/momenta}}\BibitemShut {NoStop}%
\bibitem [{\citenamefont {Wiringa}(2014{\natexlab{b}})}]{webmomenta2}%
  \BibitemOpen
  \bibfield  {author} {\bibinfo {author} {\bibnamefont {Wiringa}, \bibfnamefont
  {R.~B.}}} (\bibinfo {year} {2014}{\natexlab{b}}),\ \href@noop {} {\emph
  {\bibinfo {title} {Two-Nucleon Momentum Distributions}}},\ \bibinfo {note}
  {\url{http://www.phy.anl.gov/theory/research/momenta2}}\BibitemShut {NoStop}%
\bibitem [{\citenamefont {Wiringa}(2015)}]{Wiringa:unpub}%
  \BibitemOpen
  \bibfield  {author} {\bibinfo {author} {\bibnamefont {Wiringa}, \bibfnamefont
  {R.~B.}}} (\bibinfo {year} {2015}),\ \href@noop {} {\bibinfo  {journal}
  {unpublished}\ }\BibitemShut {NoStop}%
\bibitem [{\citenamefont {Wiringa}\ and\ \citenamefont
  {Brida}(2014)}]{weboverlaps}%
  \BibitemOpen
\bibfield  {journal} {  }\bibfield  {author} {\bibinfo {author} {\bibnamefont
  {Wiringa}, \bibfnamefont {R.~B.}}, \ and\ \bibinfo {author} {\bibfnamefont
  {I.}~\bibnamefont {Brida}}} (\bibinfo {year} {2014}),\ \href@noop {} {\emph
  {\bibinfo {title} {Spectroscopic Overlaps}}},\ \bibinfo {note}
  {\url{http://www.phy.anl.gov/theory/research/overlap}}\BibitemShut {NoStop}%
\bibitem [{\citenamefont {Wiringa}\ \emph {et~al.}(2013)\citenamefont
  {Wiringa}, \citenamefont {Pastore}, \citenamefont {Pieper},\ and\
  \citenamefont {Miller}}]{wiringa2013}%
  \BibitemOpen
  \bibfield  {author} {\bibinfo {author} {\bibnamefont {Wiringa}, \bibfnamefont
  {R.~B.}}, \bibinfo {author} {\bibfnamefont {S.}~\bibnamefont {Pastore}},
  \bibinfo {author} {\bibfnamefont {S.~C.}\ \bibnamefont {Pieper}}, \ and\
  \bibinfo {author} {\bibfnamefont {G.~A.}\ \bibnamefont {Miller}}} (\bibinfo
  {year} {2013}),\ \href@noop {} {\bibfield  {journal} {\bibinfo  {journal}
  {Phys. Rev. C}\ }\textbf {\bibinfo {volume} {88}},\ \bibinfo {pages}
  {044333}}\BibitemShut {NoStop}%
\bibitem [{\citenamefont {Wiringa}\ and\ \citenamefont
  {Pieper}(2002)}]{wiringa2002}%
  \BibitemOpen
  \bibfield  {author} {\bibinfo {author} {\bibnamefont {Wiringa}, \bibfnamefont
  {R.~B.}}, \ and\ \bibinfo {author} {\bibfnamefont {S.~C.}\ \bibnamefont
  {Pieper}}} (\bibinfo {year} {2002}),\ \href {\doibase
  10.1103/PhysRevLett.89.182501} {\bibfield  {journal} {\bibinfo  {journal}
  {Phys. Rev. Lett.}\ }\textbf {\bibinfo {volume} {89}}~(\bibinfo {number}
  {18}),\ \bibinfo {pages} {182501}}\BibitemShut {NoStop}%
\bibitem [{\citenamefont {Wiringa}\ \emph {et~al.}(2000)\citenamefont
  {Wiringa}, \citenamefont {Pieper}, \citenamefont {Carlson},\ and\
  \citenamefont {Pandharipande}}]{Wiringa:2000}%
  \BibitemOpen
  \bibfield  {author} {\bibinfo {author} {\bibnamefont {Wiringa}, \bibfnamefont
  {R.~B.}}, \bibinfo {author} {\bibfnamefont {S.~C.}\ \bibnamefont {Pieper}},
  \bibinfo {author} {\bibfnamefont {J.}~\bibnamefont {Carlson}}, \ and\
  \bibinfo {author} {\bibfnamefont {V.~R.}\ \bibnamefont {Pandharipande}}}
  (\bibinfo {year} {2000}),\ \href {\doibase 10.1103/PhysRevC.62.014001}
  {\bibfield  {journal} {\bibinfo  {journal} {Phys. Rev. C}\ }\textbf {\bibinfo
  {volume} {62}},\ \bibinfo {pages} {014001}}\BibitemShut {NoStop}%
\bibitem [{\citenamefont {Wiringa}\ and\ \citenamefont
  {Schiavilla}(1998)}]{Wiringa:1998}%
  \BibitemOpen
  \bibfield  {author} {\bibinfo {author} {\bibnamefont {Wiringa}, \bibfnamefont
  {R.~B.}}, \ and\ \bibinfo {author} {\bibfnamefont {R.}~\bibnamefont
  {Schiavilla}}} (\bibinfo {year} {1998}),\ \href {\doibase
  10.1103/PhysRevLett.81.4317} {\bibfield  {journal} {\bibinfo  {journal}
  {Phys. Rev. Lett.}\ }\textbf {\bibinfo {volume} {81}},\ \bibinfo {pages}
  {4317}}\BibitemShut {NoStop}%
\bibitem [{\citenamefont {Wiringa}\ \emph {et~al.}(2014)\citenamefont
  {Wiringa}, \citenamefont {Schiavilla}, \citenamefont {Pieper},\ and\
  \citenamefont {Carlson}}]{wiringa2014}%
  \BibitemOpen
  \bibfield  {author} {\bibinfo {author} {\bibnamefont {Wiringa}, \bibfnamefont
  {R.~B.}}, \bibinfo {author} {\bibfnamefont {R.}~\bibnamefont {Schiavilla}},
  \bibinfo {author} {\bibfnamefont {S.~C.}\ \bibnamefont {Pieper}}, \ and\
  \bibinfo {author} {\bibfnamefont {J.}~\bibnamefont {Carlson}}} (\bibinfo
  {year} {2014}),\ \href@noop {} {\bibfield  {journal} {\bibinfo  {journal}
  {Phys. Rev. C}\ }\textbf {\bibinfo {volume} {89}},\ \bibinfo {pages}
  {024305}}\BibitemShut {NoStop}%
\bibitem [{\citenamefont {Wiringa}\ \emph {et~al.}(1984)\citenamefont
  {Wiringa}, \citenamefont {Smith},\ and\ \citenamefont
  {Ainsworth}}]{wiringa1984}%
  \BibitemOpen
  \bibfield  {author} {\bibinfo {author} {\bibnamefont {Wiringa}, \bibfnamefont
  {R.~B.}}, \bibinfo {author} {\bibfnamefont {R.~A.}\ \bibnamefont {Smith}}, \
  and\ \bibinfo {author} {\bibfnamefont {T.~L.}\ \bibnamefont {Ainsworth}}}
  (\bibinfo {year} {1984}),\ \href {\doibase 10.1103/PhysRevC.29.1207}
  {\bibfield  {journal} {\bibinfo  {journal} {Phys. Rev. C}\ }\textbf {\bibinfo
  {volume} {29}}~(\bibinfo {number} {4}),\ \bibinfo {pages} {1207}}\BibitemShut
  {NoStop}%
\bibitem [{\citenamefont {Wiringa}\ \emph {et~al.}(1995)\citenamefont
  {Wiringa}, \citenamefont {Stoks},\ and\ \citenamefont
  {Schiavilla}}]{wiringa1995}%
  \BibitemOpen
  \bibfield  {author} {\bibinfo {author} {\bibnamefont {Wiringa}, \bibfnamefont
  {R.~B.}}, \bibinfo {author} {\bibfnamefont {V.~G.~J.}\ \bibnamefont {Stoks}},
  \ and\ \bibinfo {author} {\bibfnamefont {R.}~\bibnamefont {Schiavilla}}}
  (\bibinfo {year} {1995}),\ \href {\doibase 10.1103/PhysRevC.51.38} {\bibfield
   {journal} {\bibinfo  {journal} {Phys. Rev. C}\ }\textbf {\bibinfo {volume}
  {51}}~(\bibinfo {number} {1}),\ \bibinfo {pages} {38}}\BibitemShut {NoStop}%
\bibitem [{\citenamefont {Wlaz\l{}owski}\ \emph {et~al.}(2014)\citenamefont
  {Wlaz\l{}owski}, \citenamefont {Holt}, \citenamefont {Moroz}, \citenamefont
  {Bulgac},\ and\ \citenamefont {Roche}}]{Wlazlowski:2014}%
  \BibitemOpen
  \bibfield  {author} {\bibinfo {author} {\bibnamefont {Wlaz\l{}owski},
  \bibfnamefont {G.}}, \bibinfo {author} {\bibfnamefont {J.~W.}\ \bibnamefont
  {Holt}}, \bibinfo {author} {\bibfnamefont {S.}~\bibnamefont {Moroz}},
  \bibinfo {author} {\bibfnamefont {A.}~\bibnamefont {Bulgac}}, \ and\ \bibinfo
  {author} {\bibfnamefont {K.~J.}\ \bibnamefont {Roche}}} (\bibinfo {year}
  {2014}),\ \href {\doibase 10.1103/PhysRevLett.113.182503} {\bibfield
  {journal} {\bibinfo  {journal} {Phys. Rev. Lett.}\ }\textbf {\bibinfo
  {volume} {113}},\ \bibinfo {pages} {182503}}\BibitemShut {NoStop}%
\bibitem [{\citenamefont {Wuosmaa}\ \emph {et~al.}(2005)\citenamefont {Wuosmaa}
  \emph {et~al.}}]{wuosmaa2005}%
  \BibitemOpen
  \bibfield  {author} {\bibinfo {author} {\bibnamefont {Wuosmaa}, \bibfnamefont
  {A.~H.}},  \emph {et~al.}} (\bibinfo {year} {2005}),\ \href {\doibase
  10.1103/PhysRevLett.94.082502} {\bibfield  {journal} {\bibinfo  {journal}
  {Phys. Rev. Lett.}\ }\textbf {\bibinfo {volume} {94}},\ \bibinfo {pages}
  {082502}}\BibitemShut {NoStop}%
\bibitem [{\citenamefont {Wuosmaa}\ \emph {et~al.}(2008)\citenamefont {Wuosmaa}
  \emph {et~al.}}]{wuosmaa2008}%
  \BibitemOpen
  \bibfield  {author} {\bibinfo {author} {\bibnamefont {Wuosmaa}, \bibfnamefont
  {A.~H.}},  \emph {et~al.}} (\bibinfo {year} {2008}),\ \href {\doibase
  10.1103/PhysRevC.78.041302} {\bibfield  {journal} {\bibinfo  {journal} {Phys.
  Rev. C}\ }\textbf {\bibinfo {volume} {78}},\ \bibinfo {pages}
  {041302}}\BibitemShut {NoStop}%
\bibitem [{\citenamefont {Zhang}\ and\ \citenamefont
  {Krakauer}(2003)}]{Zhang:2003}%
  \BibitemOpen
  \bibfield  {author} {\bibinfo {author} {\bibnamefont {Zhang}, \bibfnamefont
  {S.}}, \ and\ \bibinfo {author} {\bibfnamefont {H.}~\bibnamefont {Krakauer}}}
  (\bibinfo {year} {2003}),\ \href {\doibase 10.1103/PhysRevLett.90.136401}
  {\bibfield  {journal} {\bibinfo  {journal} {Phys. Rev. Lett.}\ }\textbf
  {\bibinfo {volume} {90}},\ \bibinfo {pages} {136401}}\BibitemShut {NoStop}%
\bibitem [{\citenamefont {Zhu}\ \emph {et~al.}(2005)\citenamefont {Zhu},
  \citenamefont {Maekawa}, \citenamefont {Holstein}, \citenamefont
  {Ramsey-Musolf},\ and\ \citenamefont {van Kolck}}]{Zhu05}%
  \BibitemOpen
  \bibfield  {author} {\bibinfo {author} {\bibnamefont {Zhu}, \bibfnamefont
  {S.~L.}}, \bibinfo {author} {\bibfnamefont {C.~M.}\ \bibnamefont {Maekawa}},
  \bibinfo {author} {\bibfnamefont {B.~R.}\ \bibnamefont {Holstein}}, \bibinfo
  {author} {\bibfnamefont {M.~J.}\ \bibnamefont {Ramsey-Musolf}}, \ and\
  \bibinfo {author} {\bibfnamefont {U.}~\bibnamefont {van Kolck}}} (\bibinfo
  {year} {2005}),\ \href@noop {} {\bibfield  {journal} {\bibinfo  {journal}
  {Nucl. Phys. A}\ }\textbf {\bibinfo {volume} {748}},\ \bibinfo {pages}
  {435}}\BibitemShut {NoStop}%
\end{thebibliography}

%

\end{document}